\documentclass[aa, twocolumn, tighten, twocolappendix,numberedappendix, iop]{openjournal}

\makeatletter
\renewcommand\frontmatter@above@affilgroup{\vspace*{-0.3in}}  
\makeatother

\usepackage{amsmath}

\usepackage[varg]{txfonts}
\usepackage{svg}
\usepackage{orcidlink}
\usepackage{hyperref}
\hypersetup{colorlinks, allcolors=blue}
\usepackage{xcolor}
\usepackage{placeins}

\newcommand{\Msun}{{\rm\,M_\odot}}
\newcommand{\kpc}{{\rm\,kpc}}
\newcommand{\Gyr}{{\rm\,Gyr}}

\newcommand{\Fsum}{{F_{\rm{sum}}}}
\newcommand{\mFsum}{{\langle F_{\rm{sum}} \rangle}}
\renewcommand{\theequation}{\arabic{equation}}

\shorttitle{Mutliple Mode Spirals}
\shortauthors{Kwak et al.}

\begin{document}

\title{ Effects of Resolution and Local Stability on Galactic Disks:\\  Multiple Spiral Mode Formation via Swing Amplification}
\author{SungWon Kwak$^{1}$ \orcidlink{0000-0003-0957-6201}}
\author{Ivan Minchev$^{1}$ \orcidlink{0000-0002-5627-0355}}
\author{Christoph Pfrommer$^{1}$ \orcidlink{0000-0002-7275-3998}}
\author{Matthias Steinmetz$^{1,2}$ \orcidlink{0000-0001-6516-7459}}
\author{Sukyoung K. Yi$^{3}$ \orcidlink{0000-0002-4556-2619}}
\affiliation{$^1$Leibniz-Instit\"ut f\"ur Astrophysik Potsdam (AIP), An der Sternwarte 16, 14482, Potsdam, Germany}
\affiliation{$^2$Universit\"at Potsdam, Institut f\"ur Physik und Astronomie, Karl-Liebknecht-Str. 24-25, 14476, Potsdam, Germany}
\affiliation{$^3$Department of Astronomy and Yonsei University Observatory, Yonsei University, Seoul, 03722, Republic of Korea}

\email[Email: ]{skwak@aip.de}

\begin{abstract}
We investigate the formation of multiple spiral modes in Milky Way–like disk–halo systems without explicitly exciting perturbations. We explore how numerical resolution, the level of local disk stability, and the presence of a live halo influence both the initial appearance and the subsequent evolution of these modes. To characterize spiral structure, we compute Fourier amplitudes for modes $m=1$–$6$. In marginally unstable, lower-resolution disks ($N_\star = 5\times10^6$, $N_{\rm DM}=1.14\times10^7$), faint features appear within the first $0.5$~Gyr due to numerical noise, in contrast to high-resolution ($N_\star = 5\times10^7$, $N_{\rm DM}=1.14\times10^8$) models where perturbations emerge later. We find that the delay in the spiral and bar formation epoch due to the enhanced resolution is more significant in the more stable models. Across all sufficiently resolved, live-halo models with $m_{\rm DM}/m_\star \le 10$, the spirals exhibit a cascading sequence in both mode number and radius: higher-$m$ modes form and decay first, followed by the delayed emergence of lower-$m$ modes, with an inward drift of the activity’s epicenter. This behavior reflects a combination of local swing amplification, which explains the initial growth of short-wavelength modes, and interference between coexisting coherent spiral components with different pattern speeds, which accounts for the recurrent short-timescale amplitude modulations. In contrast, models with a fixed halo potential or coarse halo resolution ($N_{\rm DM}=1.14\times10^6$ and $m_{\rm DM}/m_\star=100$) show strong early spirals but lack this coherent cascading behavior, owing to excessive shot noise and insufficient halo responsiveness. The $m=3$ mode plays a transitional role, marking the onset of angular-momentum transport in the inner disk that precedes bar formation, a process absent in fixed-potential models. Our results show that a live halo with appropriate mass resolution provides the gravitational response needed to sustain and regenerate multi-mode spiral structure, even though the total angular-momentum exchange remains small.
\end{abstract}

\maketitle

\section{Introduction}
Spiral arms represent the most commonly observed features in disk galaxies across the Hubble time, offering long-standing clues about their classifications and evolutionary stages \citep{hubble26a, hubble26b, sandage61}.
Despite their prominence in galaxy morphology \citep{schweizer76, gnedin95, grosbol04, zibetti09, lintott11, hart16, hart17a, hart17b, hart18}, the physical origins and dynamical evolution of these structures are not fully understood \citep{binney08, sellwood22a}. 
Indeed, spiral arms vary in type, displaying distinct kinematic and structural characteristics depending on disk properties, formation mechanism, and environments.

Grand-design spiral arms are commonly attributed to tidal forcing \citep{byrd90, byrd92, donner94}, though strong tidal interactions can also trigger bar formation in disk galaxies \citep{lokas14, kwak19, peschken19,frosst26b}. 
To prevent bar instability, \cite{oh08,oh15} included a massive bulge in their models to investigate tidally induced spiral formation because of the stabilizing influence of a massive bulge against the bar instability \citep{hohl76, toomre81, hozumi87,sellwood01, athanassoula03, shen04, jang23}.
\cite{donghia13} proposed that persistent self-perturbed spirals driven by internal perturbations, such as giant molecular clouds, can explain the origin of hidden two-armed spirals in the bulgeless dwarf galaxy IC3328 (VCC856) \citep{jerjen00}.
Later, \cite{kwak19} reproduced the same spirals of IC3328 by tidal force, showing $m=2$ spirals are tidally induced regardless of arm strength. 
Tidal forces are ubiquitous in the universe and generate various non-axisymmetric structures via galaxy-galaxy encounters, cluster tidal fields, and galaxy harassment \citep{moore96, moore98, gnedin03a, gnedin03b, lokas14, semczuk17, kwak19, grand26}.

Aside from tidal origins, one mechanism that drives the formation of barred spirals is the manifold-driven process \citep{athanassoula12, kwak17, efthymiopoulos19, grand26}. Additionally, stochastic self-propagating star formation (SSPSF) also triggers the development of multiple spiral arms \citep{mueller76, gerola78}. 
These various mechanisms show the complex interplay of internal and external factors shaping spiral arm formation.

Apart from externally forced or internally seeded perturbations, spiral arms may also form spontaneously through local gravitational instabilities in the disk.
A widely used measure of these instabilities is the Toomre \(Q\) parameter \citep{toomre64}, defined as
\begin{equation}
Q = \frac{\kappa \sigma_R}{3.36\,G\,\Sigma_\star},
\label{eq:Toomre}
\end{equation}
where \(\kappa\) is the epicyclic frequency, \(\sigma_R\) is the radial velocity dispersion of the stars, \(G\) is Newton's gravitational constant and \(\Sigma_\star\) is the stellar surface density. Disks with \(Q\lesssim1\) experience strong gravitational instabilities that can trigger spiral features, whereas disks with \(Q>2\) are considered stable against large-scale modes \citep{athanassoula86, binney08}. In marginally stable cases (e.g., \(Q \approx 1-1.5\)), small perturbations can still be amplified through \textit{swing amplification} \citep{goldreich65, julian66, toomre81}, in which differentially rotating stellar orbits shear leading wave packets into trailing arms while self-gravity rapidly boosts their amplitudes. This mechanism has been widely applied to explain the prevalence of multi-armed structures in galaxies with moderate or low Toomre \(Q\), indicating that the local dynamics of the disk strongly dictate both the onset and the maintenance of spiral patterns. N-body simulations also show that higher galactic shear rates produce more tightly wound spiral arms, linking spiral morphology to the local dynamical state of the disk \citep{grand13}. Other simulations have found that spiral arms can be transient and recurrent features with pattern speeds that decline with radius and can drive efficient radial migration over a broad radial range \citep{grand12a, grand12b}.

However, swing amplification alone does not determine the lifetime or recurrence of these features. Many numerical experiments have shown that multiple spiral modes may coexist in the disk and overlap in radius \citep{tagger87, masset97, rautiainen99, quillen11}. Their constructive and destructive interference can produce short-lived density enhancements even when the underlying coherent patterns persist for much longer, in some simulations for several hundred Myr and up to roughly 1 Gyr \citep{comparetta12, minchev12a, hilmi20, marques25}. Observational evidence for such multi-mode structure has also accumulated. In their Fourier expansion of near-infrared images of M51, \citet{rix93} identified significant $m=1$ and $m=3$ components alongside the dominant two-armed pattern. More generally, galaxy images expanded into Fourier components reveal hidden three-armed structure in many systems \citep{elmegreen92}, and asymmetries in the stellar spirals (e.g., \citealt{henry03, meidt08}) further suggest the presence of multiple overlapping patterns. Using the radial Tremaine–Weinberg method \citep{merrifield06}, \citet{meidt09} found direct evidence for resonant coupling among several distinct patterns, including spiral-spiral and spiral-bar components, in high-quality H I and CO data cubes of external galaxies. Together, these theoretical and observational results support a picture in which multiple spiral modes coexist and interact, producing recurrent density enhancements through constructive and destructive interference.

Numerical resolution plays a pivotal role in capturing the faint transient density features produced by multiple spiral modes. 
In lower-resolution simulations, Poisson noise from a low number of particles can artificially amplify seed perturbations, triggering spirals too early and causing them to fade within a short timescale due to excessive heating \citep{fujii11}. 
Conversely, at sufficiently high resolution, the evolution of spiral arms proceeds more gradually, delaying the onset of detectable patterns while preserving them over several dynamical times without rapid dissipation \citep{donghia13, sellwood12}. 
Enhanced resolution also reduces stochastic particle scattering, allowing the system to maintain cooler kinematics and a lower Toomre \(Q\) for longer. 
As a result, the multi-arm phase can persist over several dynamical times, with swing amplification providing the initial growth of short-wavelength perturbations and interference among coexisting modes manifesting as recurrent spiral enhancements even in unperturbed disks.

In fully \emph{live} disk-halo systems, non-axisymmetrical structures such as spirals and bars continuously interact with nearby responsive halo particles. 
Angular momentum exchange with the dark matter halo (DM), as well as with any spheroidal bulge components, can significantly reshape non-axisymmetric structures \citep{athanassoula02, sellwood16, kwak17, jang23, chiba24, kwak26a}. 
While bars transfer angular momentum efficiently to the halo, which  simultaneously accelerates bar growth and slows its pattern speed  \citep{debattista00, dubinski09}, multiple spiral arms generally produce a much smaller net halo torque; for example, \cite{sellwood21b} reported that spiral arms transfer less than $0.2\%$ of total stellar angular momentum.
In relatively stable disks, the halo may impede early bar formation, giving spiral modes a prolonged window in which to dominate the disk. 
In more unstable configurations, however, strong spiral activity can quickly reinforce global asymmetries, hastening the transition to a bar phase. 
These processes imply the central role of a live halo in modulating disk instabilities: by acting as an angular momentum reservoir, it governs the timescales and amplitudes of both spirals and bars, ultimately driving the disk’s long-term secular evolution.

In this paper, we investigate the formation and evolution of multi-armed spiral modes in three-dimensional exponential disks embedded in \textit{live} DM halos with varying concentrations. 
By adjusting the halo concentration parameter, we simultaneously change the mass distribution that influences the local disk stability (Toomre \(Q\)) and the growth rates. 
In addition, we vary the disk and halo particle number to assess resolution effects on the spiral evolution. 
Our goals are as follows:
\begin{enumerate}
    \item \textbf{Characterizing Spiral Formation without External Perturbations}: We adopt equilibrium initial conditions and let local instabilities develop naturally, rather than imposing perturbations to initiate spirals \citep[e.g.,][]{sellwood21b}. We therefore follow the full, unsuppressed evolution of non-axisymmetric structure, including noise-seeded perturbations, transient shearing features, and mode interactions, rather than isolating a single prescribed spiral response.
    \item \textbf{Quantifying Resolution Effects}: We compare models of identical structure yet differing disk and halo particle numbers, thereby gauging how Poisson noise influences the amplitude and longevity of spiral patterns.
    \item \textbf{Connecting Spirals and Bars}: We identify when and how the spiral structure dominated phase transitions into bar formation, emphasizing the role of halo concentration in regulating these instabilities.
    \item \textbf{Comparing Live Halos and Fixed Potentials}: We highlight differences in spiral behavior between live DM halos and fixed potential models, underlining the dynamical friction and angular momentum exchange that only arise in the responsive halo case.
\end{enumerate}

We begin Section 2 by providing details of our numerical setup, including the disk-halo models, halo structure, and $N$-body technique. 
In Section 3, we present the principal results on spiral formation, investigating how the amplitude of different Fourier modes evolves in time for distinct halo concentrations and resolutions. 
Section 4 provides detailed discussions of the physical mechanisms, linking our simulations to the broader theoretical framework of swing amplification and non-linear mode coupling. 
In Section 5, we summarize our conclusions, highlighting implications for future high-resolution studies of galactic disks embedded in live halos.

\section{Numerical Setup}
\subsection{Initial Conditions}\label{sec:ic}

The initial conditions of our galaxies are generated using the publicly accessible \textsc{galic} code \citep{yurin14}. This initial condition generator modifies the velocities of particles through an iterative process to achieve equilibrium for a specified density profile. \textsc{galic}'s flexibility such as its ability to control the average streaming motion and the ratio of radial to vertical velocity dispersions makes it well-suited for producing the specific initial conditions. Although our initialized disk--halo models are close to equilibrium, they still exhibit weak axisymmetric ringing associated with their initial relaxation, which appears only temporarily as phase-wrapped ripples in the disk. In Appendix~\ref{appendix:r2c24}, we demonstrate that the initial relaxation process neither modifies the initial assignment of density and velocities nor triggers the formation of spirals and bars.

Our models consist of a stellar disk and a dark matter (DM) halo with mass and radius comparable to the Milky Way.
The DM halo follows a \cite{hernquist90} profile
\begin{equation}\label{eq:hernquist}
\rho_\text{DM} (r) = \frac{M_\text{DM}}{2\pi} \frac{a}{r(r+a)^3} ,
\end{equation}
where $M_\text{DM}$ is the total mass and $a$ is the scale length of the halo. The scale length of the DM halo changes by the concentration parameter $c$ as
\begin{equation}\label{eq:cc}
a = \frac{r_{200}}{c} \left[2 \ln(1+c)-\frac{c}{1+c}\right]^{1/2},
\end{equation}
where $r_{200}$ is the virial radius  \citep{springel05}. 
We fix the total mass of the DM as $M_\text{DM} = 1.14  \times 10^{12} \Msun$ while changing $c$ within the reasonable range from the Aquarius and TNG simulations \citep{springel08, vogelsberger14, bose19}.

The stellar disk follows the distribution of  
\begin{equation}\label{eq:expdisk}
\rho_{\star} (R, z) = \frac{M_{d}}{4\pi z_{d} R_{d}^2} \exp \left( -\frac{R}{R_{d}} \right)  \text{sech}^2 \left(\frac{z}{z_{d}}\right),
\end{equation}
where $M_{d}$ is the total mass of the stellar disk, $z_{d}$ is the vertical scale height, and $R_{d}$ is the scale length that is controlled by the spin parameter introduced in \cite{mo98}.
$R$ and $z$ stand for the radial and vertical distances in the cylindrical coordinates, respectively.
For our study, we fix the stellar mass, $M_{d} = 5  \times 10^{10} \Msun$, disk scale length, $R_d = 3\kpc$, and vertical scale height, $z_d = 0.3\kpc$.

\begin{table}[h!]
\caption{\label{ic} Initial Conditions}
\centering
\begin{tabular}{lccccc}
\hline\hline
Model & $c$  & $N_{\star}$ & $N_{DM}$ & $m_{\star}:m_{DM}$ \\
\hline
r1c14 & 14  & 5e6 & 1.14e7 & 1:10  \\
r1c16 & 16  & 5e6 & 1.14e7 & 1:10  \\
r1c18 & 18  & 5e6 & 1.14e7 & 1:10  \\
r1c20 & 20  & 5e6 & 1.14e7 & 1:10  \\
\hline
r2c14 & 14  & 5e7 & 1.14e8 & 1:10  \\
r2c16 & 16  & 5e7 & 1.14e8 & 1:10   \\
r2c18 & 18  & 5e7 & 1.14e8 & 1:10  \\
r2c20 & 20  & 5e7 & 1.14e8  & 1:10  \\
\hline
r1c16hdm & 16  & 5e6 & 1.14e8 & 1:1  \\ 
r1c16ldm & 16  & 5e6 & 1.14e6 & 1:100  \\ 
r1c16fdm & 16  & 5e6 & Fixed &      \\ 
\hline
r1c14fdm     & 14  & 5e6   & Fixed  & \\
r1c16ldmsf06 & 16  & 5e6   & 1.14e6 & 1:100\\
r1c22 & 22   & 5e6 & 1.14e7 & 1:10 \\
r2c24 & 24   & 5e7 & 1.14e8 & 1:10  \\
\hline
\end{tabular}
\tablecomments{Column (1) is the model name. Column (2) is the halo concentration. Columns (3) and (4) are the total number of star and dark matter halo particles, and the last column gives the mass ratio between star and DM particle.}
\label{table:model}
\end{table}

The list of our models is presented in Table \ref{table:model}.
Our models are categorized into three groups: local stability, resolution, and halo features. 
Each `r1' model consists of $5 \times 10^{6}$ particles for the stellar disk and $1.14 \times 10^{7}$ particles for the DM halo, with a mass ratio of 1:10 between the stellar and DM particles.
To study the effects of resolution, we increase the number of particles by a factor of 10 for the `r2' models while maintaining the same mass ratio between stellar and halo particles.
To investigate the gravitational stability of the models, we vary the halo concentration parameter, $c$, from 14 to 20 in the `r1' and `r2' models. We define disk stability by the Toomre $Q$ parameter \citep{toomre64}, see Eq.~\eqref{eq:Toomre}. 
The radial velocity dispersion of all models is set to be $\sigma_{R}/\sigma_{z}=1$. The corresponding minimum $Q$ values are 0.830, 0.875, 0.923, and 0.965 from c14 to c20.

The Toomre $Q$ parameter is often referred to as a stability criterion, with $Q<1$ considered indicative of very cold conditions that rapidly undergo axisymmetric rearrangement in 2D simulations of razor-thin disks \citep{athanassoula86}. However, the degree of stability measured in 2D is not directly comparable to 3D systems. For example, the 3D isolated galaxy model DP4 in \cite{kwak17}, which has a Toomre parameter $  Q \sim 2  $, experiences vertical heating due to its initially large velocity anisotropy, even without forming a bar over 10 Gyr, whereas $  Q > 2  $ is required to suppress instabilities in 2D simulations \citep{athanassoula86}. 
\cite{romeo92} showed that finite disk thickness provides stabilizing effects. Thus, the same range of Toomre $Q$ values cannot be directly applied or compared in the same manner for 3D models: the effect of thickness is to increase the effective stability parameter, depending on the ratio of vertical to radial velocity dispersion \citep{romeo11,romeo13}. The effects of disk thickness include delaying bar formation, modulating angular momentum exchange, and ultimately altering bar properties such as length and even buckling instabilities, as demonstrated in \cite{klypin09} and \cite{kwak17}. Rather than constructing galaxy models fine-tuned to achieve $Q>1$, we select structural parameters based on observations and examine their natural processes of non-axisymmetric structure formation. In this study, we primarily employ $Q$ to denote relative stability among our models.

Our models do not include a classical bulge. We seek to examine the multi-mode spiral arms that form naturally prior to bar formation in 3D exponential disks exhibiting different stability levels, modified solely by resolution and halo concentration. 
Varying the halo concentration is our means of varying disk stability: a lower $c$ reduces the dark-matter contribution to the circular speed and hence the epicyclic frequency $\kappa$, lowering the minimum Toomre $Q$ of the disk (Table~\ref{table:model}). The resulting dependence of the onset and strength of spiral activity on $Q_{\rm min}$ is analysed in Sec.~\ref{sec:halocon}.
Including a bulge influences the kinematics and density structures of the disk within the bulge's effective radius, which in turn impacts the properties of spiral arms, particularly in the inner regions. Indeed, the introduction of random kinetic energy, characterized by the ratio of rotational kinetic energy to total gravitational potential energy, affects the system's stability \citep{ostriker73, kwak17}. Therefore, rather than incorporating a massive bulge to entirely prevent bar formation and alter disk dynamics, we focus on the characteristics of spirals and their evolution in exponential disks before and briefly after bar formation.

Finally, to examine the influence of the DM halo on spiral structure formation, we construct two models with different star-to-halo particle mass ratios and one model with a fixed potential halo instead of a live halo. 
In the r1c16hdm model (where `hdm' stands for high DM resolution), we increase the number of halo particles by a factor of 10 while keeping the number of disk particles unchanged. 
To examine the impact of the star-to-halo particle mass ratio, we reduce the number of halo particles by a factor of 10 in the r1c16ldm model (where `ldm' stands for low DM resolution). 
In the r1c16fdm model, `fdm' refers to a `fixed DM' halo potential. We refer to these three models as \emph{halo models}.

We use the same softening length for both the disk and halo to account for the large variation in the number of DM halo particles across some of our models. 
The softening lengths are determined based on the mean particle separation in the disk within the effective radius.
For the r1 and r2 models, we adopt softening lengths of 0.03 kpc and 0.01 kpc.
However, Model r1c16ldm has $N_{\rm{DM}}=1.14\times10^6$ particles with $m_{\rm{DM}}/m_\star = 100$, for which a larger DM softening value is often considered, so we additionally evolve the same model with a 0.6 kpc DM softening length, $\epsilon_{\rm DM}$, and name it  r1c16ldmsf06 (see Appendix \ref{appendix:additional}). Finally, Model r2c24 is also evolved and analyzed in Appendix \ref{appendix:r2c24}, Model r1c14fdm in Appendix \ref{appendix:additional}, and Model r1c22 in Appendix \ref{appendix:spectrogram}.

Our models are pure N-body simulations without a gaseous component, and thus do not include realistic stellar feedback processes that could influence the development of instabilities.
The impacts of star formation, gravitational softening length, and optimal particle number will be investigated in detail in our future studies.
Instead, we focus on the total number of particles and mass ratio between stars and DM particles.
All models are evolved for 2 Gyr with 200 output snapshots, using the AREPO code \citep{springel10b, weinberger20}.

\subsection{Modelling strategy and scope}
\label{sec:strategy}

Our simulations are designed as ``hands-off'' experiments in which the disk is allowed to develop non-axisymmetric structure self-consistently after the initial conditions are specified. We do not impose external perturbers, use quiet-start techniques, or suppress selected azimuthal harmonics in order to isolate a particular spiral response. This differs from more controlled studies of spiral structure, where the distribution of seed perturbers, the initial noise spectrum, or the allowed harmonics are deliberately restricted in order to isolate specific mechanisms of spiral growth and decay \citep[e.g.,][]{donghia13,sellwood12,sellwood14,sellwood19}. Such controlled calculations are better suited for identifying individual modes or comparing directly with analytic mode spectra of a prescribed initial disk. Our goal is complementary: we follow how multiple non-axisymmetric components, transient shearing features, and the response of a live dark-matter halo interact in an evolving disk.
Compared with controlled experiments, our approach cannot attribute an individual density feature to a single mechanism. In exchange, it retains the coexistence of coherent patterns, shearing transients and phase-mixing debris that characterizes real disks, and it allows us to measure how the coupled outcome depends on disk stability, numerical resolution, halo particle mass and halo responsiveness. These are dependencies that quiet-start calculations remove by construction.

A consequence of this approach is that the simulations also retain the weak ringing associated with the initial relaxation of the disk. This ringing is not a physical spiral mode and should not be confused with the later non-axisymmetric spiral or bar structures. Instead, it appears as early phase-wrapped ripples and faint diagonal features in the angular-momentum maps. Because the disk is not mathematically identical to the initialized model after this early adjustment, our simulations should not be interpreted as predicting the exact normal-mode spectrum of the initial distribution function. Rather, we follow the subsequent evolution of the disk after this initial relaxation.

This strategy is motivated by the fact that real stellar disks are not quiet-start systems. Galactic disks are continually perturbed by molecular clouds, halo substructure, satellite passages, gas accretion, and pre-existing non-axisymmetric structure. The phase-space structure of the Milky Way provides clear evidence that the Galactic disk preserves a complex history of perturbations and phase mixing \citep[e.g.,][]{antoja18,antoja22,tremaine23,cao24,hunt24, chiba25}. Similarly, observed spiral structure is rarely a pure single-$m$ pattern, but is generally a superposition of several non-axisymmetric components whose amplitudes, phases, and pattern speeds may vary with radius and time. We therefore do not aim to suppress this complexity, but to characterize how such multi-component structure evolves in isolated live-halo disks as a function of disk stability, numerical resolution, and halo responsiveness.

Throughout this paper we use ``secular evolution'' in the broad sense: any slow, cumulative change in the structure or kinematics of the disk driven by internal non-axisymmetric torques. This includes the redistribution of angular momentum among disk stars, with the associated radial migration and heating, which arises from resonant exchange between the stars and the spiral pattern \citep{lynden-bell72} and is mediated by the angular momentum the waves transport away from the resonances \citep{goldreich78,goldreich79}, as well as the net transfer of angular momentum from the disk to the halo. We distinguish these two explicitly, because in our models the spiral modes accomplish the former while only the central $m=3$ mode and the subsequent bar accomplish the latter.

\section{Results}\label{sec:ch2}

\renewcommand{\arraystretch}{0} 

\begin{figure*}[htbp]
    \centering
    \renewcommand{\arraystretch}{0}
    \begin{tabular}{@{}c@{}c@{}c@{}c@{}c@{}c@{}c@{}}
        & \textbf{r2c14} & \textbf{r1c16} & \textbf{r2c16} & \textbf{r2c18} & \textbf{r2c20} \\

        \raisebox{1\height}{\rotatebox{90}{\quad \textbf{0.5 Gyr}}} &
        \includegraphics[width=0.18\textwidth]{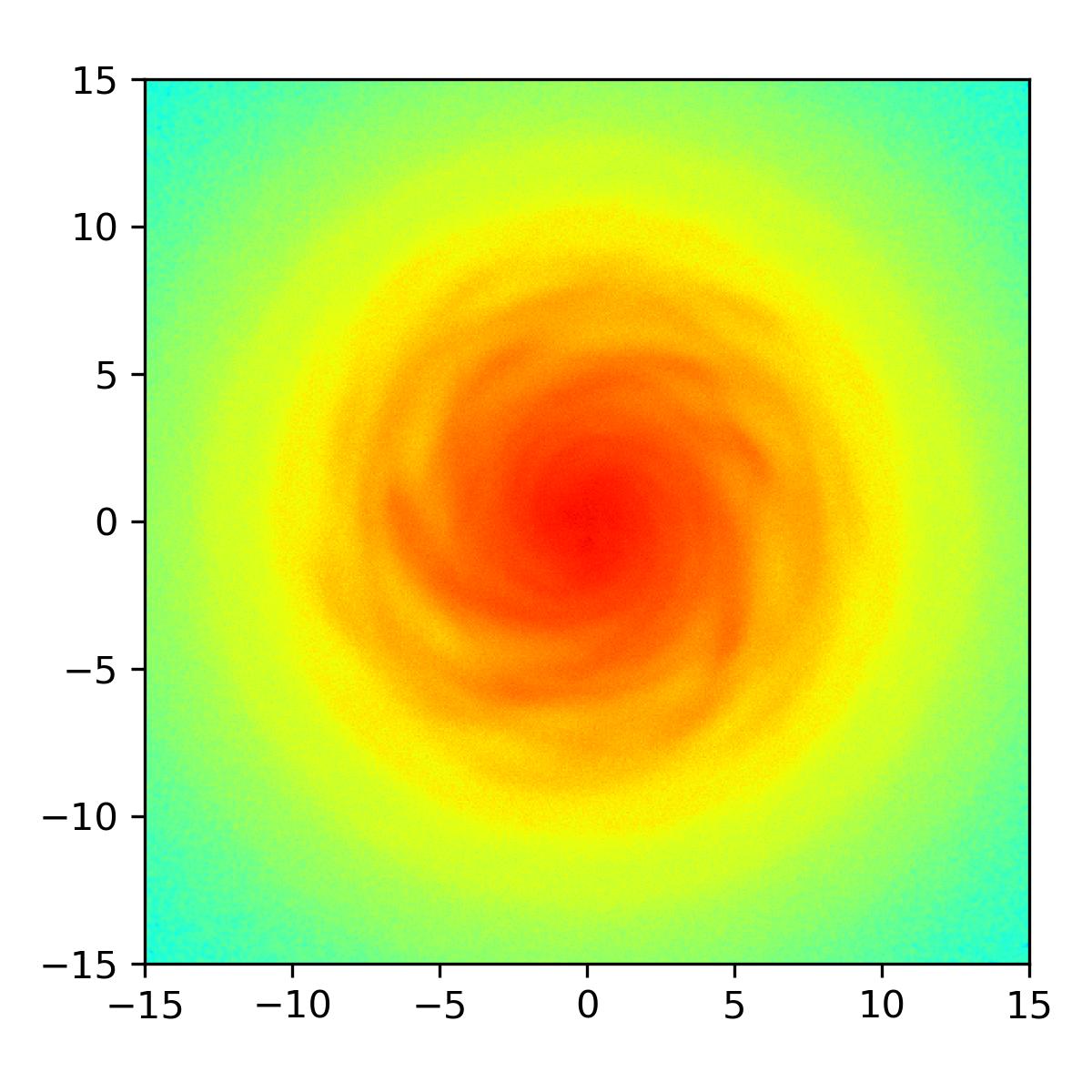} &
        \includegraphics[width=0.18\textwidth]{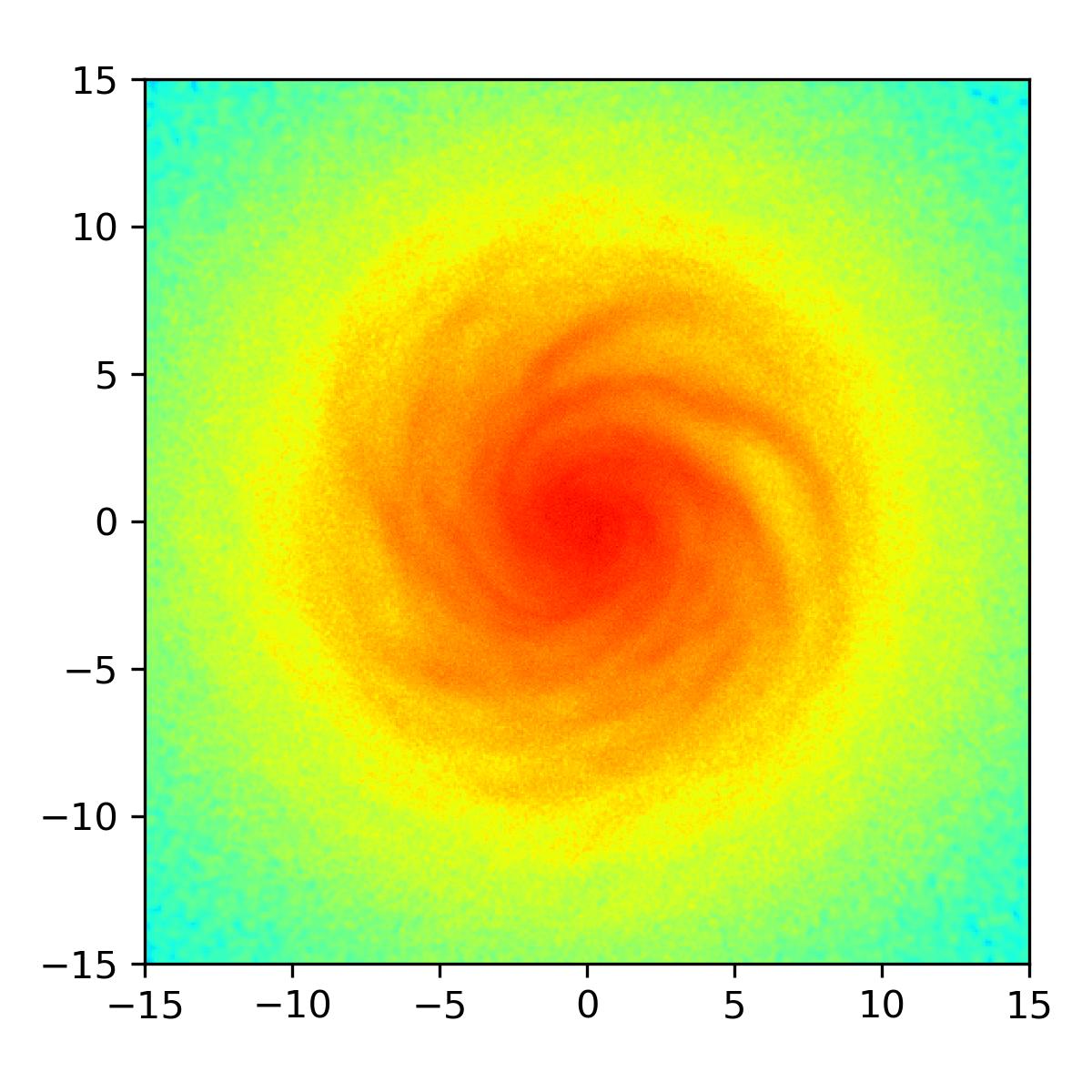} &
        \includegraphics[width=0.18\textwidth]{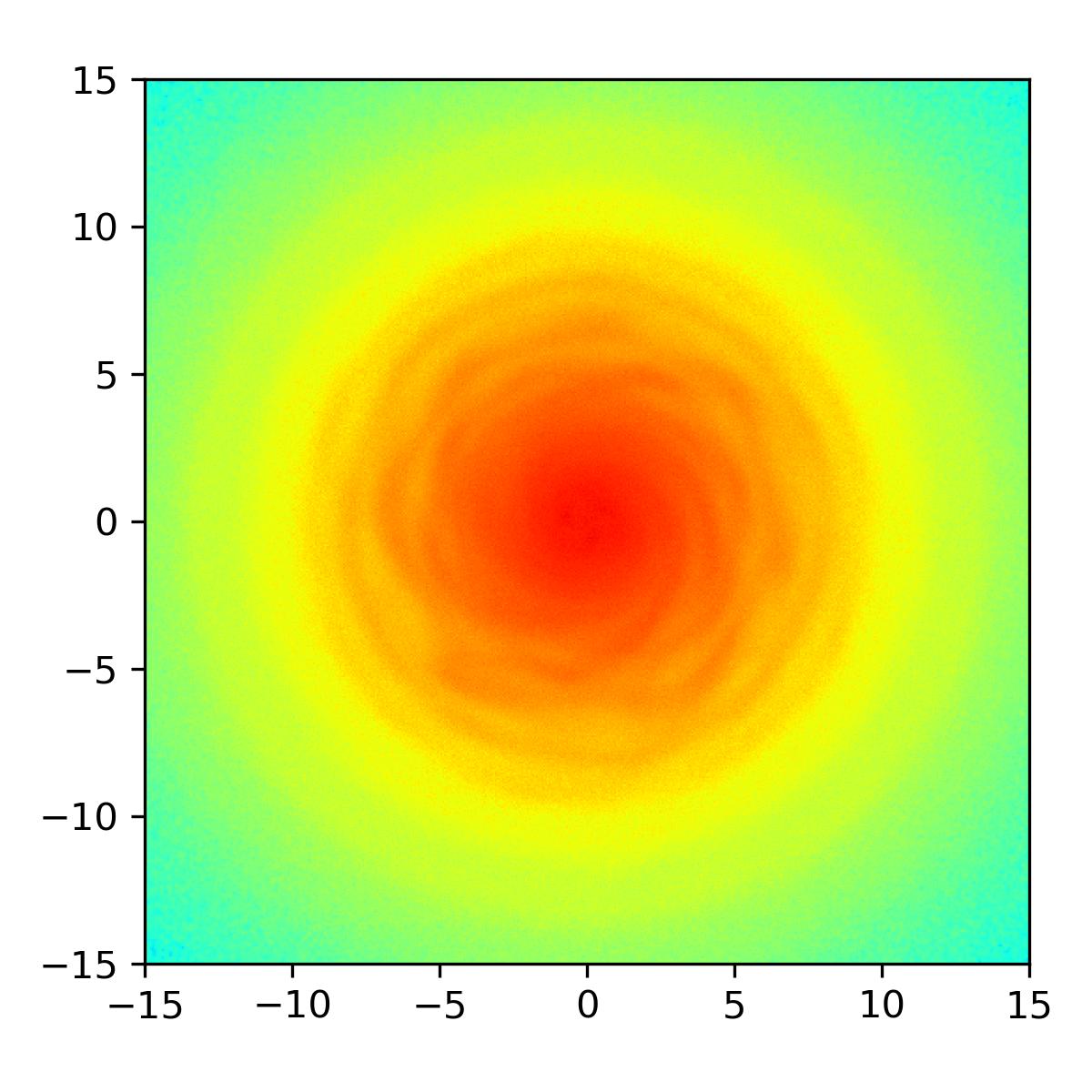} &
        \includegraphics[width=0.18\textwidth]{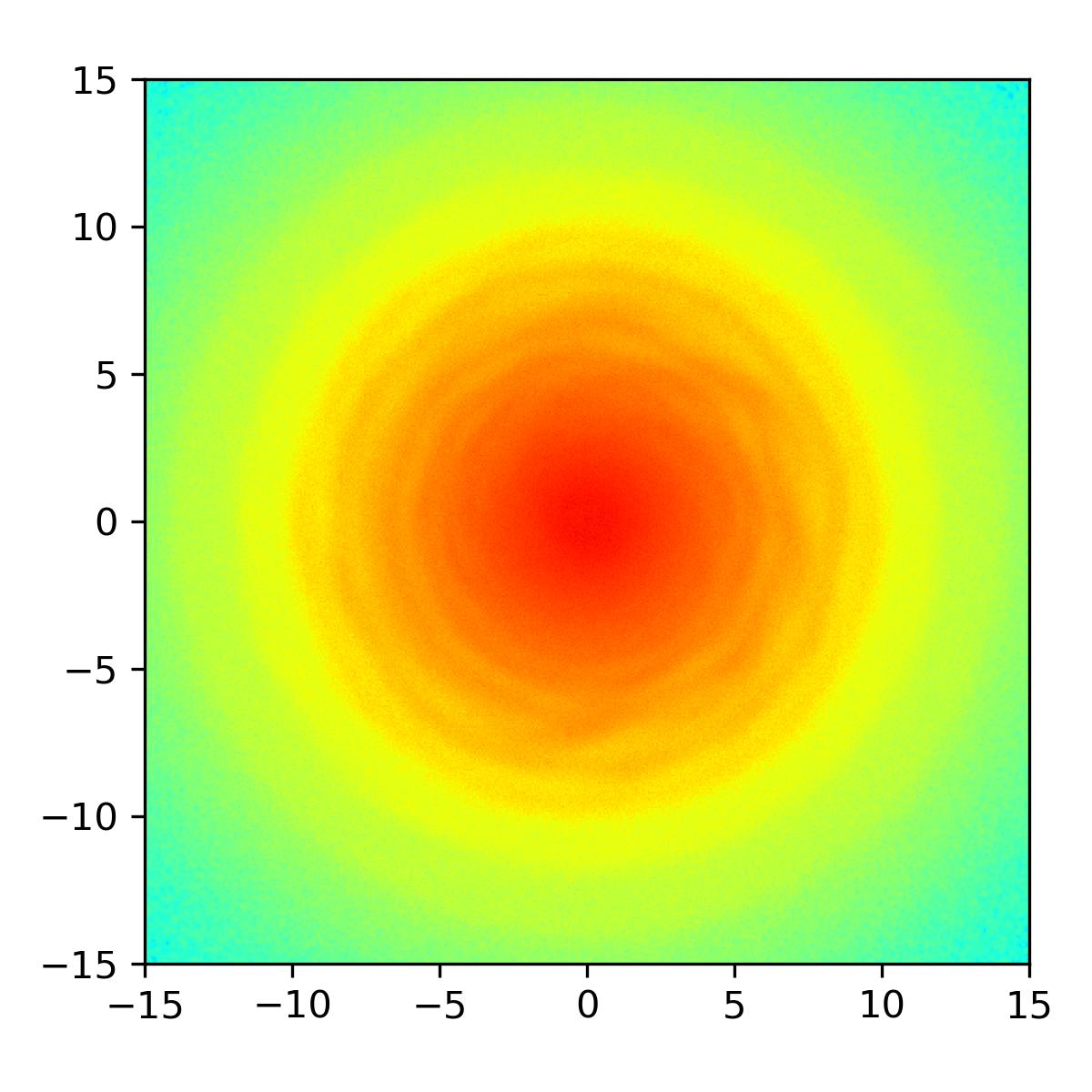} &
        \includegraphics[width=0.18\textwidth]{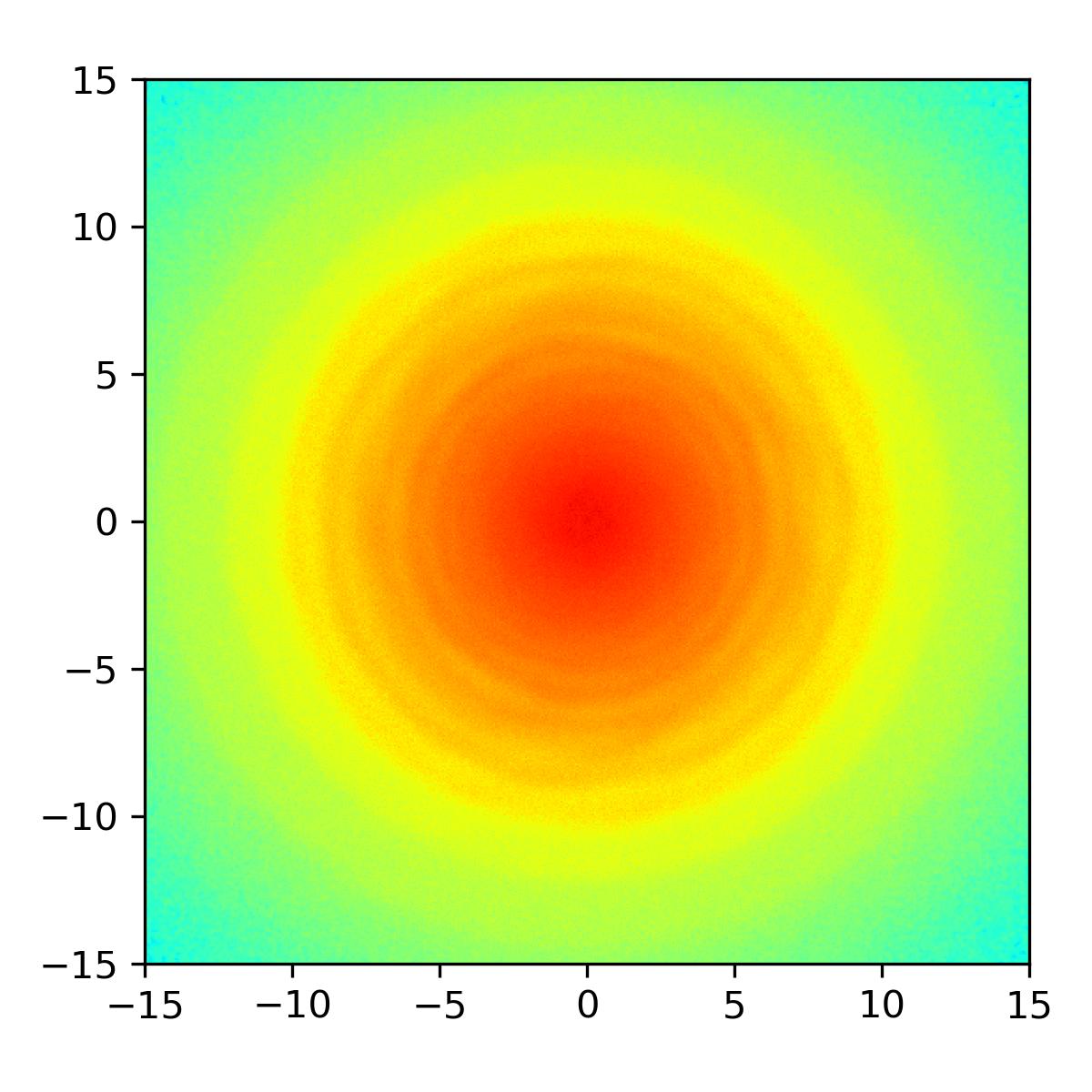} \\

        \raisebox{1\height}{\rotatebox{90}{\quad \textbf{1.0 Gyr}}} &
        \includegraphics[width=0.18\textwidth]{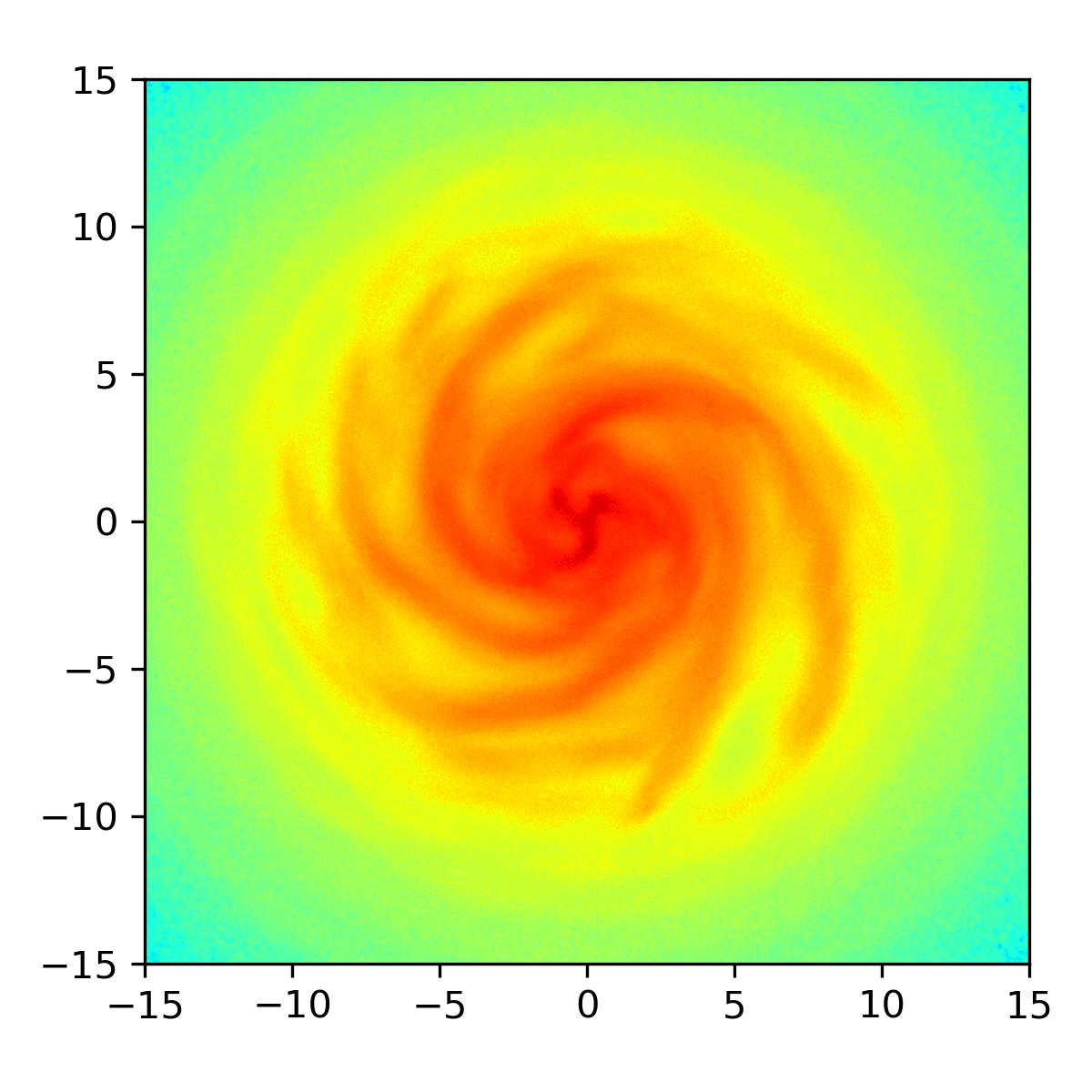} &
        \includegraphics[width=0.18\textwidth]{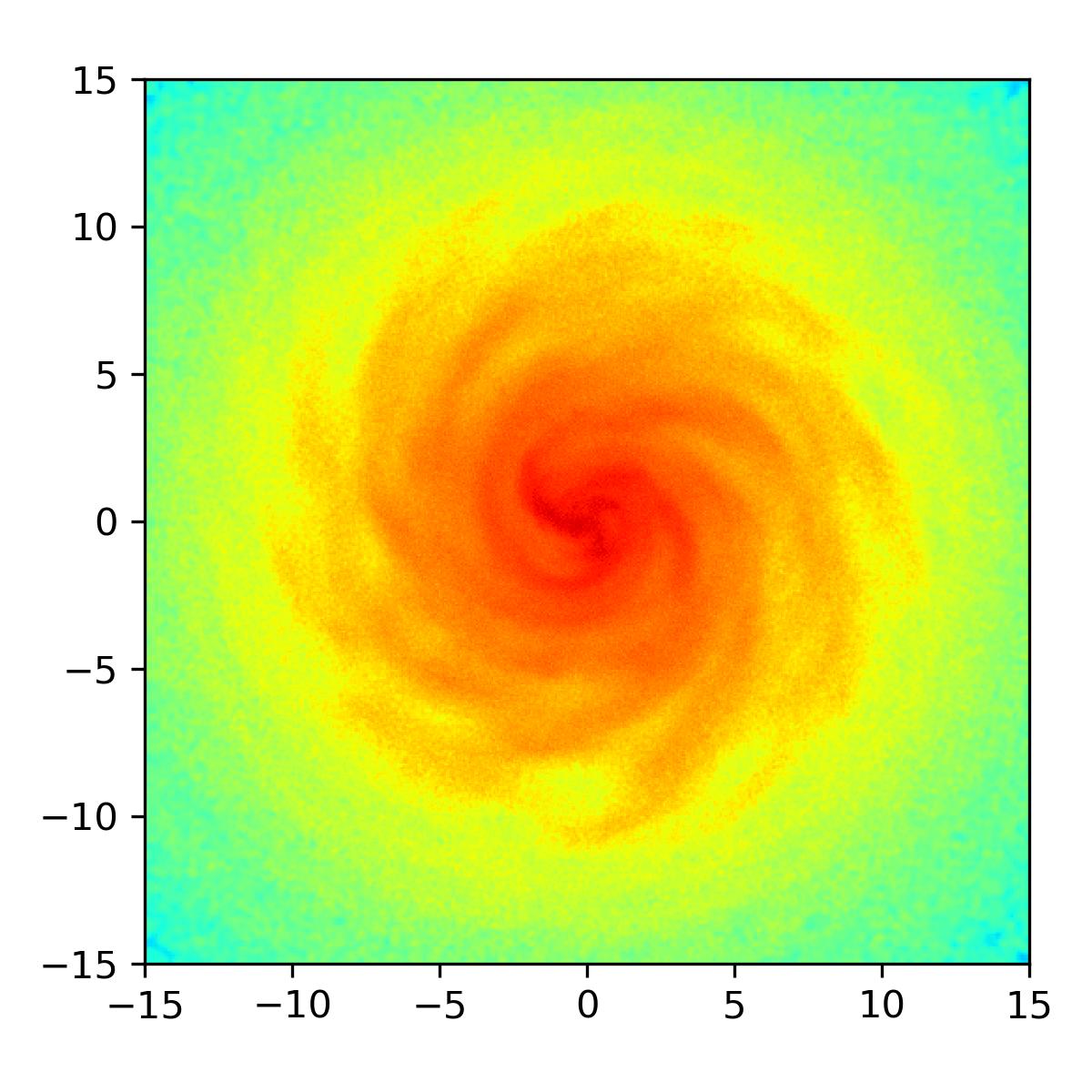} &
        \includegraphics[width=0.18\textwidth]{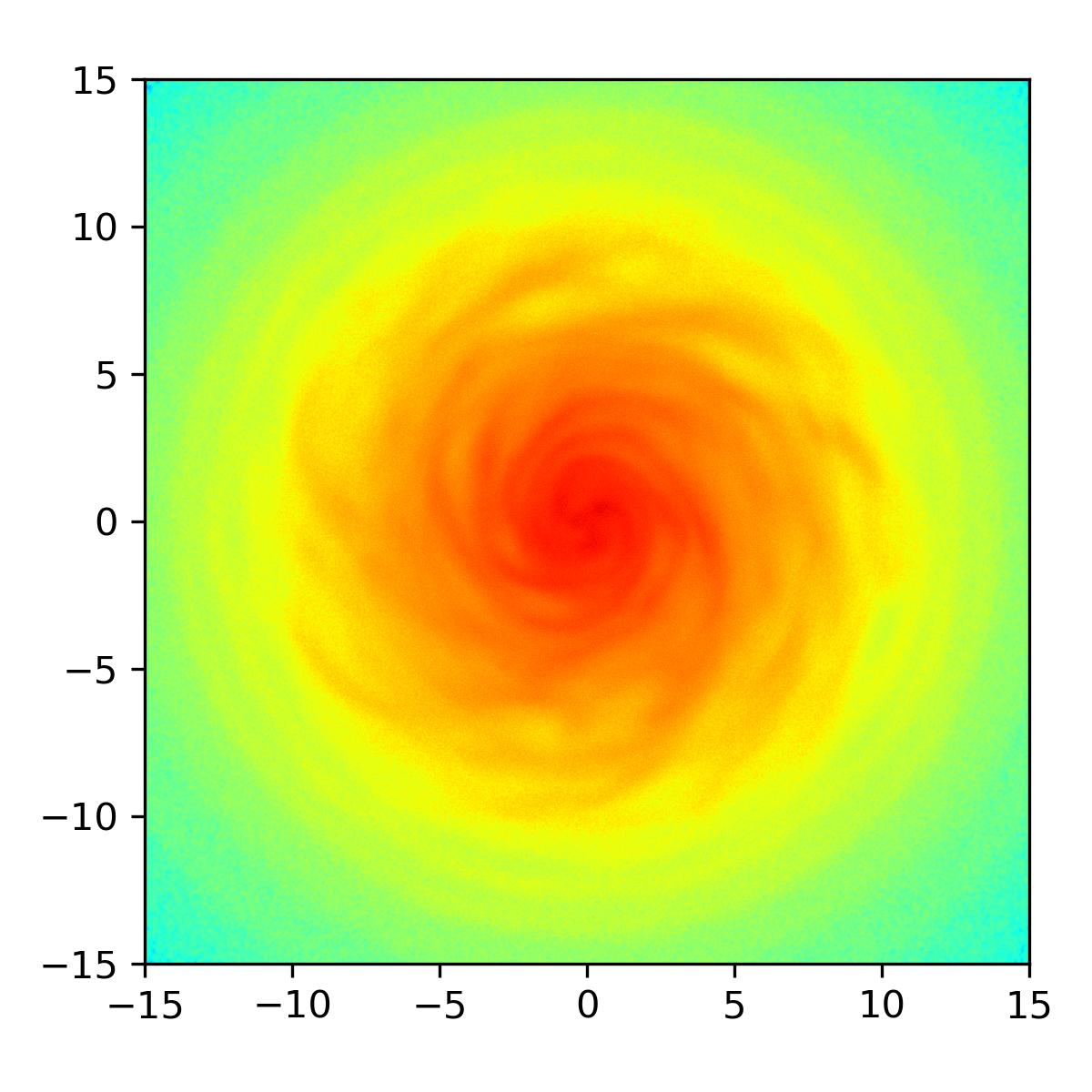} &
        \includegraphics[width=0.18\textwidth]{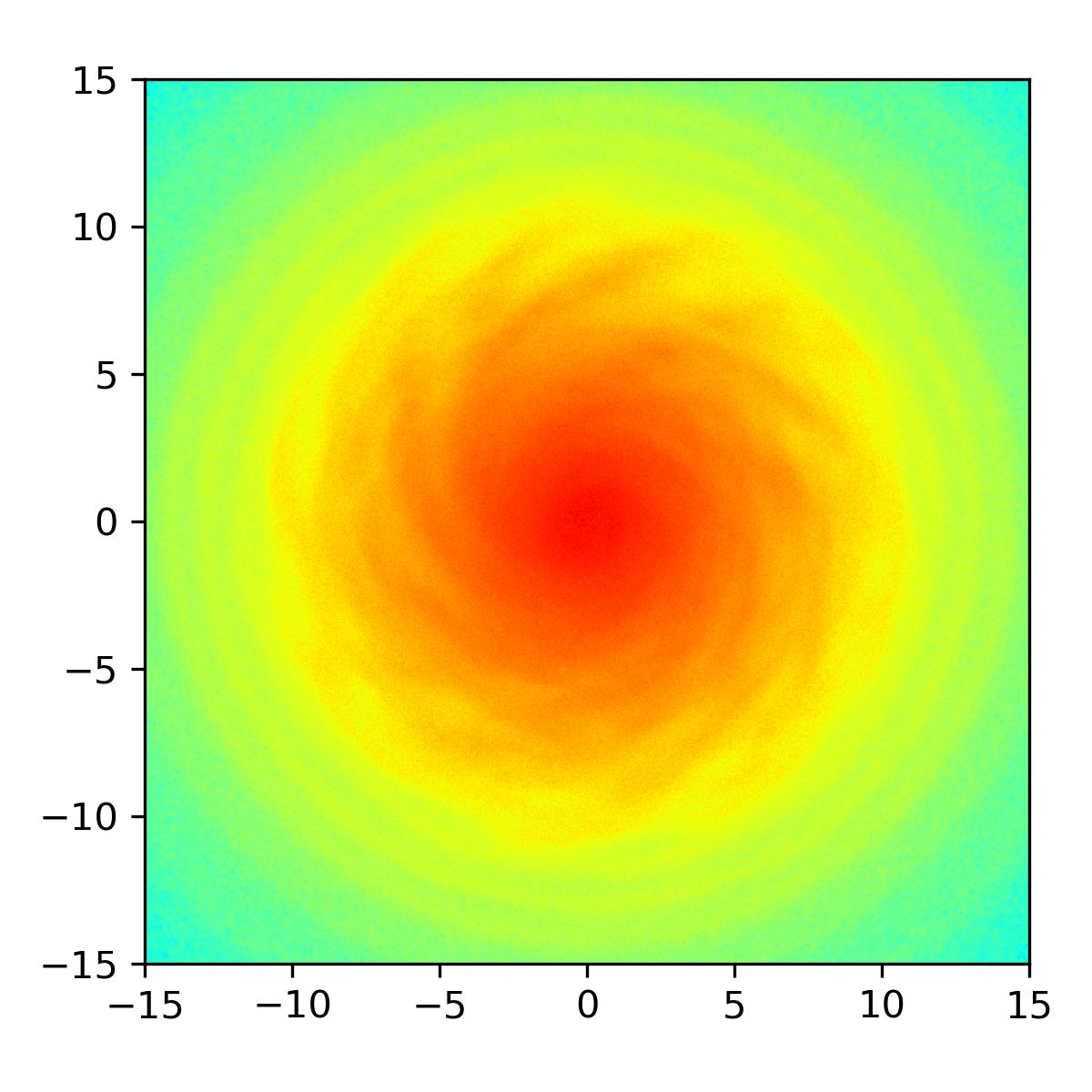} &
        \includegraphics[width=0.18\textwidth]{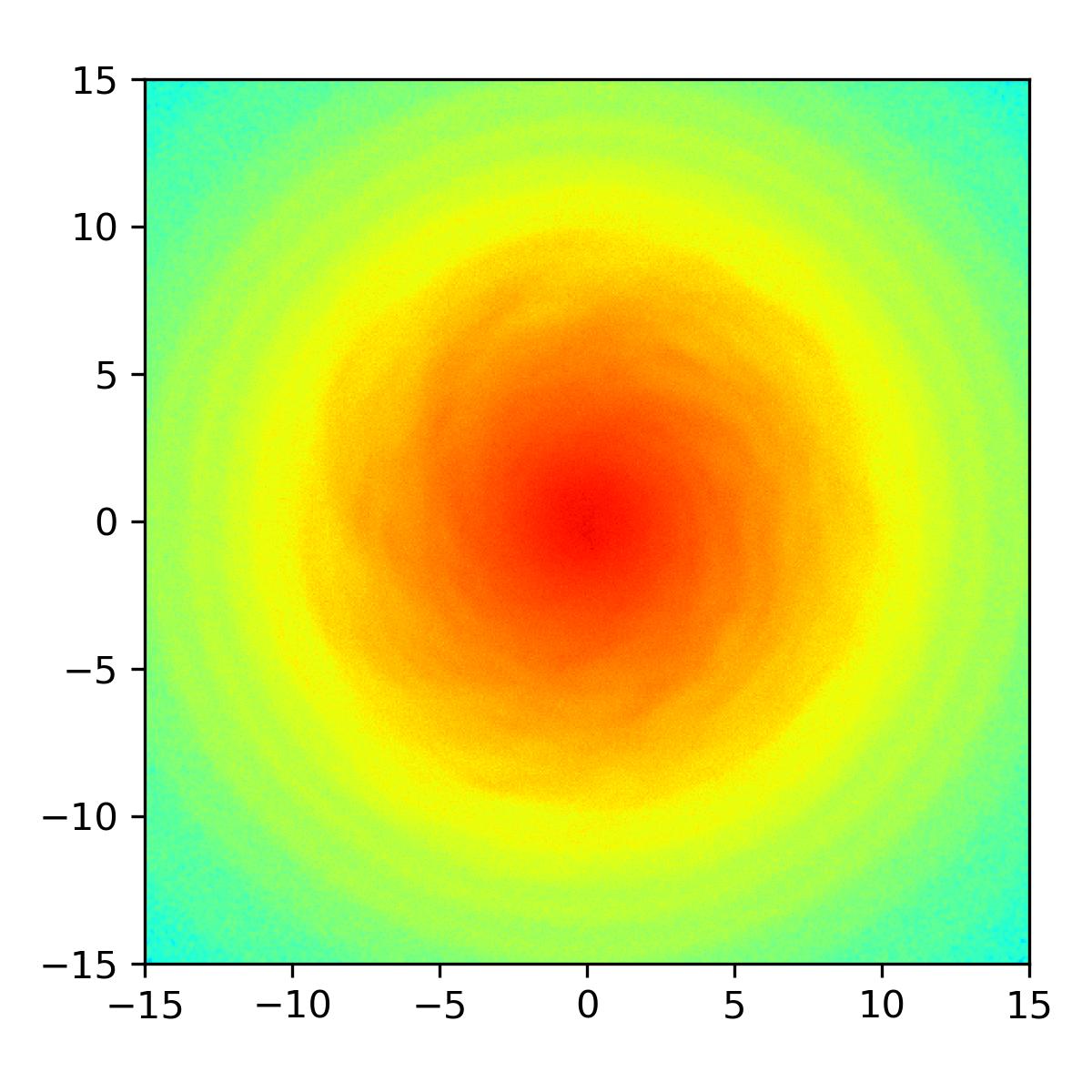} \\

        \raisebox{1\height}{\rotatebox{90}{\quad \textbf{1.5 Gyr}}} &
        \includegraphics[width=0.18\textwidth]{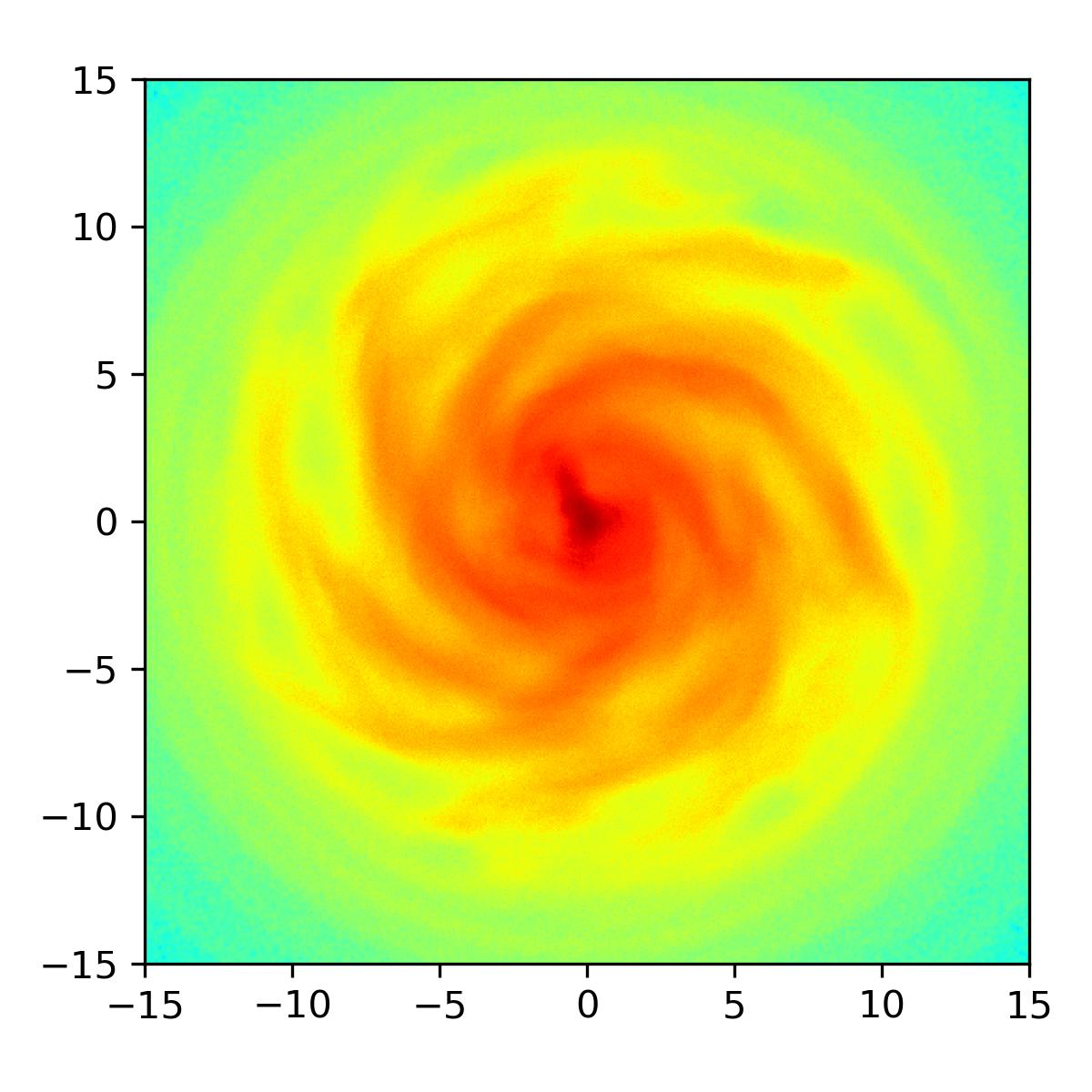} &
        \includegraphics[width=0.18\textwidth]{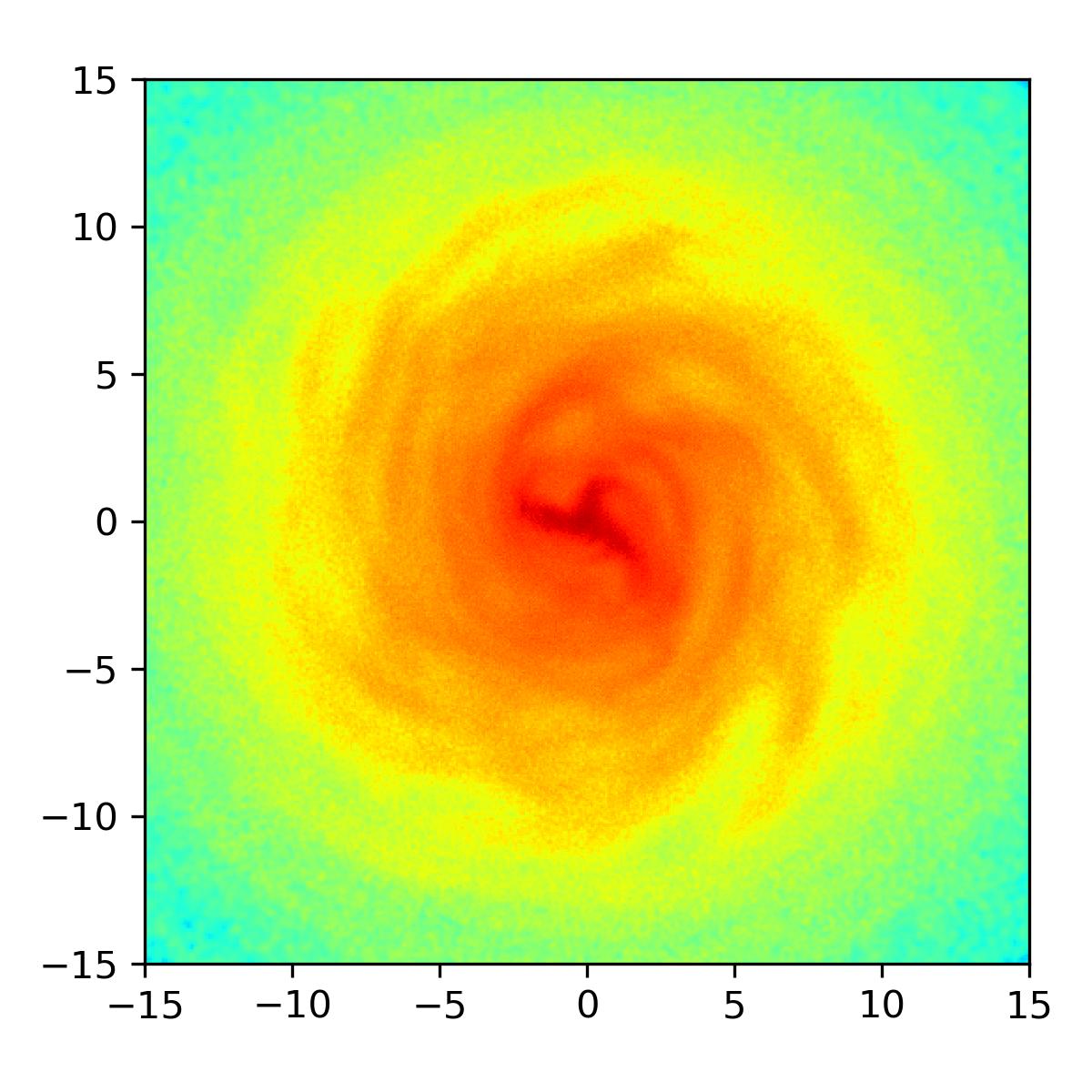} &
        \includegraphics[width=0.18\textwidth]{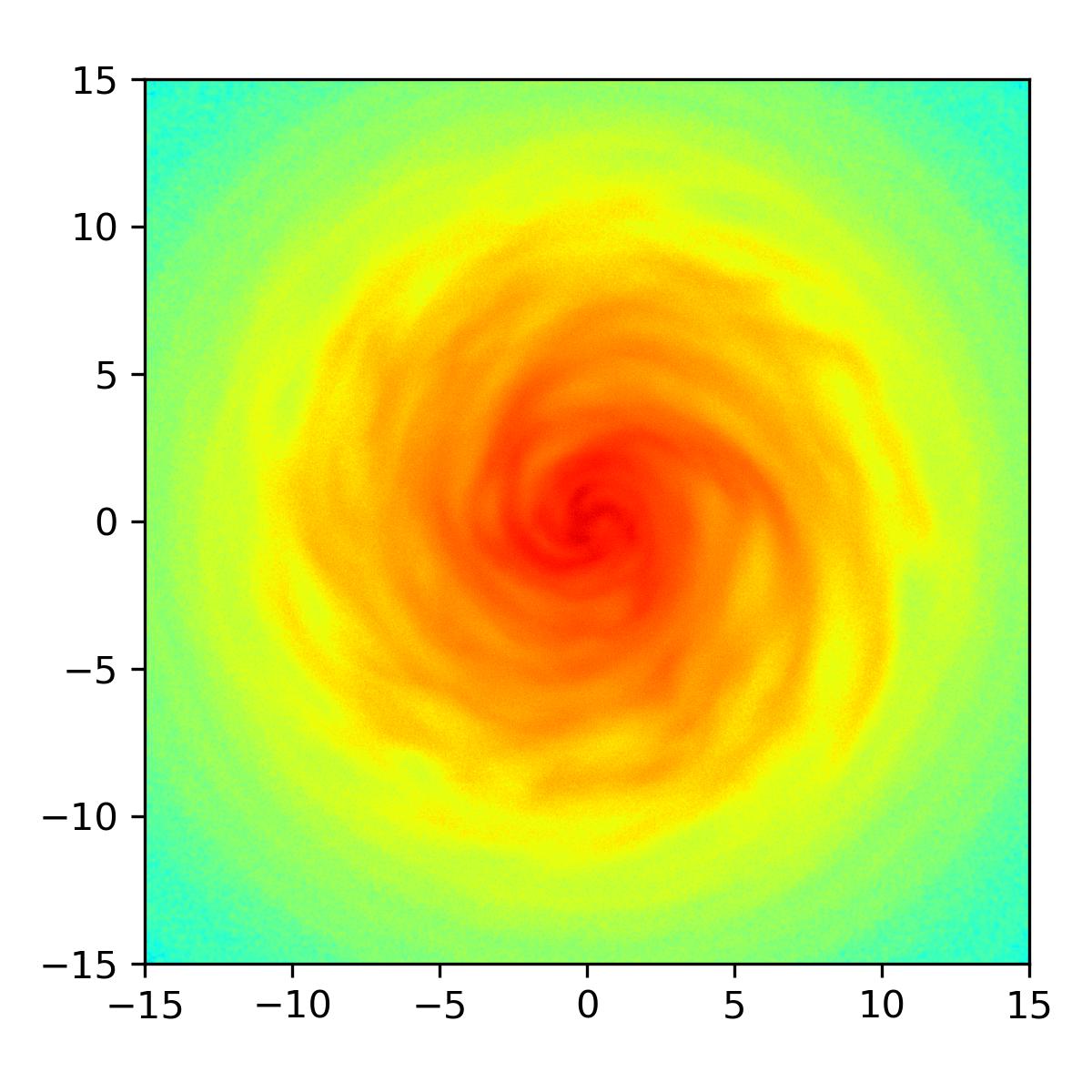} &
        \includegraphics[width=0.18\textwidth]{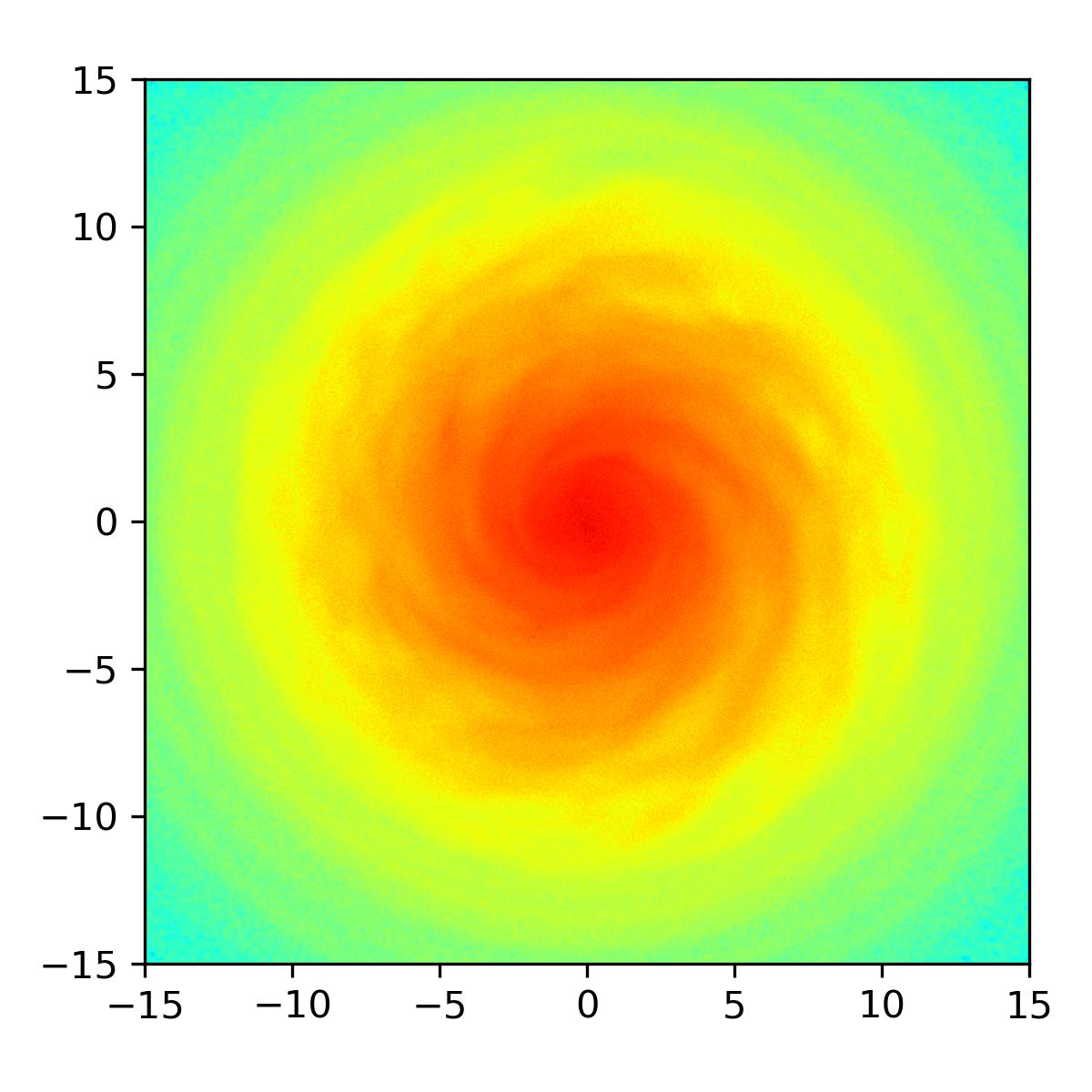} &
        \includegraphics[width=0.18\textwidth]{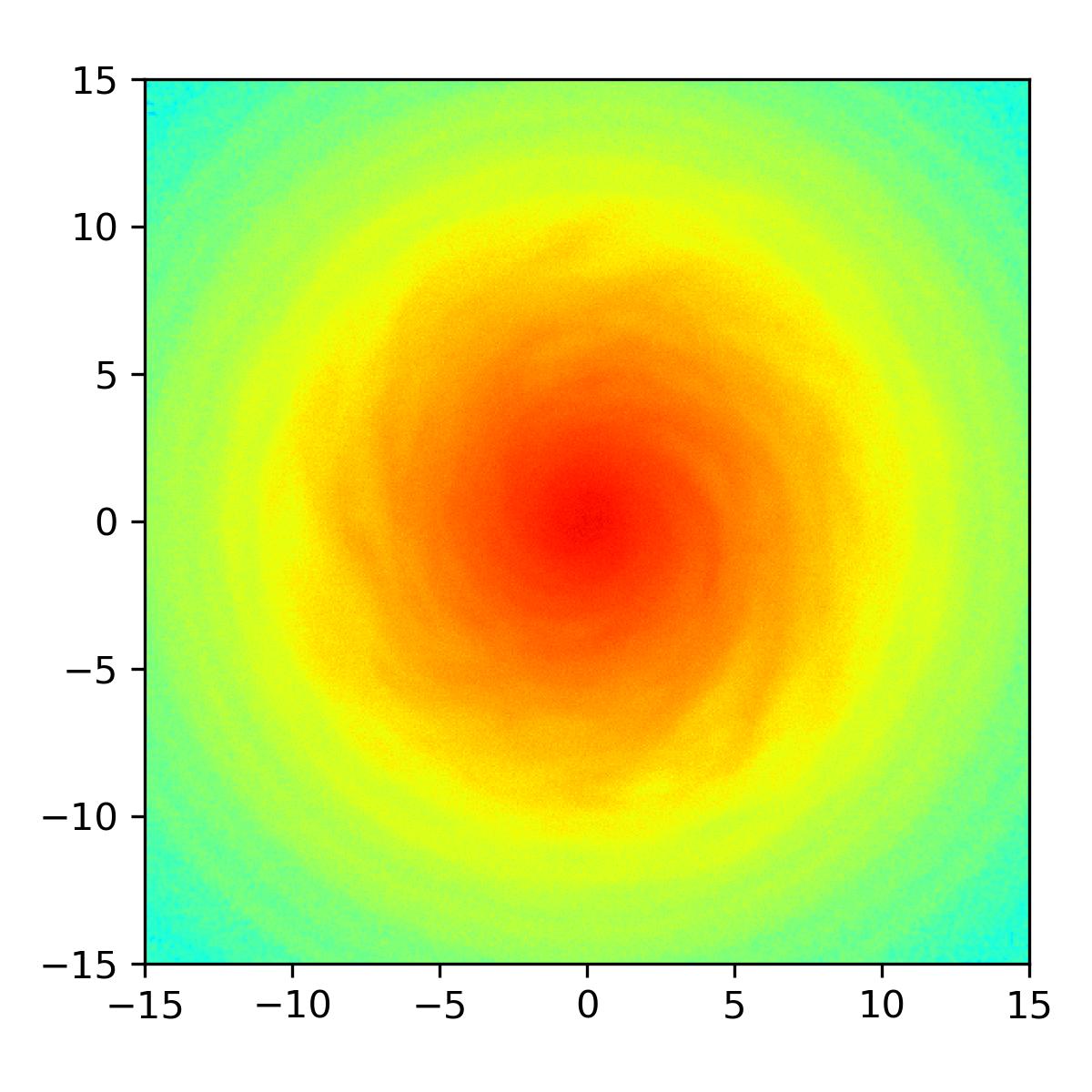} \\

        \raisebox{1\height}{\rotatebox{90}{\quad \textbf{2.0 Gyr}}} &
        \includegraphics[width=0.18\textwidth]{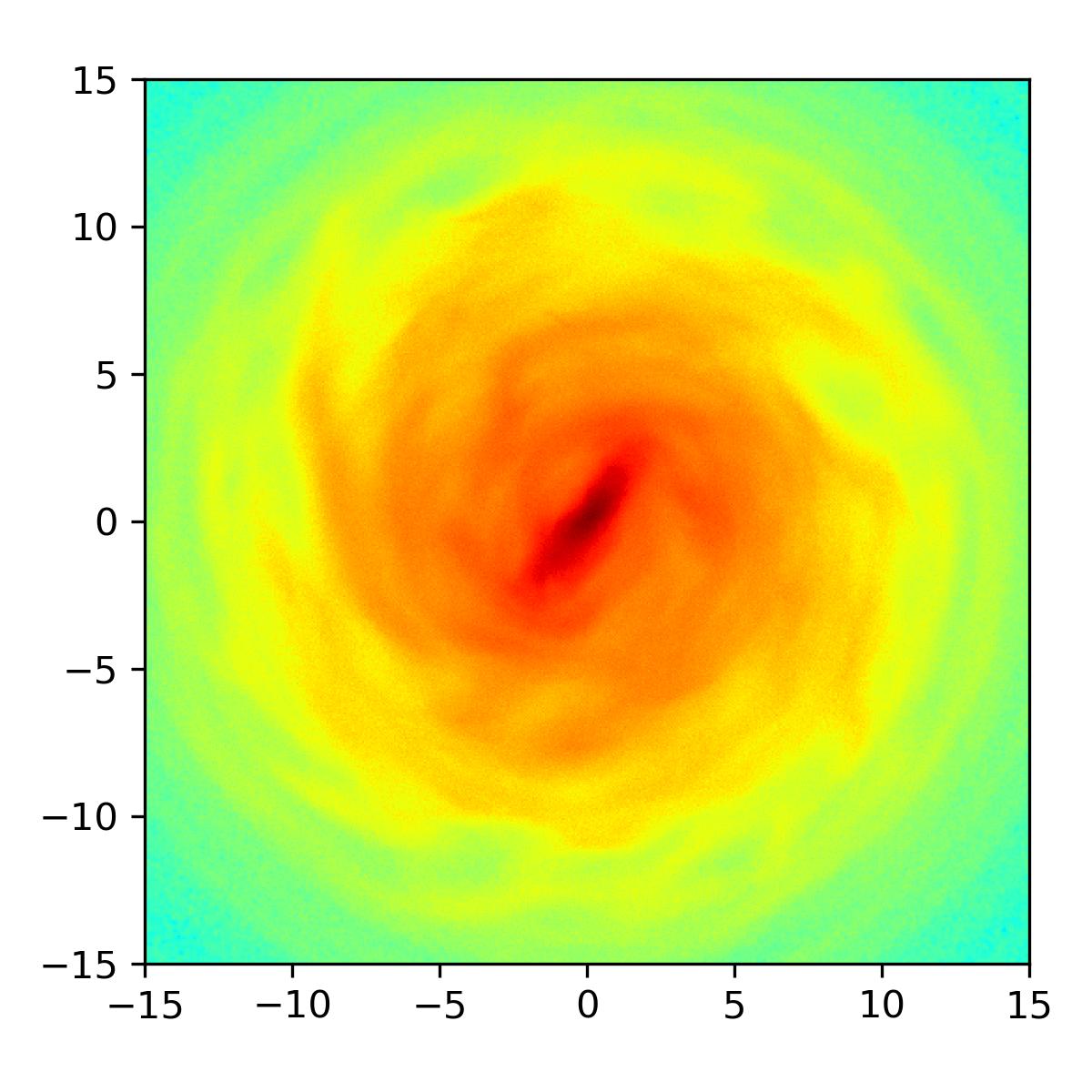} &
        \includegraphics[width=0.18\textwidth]{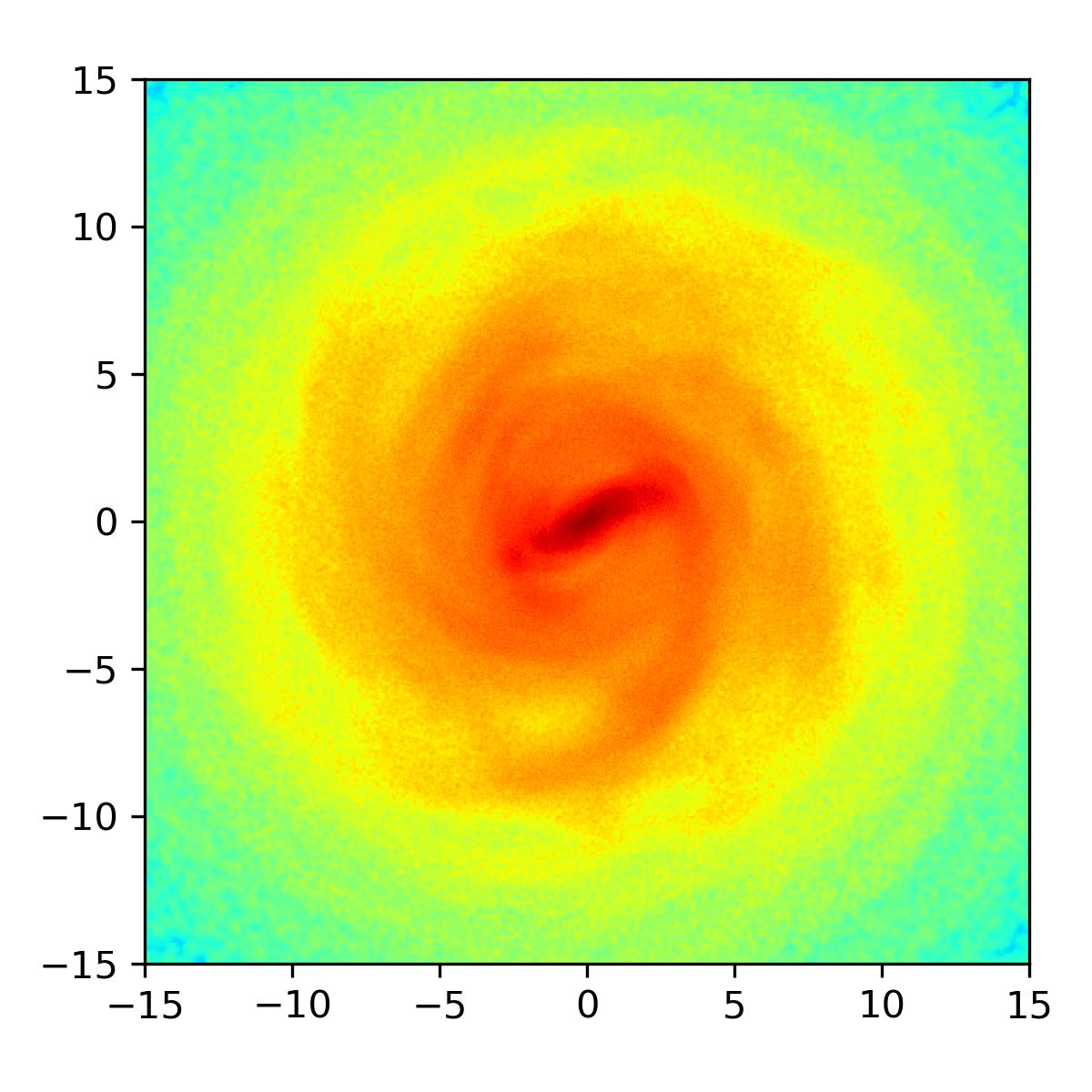} &
        \includegraphics[width=0.18\textwidth]{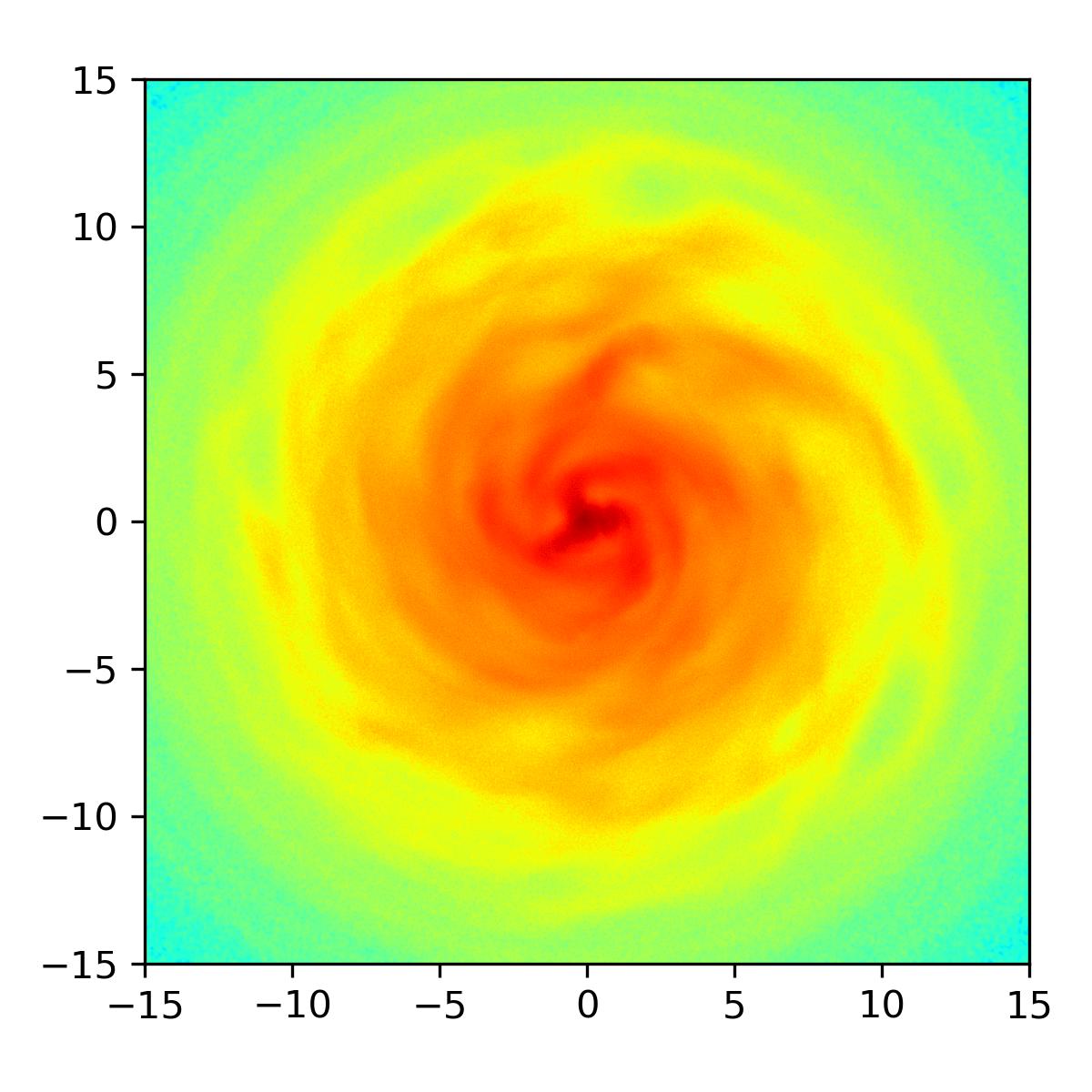} &
        \includegraphics[width=0.18\textwidth]{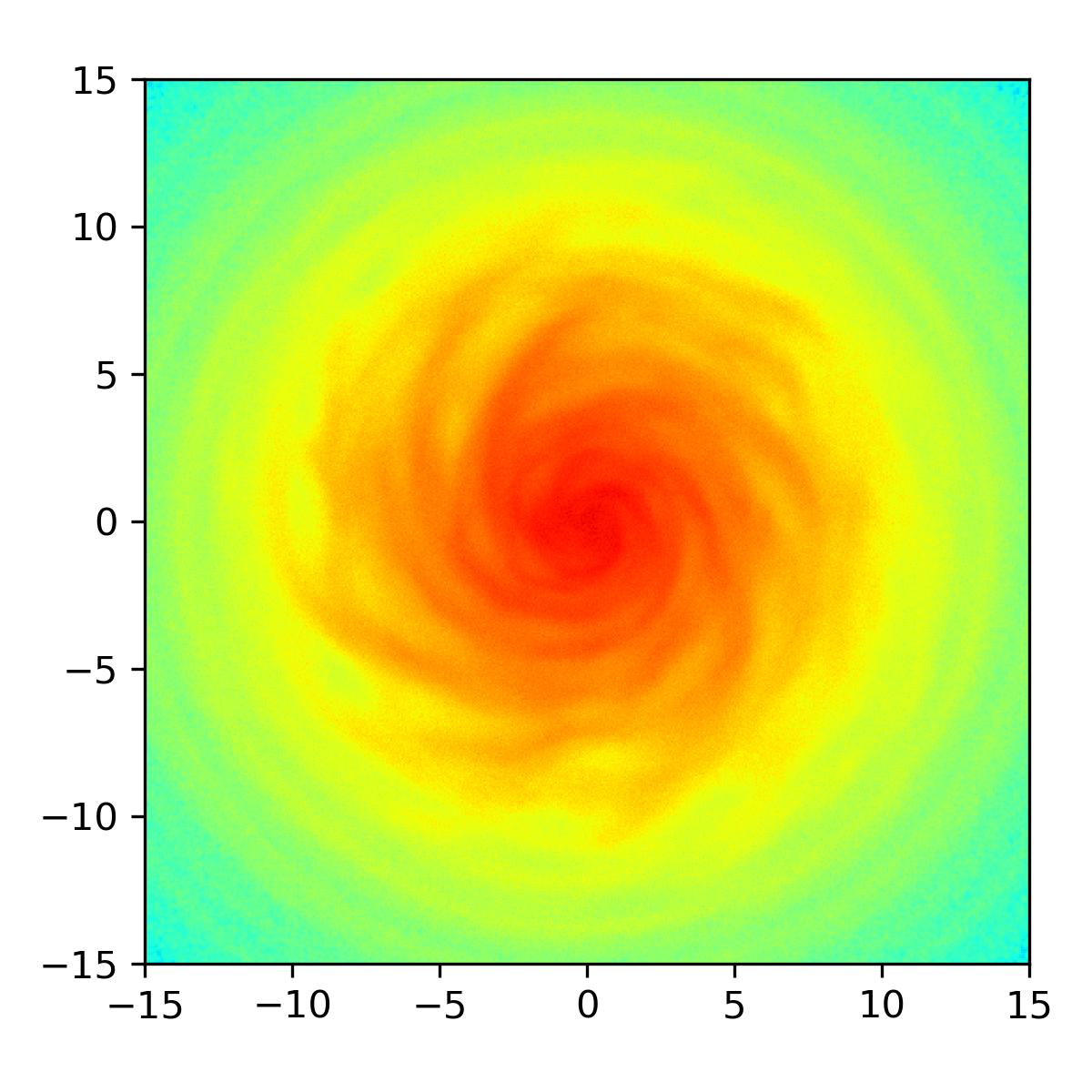} &
        \includegraphics[width=0.18\textwidth]{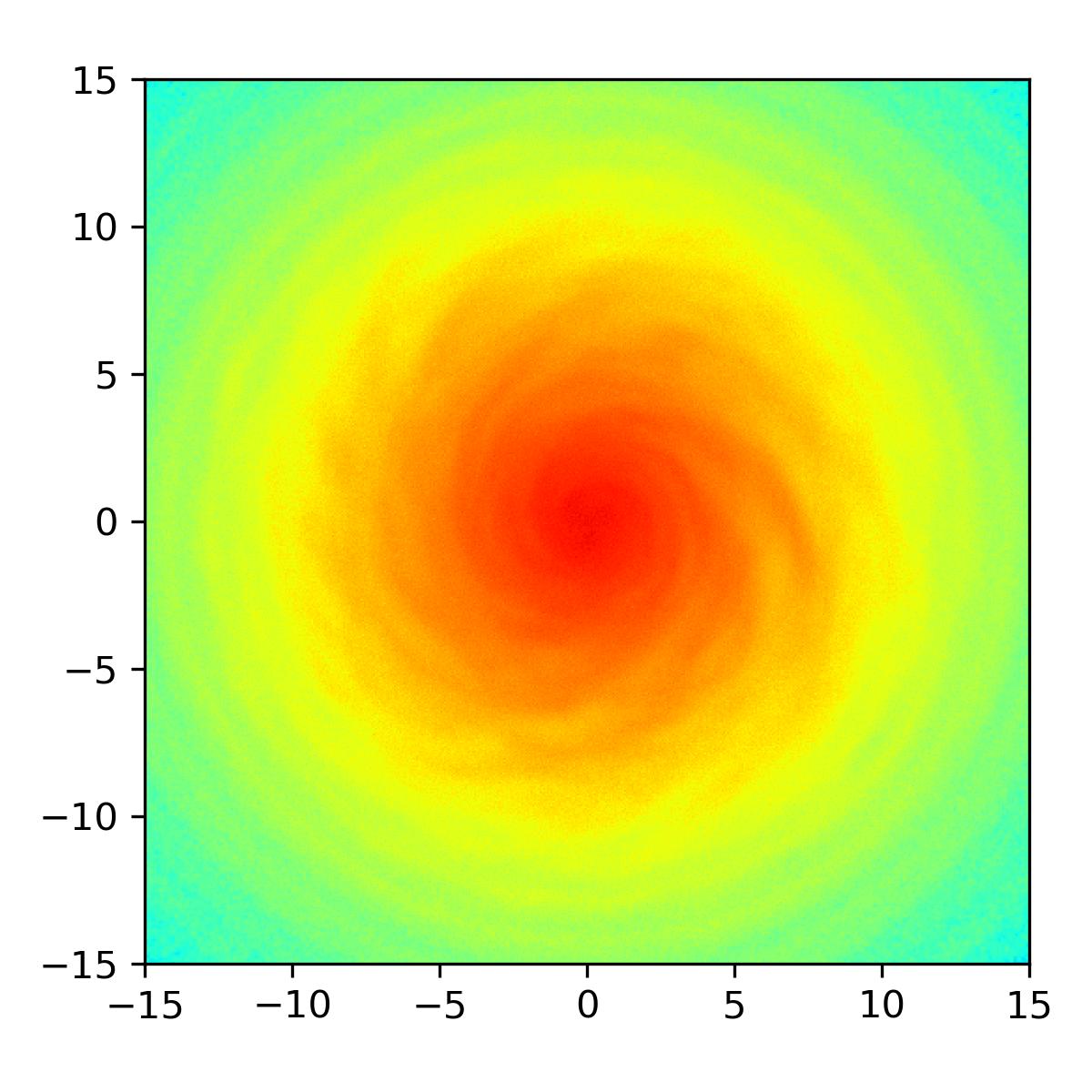} &
        \includegraphics[height=0.18\textwidth]{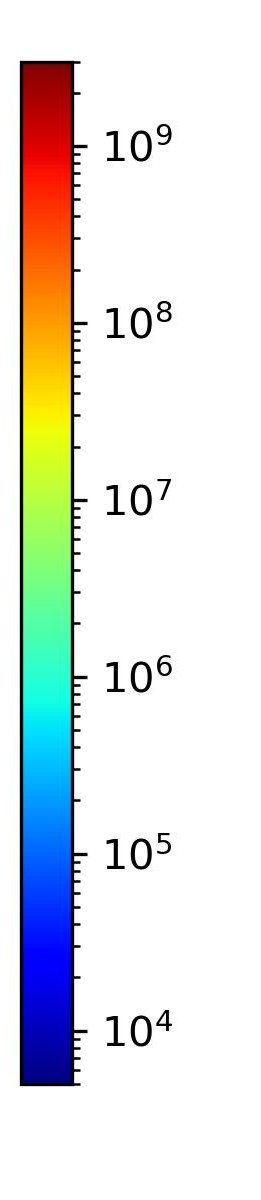} 
        \\
    \end{tabular}
    \caption{Face-on projections of surface density distribution in a $30\times30\kpc$ box at 0.5, 1.0, 1.5 and 2.0 Gyr (rows) and for model r2c14, r1c16, r2c16, r2c18, and r2c20 (columns). The color bar indicates the surface density in units of $\Msun\,\mathrm{kpc}^{-2}$.}
    \label{fig:faceon}
\end{figure*}

\begin{figure}[htbp]
    \centering
    \renewcommand{\arraystretch}{0} 
    \setlength{\tabcolsep}{2pt}     

    \begin{tabular}{cc} 
        \multicolumn{1}{c}{\textbf{r1c16}} & 
        \multicolumn{1}{c}{\textbf{r2c16}} \\[1ex]

        \includegraphics[width=0.25\textwidth]{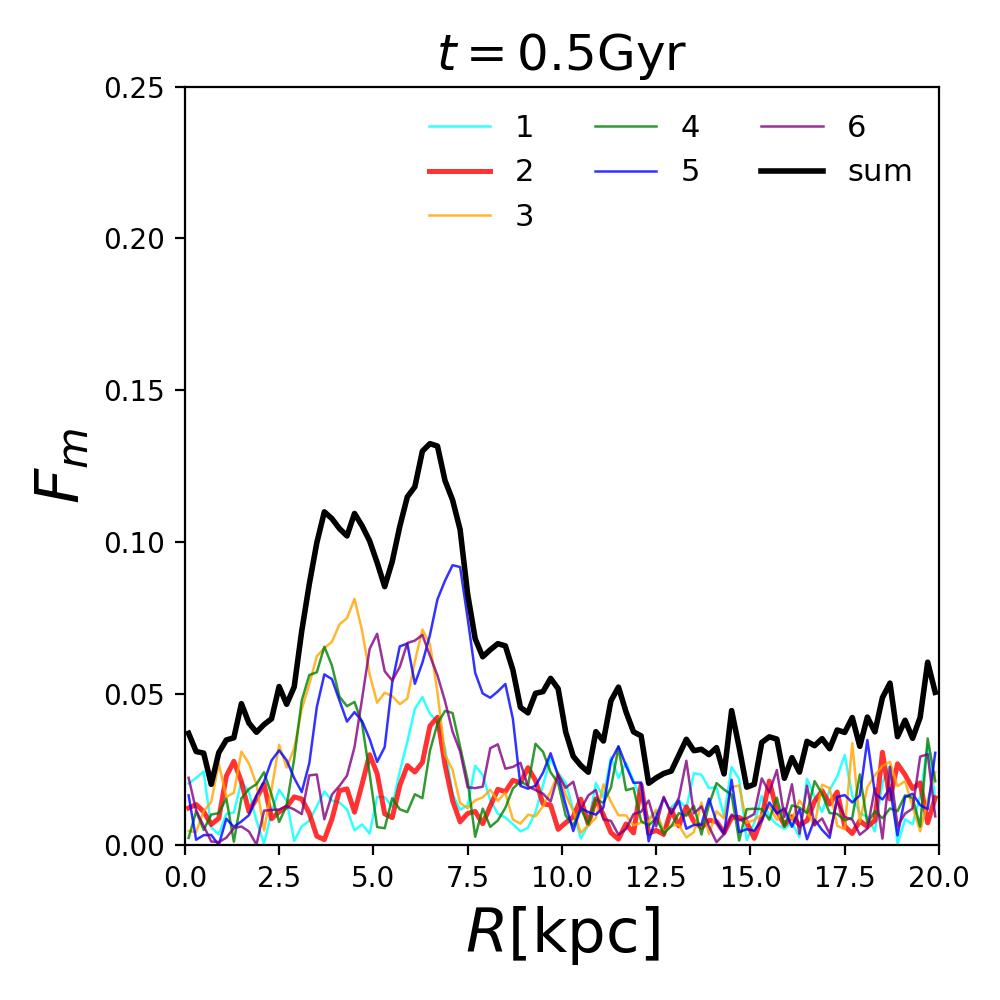} &
        \includegraphics[width=0.25\textwidth]{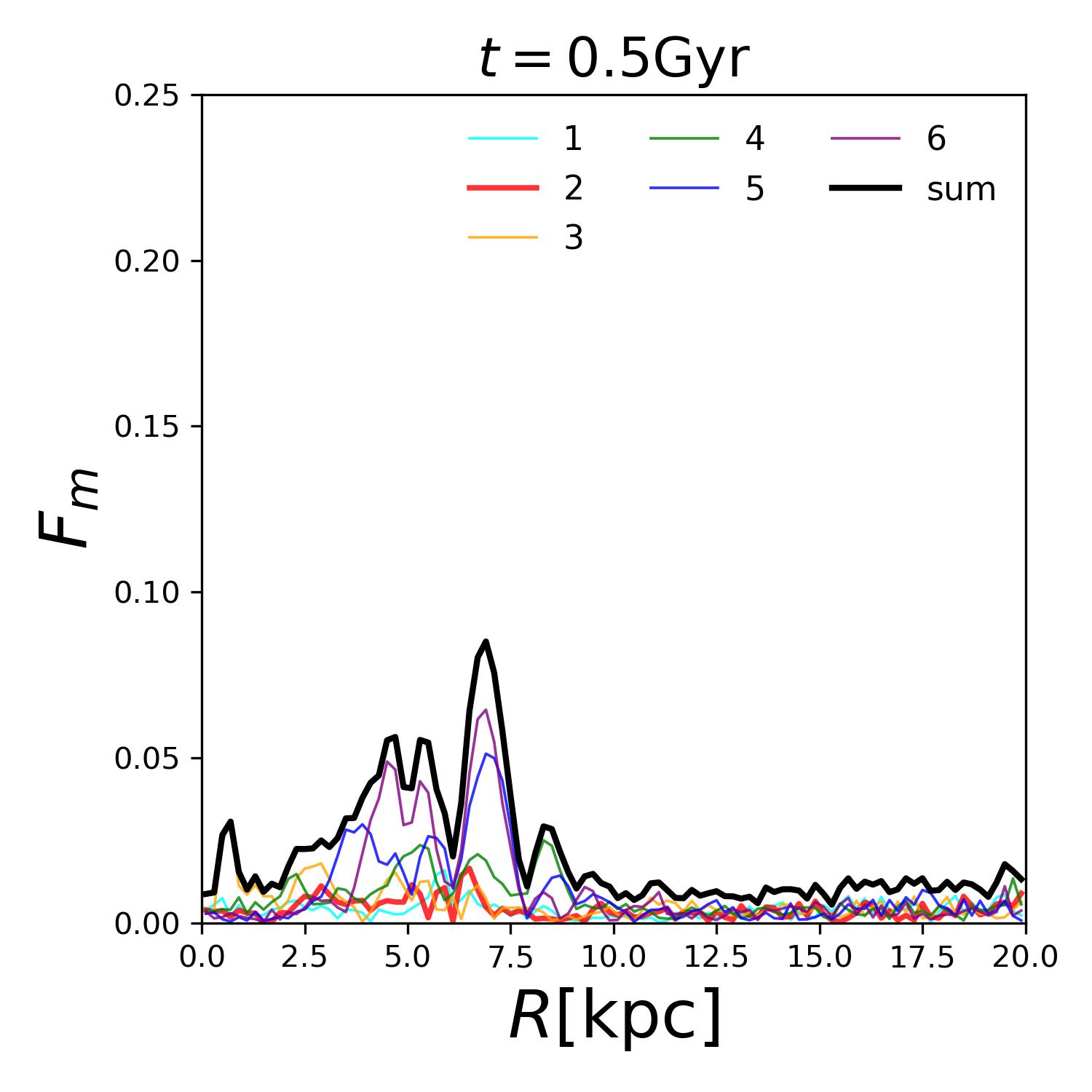} \\[-1ex]

        \includegraphics[width=0.25\textwidth]{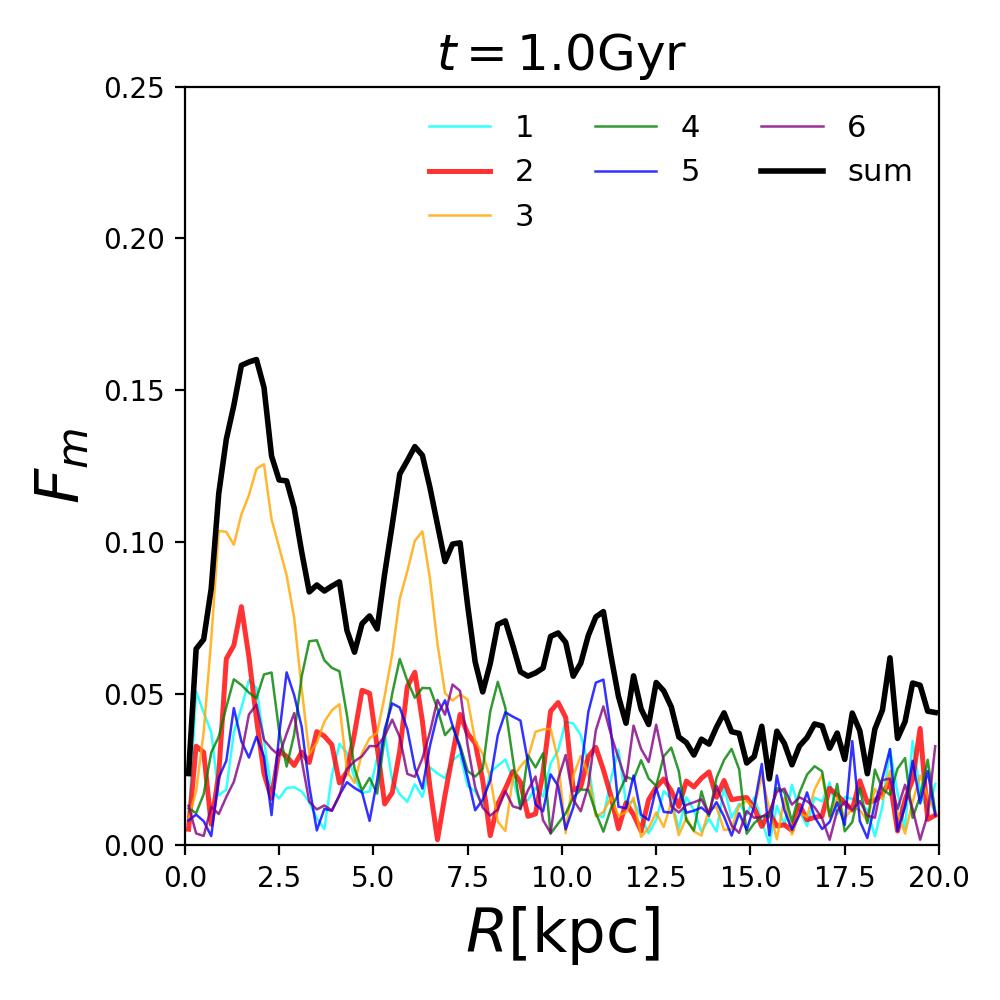} &
        \includegraphics[width=0.25\textwidth]{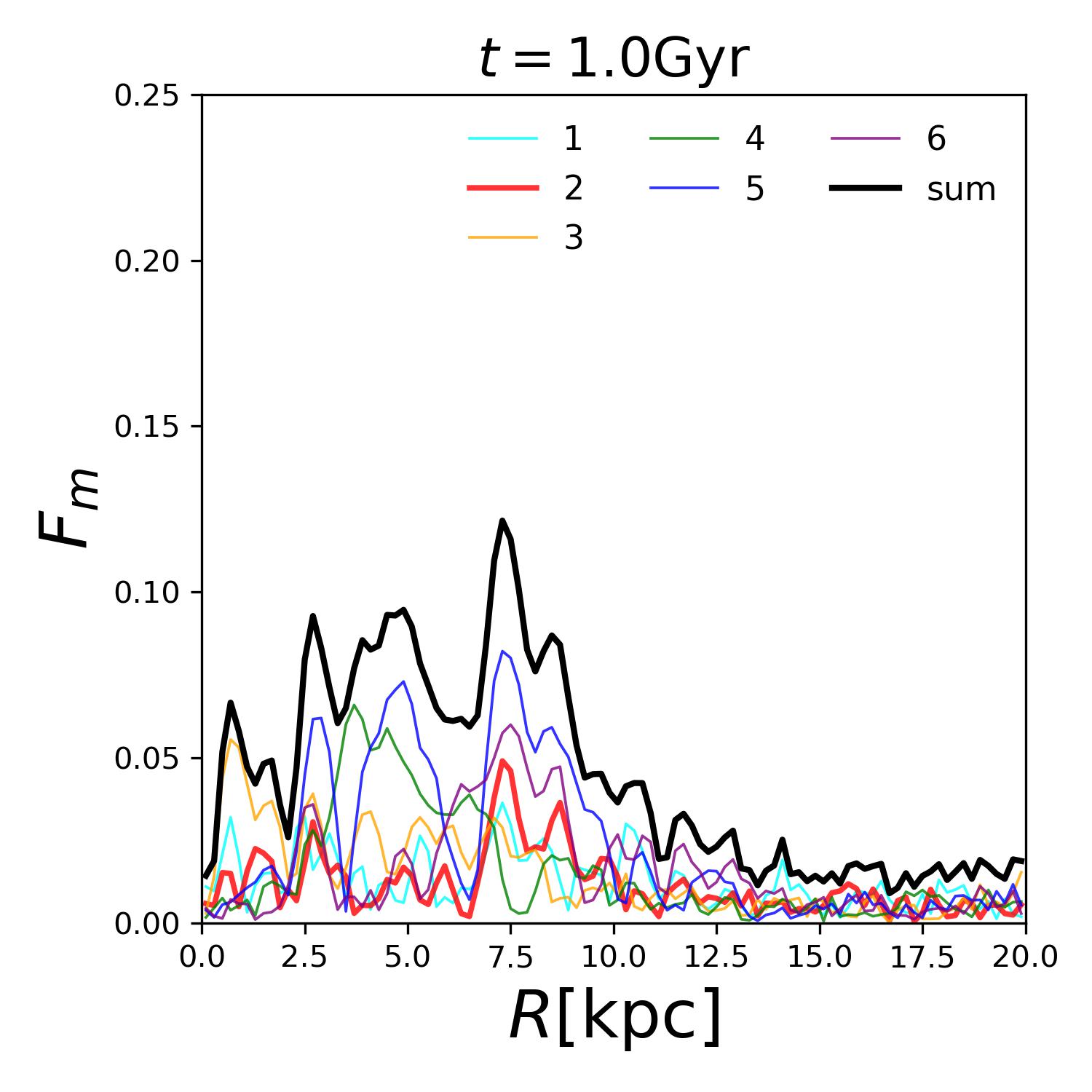} \\[-1ex]

        \includegraphics[width=0.25\textwidth]{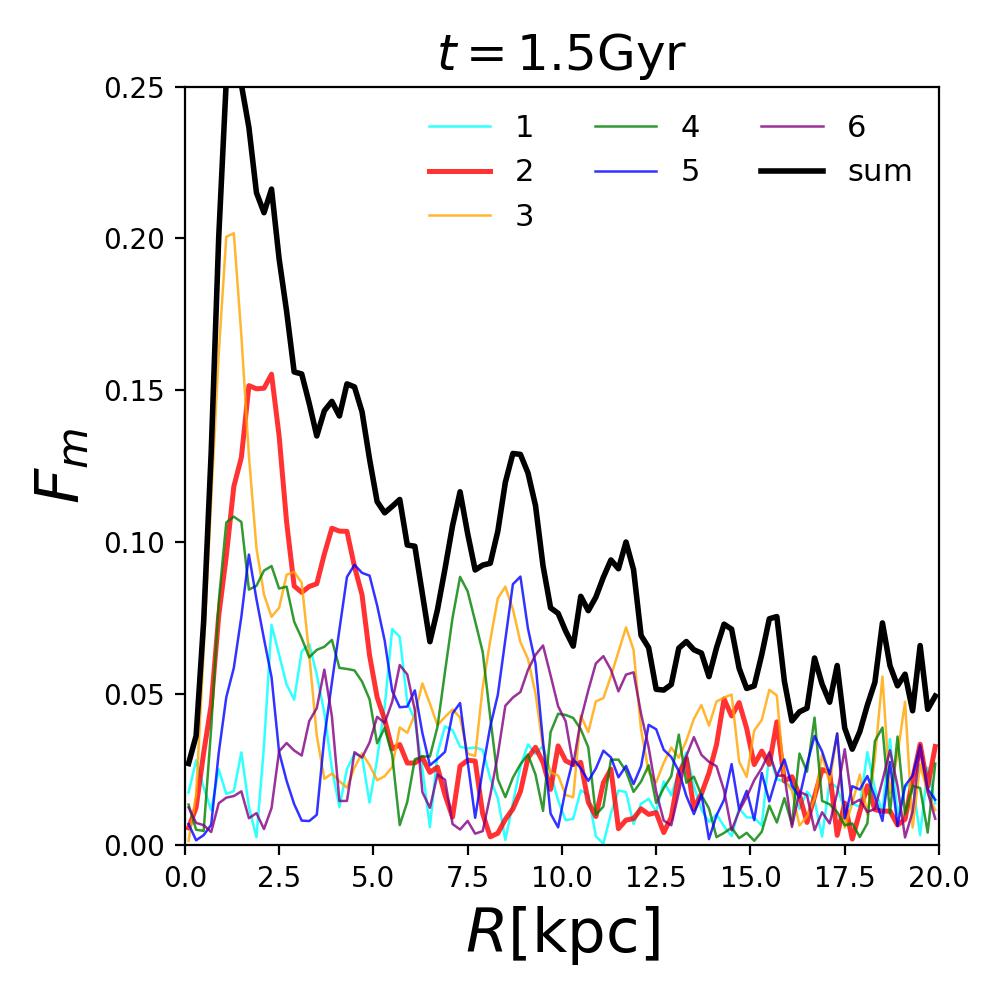} &
        \includegraphics[width=0.25\textwidth]{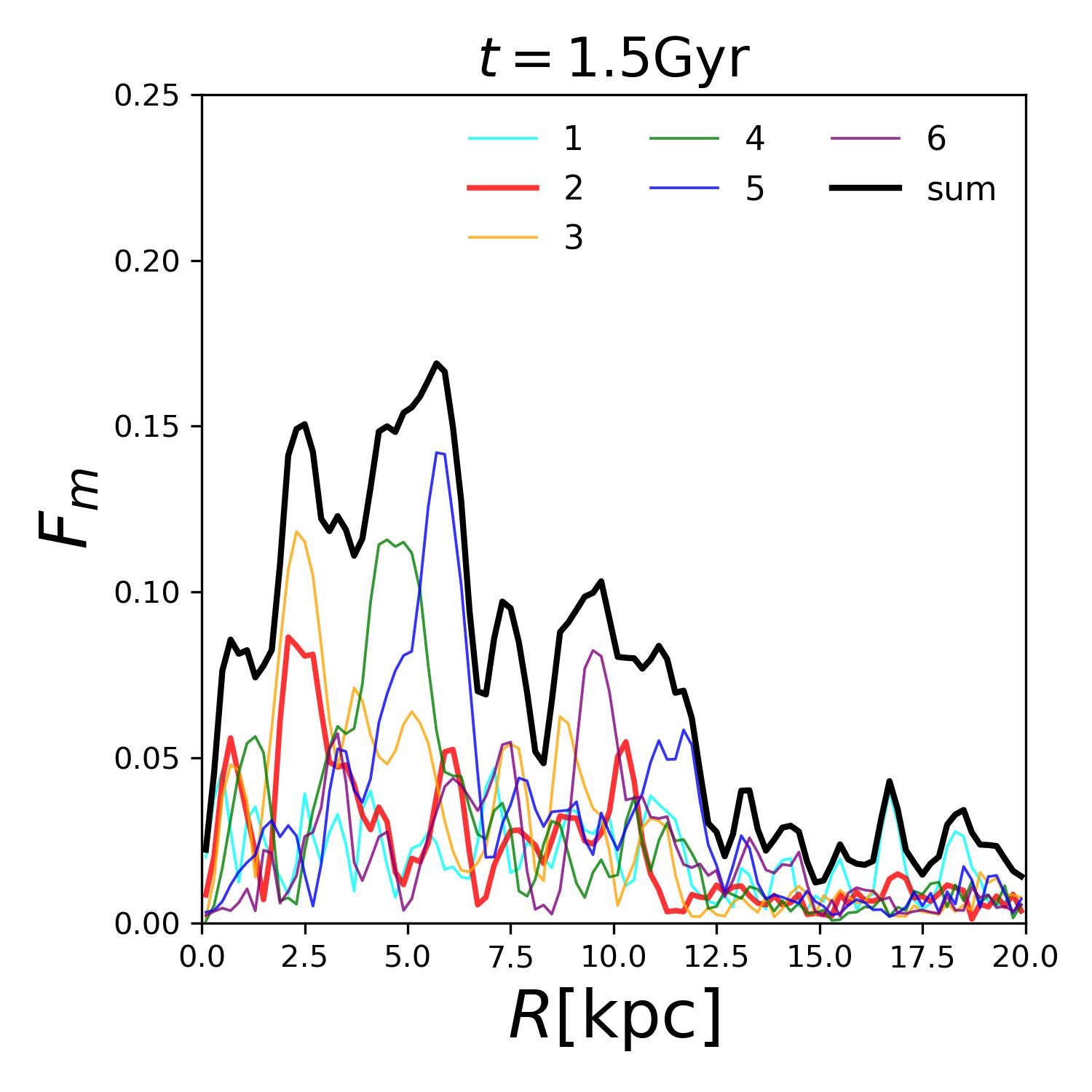} \\[-1ex]

        \includegraphics[width=0.25\textwidth]{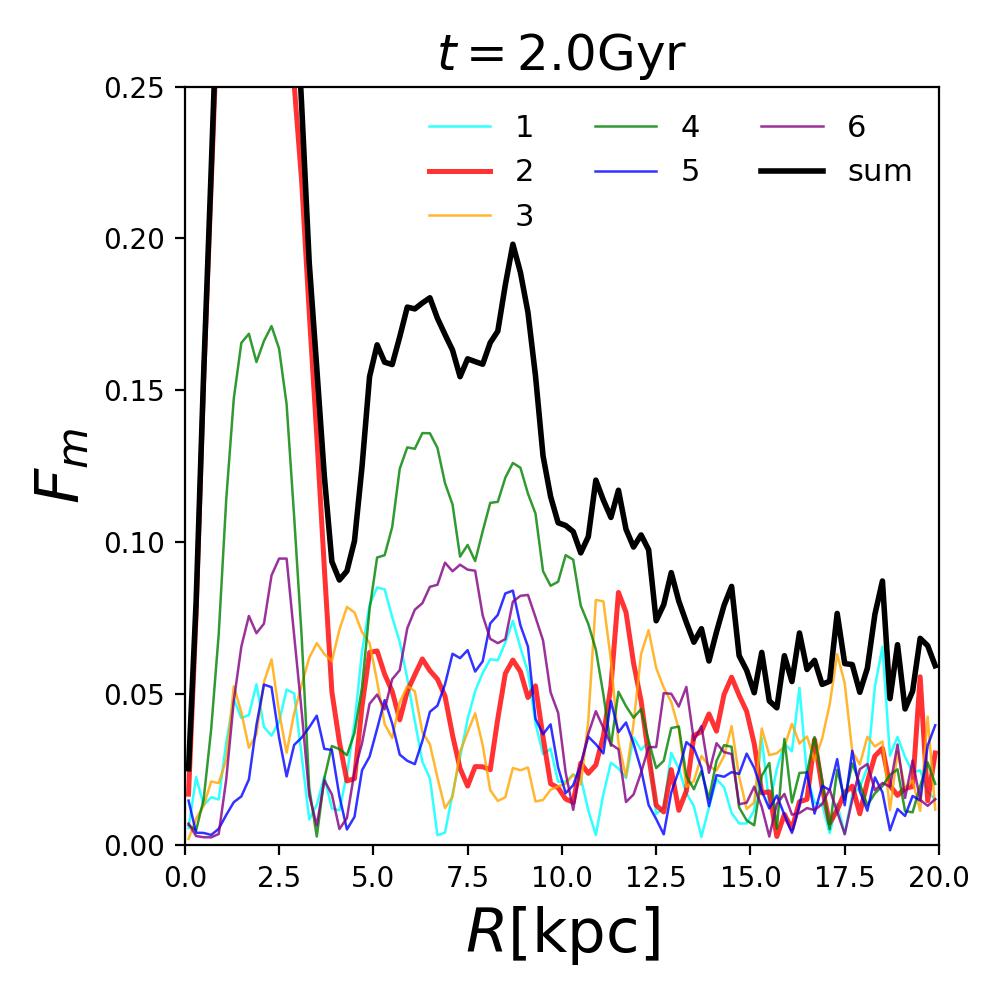} &
        \includegraphics[width=0.25\textwidth]{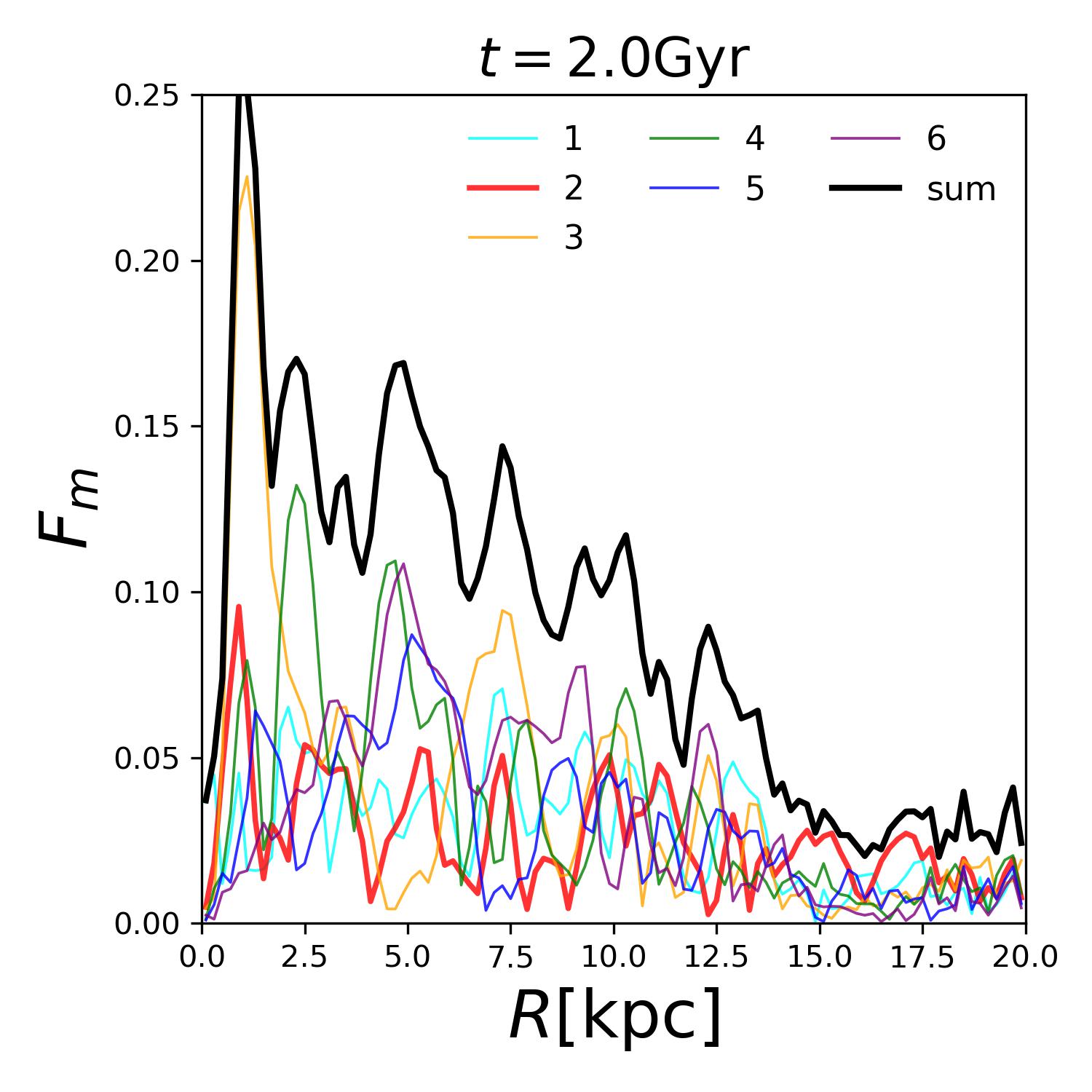} \\
    \end{tabular}

    \caption{Comparison of radial profiles of individual Fourier modes and $\Fsum$ in r2c16 (left column) and r1c16 (right column). $\Fsum$ is the root-mean-square (RMS) sum of Fourier modes from 1 to 6. Each row is obtained from a snapshot at 0.5, 1.0, 1.5, and 2.0 Gyr.}
    \label{fig:comparison_grid}
\end{figure}

\begin{figure}[htbp]
    \centering
    \includegraphics[width=0.5\textwidth]{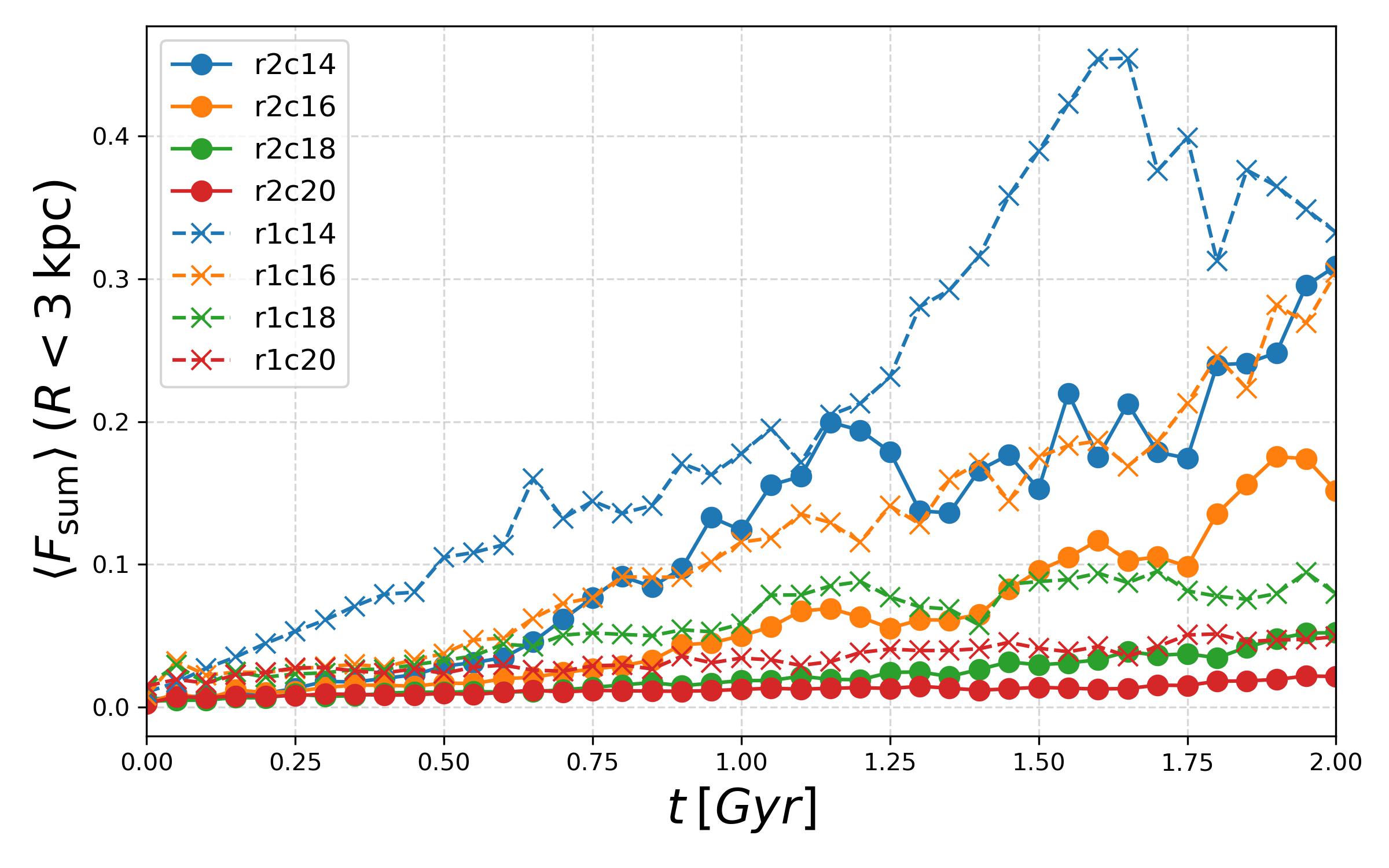}
    \includegraphics[width=0.5\textwidth]{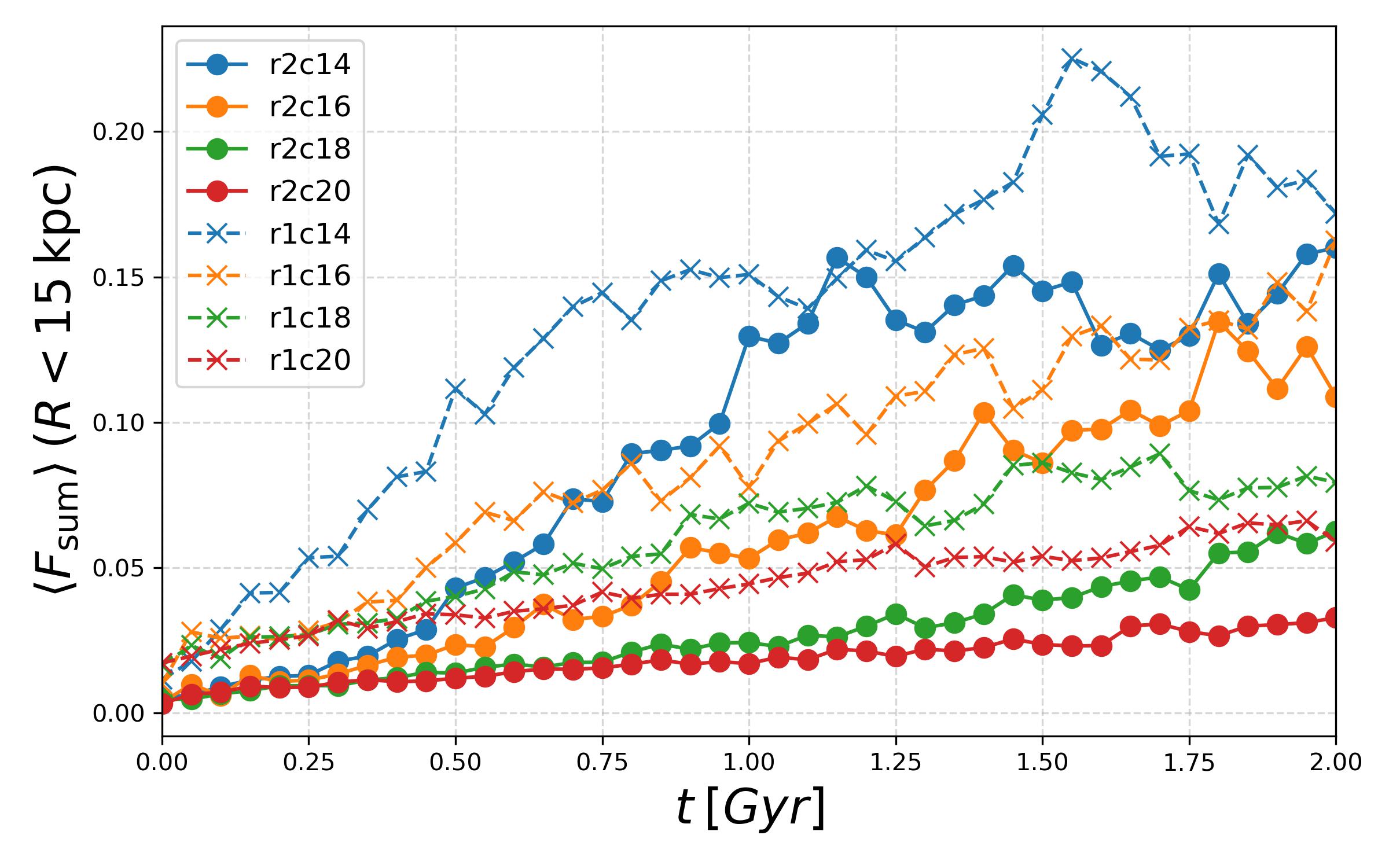}
    \caption{Temporal evolution of the radially averaged value of the RMS sum of Fourier modes, $\mFsum$, in the r1 and r2 models, as labeled, for resolution comparison. The top panel is $\mFsum$ inside $3\kpc$, and the bottom panel is the same within $15\kpc$.}
    \label{fig:fmean1}
\end{figure}

\begin{figure}[htbp]
    \centering
    \includegraphics[width=0.5\textwidth]{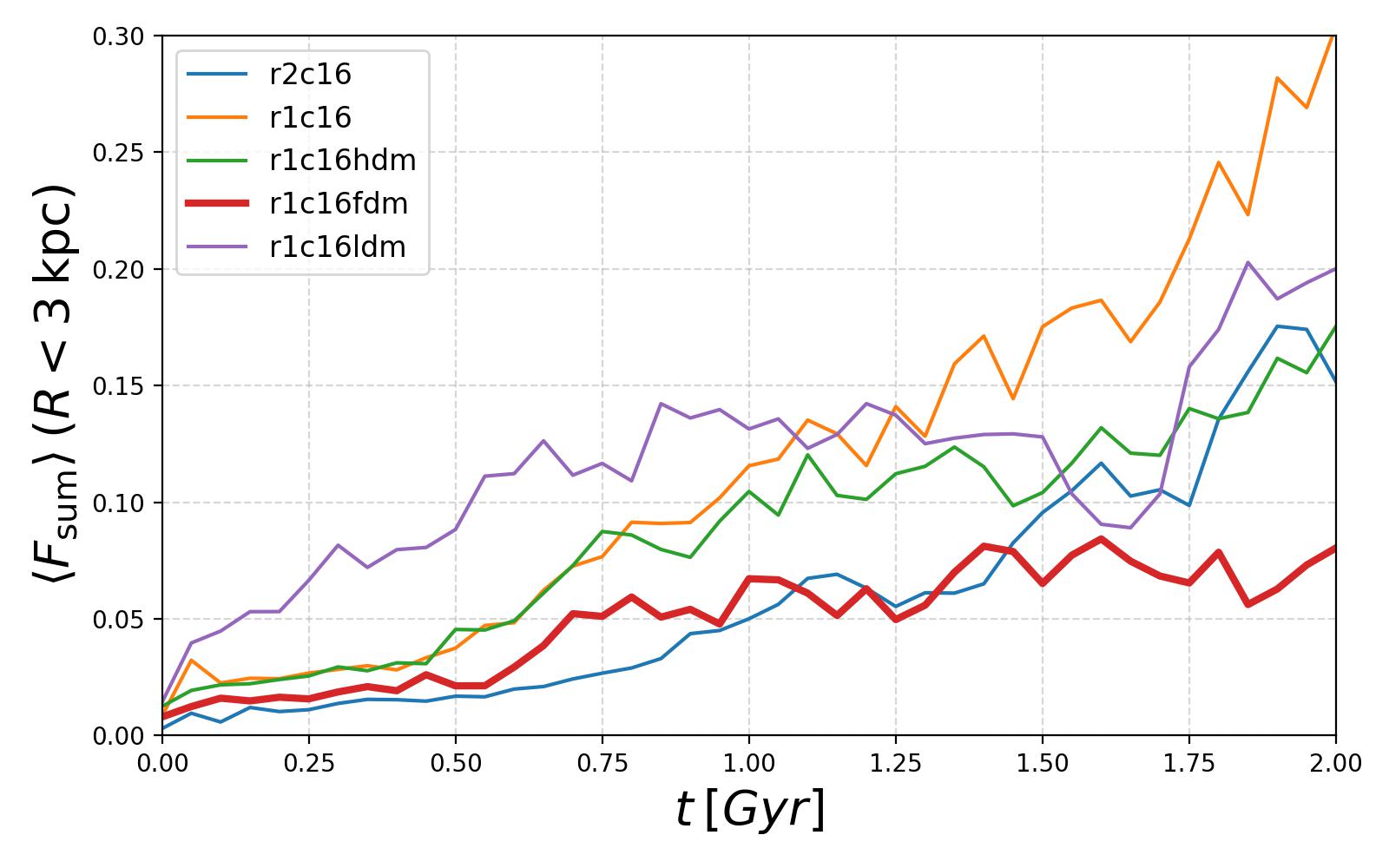}
    \includegraphics[width=0.5\textwidth]{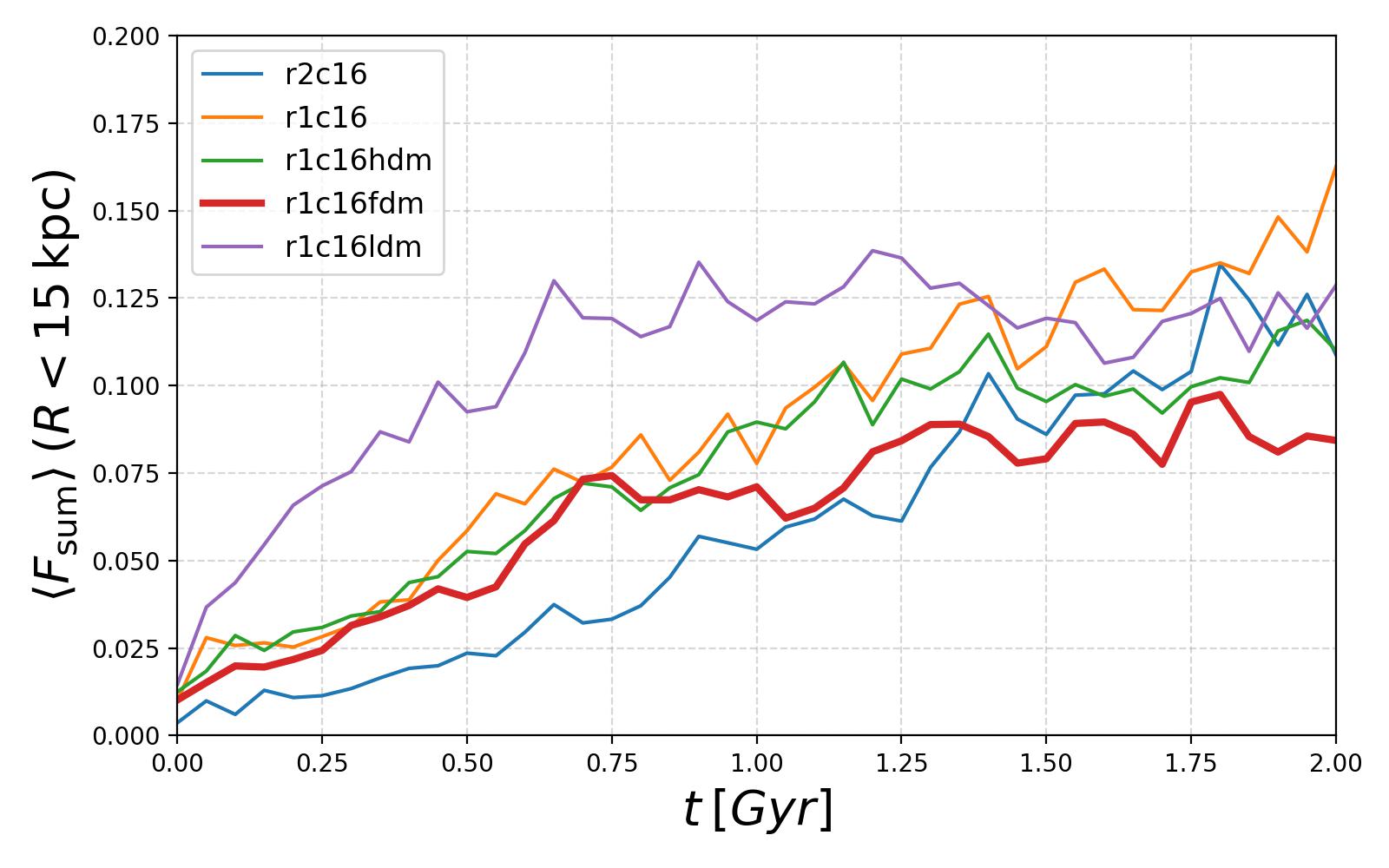}
    \caption{The same as Figure \ref{fig:fmean1}, but for the halo models.}
    \label{fig:fmean2}
\end{figure}

\begin{figure}[htbp]
    \centering
    \includegraphics[width=0.5\textwidth]{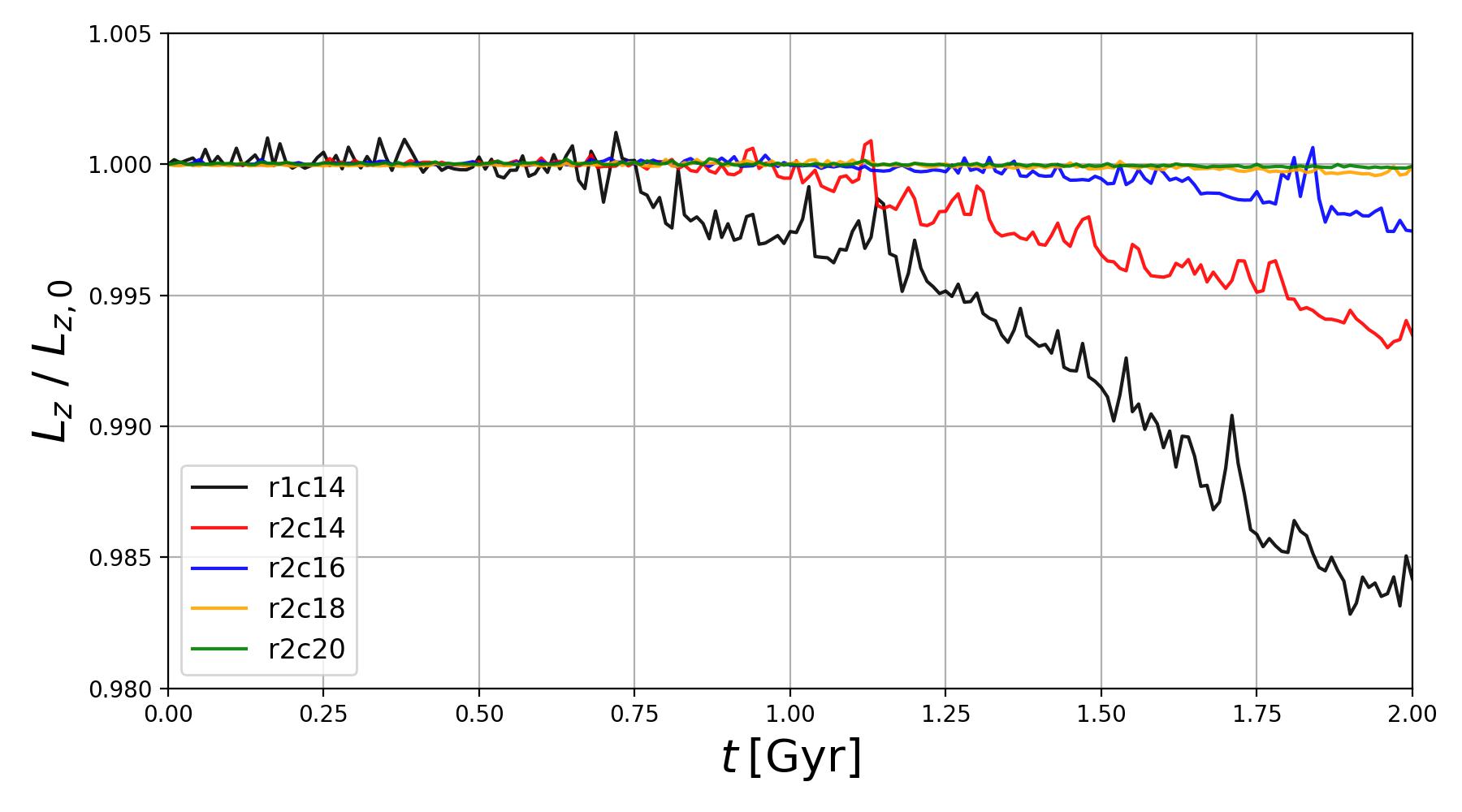}
    \includegraphics[width=0.5\textwidth]{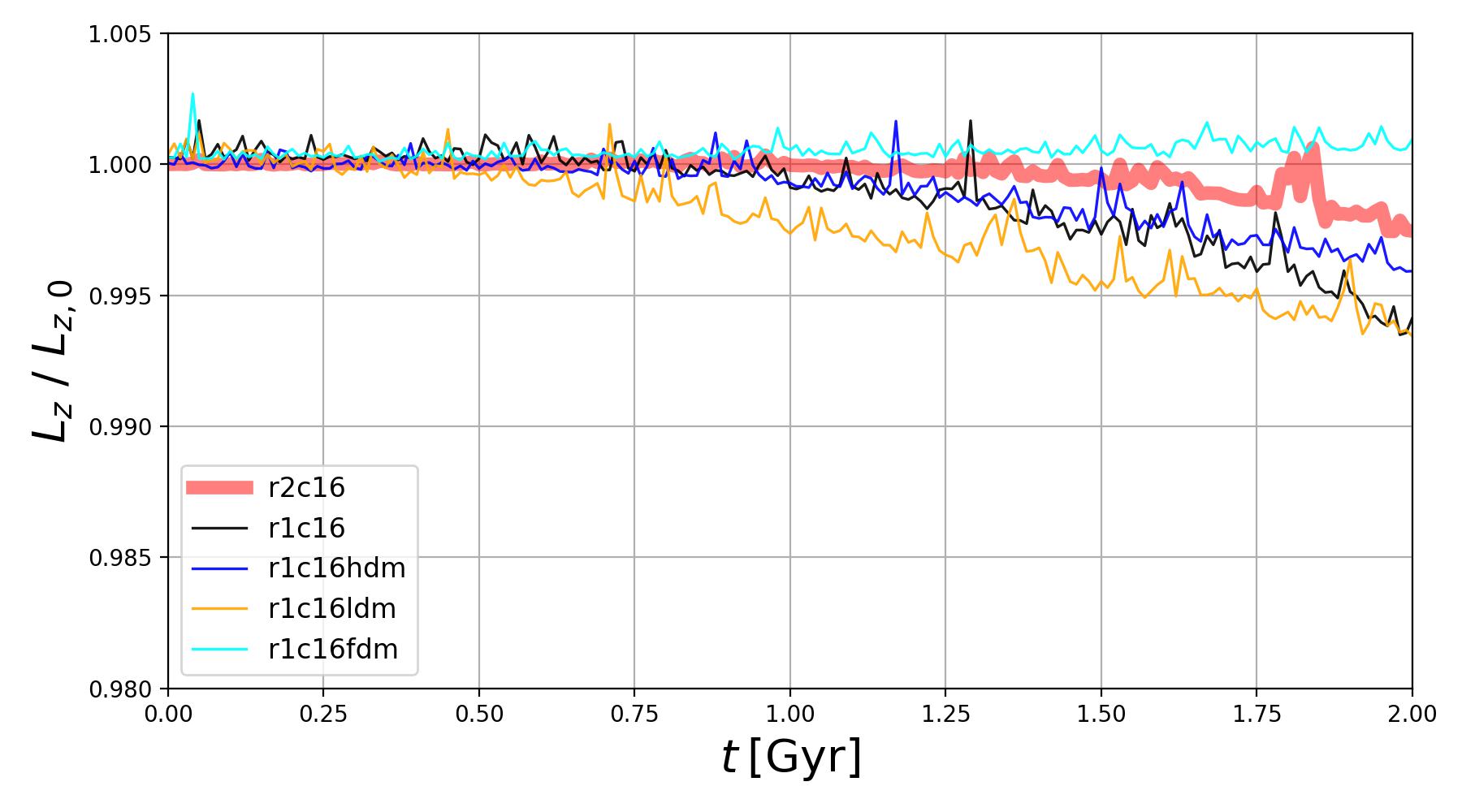}
    
    \caption{Fractional change of the total disk angular momentum divided by its initial angular momentum, $L_z/L_{z,0}$, over 2 Gyr. The top panel is for the r2 models with model r1c14 for reference, and the bottom panel is for the halo models with model r2c16 for reference, as they have the same halo concentration.}
    \label{fig:t_lz}
\end{figure}

\begin{figure*}[htbp]
    \centering
    \begin{tabular}{@{}cccc@{}}
        \textbf{r2c14} & \textbf{r2c16} & \textbf{r2c18} & \textbf{r2c20} \\

        \raisebox{0.1\height}{\rotatebox{90}
        {\textbf{
        $\Fsum \qquad\quad\quad F_6 \qquad\quad\quad\:\: F_5 \qquad\quad\quad\:\:\: F_4 \qquad\quad\quad\:\:\:\: F_3 \qquad\quad\quad\:\:\:\: F_2 \qquad\quad\quad\:\:\:\: F_1$}}}
        \includegraphics[width=0.24\textwidth]{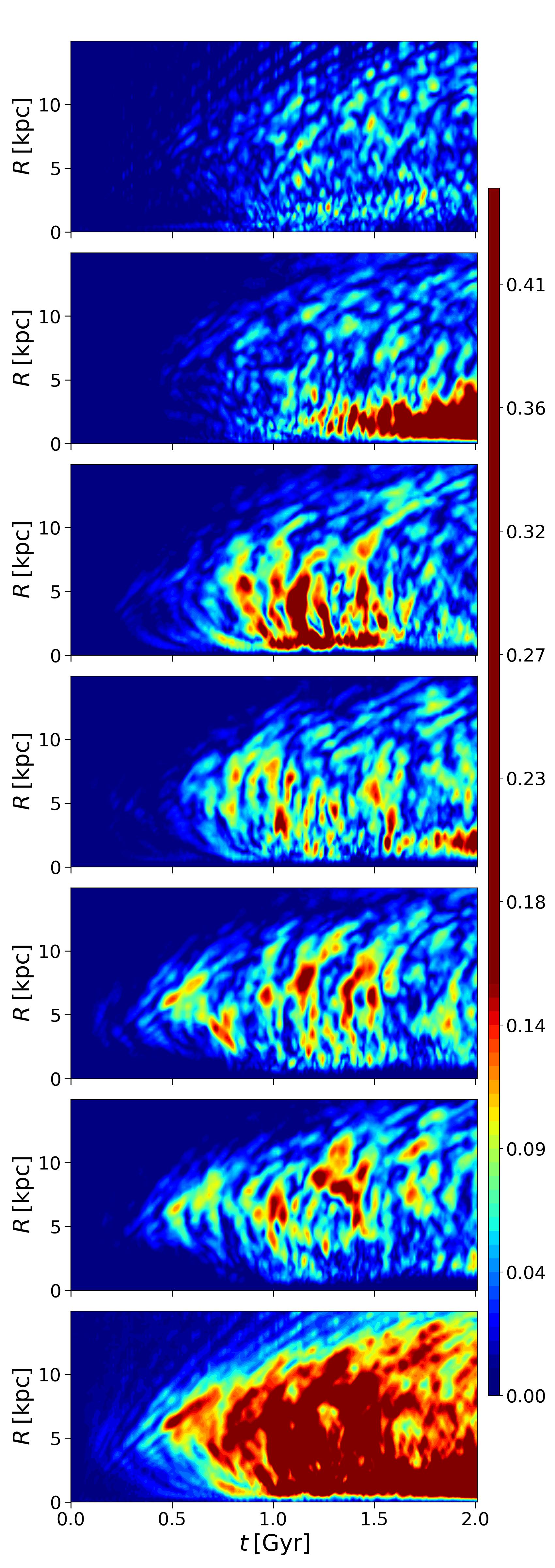} &
        \includegraphics[width=0.24\textwidth]{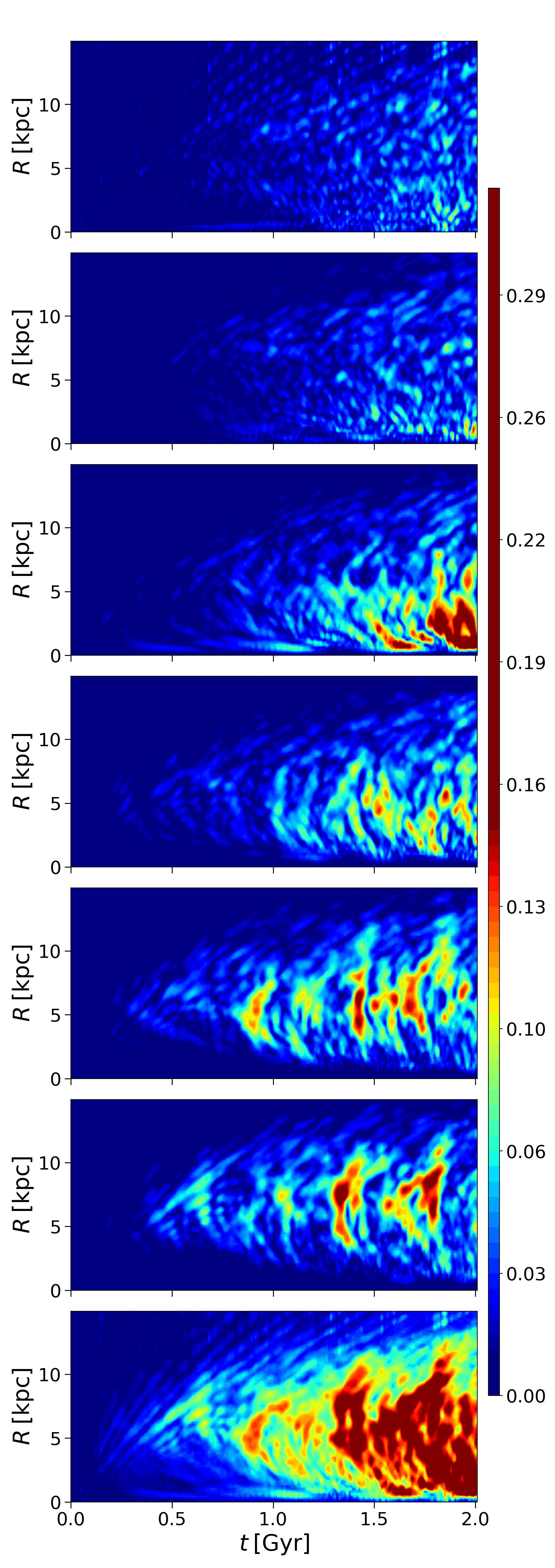} &
        \includegraphics[width=0.24\textwidth]{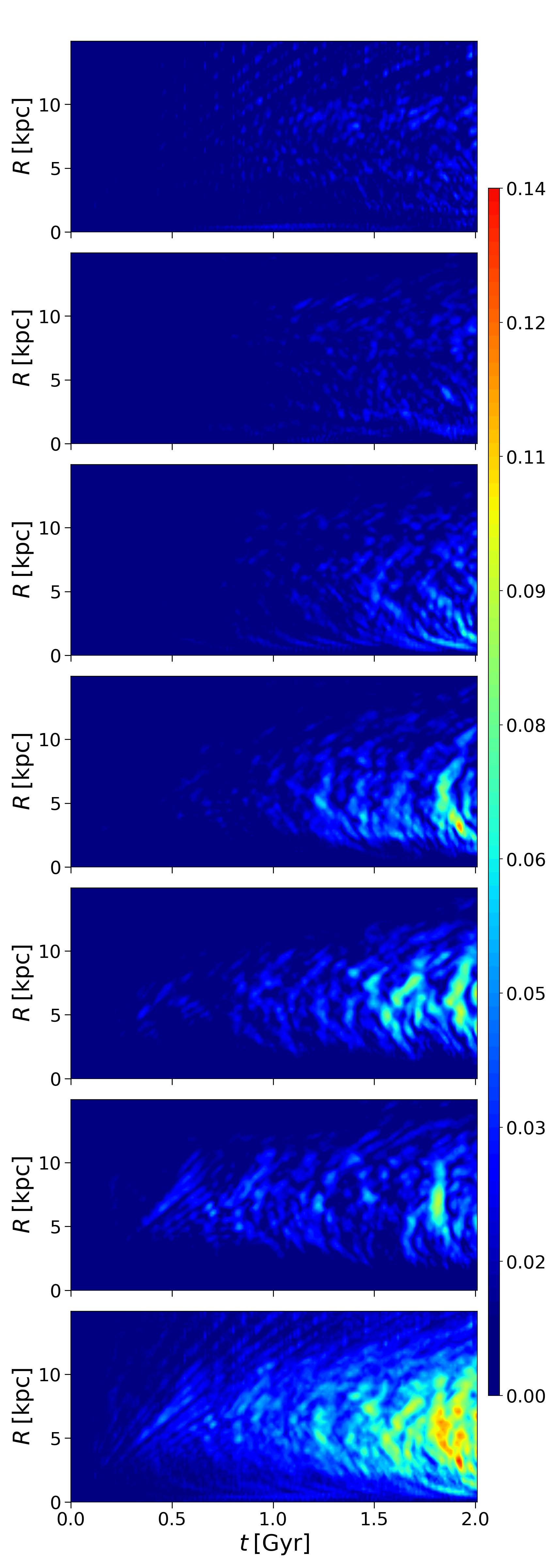} &
        \includegraphics[width=0.24\textwidth]{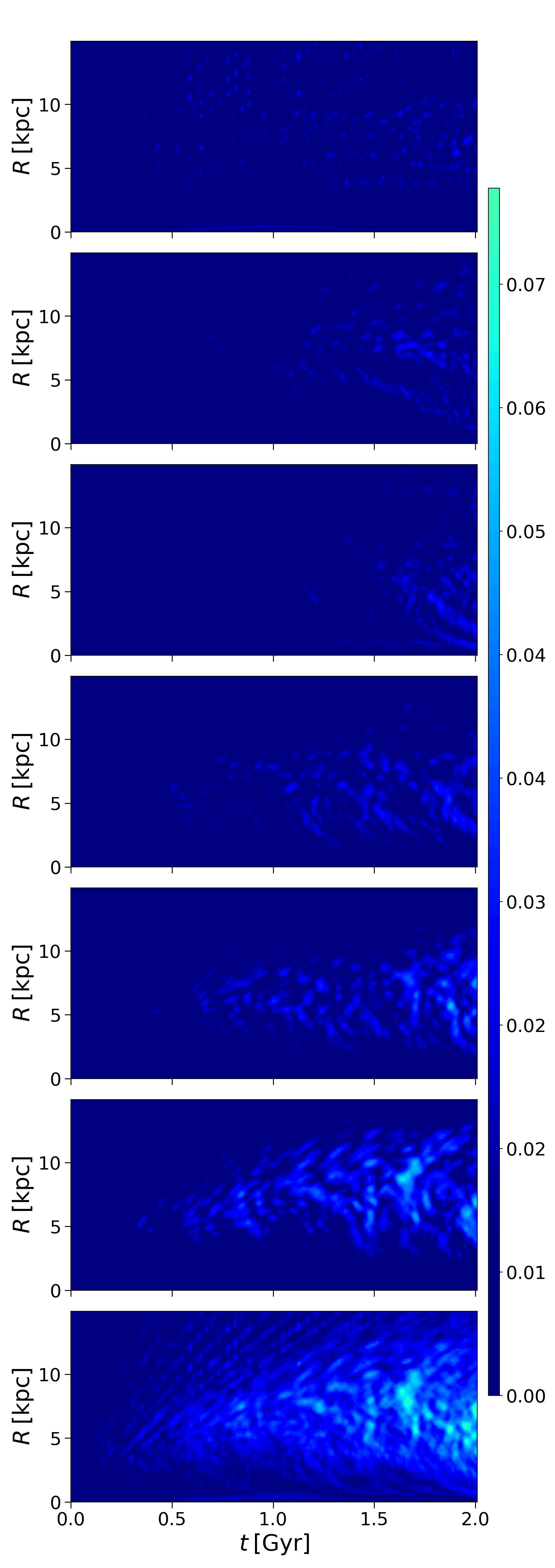}
    \end{tabular}
    \caption{Fourier contour map calculated from the time evolution of radial Fourier profiles of each mode and $\Fsum$ in model r2c14, r2c16, r2c18, and r2c20. The time interval of each snapshot used is 0.01 Gyr. The color scale of each color bar is fixed from $1\%$ to $15\%$, so that the same color indicates the same amplitude. The numbers beside each color bar indicate the full range of the data in that panel; the color mapping itself is fixed from $1\%$ to $15\%$ in all panels and saturates outside this range.}
    \label{fig:fmap1}
\end{figure*}

\begin{figure*}[htbp]
    \centering
    \begin{tabular}{@{}cccc@{}}
        \textbf{r1c16} & \textbf{r1c16hdm} & \textbf{r1c16fdm} & \textbf{r1c16ldm} \\
        
        \raisebox{0.1\height}{\rotatebox{90}
        {\textbf{
        $\Fsum \qquad\quad\quad F_6 \qquad\quad\quad\:\: F_5 \qquad\quad\quad\:\:\: F_4 \qquad\quad\quad\:\:\:\: F_3 \qquad\quad\quad\:\:\:\: F_2 \qquad\quad\quad\:\:\:\: F_1$}}}
        \includegraphics[width=0.24\textwidth]{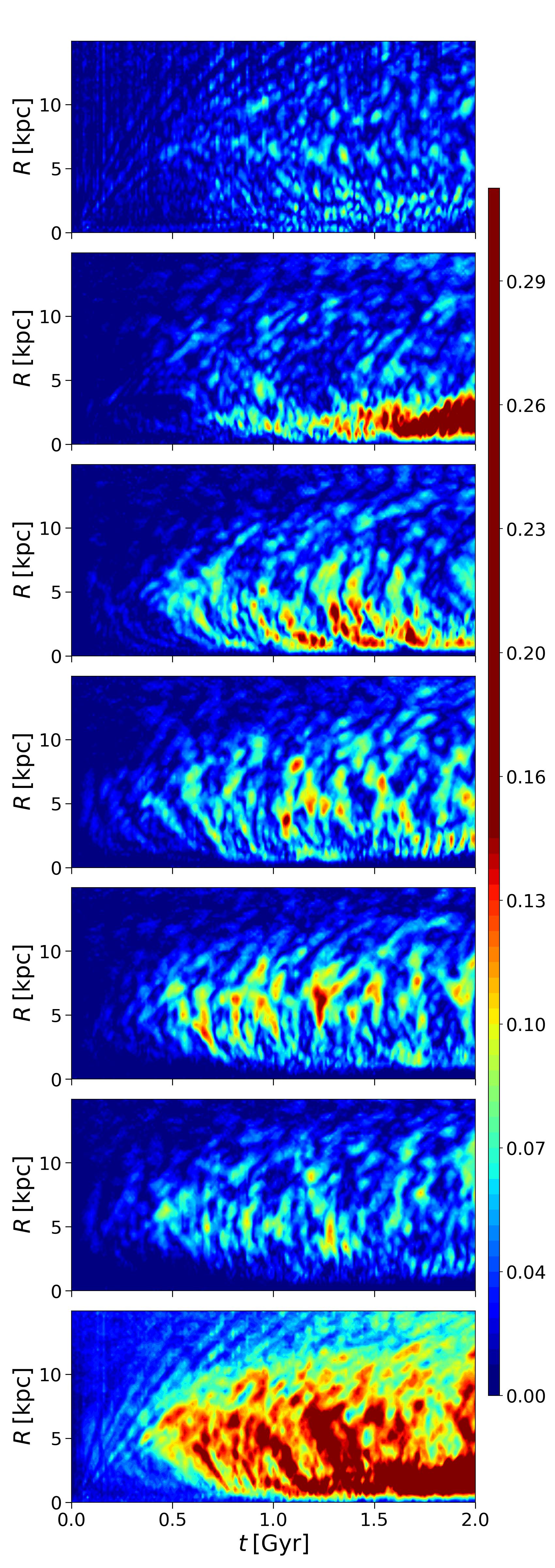} &
        \includegraphics[width=0.24\textwidth]{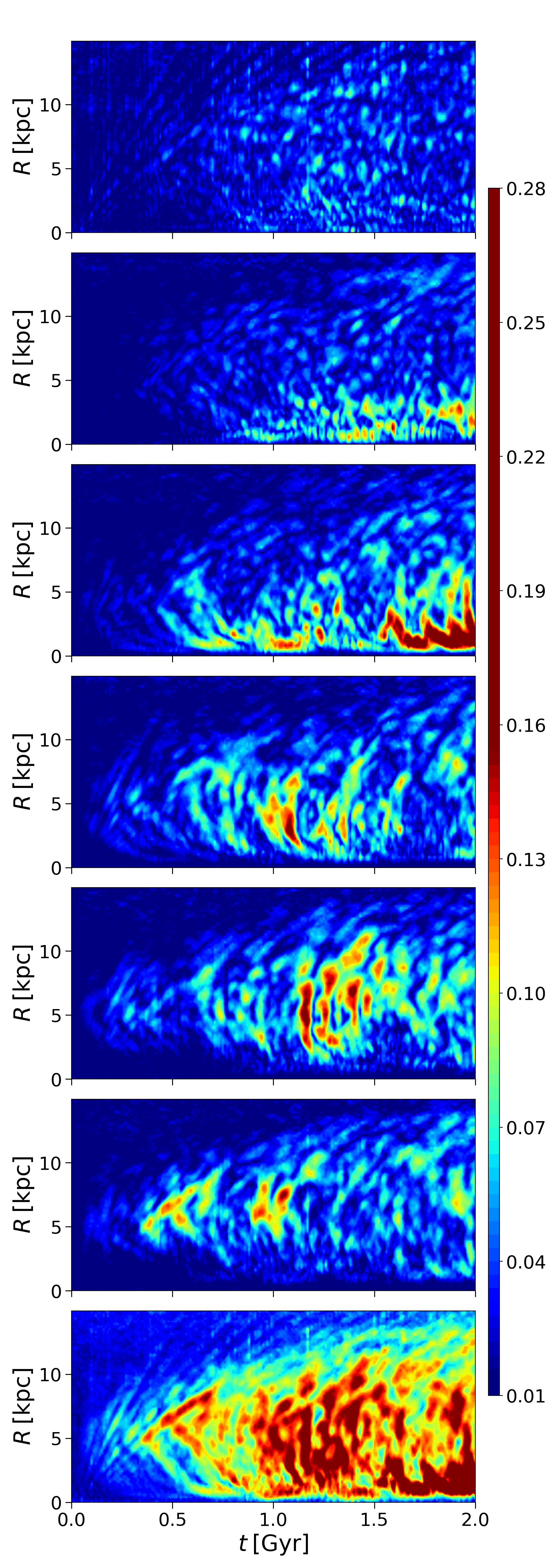} &
        \includegraphics[width=0.24\textwidth]{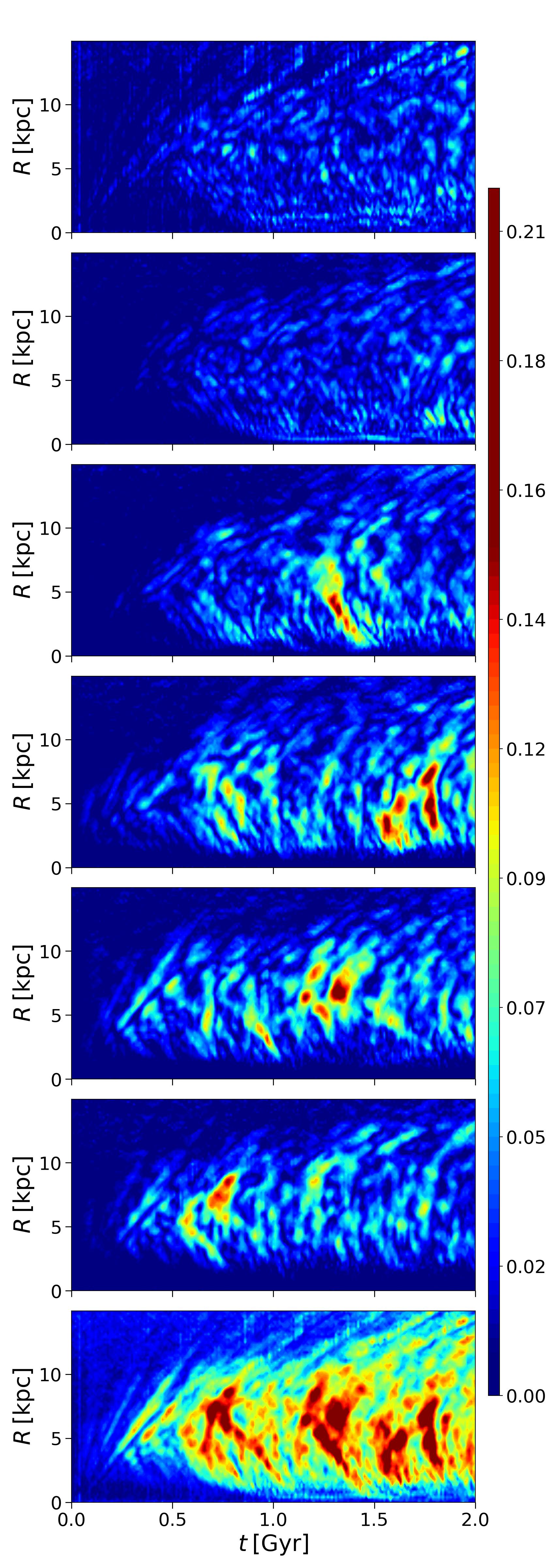} &
        \includegraphics[width=0.24\textwidth]{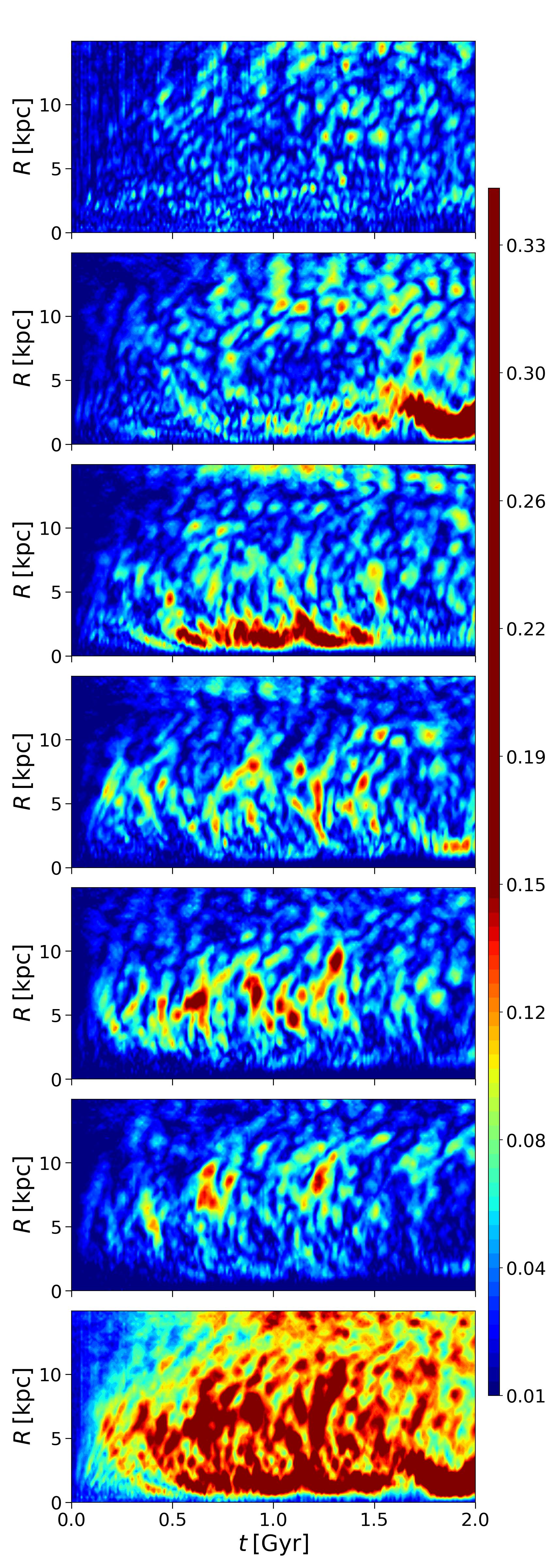}
    \end{tabular}
    \caption{The same as Figure \ref{fig:fmap1}, but for the halo models, in which the disk-halo systems have the same density structures.}
    \label{fig:fmap2}
\end{figure*}

\begin{figure}[htbp]
    \centering
    \renewcommand{\arraystretch}{0}
    \begin{tabular}{@{}c@{}c@{}}
        \raisebox{3\height}{\rotatebox{90}{\textbf{r2c14}}} &
        \includegraphics[width=0.50\textwidth]{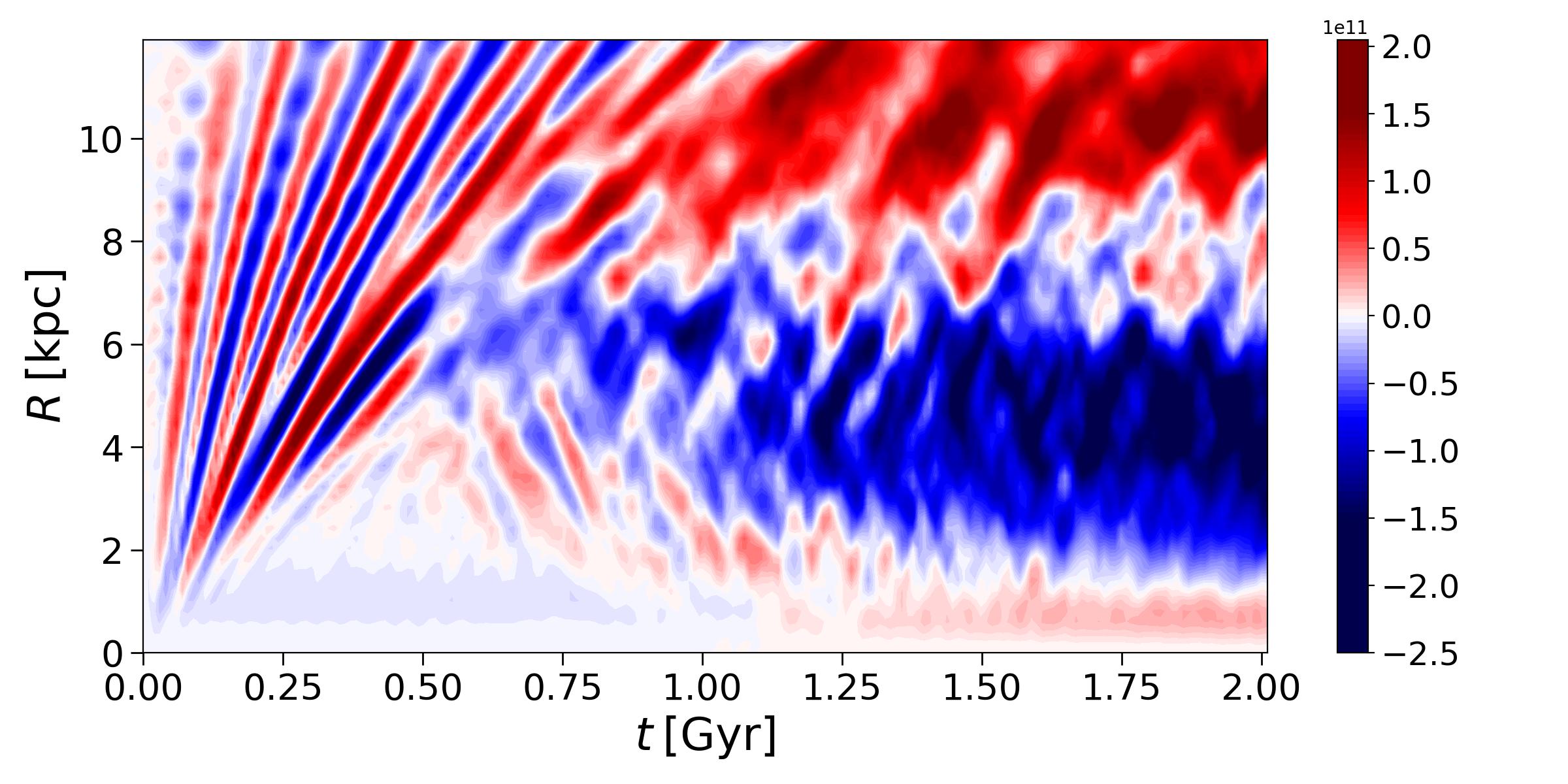} \\[1ex]

        \raisebox{3\height}{\rotatebox{90}{\textbf{r2c16}}} &
        \includegraphics[width=0.50\textwidth]{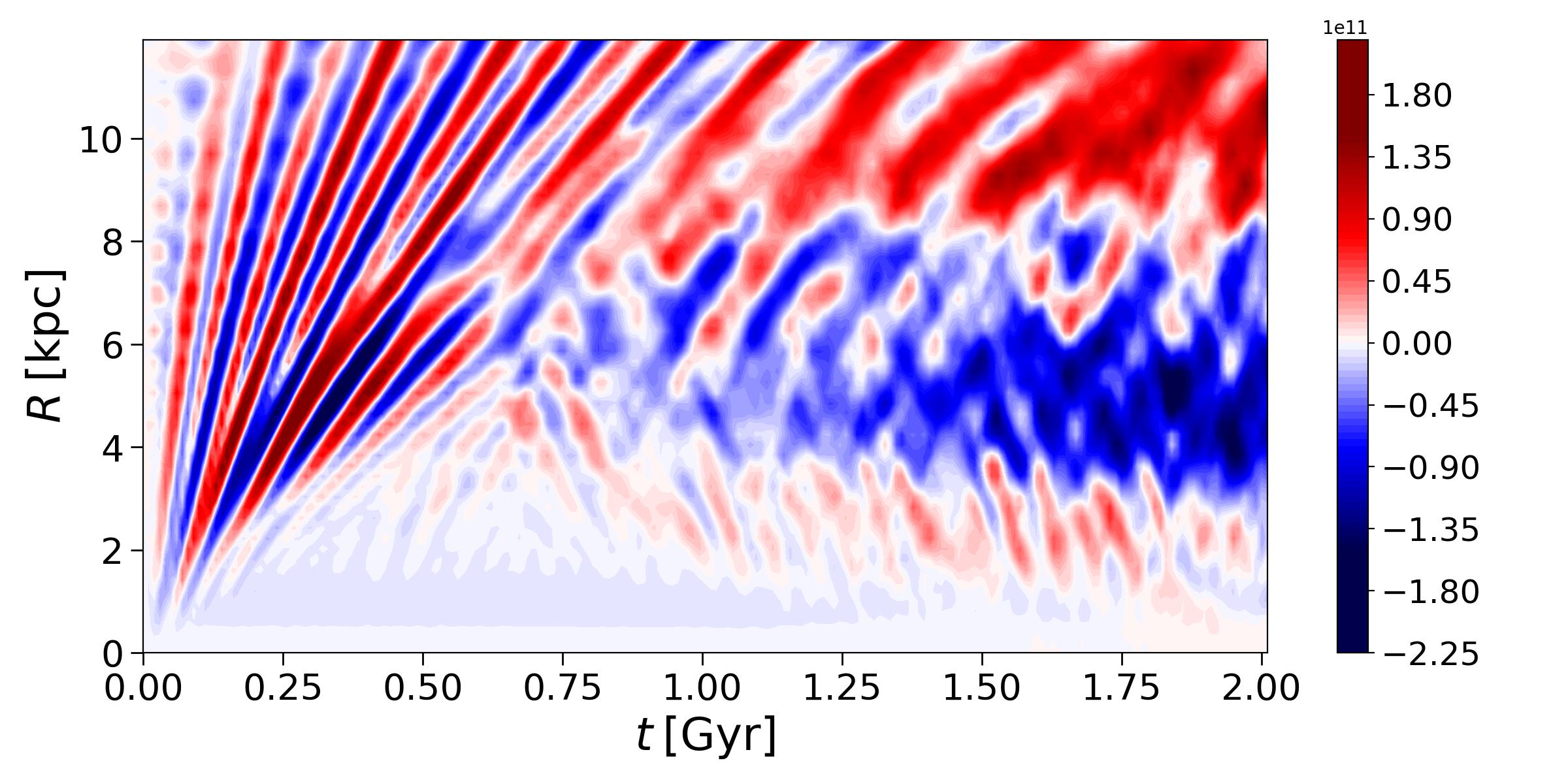} \\[1ex]

        \raisebox{3\height}{\rotatebox{90}{\textbf{r2c18}}} &
        \includegraphics[width=0.50\textwidth]{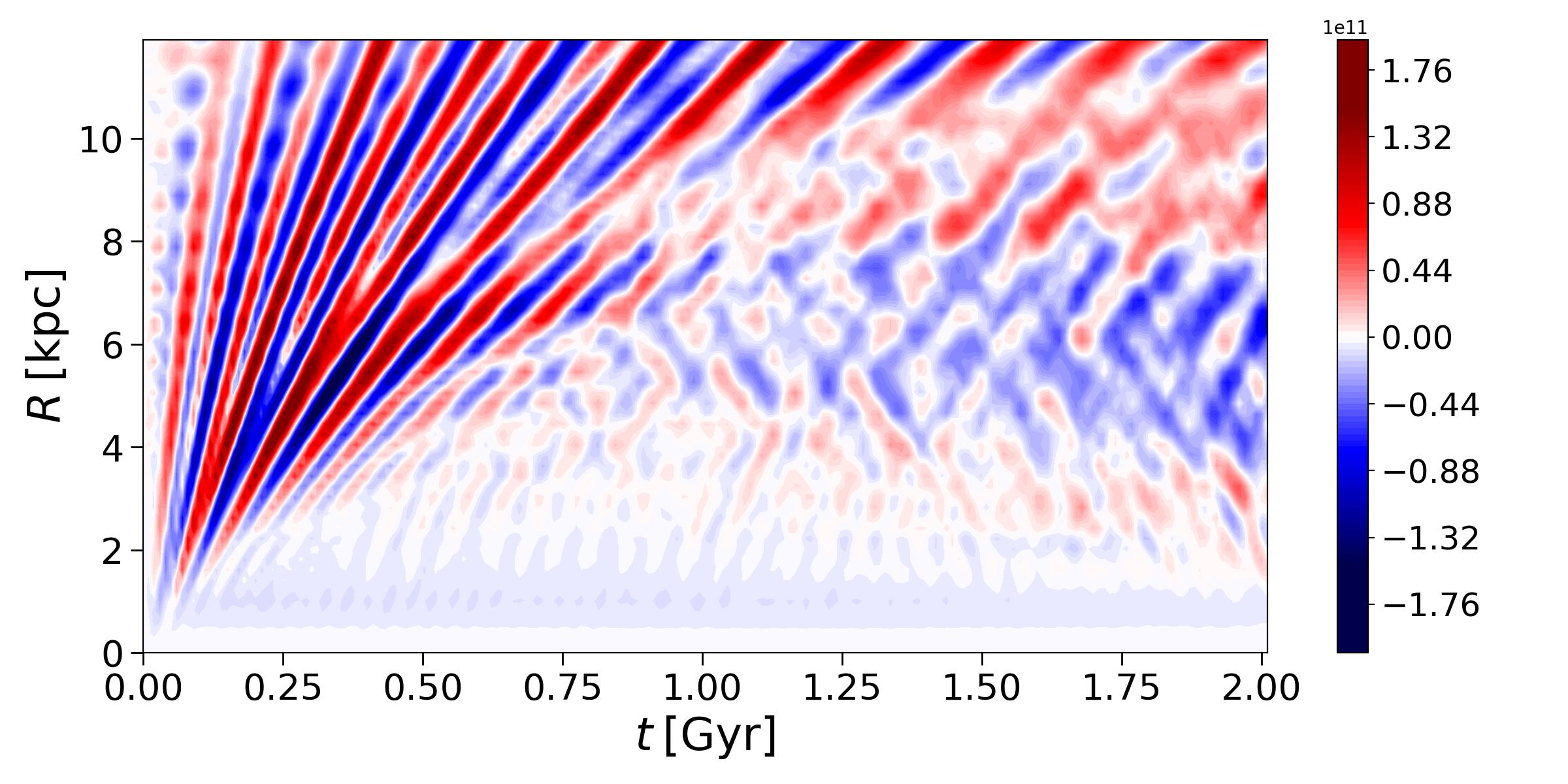} \\[1ex]

        \raisebox{3\height}{\rotatebox{90}{\textbf{r2c20}}} &
        \includegraphics[width=0.50\textwidth]{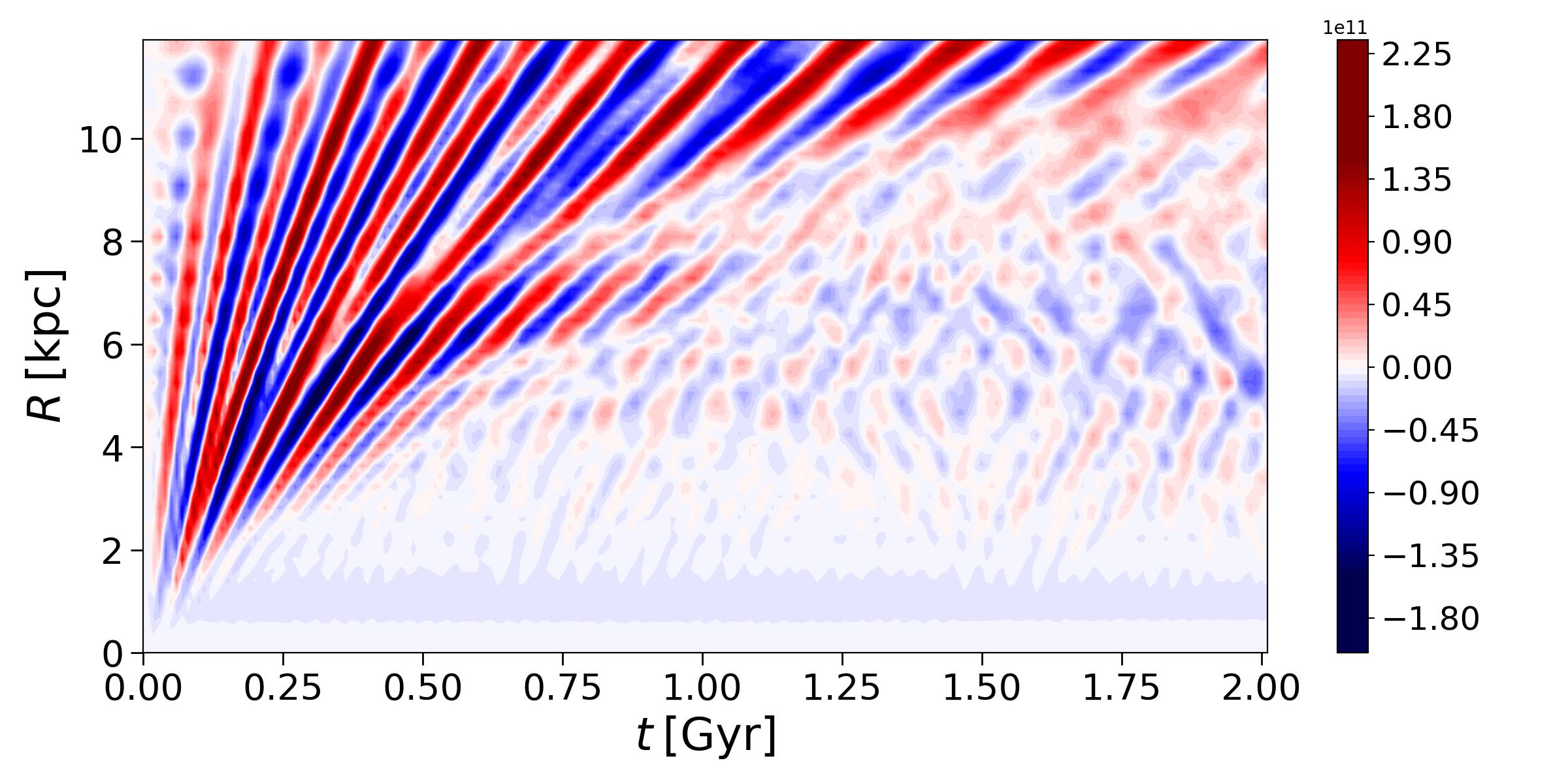} \\
    \end{tabular}
    \caption{Angular momentum map calculated from the time evolution of the radial distribution of disk angular momentum subtracted from its initial value, $L_z-L_{z,0}$, in the r2 model, as labeled. The time interval of each snapshot used is 0.01 Gyr. The range of the color bar is fixed within $\pm1.5\times10^{11} \Msun \kpc \ \rm{km/s}$, so the same color indicates the same $\Delta L_z$. Values outside this range are saturated.}
    \label{fig:lzmap1}
\end{figure}

\begin{figure}[htbp]
    \centering
    \renewcommand{\arraystretch}{0}
    \begin{tabular}{@{}c@{}c@{}}
        \raisebox{3\height}{\rotatebox{90}{\textbf{r1c16}}} &
        \includegraphics[width=0.50\textwidth]{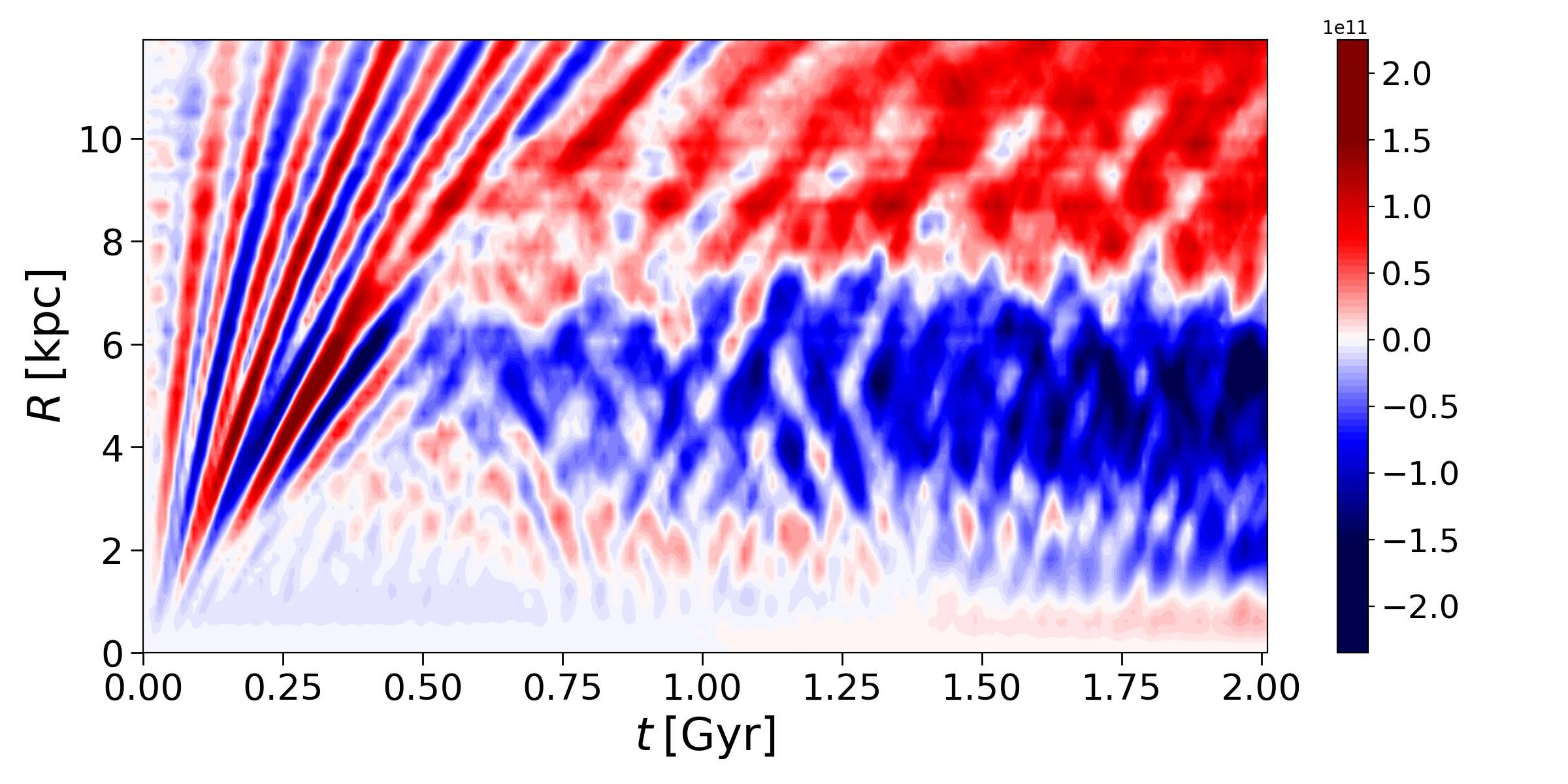} \\[1ex]

        \raisebox{1.5\height}{\rotatebox{90}{\textbf{r1c16hdm}}} &
        \includegraphics[width=0.50\textwidth]{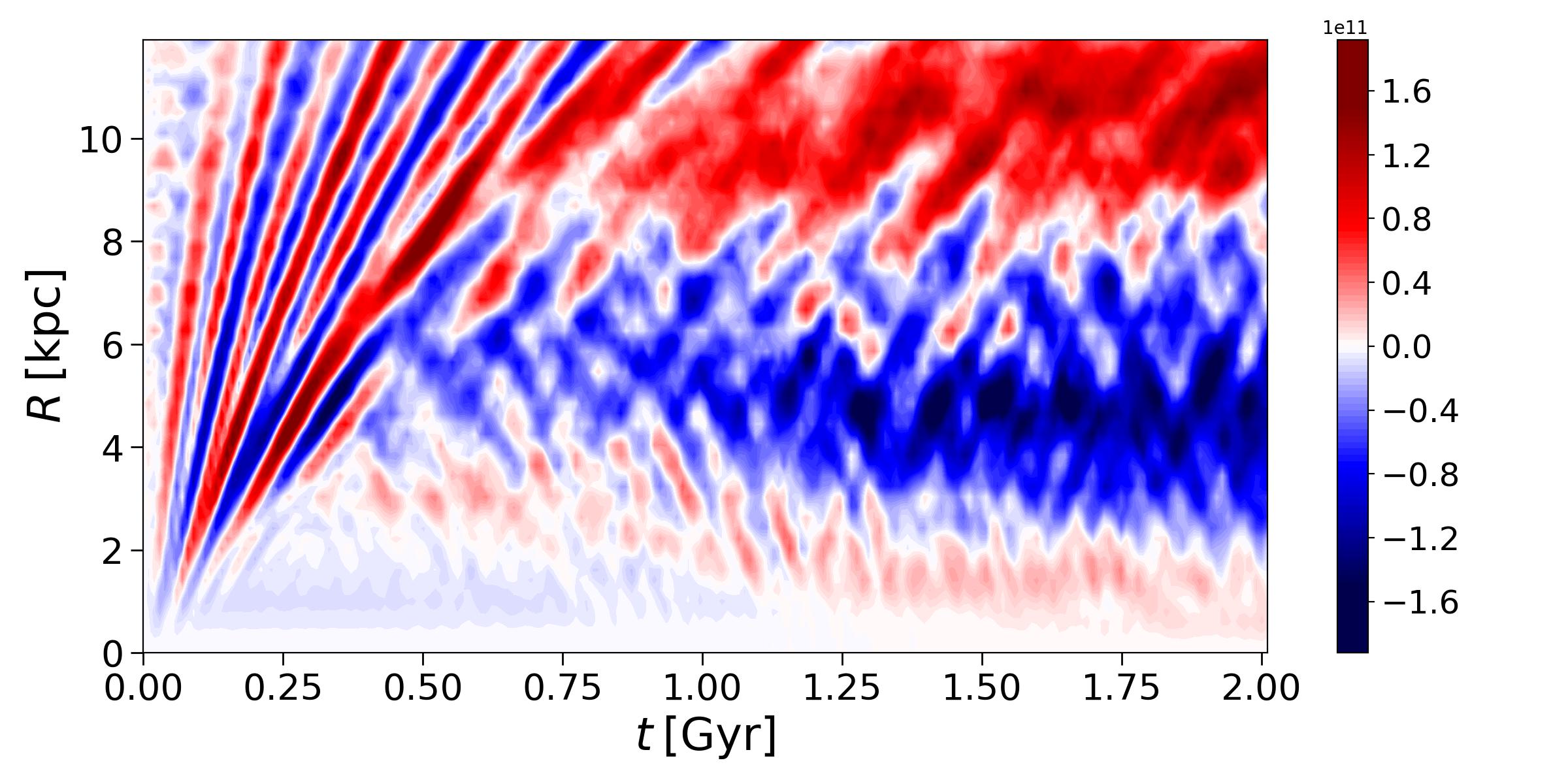} \\[1ex]

        \raisebox{1.5\height}{\rotatebox{90}{\textbf{r1c16fdm}}} &
        \includegraphics[width=0.50\textwidth]{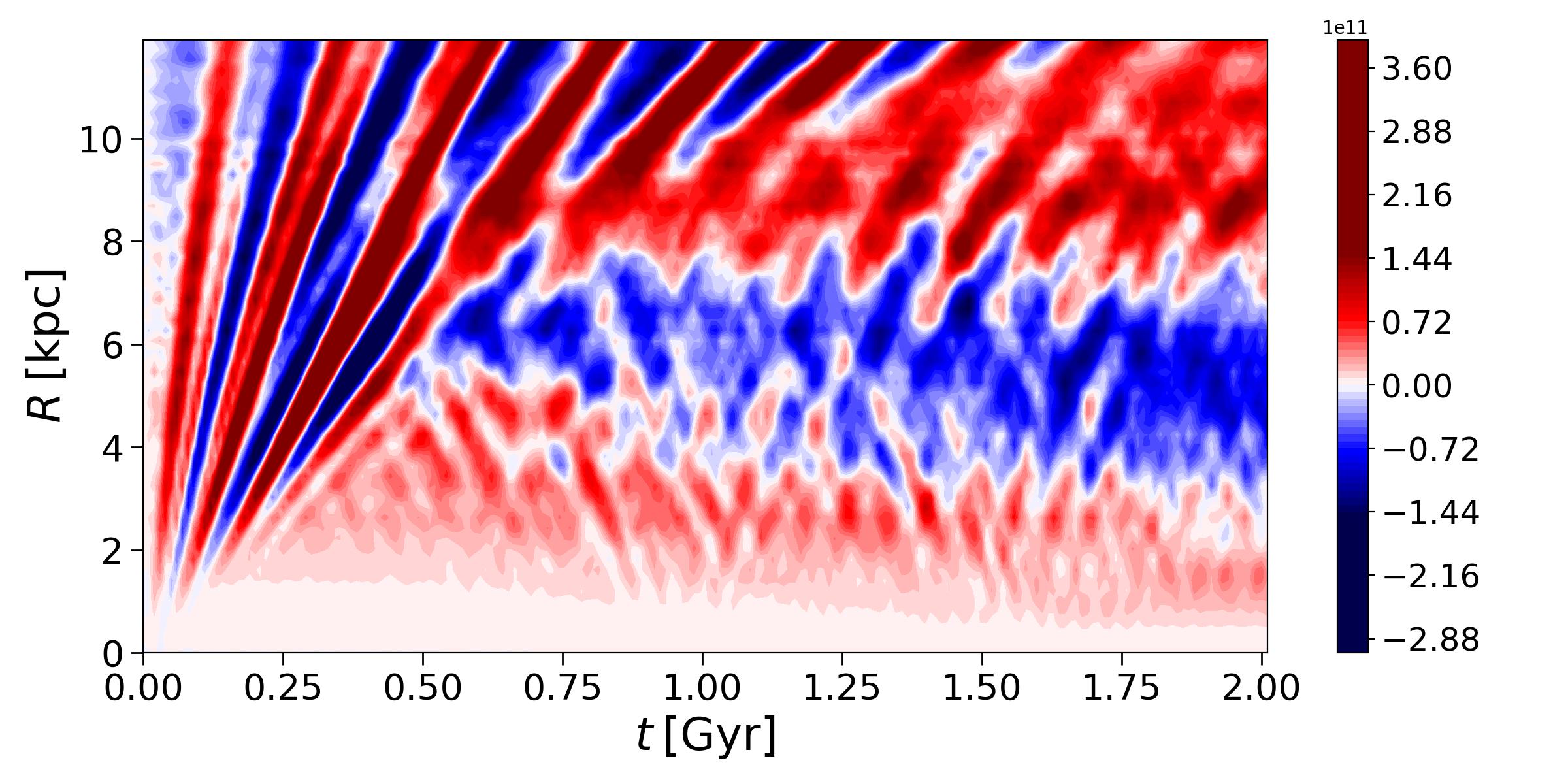} \\[1ex]

        \raisebox{1.5\height}{\rotatebox{90}{\textbf{r1c16ldm}}} &
        \includegraphics[width=0.50\textwidth]{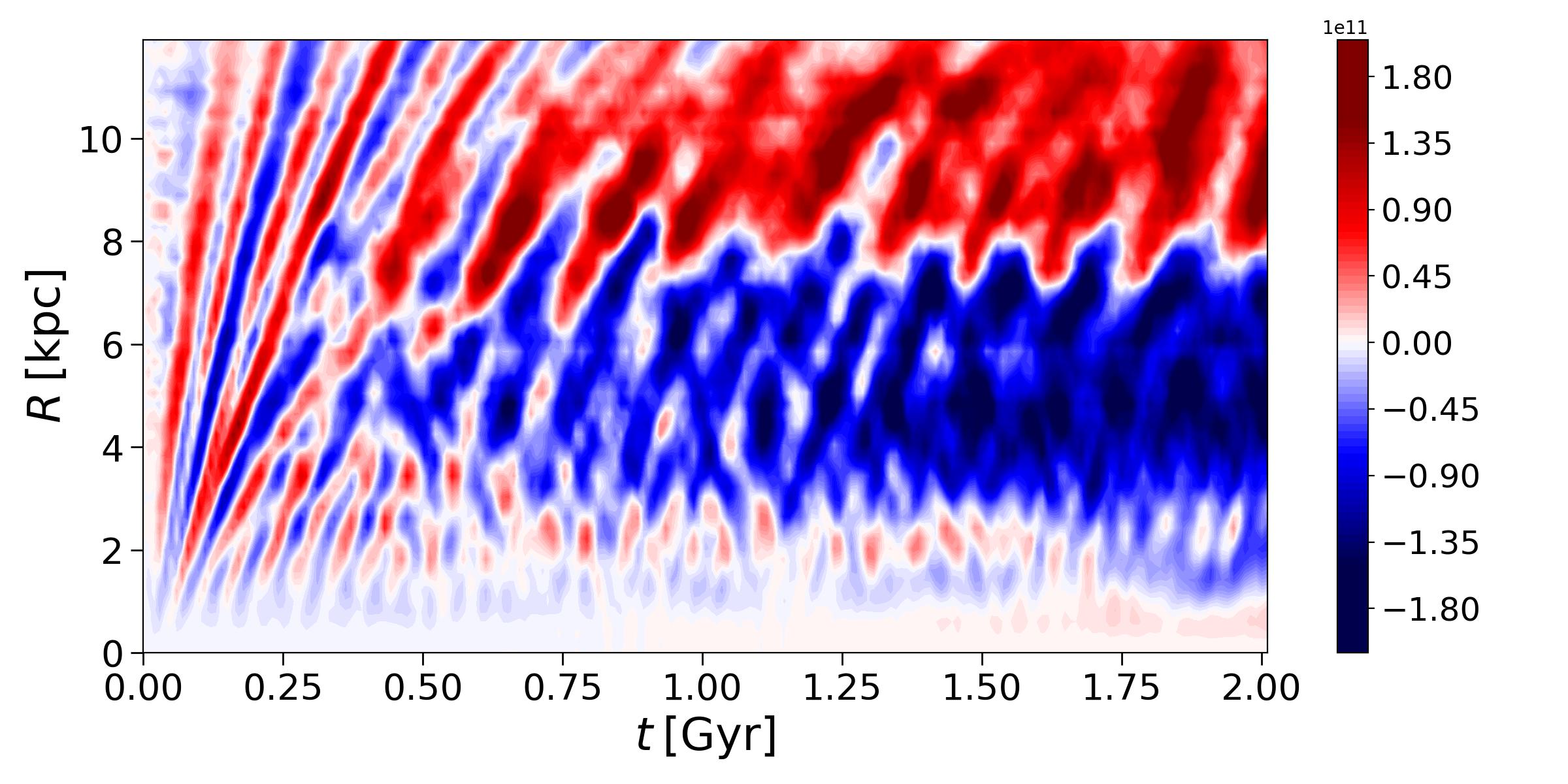} \\
    \end{tabular}
    \caption{The same as Figure \ref{fig:lzmap1}, but for the halo models, in which the disk-halo systems have the same density structures.}
    \label{fig:lzmap2}
\end{figure}

\begin{figure*}[htbp]
    \centering
    \includegraphics[width=0.45\textwidth]{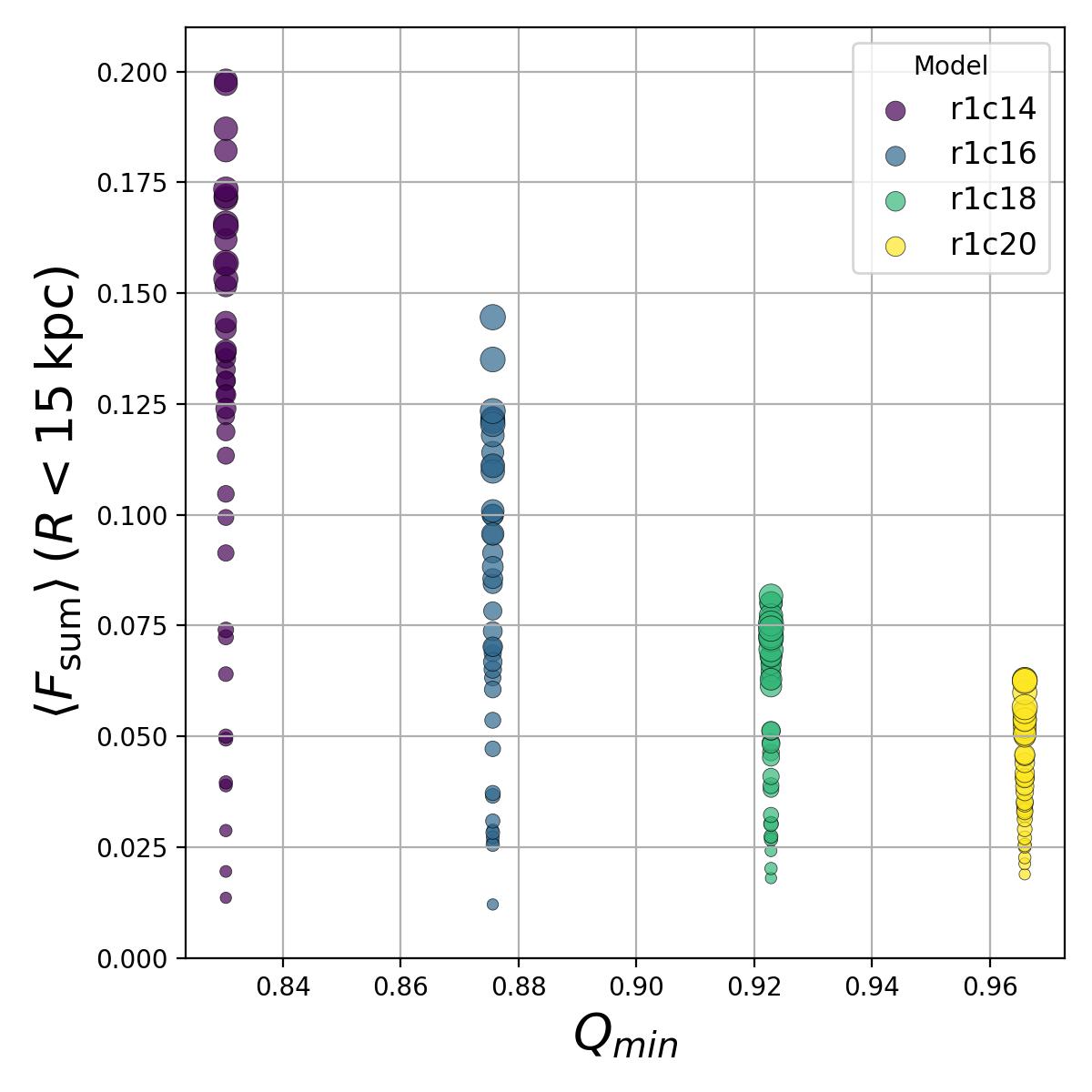} \includegraphics[width=0.45\textwidth]{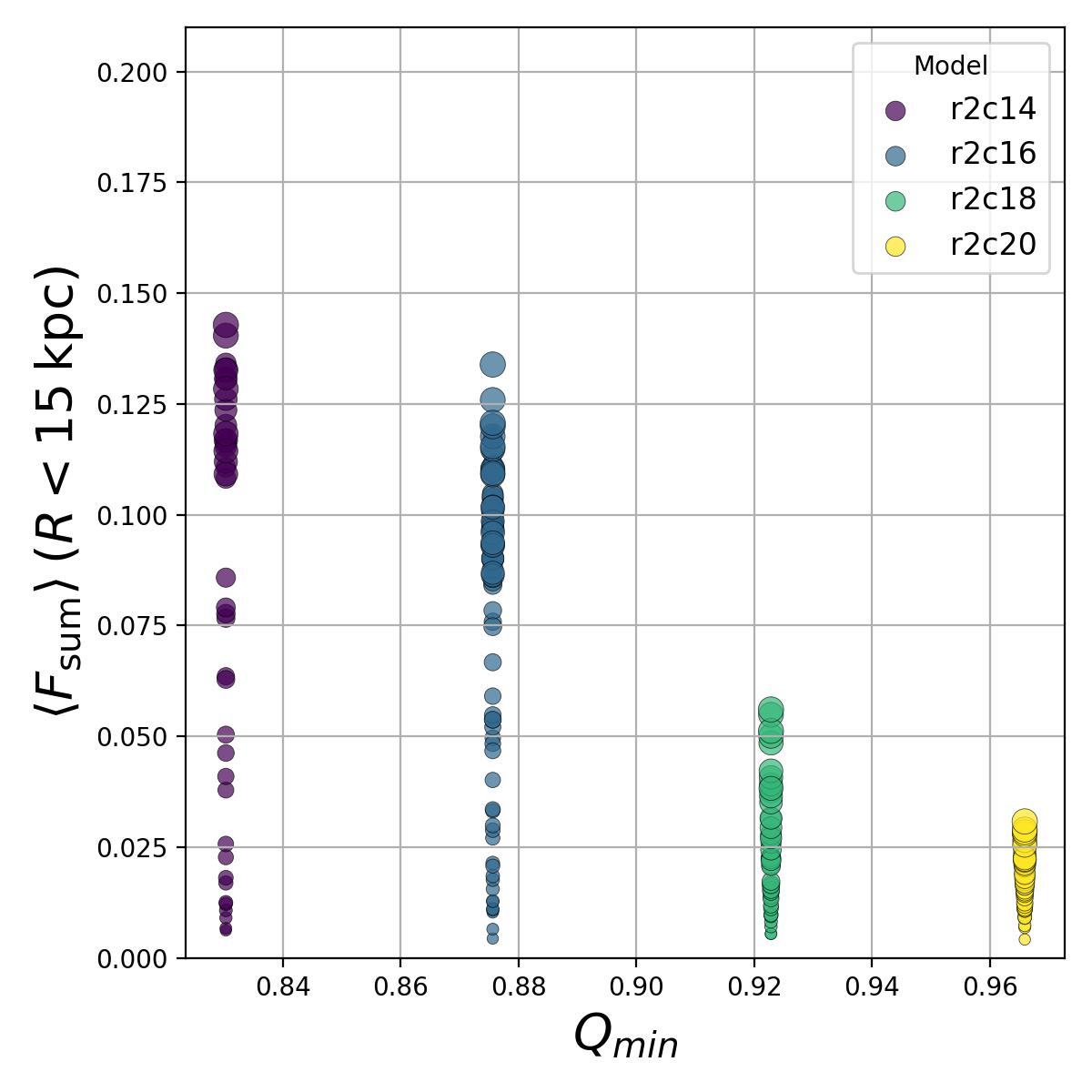}

    \caption{Amplitude of $\mFsum$ within $15 \kpc$ as a function of the minimum Toomre $Q$ value of each model, as labeled, with different halo concentration parameters  $c$. The size of a circle increases with the evolution time. The time interval of each snapshot used is 0.05 Gyr. The left panel is for the low resolution, r1, models, and the right panel is for the r2 models.  }
    \label{fig:q_fmean}
\end{figure*}

\begin{figure}[htbp]
    \centering
    \includegraphics[width=0.45\textwidth]{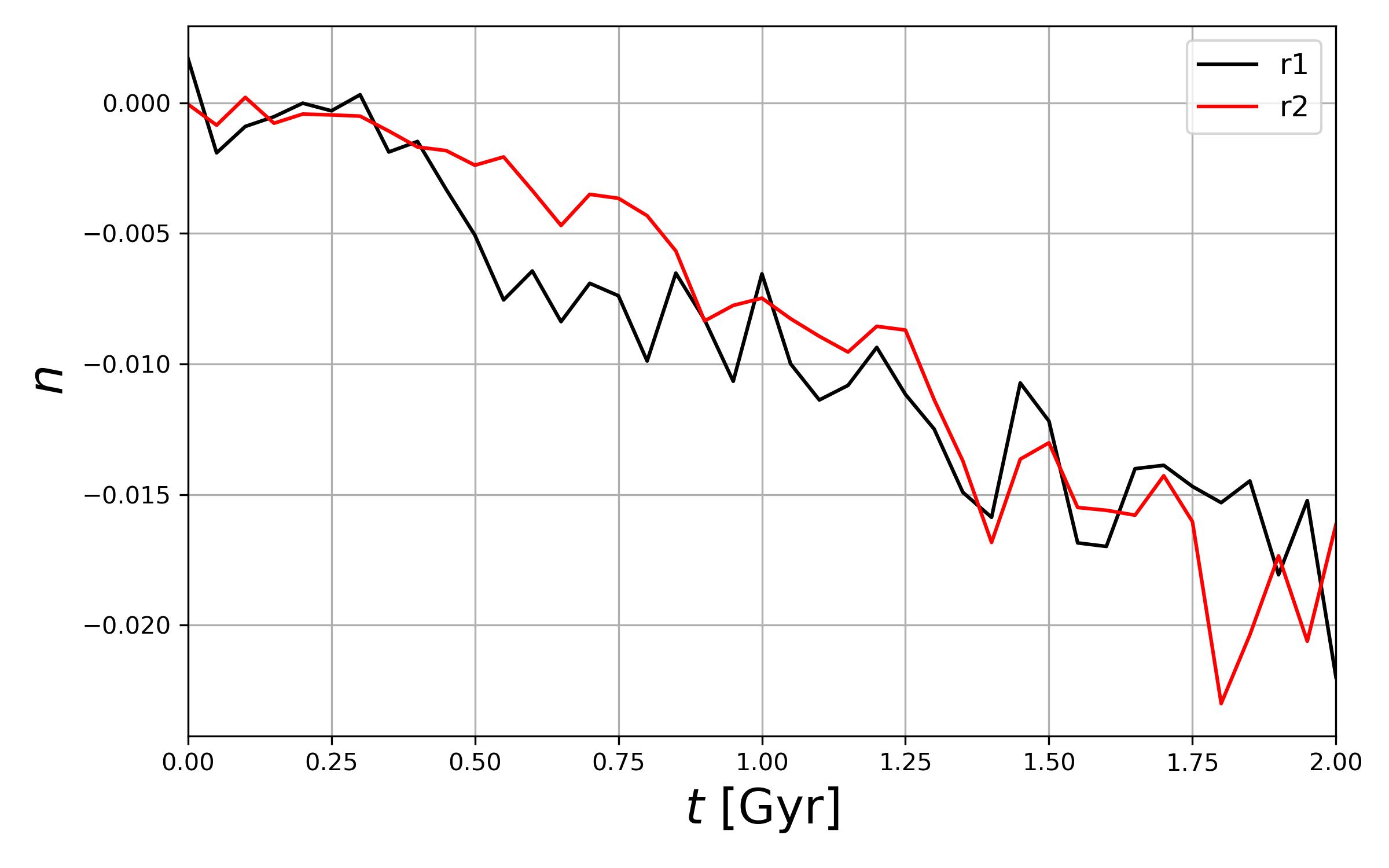}

    \caption{Temporal change of the slope obtained by performing linear fits on the circles of models in Figure \ref{fig:q_fmean} with the exception of the c14 models due to rapid bar formation. The black line is the series of linear fits on the r1 models, and the red line is the same on the r2 models.}
    \label{fig:slope}
\end{figure}

\subsection{Overview of Evolution}\label{sec:overview}
In Figure~\ref{fig:faceon}, we present the face-on surface density distributions of the r2 models with different halo concentrations $c$, along with model r1c16. 
Each column corresponds to the labeled model, and each row displays snapshots at 0.5, 1.0, 1.5, and 2.0~Gyr within a $30\times30~\kpc$ box.
Additionally, model r1c16 is included for resolution comparison and is examined in detail because the c16 models are less unstable: they form visible spirals but do not develop a strong bar by the end of the simulation.
When comparing the r2 models, those with lower halo concentrations have a smaller DM contribution within the disk region, resulting in a lower Toomre $Q$ distribution and making the disks more susceptible to spiral formation due to gravitational instabilities. 
Among these models, model r2c14 is the most unstable and rapidly develops stronger spiral arms at 0.5 and 1.0 Gyr. 
This intense spiral activity heats the disk, gradually stabilizing it and weakening subsequent spiral arm formation in the outer region of the disk.
By 1.5 Gyr, the perturbations are concentrated in the central region, initiating bar formation. 
Consequently, a bar is present at 2 Gyr. 

In the comparison between the c16 models with a resolution differing by a factor of ten, the lower-resolution model (r1c16) initially develops more visible spiral arms at 0.5 Gyr than the higher-resolution model (r2c16). 
However, at 1 Gyr, the r2c16 model also exhibits visible spiral arms, which last until the end of the simulation. 
The r2c16 model appears to be on the verge of forming a small bar at the end of the run, but overall, the evolution does not significantly differ between r1c16 and r2c16. 
We conjecture that the bar formation proceeds faster in the r1c16 model because noise-driven spiral development in its early phase eventually hastens the growth of the bar mode, $m=2$. 
Visual inspection of the face-on snapshots suggests that stability has a greater impact on spiral arm formation than resolution if the galaxy is simulated for a sufficiently long period. 

Overall, every model in this sequence develops spiral structure within 2 Gyr, later as the disks become more stable due to the increased DM contribution or enhanced resolution. At 0.5 Gyr the more stable disks show the faint ring-like features left by the initial relaxation. These are not replaced but overtaken: by 1.0 to 2.0 Gyr multi-armed spirals dominate the appearance of the inner and middle disk in every column, including r2c20, while the ring signature survives in the outermost disk, where the longer dynamical times make phase wrapping much slower (Fig.~\ref{fig:lzmap1}). What changes along the sequence is the amplitude and the onset time of the spirals, not the character of the structure.

\subsection{Resolution Effects}\label{sec:resolution}
To investigate the effects of resolution on the formation and evolution of spiral arms, we classify spirals by their number of arms and calculate their strength using the Fourier analysis:

\begin{equation}
F_{m}(R) = \frac{\sum_{j} \mu_{j} \, e^{i m \phi_{j}}}{\sum_{j} \mu_{j}},
\label{eq:fourier}
\end{equation}
where \(m\) is an integer that denotes the multipole order and ranges from 0 to 6. Here, \(\mu_{j}\) and \(\phi_{j}\) represent the mass and azimuthal angle, respectively, of the \(j\)th particle in an annulus of radial bin width \(\Delta R = 0.2\,\mathrm{kpc}\).
Since each spiral mode changes rapidly, we define a global perturbation parameter by summing all the modes as follows:
\begin{equation}
\Fsum(R)
= \sqrt{\sum_{m=1}^{6} \bigl[F_{m}(R)\bigr]^{2}}.
\end{equation}

Figure \ref{fig:comparison_grid} illustrates how the resolution affects the amplitude of $\Fsum$ and each mode over time.
The left column corresponds to model r1c16, and the right column to model r2c16. 
From top to bottom, the rows represent snapshots at 0.5, 1.0, 1.5, and 2.0 Gyr. 
At 0.5 Gyr, model r1c16 forms spirals in multiple modes due to initial noise arising from its lower resolution, whereas only the $m=5$ and $6$ modes show signs of development in model r2c16. 
In general, higher modes begin to appear and recur around $2\,R_{d}$, while the $m=2$ and $m=3$ modes dominate within $1\,R_{d}$. 
After 1 Gyr, $\Fsum$ in the higher-resolution model starts to catch up with the $\Fsum$ of the lower-resolution model. 
By the end of the simulation, their $\Fsum$ values become comparable except in the central region, as model r1c16 begins to form a small bar by exceeding $F_{2} = 0.30$.
Notably, about $5\%$ noise is present in the region beyond $15 \kpc$ in model r1c16.

Overall, even a tenfold increase in the number of particles does not appear to significantly suppress arm formation if evolved for a sufficiently long period. 
We attribute the earlier onset of small bar formation in model r1c16 to higher initial noise, which produces faster amplification of perturbations and, consequently, earlier bar formation. 
No bar has formed in model r2c16 by 2.0 Gyr. 
However, such a high resolution does not halt the formation of non-axisymmetric structures but delays it, in inverse proportion to the level of initial noise. 
Interestingly, after evolving model r2c16 for another 2 Gyr, we find that it eventually forms a bar about 1.9 Gyr later than model r1c16 by surpassing $F_{2,\mathrm{max}}=0.30$ at $t=3.56 \Gyr$. 

In the c14 models, we calculate that the difference in the bar formation epoch ($F_{2,\mathrm{max}}=0.30$) is smaller, about 0.5 Gyr, due to stronger disk instability arising from the lower Toomre $Q$ distribution, which leads to more rapid arm and bar formation. 
Note that the c16 models are an exceptional case, placed in the middle of the stability spectrum: they are unstable enough to form multiple spirals regardless of resolution, yet the disk’s intrinsic instability lies near the threshold for bar formation. 
Consequently, initial noise boosts amplification of perturbations and results in a considerable divergence in the timing of bar formation as a result of the different resolution.
Compared to the c16 models, the effect of resolution on the bar instability is smaller in the c14 models, where disk instability dominates over resolution effects.

To quantify the overall growth of perturbations due to resolution and gravitational instability, in Figure~\ref{fig:fmean1}, we illustrate the time evolution of the mean value of $\Fsum$ for models with the parameter $c$ ranging from 14 to 20.  
We calculate $\mFsum$ within $R<3\kpc$ and $R<15 \kpc$, using time snapshots taken every $0.05 \Gyr$. 
Because the central region is largely dominated by $m=2$ and $m=3$ modes, and some of our models form bars, we examine spiral amplification in both the central region and the entire disk region within $5R_d$.

As shown in Figure~\ref{fig:fmean1}, compared to the r2 models, the r1 models start with $\mFsum$ at approximately 1--2$\%$, leading to divergences in $\mFsum$ over time.
In model r1c14, noise-driven spirals form and amplify rapidly from the beginning.
In the outer region, perturbations reach a plateau at $\mFsum \sim 0.15$ by 0.75 Gyr, which is later breached as the inner $\mFsum$ surpasses 0.2 and $F_2$ exceeds 0.3 at 1.26 Gyr.
Similar behavior occurs in model r2c14 but with a time delay of 0.5 Gyr: perturbation growth begins later due to reduced noise, reaches a plateau at 1.26 Gyr, and is breached as the inner $\mFsum > 0.2$ and $F_2 > 0.3$ at 1.77 Gyr.
Overall, models with the same density structure do not converge by the end of the simulations due to resolution differences.
Note that the decrease of $\mFsum$ in model r1c14 at 1.6 Gyr results from mild buckling instability, which temporarily weakens the bar.

Due to the presence of initial noise, the evolution of $\mFsum$ in model r2c14 is rather close to that of model r1c16, especially within 3 kpc.
Indeed, model r1c16 forms a bar as $F_2$ exceeds 0.3 at 1.6 Gyr, which is comparable to that of model r2c14.
The divergence in the bar formation epoch is larger in the c16 models, with a time delay of 1.9 Gyr, because their disks are marginally unstable, making the resolution effect relatively more significant than in the c14 models.
Within 15 kpc, spiral arms begin to emerge earlier in model r1c16 due to the initial noise, but $\mFsum(R < 3 \kpc)$ in model r2c14 catches up and exceeds it around 0.75 Gyr with a higher growth rate due to the greater intrinsic disk instability. 

All the models with the same halo concentration also do not converge until the end of the evolution, and the divergence in $\mFsum$ becomes large when bar formation begins.
When a bar is not present, the initial divergence due to noise in $\mFsum(R < 15~\kpc)$ is comparable to the divergence by the end of the evolution.
For example, the divergence in the inner $\mFsum$ in the c16 models becomes larger as soon as a bar forms, while it is mild in the outer region where the effects of the bar are reduced.
In more stable models with $c=18$ and $c=20$, the divergence is only 2--3$\%$ in $\mFsum(R<15~\kpc)$ when comparing the initial and final values.

Overall, more unstable models tend to form stronger non-axisymmetric structures, undergoing faster phase changes with relatively less impact from resolution in terms of formation epoch.
Throughout the disk region, the final perturbation levels in stable models without bars do not diverge significantly from the initial noise levels.

\subsection{Role of Halo}\label{sec:halo}
As the angular momentum exchange between stars and DM particles plays an important role in bar formation and growth \citep{athanassoula02, sellwood16, kwak17, seo19, frosst24, baek25, kwak26a}, we extend our study of the DM halo's effect on spiral structures by constructing three models: two models with different halo resolutions and one model with a fixed potential halo instead of adopting a live halo (see Table~\ref{table:model}). 

Figure~\ref{fig:fmean2} is plotted in the same manner as Figure~\ref{fig:fmean1}, but for the halo models in comparison to model r2c16, allowing us to focus on halo resolution and its effects rather than differences in gravitational stability. 
The fixed potential model, r1c16fdm,
slowly develops spirals so that the model gradually diverges from model r1c16.
Especially in the inner region, the power of the summed Fourier modes does not go beyond $\mFsum=0.10$.

The disk in a fixed-potential halo is intrinsically more stable in terms of spiral formation, consequently leading to less prominent spirals in model r1c16fdm.
\cite{sellwood16} showed that disks in rigid halos are more stable against bar formation than those in responsive halos.
Our fixed-potential model forms weaker spirals with a lower growth rate due to the absence of angular momentum exchange between stars and live halo particles, unlike the other models that adopt a live halo.
For example, model r1c14 also undergoes rapid spiral formation with continuous growth in $\Fsum$, which eventually leads to bar formation without external forcing.
In Appendix \ref{appendix:additional}, we additionally test r1c14 in a fixed-potential halo, model r1c14fdm, and find that even the same disk-halo system that rapidly forms a bar in a responsive halo does not form a bar in a fixed-potential halo.

In the spiral dominant phase, no dramatic change is shown when increasing the DM resolution by a factor of 10 in model r1c16hdm, compared to model r1c16.
However, bar formation ($F_2 > 0.3$) is delayed by 0.8 Gyr in model r1c16hdm with $m_{\star} : m_{\mathrm{DM}} = 1 : 1$.
Slightly after 1 Gyr, before bar formation in model r1c16 at 1.67 Gyr, $\mFsum$ begins to diverge, and more significantly in the inner region where the bar mode is dominant.
We revisit and explain this earlier divergence prior to bar formation in a later section, in light of the amplification of different Fourier modes.

However, when decreasing the DM resolution by a factor of 10 in model r1c16ldm with $m_{\star} : m_{\mathrm{DM}} = 1 : 100$, the stellar disk is quickly disturbed by massive DM particles, resulting in an earlier increase in $\mFsum$.
These stronger spiral arms cause more disk heating (e.g., \citealt{minchev06}), and disk heating stabilizes the system, thus weakening subsequent spiral arms \citep{sellwood19}. 
For instance, the \textsc{NewHorizon} simulation, which adopts a comparable mass resolution ($m_{\mathrm{DM}}=1.2\times10^6$) and mass ratio ($m_{\mathrm{DM}}/m_{\star}\approx92$), is expected to experience numerical heating that eventually affects the formation and lifetime of non-axisymmetric features \citep{dubois21, reddish22}.
Indeed, $\mFsum(R < 15 \kpc)$ in model r1c16ldm peaks at 0.125 and then weakens until the end of the evolution, while after $\mFsum$ peaks at 0.15 within 3 kpc, the amplitude rapidly grows further as it forms a bar at 1.89 Gyr.

\subsection{Angular Momentum and Spiral Arms}\label{sec:angmom}

We extend our investigation of the role of the DM halo in the redistribution of angular momentum via spiral arms.
In Figure \ref{fig:t_lz}, we illustrate the temporal evolution of the total disk angular momentum defined as $L_z/L_{z,0}$, where $L_{z,0}$ is the initial angular momentum. Here $L_z$ is the $z$-component of the angular momentum summed over all disk particles, integrated over the whole stellar system, so that it measures only the net exchange with the halo. It is distinct from the radially resolved quantity $\Delta L_z(R,t)$ used in Figs.~\ref{fig:lzmap1} and \ref{fig:lzmap2}, which is the change in the mean angular momentum of stars in a radial bin and therefore also registers redistribution within the disk at fixed total.
We use this fractional term to quantify the percentage change, since the total angular momentum loss is very small (e.g., less than $0.2\%$ in \citealt{sellwood21b}).
In the top panel of Figure \ref{fig:t_lz}, we plot the higher-resolution models to investigate the correlation between arm strength and angular momentum loss.
Overall, the loss of angular momentum is minor in all bar-stable models.
Model r2c16 forms evident spirals with $\mFsum > 0.1$ without forming a bar within 2 Gyr, and it loses about $0.25\%$ of its disk angular momentum.
Model r2c14 forms a bar, which drives more active angular momentum exchange between the stellar bar and the live halo.
Additionally, we include model r1c14 in the same panel to examine how different resolutions affect angular momentum loss during bar formation.
At the end of the evolution, model r1c14 loses more angular momentum than model r2c14, but considering the time delay of 0.5 Gyr in bar formation, the loss of angular momentum in model r2c14 at 2 Gyr is comparable to that in model r1c14 at 1.5 Gyr.

In the bottom panel of Figure \ref{fig:t_lz}, we plot the same $L_z/L_{z,0}$ for the halo models, with model r2c16 overlaid to highlight the resolution effects.
Throughout the evolution, the angular momentum shows occasional spikes of about 0.1--0.2$\%$ in the lower-resolution models ($N_\star = 5\times 10^6$).
Model r1c16fdm, which does not have a responsive halo, does not lose its disk angular momentum.
As discussed, model r1c16ldm forms spirals earlier due to strong gravitational kicks driven by interactions of massive DM particles, so it loses angular momentum earlier as well, but the further growth is not properly resolved due to lacking proper interaction between stars and DM particles, so the angular momentum loss does not decay further and the bar formation is slightly delayed compared to model r1c16.
In the remaining models, the amount of disk angular momentum loss follows the order of the disk stability caused by resolution effects.
Overall, the loss of angular momentum is less than 0.6$\%$ in all the c16 models, indicating that the spiral arms do not actively transfer their angular momentum to the halo, while this little interaction still diverges the growth of $\mFsum$.

In Figure \ref{fig:fmap1}, we visualize the time evolution of Fourier modes and $\Fsum$. 
The range of each Fourier amplitude varies greatly among the models, so the color scale is fixed from $1\%$ to $15\%$ for all models in the color bar to ensure that the same amplitude is indicated by the same color.
We choose 15$\%$ as the upper limit because the spiral amplitudes of different modes do not exceed this level, as shown in Figure~\ref{fig:comparison_grid}.
Each column is labeled with the names of the r2 models, and each row displays different Fourier modes (from 1 to 6) as well as $\Fsum$ in the bottom row. 
Each mode carries significant amplitude over a limited range of radii at any one time, but several modes are active simultaneously at different radii, so their superposition in the density field appears as arms spanning a much broader radial range (cf. Fig.~\ref{fig:faceon}).

Overall, the more gravitationally unstable models experience more active spiral arm formation over time. 
The spiral structure originates with a larger number of arms in the outer region and transitions to fewer arms toward the center. 
Model r2c14 begins forming spiral arms at around $2R_d$ in the $m=5$ and $m=6$ modes at approximately 0.5 Gyr, and these arms partially propagate both inward and outward. 
The growth in the $m=4$ mode starts at around $0.75 \Gyr$ at smaller radii than the $m=5$ and $m=6$ modes. 
Its epicenter eventually decays and becomes dominant in the central region ($1R_d$) after 1.5 Gyr, at which point the $m=2$ bar mode becomes dominant. 
The $  m=3  $ mode is concentrated in the central few kpc, though it occasionally contributes to the outer spiral arms before 1.5 Gyr.
Spiral arms in multiple modes become weakened after the active formation transitions to the bar mode at around 1.5 Gyr.

In general, spiral arms driven by local instabilities are faint and recur throughout the evolution, with modes and their radii decreasing from higher to lower values over time. The cascading pattern in the Fourier modes with an inward-drifting epicenter is more evident in model r2c16. Because model r2c16 is marginally unstable and does not form a bar until the end of its evolution, we can observe the gradual evolution of each spiral mode.
Fig.~\ref{fig:faceon_r2c16_all} shows the same sequence in real space. Higher-$m$ arms first become visible in the surface density in the outer disk at $\sim0.6\Gyr$, while the central region within a few kpc remains nearly featureless. By $\sim0.8$--$0.9\Gyr$ the $m=4$ and $m=5$ arms dominate the visible pattern at $R\approx5$--$6\kpc$. As the modes cascade and the activity moves inward, the inner region develops spiral overdensities. The central $m=3$ ridge is already present in the Fourier map after $\sim1\Gyr$ (Fig.~\ref{fig:fmap1}), but it becomes discernible in the surface density itself only around $1.5$--$1.6\Gyr$, the epoch at which it also appears as a coherent feature in the spectrogram (Fig.~\ref{fig:spec}). As the $m=3$ mode amplifies further, the central spiral features become more prominent. Notably, repetitive spiral formation with similar strength in the $m=6$ mode is clearly visible in Fig.~\ref{fig:fmap1}. The $m=3$ mode exhibits predominant growth in the central region, which notably coincides with the epoch at which the disk angular momentum in model r2c16 begins to decrease at around 1.5 Gyr in Figure \ref{fig:t_lz}. The central $m=3$ mode is evident not only in the Fourier map and in the disk angular momentum, but is also morphologically visible in Figure \ref{fig:faceon} before the bar formation.

While most spiral modes remain too transient to generate a sustained gravitational wake in the halo, the centrally concentrated $m=3$ mode represents an important exception. Once it grows to appreciable amplitude, it persists long enough in the inner few kpc, where the DM density is highest, to act as a coherent non-axisymmetric perturber. In this regime, the pattern no longer behaves like a short-lived spiral feature but rather as a quasi-stationary mode that exerts measurable torques on the halo. This interaction produces genuine dynamical friction, leading to a net transfer of angular momentum from the disk to the halo and enabling the subsequent amplification of the $m=2$ mode. Thus, the emergence of the $m=3$ component marks the point at which spiral activity begins to drive net angular-momentum transfer to the halo, effectively seeding the transition from multi-armed structure to bar formation. 
Appendix~\ref{appendix:spectrogram} presents spectrograms of extended runs that support this interpretation, showing a coherent $m=3$ phase followed by the later dominance of the $m=2$ mode.

Figure~\ref{fig:fmap2} presents the same distributions of Fourier modes as in Figure~\ref{fig:fmap1}, but for the halo models that share the same stellar disk with $N_{\star} = 5 \times 10^6$.
As shown in Figure \ref{fig:fmean2}, the initial noise in these lower-resolution models is about 1--2$\%$ higher, which is readily apparent in Figure \ref{fig:fmap2}.
Similar to models r2c14 and r2c16, recurrent spiral arms of comparable strength and morphology are also observed in model r1c16, albeit obscured and enhanced by numerical noise.
Due to the augmented instability caused by the initial noise in model r1c16, the formation of multi-armed spiral structures appears earlier, which causes the cascading pattern in the Fourier modes to appear fainter.
However, the cascading pattern is more pronounced in model r1c18 (see Fig. \ref{fig:additional}), as the greater stability of the latter permits a more gradual progression of the spiral modes.

A comparison between models r1c16 and r1c16hdm reveals that the DM halo resolution substantially influences the timing of bar formation (time delay of 0.8 Gyr), whereas the development of higher Fourier modes appears largely unaffected.
In both models, the $m=3$ mode gradually emerges between 0.5 and 0.75 Gyr, although this prompts earlier amplification of the $m=2$ mode in model r1c16.
This suggests that a mass ratio of $m_{\star}:m_{\rm{DM}}=1:10$ between stellar and DM particles may exert minimal influence during the spiral phase but has considerable effects during the transitional phase from spiral modes to bar formation with central amplification driven by the $m=3$ mode, as also evidenced by the divergence in $\mFsum$ shown in Figure \ref{fig:fmap2}.
A tenfold increase in total resolution (as in model r2c16) results in a more substantial suppression of noise-driven spiral amplification, but a tenfold increase in DM resolution (as in model r1c16hdm) yields a considerable influence in the transition of central bar modes from $m=3$ to $m=2$.
Note that the resolution effects diminish in more unstable galaxies, such as those represented by the c14 models, where disk instability predominates over resolution effects.

As shown in Section~3.4, the early, high-$m$ spiral patterns primarily redistribute angular momentum among neighboring stars, and the total disk $L_z$ remains nearly conserved during the spiral-dominated phase. In fixed-potential models, the absence of halo back-reaction therefore does not strongly affect the initial spiral-driven redistribution, but it does prevent the subsequent growth of the centrally concentrated $m=3$ mode and the associated bar.

While model r1c16 forms a bar, model r1c16fdm does not. We additionally simulate model r1c14, which is more unstable and forms a bar at 1.26~Gyr, by replacing its live halo with a rigid halo. We designate this variant as r1c14fdm, shown in Figure~\ref{fig:additional} (see Appendix~\ref{appendix:additional}).
Shortly after the onset of evolution, model r1c14fdm rapidly develops strong spiral arms.
However, despite this rapid spiral formation in r1c14fdm, bar formation is absent. Neither the $m=2$ nor the $m=3$ mode exhibits amplification within $1~R_d$, owing to the lack of angular momentum exchange.
Instead, we observe faint, recurrent $m=2$ perturbations in the central region.

The low DM resolution in model r1c16ldm produces an effect opposite to that in r1c16fdm, with spiral arm formation being notably amplified over time.
Although the total angular momentum exchange between the spirals and the DM halo is small in general (e.g., \citealt{sellwood21b}), this interaction nevertheless substantially influences the morphology, lifetime, and recurrence of spiral structures.
We compute the average number of DM particles within a $1~\kpc$ cubic volume at $r = 6.5~\kpc$, yielding approximately 30 particles for model r1c16ldm ($N_{\rm DM} = 1.14 \times 10^6$).
The appropriate representation of the halo’s gravitational response to the disk, which regulates the subtle torques on multi-armed spirals and the subsequent angular-momentum transfer during bar growth, is significantly polluted by the insufficient number of interactions.
Meanwhile, each interaction imparts strong gravitational perturbations owing to the halo particles being 100 times more massive, thereby exciting spiral arms across all modes immediately upon the beginning of the evolution.
Consequently, the cascading mode pattern is not observed (Fig. \ref{fig:fmap2}).
Furthermore, the $m=3$ mode becomes over-amplified starting from 0.5~Gyr, coinciding with the onset of disk angular momentum loss (see Fig.~\ref{fig:t_lz}). Subsequently, a bar forms at 1.89~Gyr when $F_2$ exceeds 0.3.
Overall, the disk region, extending even beyond $4 R_d$, exhibits greater amplification and blobs compared to models with $m_{\rm DM}/m_\star < 10$.
As \cite{diemand04} demonstrated that the small DM softening length reduces the two-body relaxation time, but rerunning model r1c16ldm with 20 times higher DM softening does not result in visible differences (see model r1c16ldmsf06 in Fig. \ref{fig:additional}). 

As the results of models r1c16fdm and r1c16ldm underscore the importance of physical interactions between stars and halo particles, we proceed by examining how spiral arms and bars redistribute the disk's angular momentum over time. 
In Figures~\ref{fig:lzmap1} and \ref{fig:lzmap2}, we show the time evolution of the stellar disk angular momentum relative to its initial value, $L_z - L_{z,\,0}$, within $15\,\textrm{kpc}$, with blue and red regions indicating angular momentum loss and gain, respectively.
Each panel is labeled with the corresponding model name, and the same color scale is used for comparison. 
Notably, the disk angular momentum distributions exhibit wave-like patterns throughout the disk region over time.

In model r2c14, shortly after the emergence of prominent spiral arms at approximately 0.5 Gyr, the disk's angular momentum is transported from the inner to the outer region with a fix point near $2R_d$, where loss and gain balance each other. 
As the radii of initial formation and recurrence decrease over time---in accordance with the mode cascading from $m=6$ to $m=2$---the radial extent of angular momentum loss expands and shifts inward until the end of the evolution.
A notable feature is the recurrent inward transport of angular momentum toward the galactic center during phases dominated by higher Fourier modes near $2R_d$, preceding the central amplification of the $m=3$ and $m=2$ mode.
As illustrated in Figure~\ref{fig:t_lz}, the total disk angular momentum remains nearly conserved until approximately 1~Gyr, suggesting that the spiral arms of the higher Fourier modes primarily redistribute angular momentum among neighboring stars rather than transferring it to halo particles.
Indeed, Figure~\ref{fig:fmap1} demonstrates that, between 0.5 and 1.0~Gyr, the $\Fsum$ in model r2c14 reveals a portion of the spiral arms being sheared inward and outward from $2 R_d$.
This pattern reflects a recurrent rearrangement of angular momentum across the spiral epicenters within the disk, driven by the interference of multiple modes and the gradual drift of the activity's epicenter, while the halo response remains weak until the centrally concentrated $m=3$ mode begins to grow. Throughout, we use \emph{interference} (equivalently linear superposition, or mode beating) for the addition of the density fields of two or more independently existing patterns. This produces a modulation on the beat period $2\pi/|m_1\Omega_{p,1}-m_2\Omega_{p,2}|$ but transfers no power between the patterns, and it operates at any amplitude. We reserve \emph{non-linear mode coupling} for a genuine transfer of power between modes, which requires finite amplitude and obeys selection rules on both mode number and frequency \citep{tagger87,masset97}.

Since model r2c16 develops spiral arms at a slower growth rate than r2c14, the redistribution of angular momentum can be observed in greater detail. 
The coexistence of several spiral modes, together with their constructive and destructive interference, gradually lowers the stellar angular momentum around $2R_{\rm d}$ and produces the net-like or cross-hatched pattern visible in Figure~\ref{fig:lzmap1}. 
As the epicenters of spiral activity drift inward in radius, the loci of strongest interference move as well, shuffling angular momentum inward and outward among disk stars and thereby creating intersecting tracks of angular-momentum flow. The same net-like feature appears in r2c18 and r2c20, but later due to the increased stability and ensuing delayed emergence of spirals. This pattern is distinct from the diagonal phase-wrapping bands discussed in Sec.~\ref{sec:lzredist} and is absent when spirals are not present (e.g., the early phase of r2c18, r2c20 and r2c24 in Fig.~\ref{fig:lzmap_r2c24}).

In model r1c16 and r1c16hdm (Fig.~\ref{fig:lzmap2}), the net-like pattern induced by spirals remains evident, despite their lower resolution, accompanied by observable inward and outward redistribution of angular momentum.
In model r1c16hdm, the details of this redistribution stand out more clearly, especially the repeated inflows caused by the spiral arms.
Nevertheless, the differences in $\Fsum(t)$ and total angular momentum loss during the spiral phase are minimal (Figs.~\ref{fig:fmean2} and \ref{fig:t_lz}).
Owing to bar formation in model r1c16 at 1.67~Gyr, the angular momentum loss within 3 kpc occurs earlier than in model r1c16hdm.

The net-like pattern induced by spiral arms is also evident in the fixed potential model.
However, by the absence of dynamical friction from DM particles, weaker spirals form, particularly in the central region, resulting in a narrower range of disk angular momentum loss compared to model r1c16.
In bar-forming models such as r2c14 and r1c16, angular momentum inflow via spirals is transferred to the very central region ($\sim 1 \kpc$) during amplification of the $m=3$ mode, while angular momentum beyond 1 kpc diminishes.
This process is not observed in the angular momentum distribution of model r1c16fdm, as the fixed potential halo lacks the responsive halo that absorbs disk angular momentum during central amplification of the $m=3$ mode.

In model r1c16ldm, which employs lower DM resolution, the redistribution of disk angular momentum is not accurately captured.
Owing to non-physical interactions between the spiral arms and the DM halo, the disk develops over-amplified spiral arms accompanied by irregular overdensities (see Fig. \ref{fig:fmap2}), resulting in the absence of a clear net-like pattern in Figure \ref{fig:lzmap2}.
Upon bar formation at 1.89 Gyr, the angular momentum contained within 1 kpc is lower than that in models r1c16 and r2c14 at their respective bar-formation epochs, suggesting that angular momentum transfer from the outer spiral arms to the center may be inefficient due to the low number of halo particles and the mass ratio $m_{\rm DM}/m_{\star} = 100$.

\subsection{Toomre Q vs. Resolution}
In Figure~\ref{fig:q_fmean}, we compare the effects of resolution and gravitational disk instability over time. 
We show the growth of gravitational instabilities, defined by $\langle F_{\mathrm{sum}}\rangle(R<15\kpc)$, for the r1 and r2 models, along with the corresponding initial minimum Toomre $Q$ value. 
The size of each circle increases with time. 
The lower-resolution (r1) models appear in the left panel, while the higher-resolution (r2) models are presented in the right panel.
The initial circles in r1 models start from $>1\%$ as already presented in Figure \ref{fig:fmean1}.
Despite the initial Toomre $Q < 1$, the $\mFsum$ in our models increases \emph{gradually} over time, in contrast to razor-thin disk models where perturbations develop rapidly with $Q<1$. Indeed, disk thickness provides stabilizing effects, rendering the range of Toomre $Q$ values from 2D simulations not directly comparable to those in 3D models \citep{romeo92, romeo11, romeo13}. Moreover, the resolution clearly exerts a considerable influence on stability and spiral formation, as evidenced by comparing the temporal evolution of $\mFsum$ in the `r1' and `r2' models.

The growth trend of non-axisymmetric structures over time is measured using the slopes from these models except the c14 model because model r1c14 rapidly forms a bar and undergoes a mild buckling instability, which significantly fluctuates the amplitude of Fourier modes.
We apply linear fitting to the circles at each time snapshot from 0 to $2$ Gyr.
The resulting slopes appear in Figure \ref{fig:slope}.
Overall, the growth trend does not differ dramatically between the two resolutions. 

\section{Discussion}
We studied the nature of multiple spiral arms arising from a three-dimensional exponential disk in a live halo using a series of $N$-body simulations without gaseous components and star formation.
Without explicitly exciting spiral instabilities (e.g., \citealt{sellwood21b}), we varied the local stability of the disks by adjusting the halo concentration parameter, which simultaneously changes the mass-to-light ratio in the disk region.
The central fraction of the dark matter (DM) halo and the spheroidal stellar component is known to affect disk stability and bar formation \citep{kwak17, jang23}, so we observed how multiple spiral arms form and evolve naturally from the local instability before bar formation begins.
In addition to these stability effects, we explored the impact of total resolution and DM halo resolution and compared the results with a fixed potential halo model.

\subsection{Halo Concentration and Local Stability}
\label{sec:halocon}
Disks embedded in less concentrated halos exhibit an earlier and stronger multi-armed spiral phase in our models (Fig.~\ref{fig:fmap1}). 
This is a monotonic sequence rather than a division into active and inactive models. As $Q_{\rm min}$ rises from $0.830$ in c14 to $0.965$ in c20 (Sec.~\ref{sec:ic}), every signature weakens together: the spiral amplitude in the face-on maps (Fig.~\ref{fig:faceon}), the Fourier amplitudes (Fig.~\ref{fig:fmap1}), and the net-like pattern that the overlapping modes leave in $\Delta L_z$ (Fig.~\ref{fig:lzmap1}).
A smaller halo concentration, $c$, produces a shallower inner halo density profile and a weaker contribution of the DM component to the circular speed, thereby lowering the epicyclic frequency $\kappa$. 
Consequently, the Toomre parameter, $Q = \kappa \sigma_R / (3.36\,G\,\Sigma_{\star})$, decreases to a threshold where gravitational instabilities can grow even without external perturbations \citep{toomre64, binney08}. 
Theoretically, $Q < 1$ is considered unstable, while a higher $Q > 2$ may be required to suppress bisymmetric instabilities \citep{athanassoula86}. 
All our models have $Q < 1$, meaning they are susceptible to the formation of spiral arms via swing amplification \citep{goldreich65, julian66, toomre81}. However, due to the stabilizing effects of disk thickness \citep{romeo92,romeo11,romeo13}, our models with $Q < 1$ do not experience such rapid axisymmetric rearrangement and the formation of very strong spirals, unlike 2D simulations of razor-thin disks that employ much lower resolution.

Controlling the gravitational stability by adjusting the halo concentration has a two-sided effect on bar formation and growth via angular momentum exchange between stars and halo particles. 
However, in our higher-resolution models without a bar, the angular momentum transfer from spirals to the halo remains insignificant, at less than $0.1\%$.
\citet{sellwood21b} also reported that only $0.2\%$ of the disk's angular momentum was transferred to the halo by spiral arms.

In the context of bars, varying the DM concentration alters not only the timing of bar formation but also the rate at which bars lose angular momentum to the halo: a larger DM fraction in the disk region stabilizes the disk--halo system against bar formation, yet once a bar forms, it grows faster due to enhanced angular momentum exchange, leading to a more rapid increase in bar strength and a decrease in pattern speed \citep{athanassoula02, athanassoula14, sellwood16, kwak17}. 
For instance, the model DM1 in \citet{kwak17} quickly develops a bar in a low-concentration DM halo, but the bar does not grow stronger and maintains a nearly constant pattern speed due to its low angular momentum exchange. 
Unlike a bar, which behaves like a rigid structure (even though its shape changes over time) and exerts torque on surrounding halo particles, spiral arms are faint and recurrent in the stellar density, forming in multiple modes at different radii. 
Spirals formed in halos of \emph{low or high} concentration do not experience correspondingly \emph{slower or faster} growth through \emph{reduced or enhanced} angular momentum exchange. 
Indeed, our models with higher DM concentration do not show a higher growth rate of spirals.
Hence, in spiral studies, we suggest that adjusting the halo concentration to control disk stability does not yield the two-sided effect seen in bars, which significantly alters the growth rate.

\subsection{Disk Resolution Effects}
Multi-armed spiral structure in galactic disks typically appears as low-contrast, recurrent density enhancements produced by the superposition of several modes, and its detailed morphology is particularly sensitive to numerical resolution. 
\citet{fujii11} showed that in higher-resolution $N$-body simulations of pure stellar disks (with $N_{\star}\gtrsim 3 \times 10^6$ particles), the disks can maintain spiral features for several tens of rotations even without additional cooling. 
By contrast, lower-resolution simulations produce greater disk heating (increasing the Toomre $Q$ parameter more rapidly), and thus the spirals become faint within a few Gyr. 
For instance, in simulations with a smaller number of particles ($N_{\star} = 3 \times 10^5$), the spiral arms fade away quickly: they initially form through swing amplification from the seed density perturbations caused by local instabilities and Poisson noise in the initial conditions. 
Since fewer particles amplify the noise, such disks are heated more rapidly and consequently lose their spiral structures in a shorter timescale.
Our lower-resolution models contain $5\times10^6$ disk particles, which is sufficiently high to prevent the spirals from disappearing rapidly due to noise-driven heating \citep{fujii11}.
Indeed, our models sustain multiple spiral arms until the end of the evolution without requiring external cooling (Fig. \ref{fig:fmap2}), though the spirals tend to weaken in the bar forming region.

Increasing the particle number, and thereby reducing Poisson noise in the initial conditions, systematically postpones the onset of spiral activity, in agreement with \citet{sellwood12,sellwood20}.  
In our marginally unstable disk embedded in a halo of concentration $c=16$, the first coherent spiral pattern appears ${\sim}0.5 \Gyr$ later when the total mass resolution is raised by an order of magnitude.  
The lower-resolution model r1c16 ($N_{\star}=5\times10^{6}$, $N_{\mathrm{DM}}=1.14\times10^{7}$), which has an initial noise level of $\mFsum\approx1\%$ inside $R=15\;\mathrm{kpc}$, already exhibits visible spiral arms at 0.5 Gyr.  
The high-resolution counterpart r2c16 requires an additional ${\sim}0.5\;\mathrm{Gyr}$ to reach comparable spiral amplitudes (Fig. \ref{fig:fmean1}).  
A similar delay was reported by \citet{donghia13}, who increased the particle number from $10^{6}$ to $10^{8}$ and evolved their models for only $0.5\;\mathrm{Gyr}$. 
We concur with \citet{sellwood20} that such a short integration time can mask the eventual growth of spirals at high resolution.

The resolution-driven delay in the formation of non-axisymmetric structures shortens as the system becomes more unstable through a reduction in halo concentration.
In such scenarios, the intrinsic stability and perturbation growth rate predominate over the suppression of seed noise.
As spirals form and undergo physical amplification within a live halo, the Fourier modes cascade from $m=6$ to $m=2$, with this process occurring more rapidly in increasingly unstable disks.
For instance, the time lag to attain $\mFsum(R < 15~\kpc) \approx 0.05$ is approximately 0.3 Gyr in the c14 models, compared to 0.5 Gyr in the c16 models (Fig. \ref{fig:fmean1}).
Once the $m=2$ Fourier mode surpasses $F_2 = 0.30$, a bar forms, and the resolution difference induces an even greater delay in bar formation: 1.9 Gyr in the c16 models versus 0.5 Gyr in the c14 models (see Fig. \ref{fig:fmean1}).
Owing to the rapid bar formation and associated angular momentum exchange, the c14 models exhibit further divergence in $\mFsum$ beyond the difference arising from noise-driven spirals.
The models with $c = 18$ and $c = 20$ also fail to converge by the end of the evolution in terms of $\mFsum$ because their minimum $Q$ values approach unity and are thus more stable (cf.\ Sec.~\ref{sec:ic}).

\citet{dubinski09} demonstrated that spiral and bar instabilities depend \emph{wholly} on initial Poisson noise, using $N$-body models with $N_\star=$18K, 180K, 1.8M, 18M and $N_{\rm{DM}}=$100K, 1M, 10M, 100M (see Fig. 15 in \citealt{dubinski09}). 
Their two highest-resolution models showed a 0.2 Gyr delay in bar formation. 
We compared 5M and 50M disk resolutions, observing formation delays of 0.5 Gyr and 1.9 Gyr for spirals and bars in a marginally unstable model, and 0.3 Gyr and 0.5 Gyr for the same in the more unstable model. 
Our findings confirm that higher resolution reduces Poisson noise, delaying non-axisymmetric structure formation. 
This effect is more pronounced for spirals and in marginally unstable disks. 
In our 5M lower-resolution model, initial noise ($\mFsum\approx1\%$) triggers early and faint spiral arms.
However, in more unstable disks, the resolution impact on spirals diminishes due to their rapid growth. 
For bars, once perturbations amplify to bar-forming levels, the bar emits angular momentum actively to the DM halo, with growth rates remaining resolution-independent (see Fig. 16 in \citealt{dubinski09}). 
Our results emphasize that disk stability also significantly influences the formation of non-axisymmetric structures and their amplification.

\subsection{Mode Cascade with Decaying Epicenter}
\label{sec:cascade}

In our simulations, spiral structure appears as a sequence of recurrent density enhancements that involve multiple $m$-modes at different radii. 
Interestingly, a cascading of Fourier modes from $m=6$ down to $m=2$ is captured, along with an inward drift of the epicenter where each mode takes shape and becomes dominant (Fig. \ref{fig:fmap1}).
The high resolution of the c16 models illustrates this sequence most clearly because its disk is marginally unstable: it is cold enough to seed faint spirals but still warm enough to avoid bar formation for the first 2 Gyr. At $t\approx0.4 \Gyr$ the strongest signal lies in $m=5$ and $m=6$ around $R=6 \kpc$, by 0.8 Gyr the power has shifted to $m=4$ around $R = 5\kpc$, and after 1 Gyr an $m=3$ ridge dominates around $R=3\kpc$. As the disk evolves beyond 2 Gyr, the $m=2$ mode becomes dominant and gradually transforms into a bar. This transition from the cascading spiral modes to the bar phase is also evident in other models employing live halos with $m_{\rm DM}/m_\star \leq 10$. In more stable disks---whether due to intrinsic gravitational stability or enhanced resolution---the trend is more pronounced, as the cascade in modes and epicenter proceeds more gradually.

The inward drift of the activity’s epicenter reflects a combination of physical and apparent effects. As the outer disk heats and the local Toomre $Q$ rises, short-wavelength perturbations are no longer efficiently swing-amplified, so subsequent wave packets are preferentially seeded at smaller radii. In addition, several spiral modes with similar $m$ but different pattern speeds overlap in radius, and the location of maximum constructive interference between them shifts gradually inward as differential rotation shears their phases. This slow, secular drift produces the broad inward progression of the ridge seen in the Fourier maps.

In addition to this cascading sequence, the short-timescale recurrences in the amplitudes of individual $m$-modes (typically $100$--$200\,$Myr) do not necessarily correspond to repeated, independent episodes of swing amplification. 
Here “short-timescale” refers to the lifetime of the visible density enhancement or Fourier-amplitude peak, not necessarily to the lifetime of the underlying pattern. Instead, these modulations are consistent with interference between multiple coherent spiral components with different pattern speeds, whose overlap produces short-lived density peaks on the beat timescale. A number of works have demonstrated that even persistent spiral waves can produce short-lived density enhancements when their phases momentarily align, with the enhancement fading once the waves shear out of phase (e.g., \citealt{comparetta12, minchev12a, vislosky24, marques25}). The cyclic peaks we observe in $F_m(t)$ (Fig. \ref{fig:fmap1} and \ref{fig:fmap2}) are consistent with such constructive and destructive interference among coexisting patterns, including overlaps between spiral modes and the early bar component, implying that the apparent ``burstiness'' reflects mode beating rather than a sequence of rapid growth and decay episodes.

While the long- and short-timescale behaviors are shaped by interference and drift, the first appearance of each high-$m$ ridge on the Fourier contour map (Fig. \ref{fig:fmap1} and \ref{fig:fmap2}) is well explained by linear swing amplification \citep{goldreich65,julian66,toomre81}. A small leading disturbance is sheared into a trailing wave by differential rotation, and as long as the local Toomre $Q$ parameter remains of order unity, its surface-density contrast can grow by an order of magnitude before the wave is sheared apart. Because $Q$ is smallest in the disks embedded in low-concentration halos, the c14 and c16 models show the earliest and strongest formation, and both the onset time and the amplitude vary monotonically along the sequence: in c20 the arms appear later and remain faint, but they are clearly present over the second half of the run (Fig.~\ref{fig:faceon}). Typically, in this linear regime, the perturbation is relatively small ($\delta \Sigma / \Sigma_0 < 5$--$10\%$), so the stellar kinematics undergo minimal changes in terms of pattern speed and spiral-driven heating. For instance, model r1c18 develops faint spirals that have minimal influence on disk dynamics, resulting in quasi-periodic recurrence of spirals with comparable strength (see Fig.~\ref{fig:additional}).

Once the spiral contrast exceeds $\delta \Sigma / \Sigma_0 \sim 10\%$ and recurs repeatedly, the system transitions beyond the regime where the collisionless Boltzmann and Poisson equations can be linearized, entering the non-linear regime \citep{sellwood02,minchev11,sellwood14}.
In this phase, each large-amplitude perturbation transfers angular momentum across its epicenter, sending some stars outward, while drawing others inward.
Although this redistribution does not by itself produce significant radial heating \citep{minchev12b}, non-linear interactions among multiple modes and the early growth of the $m=2$ component progressively increase random motions in the disk. As the outer disk becomes warmer and the local Toomre $Q$ rises, short-wavelength perturbations cease to be efficiently swing-amplified, so subsequent wave packets are seeded at smaller radii.
Concurrently, non-linear mode coupling may contribute to the transfer of power from high-$m$ waves to slightly lower-$m$ components, whose Lindblad resonances reside deeper within the potential well \citep{tagger87,quillen11,fouvry15}.
The combined effects of evolving disk stability, phase shifts between coexisting spiral patterns, and possibly mode coupling, produce the observed cascade from $m=6 \to 5 \to 4 \to 3 \to 2$, together with the gradual inward drift of the activity’s epicenter. 

An alternative account of recurrent spiral activity should be noted here. In the groove-mode picture of \citet{sellwood14,sellwood19}, each saturated transient spiral leaves a narrow feature in the angular-momentum distribution of the disk at its inner Lindblad resonance, and that feature seeds the next mode; the resonance overlaps that would otherwise be read as signatures of mode coupling are then the signature of groove creation by successive transients, and the individual spirals are short-lived. The same is true of the finite-lifetime pictures of \citet{donghia13} and \citet{meidt26}. Our simulations cannot distinguish these from the interference reading adopted here: we neither resolve the phase-space grooves nor measure the lifetimes of the individual higher-$m$ components. We also cannot exclude beating between a coherent mode and a shearing transient rather than between two modes \citep{polyachenko16}.

\subsection{The Central \(m = 3\) Mode as a Precursor to Bar Formation}
\label{sec:m3_prebar}

Unlike the higher-$m$ modes, whose density signatures vary rapidly in time because several patterns with similar $m$ and different pattern speeds overlap and interfere, the $m=3$ mode exhibits a qualitatively distinct behaviour once it becomes confined to the inner few kpc. In the c14 and c16 models, the $m=3$ component reaches appreciable amplitude in this central region and, because few other modes penetrate to such small radii with comparable strength, it is largely free from destructive interference. As a result, the $m=3$ pattern appears as a coherent, slowly evolving non-axisymmetric structure and it is visible not only as a persistent ridge in the Fourier maps, but also as a relatively steady asymmetry in the stellar density itself (see Figure~\ref{fig:faceon}). Its central concentration and isolation from competing modes make $m=3$ the dominant non-axisymmetric feature prior to the onset of the bar.
A qualitatively similar appearance of temporary three-armed spirals prior to the onset of a two-armed spiral and/or bar-like flow has also been reported in collisionless N-body models \citep{jang24} and gas-rich hydrodynamical models \citep{seo19, bland-hawthorn24}.

This distinction has important dynamical consequences. Because the $m=3$ mode resides in the region of highest dark-matter density and persists for several rotation periods, it is capable of raising a gravitational wake in the halo and therefore experiences genuine dynamical friction. This frictional torque extracts angular momentum from the inner disk, making the $m=3$ mode the first pattern to drive measurable net loss of $L_z$ to the halo, although the bulk of the transfer follows only once the $m=2$ bar is established (Appendix~\ref{appendix:spectrogram}). The emergence of the $m=3$ component thus marks the transition from the multi-armed spiral regime, in which the halo response is weak and the disk’s total angular momentum is nearly conserved, to the secular evolution regime, in which the halo becomes dynamically engaged.

The subsequent growth of the $m=2$ mode is closely tied to this process. As the $m=3$ pattern transfers angular momentum to the halo, the central disk becomes increasingly susceptible to the bar instability. The $m=2$ mode therefore strengthens shortly after the rise of the $m=3$ component and eventually overtakes it, completing the transformation into a bar. This sequence explains the coincidence between the onset of net angular-momentum loss in the live-halo models and the amplification of the central $m=3$ mode, as well as the absence of both phenomena in the fixed-halo models. This finding is further supported by the spectrograms, which reveal a transition of the central mode in a bar-forming model (Fig. \ref{fig:spec}).

We do not claim that $m=3$ is dynamically privileged. Two properties make this component effective: (a) it is radially confined to the inner few kpc, where few competing patterns of comparable amplitude reach, so it is not destroyed by interference and can persist for many rotation periods; and (b) that region coincides with the highest dark-matter density, where the halo response, and hence the frictional torque, is strongest. Any coherent, centrally confined pattern satisfying (a) and (b) should play the same role, and it is the combination rather than the value of $m$ that matters. Which $m$ happens to satisfy them is set by where the cascade halts, which depends in turn on the radial run of $Q$ and on the extent of the region already heated by the outer spirals.

\subsection{Disk Angular Momentum Redistribution}
\label{sec:lzredist}

The phenomenon of mode cascading, characterized by a radially decaying epicenter, leaves distinctive signatures in the stellar angular momentum distribution over time. In model r2c16, the early appearance and interference-driven recurrence of modes $m=5$ and $m=6$ produce alternating regions of angular-momentum gain and loss, forming a persistent wedge-like pattern in the Fourier maps (Fig. \ref{fig:fmap1}). As these high-$m$ modes modulate the gravitational torque field, a net-like pattern emerges in $\Delta L_z(R,t)$, as illustrated in Figure~\ref{fig:lzmap1}. 
Furthermore, the decaying epicenter associated with lower spiral modes causes the radial extent of the blue region, indicating angular momentum loss, to widen inward over time. 

At early times all models exhibit thin, positively sloped bands in $\Delta L_z(R,t)$, which arise from epicyclic phase wrapping of the initially non-equilibrium disk. This is the well-known “disk ringing’’ phenomenon \citep[e.g.,][]{minchev09,gomez12a}, generated by the axisymmetric response of the disk to small mismatches between the initial Jeans-based distribution function and true equilibrium. Because differential rotation decreases with radius, the outer disk phase-wraps more slowly, so these diagonal features persist to progressively later times at larger radii, remaining visible in the outermost disk. In Appendix~\ref{appendix:r2c24}, we show that these signatures are axisymmetric in origin and are not sufficient to seed the spiral cascade in model r2c24, whose stability is increased due to higher DM concentration, resulting in a delay in spiral formation. The ringing remains visible at the same amplitude even though no coherent modes develop (e.g., faint spirals of amplitude less than $2\%$ in modes 5 and 6 appear after 1 Gyr). We therefore attribute the diagonal bands in Figs.~\ref{fig:lzmap1} and \ref{fig:lzmap2} to this phase wrapping, and reserve the net-like, cross-hatched pattern at intermediate radii, which is absent in r2c24, for the spiral-driven torques.

Although the $\Delta L_z(R,t)$ maps display a sequence of alternating bands of angular-momentum gain and loss, these do not necessarily indicate discrete, repeated episodes of spiral growth and decay. As discussed in Section~\ref{sec:cascade}, several spiral components can coexist and overlap in radius, and changes in their relative phases can naturally produces recurrent enhancements in density and gravitational torque. These interference-driven fluctuations can generate the repeated $\Delta L_z$ wedges even when the underlying coherent components persist longer than the visible density enhancements. The regular recurrence of these features in time makes a purely stochastic sequence of unrelated transients unlikely. However, a full spectrogram decomposition of the higher-\(m\) spiral phase would be required to distinguish uniquely between coherent mode beating, finite-lived density waves, and shearing transients. In addition, once the $m=2$ component begins to strengthen in the inner disk, overlaps between the emerging bar and the remaining spiral structure further modulate the torque pattern, reinforcing the alternating structure. Thus, the net-like appearance of the angular-momentum map reflects the combined effects of linear superposition/interference among coexisting modes and early bar–spiral coupling, rather than a sequence of independent swing-amplified events or definitive evidence for nonlinear mode coupling.

In the more unstable, bar-forming model r2c14, the early rise of the $m=2$ mode propels the system into the non-linear regime sooner than in r2c16.
As the bar strengthens, it increasingly dominates the gravitational-torque budget, producing stronger but more intermittent $\Delta L_z$ signatures compared to the more stable r2c16 model.
Following the transition to the bar phase at approximately 1.5 Gyr, the bar facilitates the transfer of a substantial fraction of the disk's angular momentum to the DM halo, resulting in a broader and deeper region of angular momentum depletion in the distribution (Figure~\ref{fig:lzmap1}).
This angular momentum loss contributes to the dissipation of multiple spiral structures in the vicinity of the $m=2$ mode-dominated region.
Models that lack mode cascading with an inward-drifting epicenter---owing to low halo resolution or a fixed potential (Figure~\ref{fig:fmap2})---still show local angular-momentum fluctuations, but the $\Delta L_z(R,t)$ patterns are more irregular and lack the coherent, diagonally organized net-like structure seen in the well-resolved live-halo models (Figure~\ref{fig:lzmap2}).
This underscores the critical role of the DM halo in regulating angular momentum evolution within disks and in sustaining spiral structures.

\subsection{Disk Stability on the Number of Spiral Arms and Bar Formation}

The transition in the number of spiral arms, driven by mode evolution, occurs more rapidly in dynamically unstable disks. 
When comparing two systems with differing degrees of instability evolved over identical time periods, the more unstable disk typically exhibits fewer spiral arms. 
This is evident in the comparison of models r2c14 and r2c16, as shown in Figure~\ref{fig:faceon}. 
The more unstable disk, characterized by stronger initial perturbations, undergoes enhanced swing amplification, leading to a higher growth rate and, consequently, fewer spiral arms compared to the more stable model. 

This behavior aligns partially with findings by \citet{donghia15}, who identified a correlation between disk stability and the number of spiral arms, where systems with lower Toomre $Q$ parameters, driven by higher disk mass fractions, tend to form fewer spiral arms. 
However, our results, as illustrated in Figure~\ref{fig:fmap1}, suggest that unstable disks do not directly form fewer arms. 
Instead, the prolonged Fourier mode transition times in more stable disks increase the likelihood of observing a higher number of spiral arms in systems with higher Toomre $Q$ values.
This mode cascade is consistent with the analytical framework of \cite{meidt26}, in which higher-$m$ spirals emerge first in cold disks and cascade to lower-$m$ as dynamical heating warms the disk.

Following the decay in the number of spiral arms, disks naturally transition to bar formation. This process is faster in the c14 models compared to the c16 models due to the greater disk instability in the c14 models. 
Additionally, the bar formation occurs earlier in the r1c16 model than in the r2c16 model, as Poisson noise-driven perturbations accelerate spiral formation and subsequent amplification. 
However, the c16 models are marginally unstable, and the resolution effect on the bar stability is minimal in unstable disks. 
These findings indicate that halo concentration mainly governs the transition from the linear to the non-linear amplification regime, which eventually leads to the bar phase. 
Such a transition is absent in models with a fixed potential halo.

\subsection{Live vs.\ Fixed Potential Halo}
Most studies on multiple spiral arm formation have relied on their dynamical behavior within a fixed potential halo. 
Consequently, some of our findings are challenging to reconcile with earlier literature. 
In a fixed potential halo, even theoretically stable disks within massive halos continue to develop mild bisymmetric spirals \citep{toomre81, athanassoula86}. However, our more stable models in live halos, with high halo concentration and enhanced resolution, show neither active spiral activity nor persistent two-armed spirals by the end of their evolution. 

Regardless of stability and resolution, all our disks embedded in live halos with $m_{\rm DM}/m_\star \leq 10$ develop spiral arms that progress from higher to lower modes and from outer to inner regions.
This trend persists, albeit faintly and with a delay, in more stable models.
In contrast, \citet{sellwood14} reported that coherent waves are strongest in the inner disk, propagating outward over time while weakening the inner region owing to rapid heating induced by spiral arms.
Even in their low mass disk ($f_d = 0.2$), coherent waves resemble those in more massive disks. Spiral activity predominantly features $m \geq 4$, with minimal power in sectoral harmonics $m \leq 3$.
Higher-multiplicity patterns are favored because waves with greater angular periodicity exhibit smaller radial extents, positioning Lindblad resonances closer to corotation.
No such features are observed in model r1c16fdm, in which weak spirals recur quasi-periodically.
Instead, in the more unstable model r1c14fdm, stronger spirals form rapidly, attaining amplitudes well beyond $\Fsum > 0.10$ (see Fig. \ref{fig:additional}), which influences the disk properties. 
As a result, higher-$m$ modes such as $m=5$ and $m=6$ move further outward, whereas the $m=3$ mode tends to propagate inward.
Owing to this active spiral activity, the inner region is heated and stabilized, resulting in no spiral formation within $1$--$2 R_d$ for the $m=5$ and $m=6$ modes.
This aligns with the findings of \citet{sellwood14}.
However, such strong spiral formation arising from substantial instability rapidly triggers bar formation when the rigid halo is replaced with a live halo.
Compared to live halo models, central amplification within $1 R_d$ is absent in fixed potential models irrespective of stability. 
This eventually prevents the mode cascade from $m=3$ to $m=2$.
Bars themselves do form in rigid halos once the disk contributes a sufficient fraction of the central mass \citep{hohl71,ostriker73}; what is absent here is not the bar instability in general but this central amplification at the disk-to-halo mass ratio we adopt.

Furthermore, we do not observe the flat saturation reported by \citet{sellwood12}. In that study, the final amplitude of non-axisymmetric structures saturates independently of particle number and the varying growth rates arising from randomized restarts.
When comparing two of our models with identical initial Toomre $Q$ values, the model with fewer stellar particles tends to amplify perturbations more rapidly. Nevertheless, the subsequent peak amplitudes of the Fourier modes in both models do not saturate.
For example, the temporal evolutions of $\mFsum$ in models r1c16 and r2c16 do not fully converge over 2 Gyr (Fig.~\ref{fig:fmean1}).
The spiral amplitudes in some halo models appear to converge with model r2c16, around $\mFsum(R < 15 \kpc) = 0.125$ (see Fig.~\ref{fig:fmean2}). 
However, the divergence remains substantial in the inner region. 
Dynamical friction between DM halo particles and stars, which is essential for the growth of non-axisymmetric structures, enables continuous amplification, presumably exceeding the saturation levels observed in fixed potential cases, particularly in the central region.

The multi-mode wave pattern and its $\mFsum$ fluctuations with time in the fixed potential model do not differ substantially from those in live halo models, except in the central region.
In the fixed-potential case, spirals form slightly weaker than those in live-halo models. However, they still recur with comparable strength in a quasi-periodic manner until divergence from the $m=3$ mode.
This divergence in central amplification is also evident in the angular momentum map (see Fig. \ref{fig:lzmap2}).
In bar-forming models, disk angular momentum driven inward by spirals to the central $1 R_d$ is absorbed. Subsequently, only the residual remains within the central 1~kpc by the time the bar mode $m=2$ becomes dominant. In contrast, the fixed-potential model does not exhibit this trend, irrespective of greater instability, as it experiences no loss of disk angular momentum in the center.
We primarily attribute this divergence to the absence of DM-star interactions. 
These interactions enable spiral growth beyond saturation levels and promote central amplification through dynamical friction and angular momentum exchange (see Fig.~\ref{fig:fmean2}). 
Such processes are indeed essential for bar formation and growth \citep{athanassoula02,athanassoula14,sellwood16,kwak17,fujii18,frosst26}.
In fact, when the $m=3$ mode begins to amplify in the center, the total disk angular momentum begins to decrease, as shown in model r2c16 around 1.5 Gyr, which is to be transferred to its responsive DM halo (see Fig.~\ref{fig:lzmap1}).
In other bar-forming models, the total disk angular momentum also begins to decrease when the $m=3$ mode appears with continuous central amplification.
This angular momentum exchange gradually hastens or initiates the growth of the $m=2$ mode in the center.

Our results highlight the critical role of physical interactions between DM particles and stars in modulating spiral instabilities and transitioning from the $m=3$ mode to the bar formation. 
However, the relative contributions of halo graininess and responsiveness remain unclear.
\cite{zemp09} showed that the structure of DM halos exhibits significant graininess in phase space with small-scale clumps, so it is unsuitable for description by a smooth distribution function.
An interesting future research would involve examining the effects of responsiveness and graininess separately.
This could be achieved by enhancing the graininess in a fixed potential model and comparing it to a live halo with equivalent graininess.

\subsection{Halo Resolution and Softening}
In typical $N$-body simulations of galaxies, the stellar disk and DM halo are modeled with distinct mass resolutions, often requiring different gravitational softening lengths for each component. 
However, our study employs a uniform softening length for both stellar and DM particles to minimize technical parameters and examine particle number effects. 
To deal with the potential need for differential softening, we simulate one model, r1c16hdm, with equal mass resolution for DM and stellar particles ($m_{\rm{DM}}/m_\star=1$), using the same softening length. 
Notably, the overall patterns in the Fourier map and angular momentum distribution for $m_{\rm{DM}}/m_\star=1$ and 10 show little difference during the spiral phase, except the bar formation phase with a time delayed of 0.8 Gyr in $m_{\rm{DM}}/m_\star=1$ case (Fig. \ref{fig:fmean2} and \ref{fig:fmap2}). 
Overall, both cases exhibit the cascading of spiral modes with a decaying epicenter and net-like patterns in disk angular momentum, consistent with the high-resolution models. 
Again, our disk resolution is sufficiently high enough to avoid artificially rapid heating by spirals \citep{fujii11}. 
Although no studies on faint, multiple spirals investigate the optimal range of particle number, gravitational softening, and mass resolution in a live halo, the literature on bars indicates that softening effects diminish with higher particle numbers \citep{gabbasov06} and that adaptive softening has minimal influence on bar formation and evolution \citep{iannuzzi13}.
Given our high enough DM resolution ($N_{\rm{DM}}\geq1.14\times10^7$), the numerical effects of uniform softening on spiral formation and growth are indeed insignificant.

Remarkably, reducing the number of DM particles by a factor of ten to $N_{\rm DM}=1.14\times10^6$ with the mass ratio $m_{\rm DM}/m_\star=100$ leads to substantial deviations in spiral formation.
Shortly after the onset of evolution in model r1c16ldm, strong spirals form rapidly across all modes without exhibiting the mode cascading pattern (Fig.~\ref{fig:fmap2}).
Once their spiral amplitudes reach an early plateau, they do not grow further until bar formation is initiated.
This discrepancy stems from an insufficient number of DM particles with 100 times greater mass, which fails to accurately capture the physical interactions between DM and stellar particles.
Concurrently, the high mass ratio ($m_{\rm DM}/m_\star=100$) likely introduces substantial numerical errors such as collisional heating shortly after the simulation begins.
The entire disk plane experiences sporadic perturbations from point-like massive DM particles, likely exacerbated by the small softening length of 0.03~kpc.
However, these perturbations do not grow continuously in $\mFsum$ due to the lack of sufficient and appropriate gravitational interactions between DM particles and stars (Fig.~\ref{fig:fmean2}).
Specifically, fewer than 30 particles reside within a $1~\kpc^3$ volume at $R=6~\kpc$ in model r1c16ldm with a mean particle separation of 0.34~kpc that exceeds the disk scale height.
Such sparse sampling substantially limits DM-star interactions, which in turn fails to mediate the appropriate dynamical friction from the DM halo that enables physical growth and winding of spiral arms.
Consequently, the angular momentum distribution map lacks a clear net-like pattern observed in other cases (see Fig. \ref{fig:lzmap2}).
\cite{ludlow21} indicated that systems with fewer than $N_{\rm{DM}}=10^6$ particles experience collisional heating of stars near the stellar half-mass radius, leading to artificial disk thickening. 
Our findings further suggest that simulating faint and small-scale structures, such as multiple spirals, requires enhanced spatial resolution in the DM halo and an appropriate DM-star mass ratio to mitigate these effects.

The DM halo resolution with $N_{\rm DM} \approx 10^6$ surpasses the threshold at which gravitational softening becomes critical.
Noise decreases with increased softening, but bias may increase, requiring careful selection \citep{merritt96, dehnen01}.
Accordingly, we evolve the same galaxy (model r1c16ldm) with a softening length 20 times larger, at 0.60 kpc, for the DM halo particles (see Appendix \ref{appendix:additional}).
However, the overall patterns in all Fourier modes do not differ substantially from those in the 0.03 kpc softening case.
Large perturbed patterns by strong gravitational kicks from massive particles persist, especially in the outer regions of the disk.
This suggests that spirals are faint structures such that a large mass ratio ($m_{\rm DM}/m_\star = 100$) artificially induces perturbations that form spirals much earlier.
Without awareness of these physical patterns or direct comparison to high-resolution simulations, randomly selecting a face-on snapshot at a given epoch may lead to visual misinterpretation, as if realistic spiral arms are forming in the disk.
Our findings demonstrate that even increasing gravitational softening cannot conceal significant numerical errors in spiral formation under low DM halo resolution.
We defer a parameter survey to identify the optimal $N_{\rm DM}$ and softening values for spiral arms originating from gravitational instability to future work.

\section{Conclusions}

We performed a suite of $N$-body simulations of Milky Way--like disk--halo systems, including two models evolved within fixed halo potentials and twelve with fully responsive dark-matter halos spanning a wide range of resolutions and disk stabilities. After evolving these systems for 2~Gyr in isolation, we examined the emergence, evolution, and interplay of multiple spiral modes. Our main conclusions are as follows:

\begin{enumerate}

\item 
\textit{Resolution affects timing but not the overall evolutionary path:} 
Increasing the number of disk and halo particles in a marginally unstable disk by a factor of ten delays the onset of visible spiral structure and bar formation by roughly $0.5$~Gyr and $1.9$~Gyr, respectively, but does not qualitatively alter the mode cascade or the eventual transition to a bar. High resolution primarily suppresses early Poisson-noise--driven features.

\item 
\textit{Disk stability regulates sensitivity to numerical noise:}  
More stable disks (higher $Q$) exhibit a delayed onset of measurable spirals, with the earliest features in such models dominated by initial Poisson fluctuations rather than physical perturbations.

\item 
\textit{A coherent mode cascade is a robust feature of well-resolved, live-halo disks:}  
In all responsive-halo models with $m_{\rm DM}/m_\star \le 10$, spiral activity proceeds systematically from higher to lower $m$ and from larger to smaller radii. This inward progression of the dominant mode reflects the combined effects of local swing amplification, evolving disk stability, and the inward drift of the locus of constructive interference among coexisting coherent patterns; non-linear mode coupling may also contribute, although our analysis does not establish it. More stable disks exhibit the same sequence but with slower evolution.

\item 
\textit{Angular-momentum redistribution reflects mode interference and bar--spiral coupling:}  
The repeated red--blue banding in $\Delta L_z(R,t)$ does not correspond to discrete episodes of spiral formation and decay. Rather, it arises from interference between multiple spiral modes with nearby pattern speeds, with additional modulation once the $m=2$ mode strengthens in the central region. The resulting net-like structure in $\Delta L_z$ is therefore a signature of multi-mode dynamics, not repeated swing-amplification events.

\item 
\textit{A fixed halo or low-resolution halo suppresses coherent mode cascading:}  
Models evolved in a rigid halo potential or with coarse halo resolution ($N_{\rm DM}=1.14\times10^6$, $m_{\rm DM}/m_\star=100$) exhibit strong early spirals driven by shot noise but do not sustain a coherent inward mode cascade. Their $\Delta L_z$ maps show irregular, vertically oriented torque patterns dominated by bar-driven or noise-driven fluctuations, reflecting the absence of a responsive halo capable of mediating the appropriate gravitational back-reaction.

\item 
\textit{Halo particle mass is a critical parameter:}  
Increasing halo resolution from $m_{\rm DM}/m_\star = 10$ to $1$ yields no significant change during the spiral-dominated phase. In contrast, a ratio of $m_{\rm DM}/m_\star = 100$ produces large-amplitude, irregular perturbations across all modes due to massive dark-matter particles acting as point-like perturbers. Even adopting a 20$\times$ larger softening length does not mitigate this behavior. These results highlight that an excessively large halo particle mass both over-amplifies noise-driven structure and fails to provide the smooth gravitational response required for physical mode growth.

\item 
\textit{The central \(m = 3\) mode provides the dynamical link between multi-armed spirals and bar formation:}
In all responsive-halo models, the inward mode cascade culminates in the growth of a centrally concentrated $m=3$ component, which becomes the dominant non-axisymmetric pattern inside $\sim 1-2$\,kpc for several rotation periods. Once established, this coherent $m=3$ distortion reorganizes the central stellar orbits and lowers the local stability threshold, enabling the rapid amplification of the $m=2$ mode. The $m=3$ component therefore promotes and accelerates the subsequent bar instability rather than being a prerequisite for it: bars form in rigid halos when the disk is sufficiently self-gravitating \citep{hohl71,ostriker73}, but at the disk-to-halo mass ratio adopted here our fixed-potential models show no central amplification of the $m=3$ mode and form no bar within $2\Gyr$.

\item
\textit{Among the spiral modes, only the central \(m = 3\) mode experiences significant dynamical friction from the halo:}
The higher-$m$ spiral patterns remain short-lived in the density field due to constructive and destructive interference among coexisting modes, and they primarily redistribute angular momentum among disk stars. As a result, they exert negligible net torque on the halo. In contrast, once the $m=3$ mode becomes confined to the inner kpc and persists for several rotation periods, it acts as a coherent non-axisymmetric perturber in a region of high dark-matter density. Under these conditions it raises a gravitational wake and experiences genuine dynamical friction, producing the first sustained loss of disk angular momentum and enabling the bar to grow. The bar that follows drives the larger part of the subsequent transfer; the role of the $m=3$ mode is to initiate it. Coarse-resolution and fixed-potential halos cannot support this coupling and therefore do not form bars in our simulations.

\end{enumerate}

Overall, our results demonstrate that a live halo with sufficiently high mass resolution is essential for capturing the coherent evolution of multiple spiral modes and their associated torque signatures. Although the total angular-momentum transfer during the spiral phase remains small, the halo's gravitational response mediates the mode cascade and enables the disk to transition naturally from multi-armed structure to bar formation. 
In fact, numerical factors play an important role in bar formation, affecting the fraction of bars in cosmological simulations as well. For instance, \cite{reddish22} reported a ``missing bar problem'' in the \textsc{NewHorizon} simulation \citep{dubois21}, which adopted $m_{\rm DM}/m_\star \approx 92$.
Minimizing numerical noise at a given resolution is indeed complicated and requires optimal softening to resolve the dynamics of galaxies \citep{white88, romeo94, romeo97, romeo98, steinmetz97, power03, rodionov05}.
In the follow-up paper, we extend our study to the effects of gravitational softening and the resolution of the halo on bar formation and the ensuing vertical instability \citep{kwak26a}.

In the process of realistic galaxy formation, internal and external perturbations arise naturally from multiple sources, driving disk heating and forming a thick disk \citep{minchev15, yi24}. Among the internal perturbations, stellar feedback mechanisms—such as stellar winds, supernovae, and radiative feedback—are expected to play a crucial role in triggering and amplifying multi-armed spirals in observed galaxies. Therefore, in addition to numerical effects, it is important to account for the impacts of gas dynamics and feedback mechanisms. In future work, we will revisit the features of multi-mode spirals using a multi-phase ISM model which includes a comprehensive treatment of stellar feedback, such as the SMUGGLE framework \citep{marinacci19, kwak26b, kwak26c}.

\bigskip
\bigskip
\bigskip

\begin{acknowledgements}
We thank the referees for their constructive comments. 
SK thanks Alessandro B. Romeo for helpful suggestions and discussions on the stability of 3D structures. 
SK appreciates Thomas R. Quinn for guidance on numerical setup during an undergraduate research project at the University of Washington, Seattle.
IM acknowledges support by the Deutsche Forschungsgemeinschaft under the grant MI 2009/2-1.
SKY acknowledges support from the Korean National Research Foundation (RS-2025-00514475; RS-2022-NR070872).
\end{acknowledgements}

\bibliographystyle{aa} 
\bibliography{ref}

\begin{appendix}
\section{Disk Ringing Effects in Model r2c24}\label{appendix:r2c24}
\begin{figure*}[htbp]
    \centering
    \begin{tabular}{@{}cccc@{}}
        \textbf{r2c24}  \\

        \raisebox{0.09\height}{\rotatebox{90}
        {\textbf{
        $\Fsum \qquad\quad\quad\: F_6 \qquad\quad\quad\:\:\:\:\: F_5 \qquad\quad\quad\:\:\:\:\:\: F_4 \qquad\quad\quad\:\:\:\:\:\: F_3 \qquad\quad\quad\:\:\:\:\:\: F_2 \qquad\quad\quad\:\:\:\:\: F_1$}}}
        \includegraphics[width=0.48\textwidth]{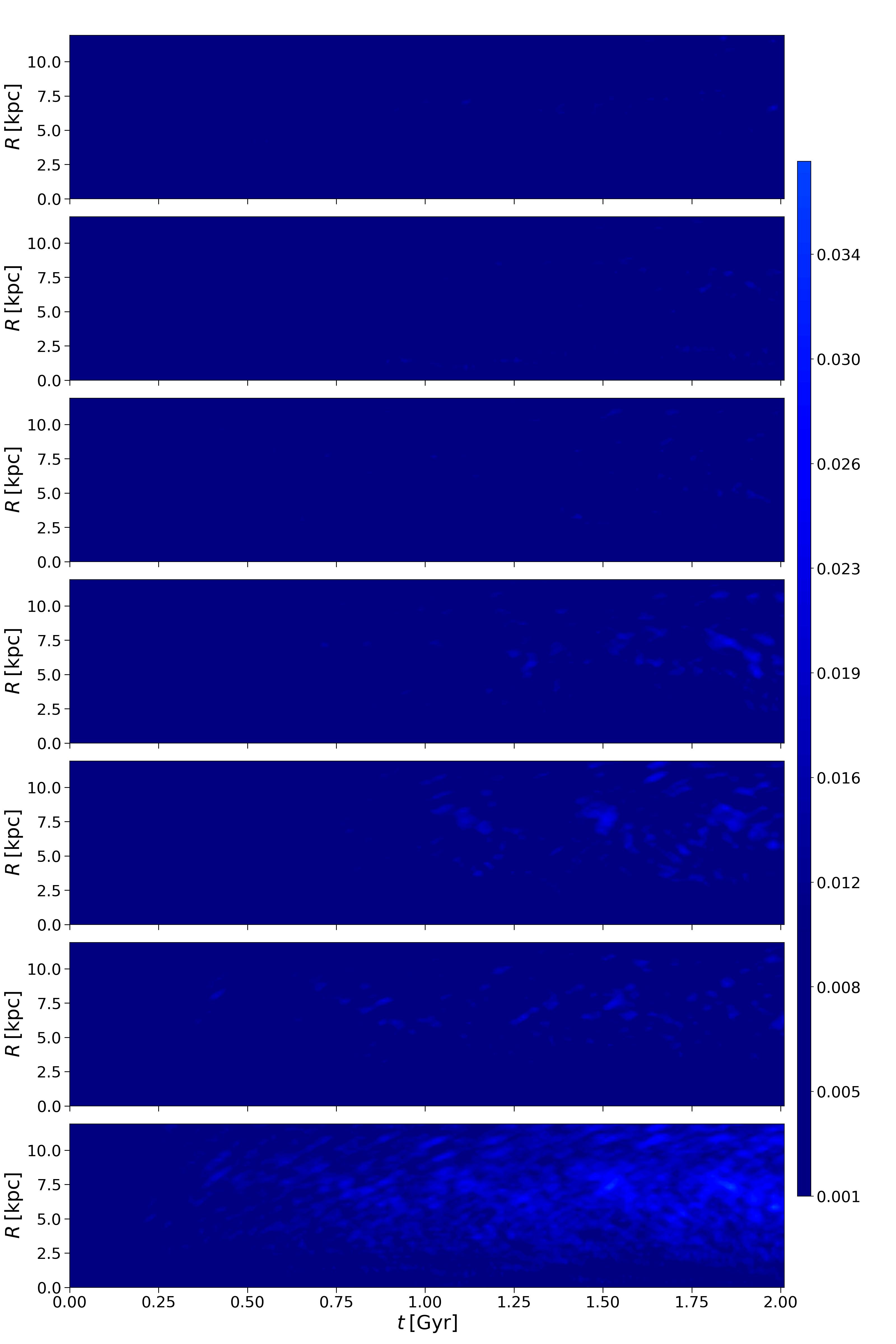}\includegraphics[width=0.48\textwidth]{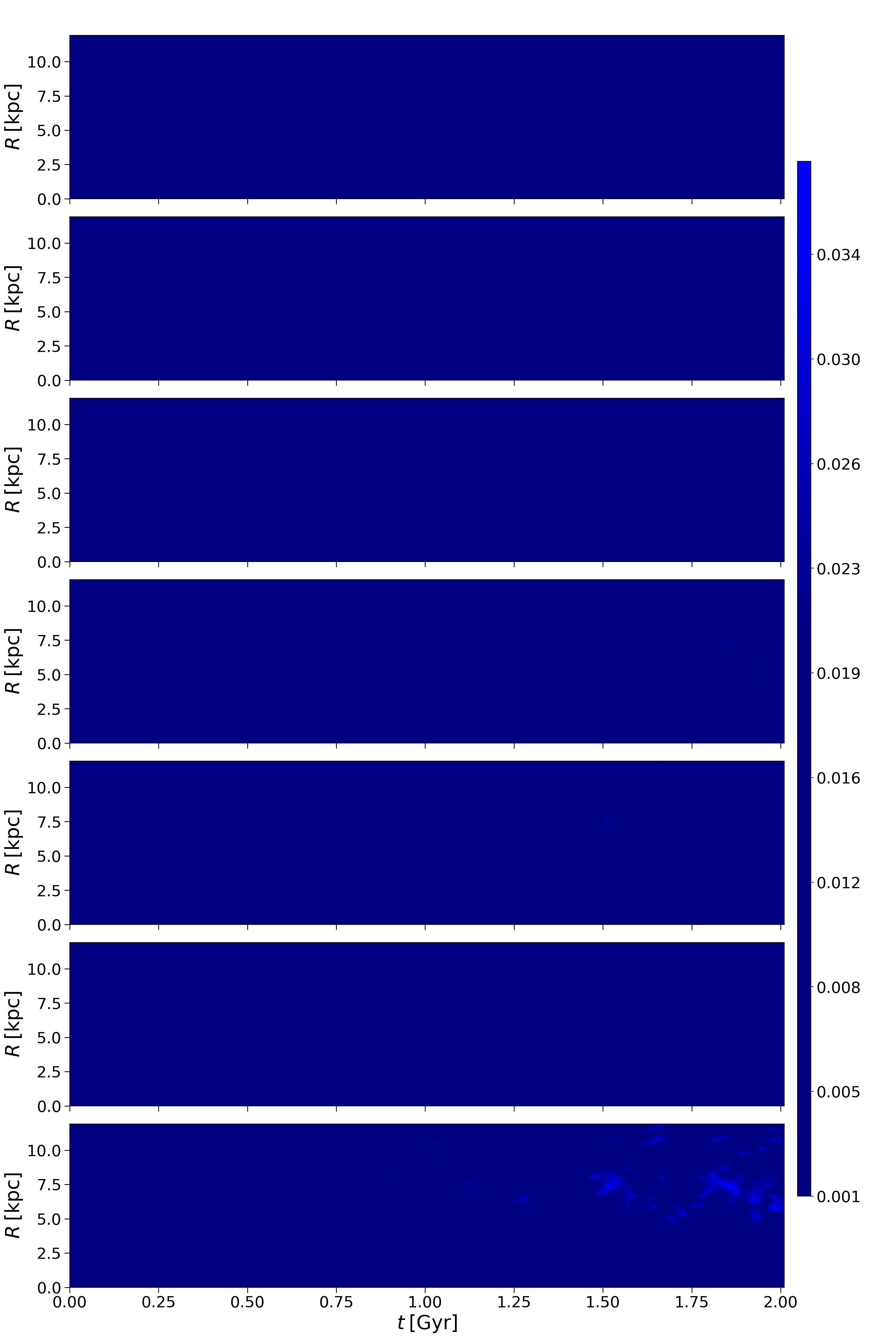}  
        
    \end{tabular}
    \caption{The same as Figure~\ref{fig:fmap1}, but for model r2c24 evolved for 2 Gyr. The minimum value of the color scale in the color bar is 0.01 and 0.02 for the left and right column, respectively.}
    \label{fig:fmap_r2c24}
\end{figure*}

\begin{figure}[htbp]
    \centering

    \includegraphics[width=0.48\textwidth]{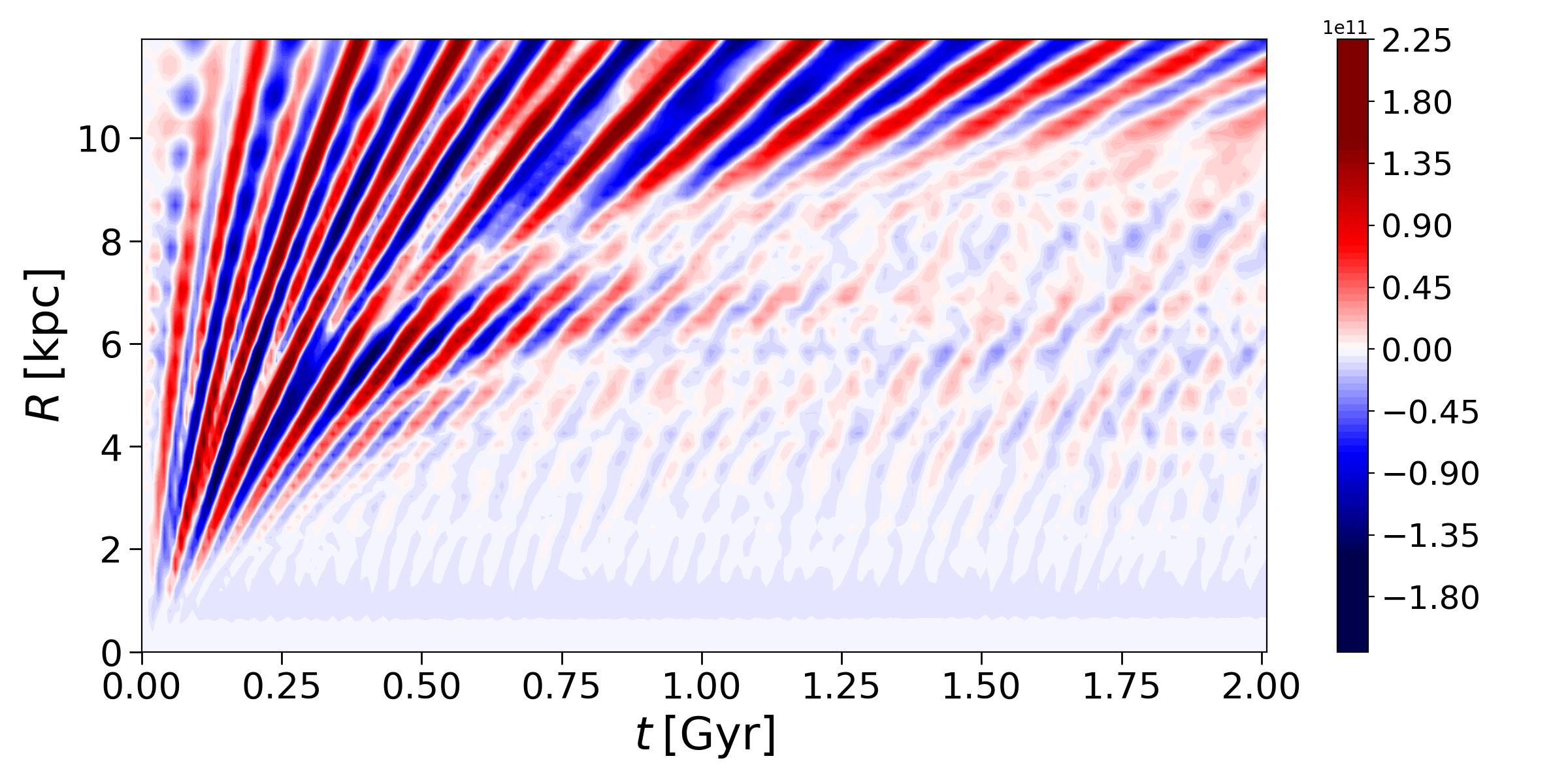} 

    \caption{The same as Figure \ref{fig:lzmap1}, but for models r2c24 evolved for 2 Gyr. \\}
    \label{fig:lzmap_r2c24}
\end{figure}

\begin{figure}[htbp]
    \centering

    \includegraphics[width=0.48\textwidth]{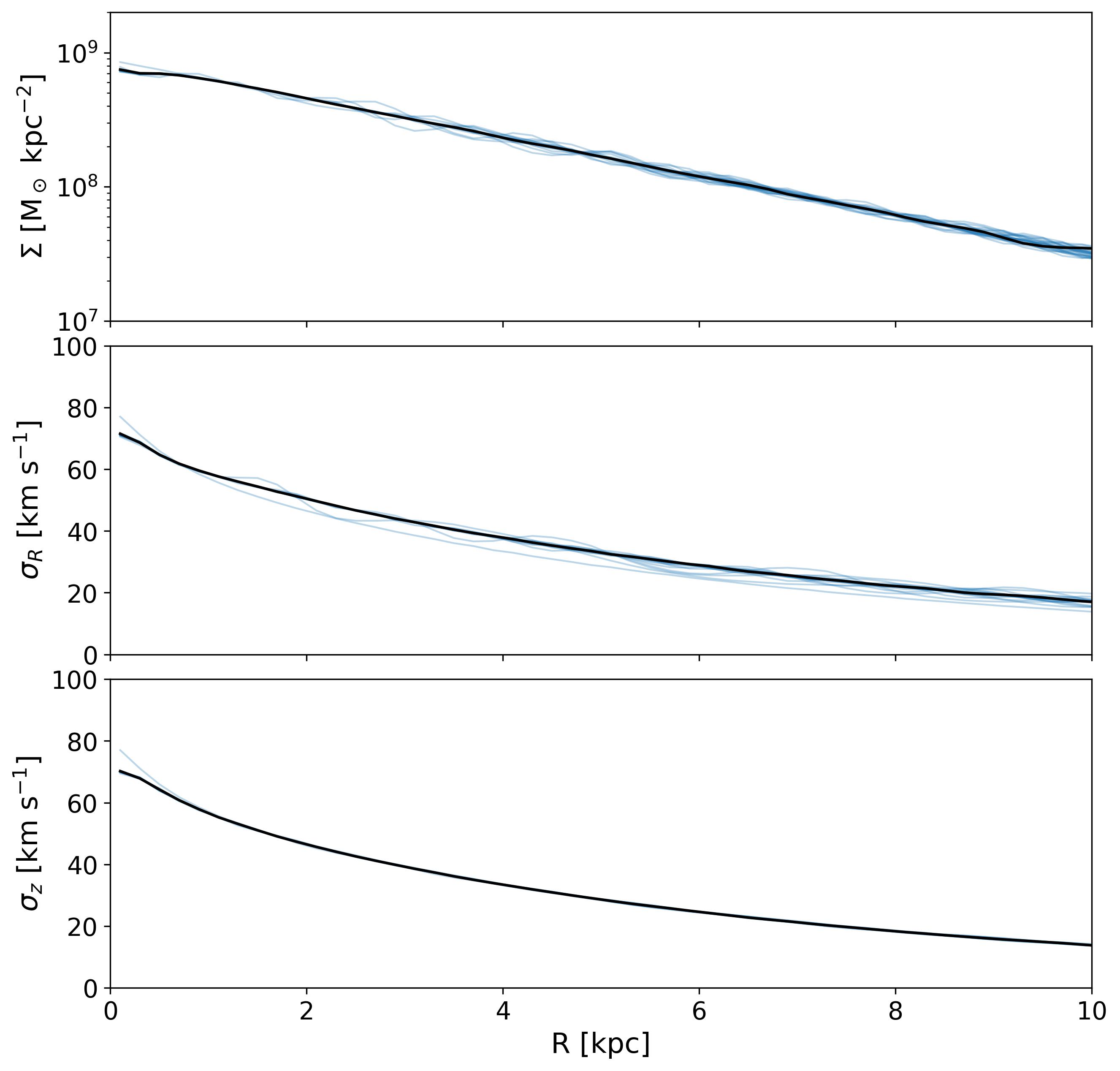} 

    \caption{Density profile, radial velocity dispersion, and vertical velocity dispersion in model r2c24. Radial profiles of these properties are drawn every 0.04 Gyr for 1 Gyr. The solid black line indicates the profiles at 1 Gyr.}
    \label{fig:den_vel_r2c24}
\end{figure}

The initialized disk-halo models are close to equilibrium but still show a small axisymmetric relaxation response shortly after the start of the simulation, which is well known as the ``disk ringing'' phenomenon (e.g., \citealt{minchev09}, \citealt{gomez12a}). This initial relaxation process is temporary and evidently does not trigger the emergence of spirals. For example, the $m=5$ and $6$ modes only gradually occur around 0.5 Gyr in model r2c20 (see Fig.~\ref{fig:fmap1}). To further support that the effect of disk ringing is negligible for spiral formation, we construct an additional model, r2c24, with higher DM contribution within the disk region that stabilizes the disk and thus delays the emergence of spirals. In Figure~\ref{fig:fmap_r2c24}, the left column shows the Fourier map of model r2c24 with the same color scale of values ranging from $1\%$ to $15\%$, consistent with other Fourier maps, while the right column uses $2\%$ to $15\%$. As any values below the minimum are saturated in dark blue, the comparison of these two columns indicates that no spiral mode larger than $2\%$ appears during 2 Gyr, and faint spiral modes below $2\%$ only appear around 0.75 and 1 Gyr for the $m=6$ and $5$ modes at $R\sim6~\kpc$. $\Fsum$ also appears faintly below $2\%$ around 0.5 Gyr. The comparison of model r2c24 with other r2 models in Figure~\ref{fig:fmap1} aligns with the sequence in which the emergence of spirals is delayed due to the enhanced disk stability, and the amplitudes of Fourier modes become weaker. This demonstrates that disk ringing causes neither rapid nor visible Fourier fluctuations. As model r2c24 mainly possesses only weak Fourier residuals below $2\%$ without showing the full sequence of inward mode cascading within 2 Gyr, the angular momentum map of model r2c24 in Figure~\ref{fig:lzmap_r2c24} does not exhibit the net-like or cross-hatched pattern produced by spiral torques in the main models (see Figs.~\ref{fig:lzmap1}). Furthermore, Figure~\ref{fig:den_vel_r2c24} shows that the corresponding density and velocity-dispersion profiles remain close to their initialized structure. We therefore conclude that the effects of the initial disk ringing are negligible for both the initial disk setup and the emergence of non-axisymmetric features in our models. Nevertheless, this should be tested over a wider range of numerical resolutions and disk stabilities, and with various initial condition generators.

\begin{figure*}[htbp]
    \centering
    \setlength{\tabcolsep}{1pt}   
    \renewcommand{\arraystretch}{0.1}  

    \begin{tabular}{@{}cccc@{}}
        \includegraphics[width=0.26\textwidth]{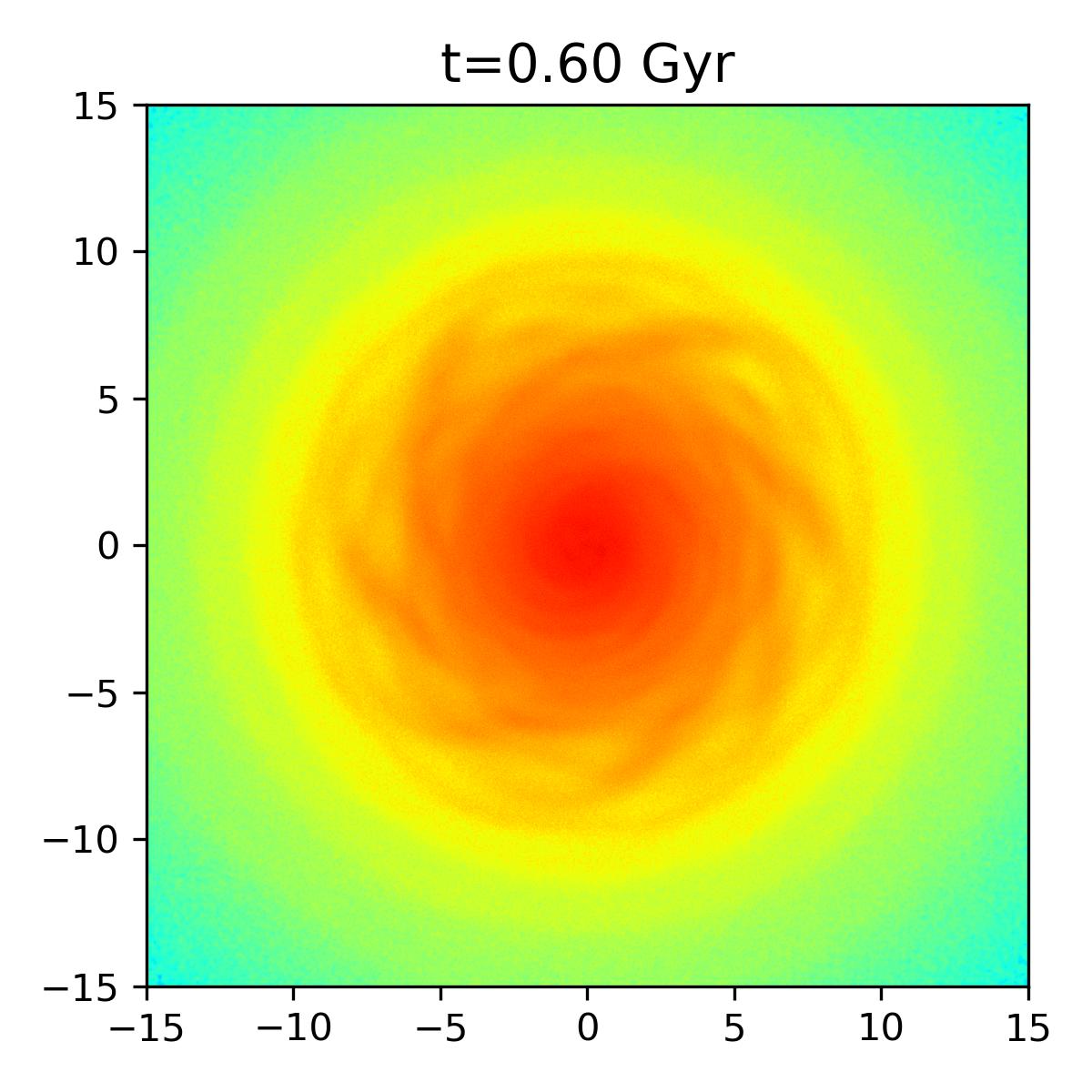} &
        \includegraphics[width=0.26\textwidth]{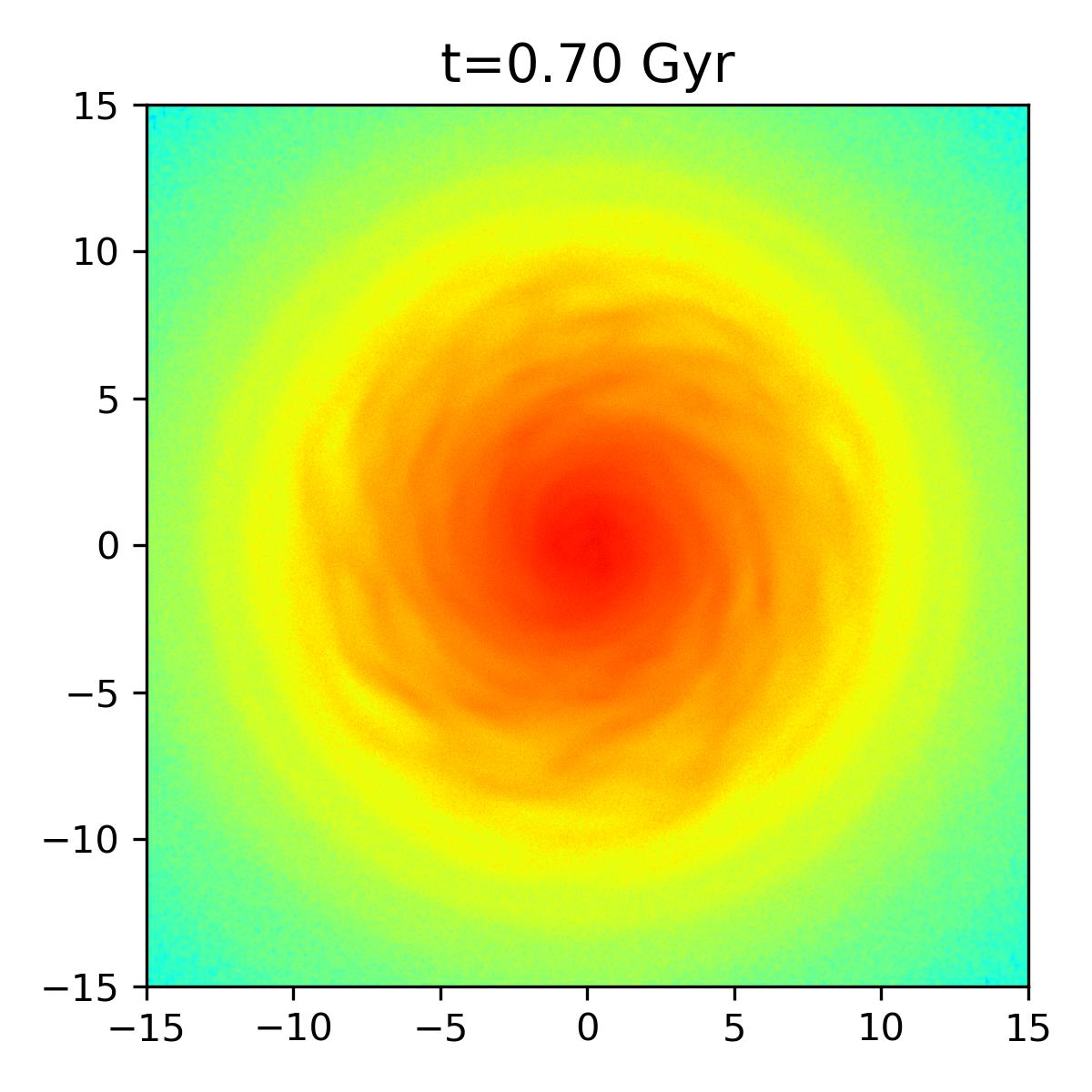} &
        \includegraphics[width=0.26\textwidth]{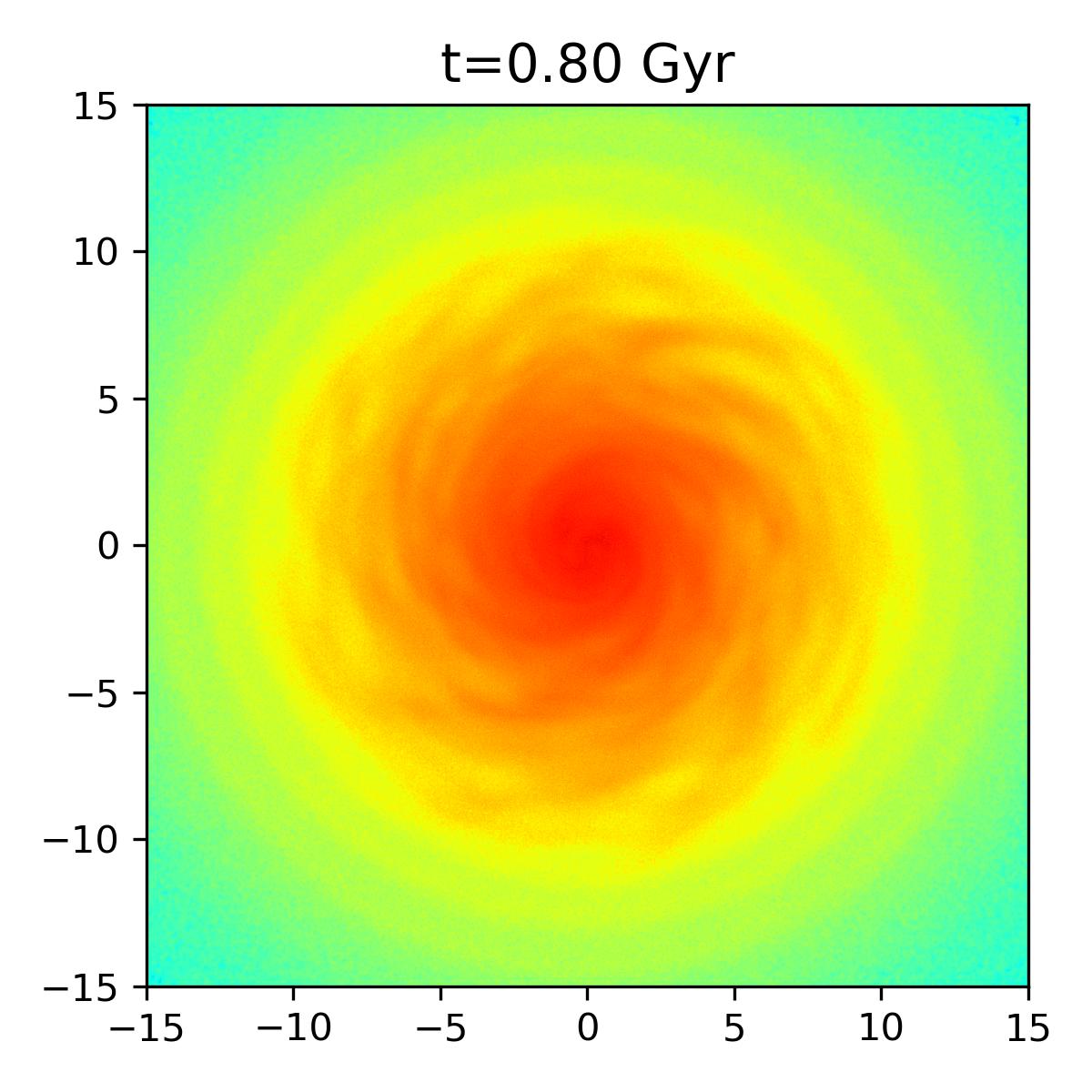} &
        \\[-5pt]

        \includegraphics[width=0.26\textwidth]{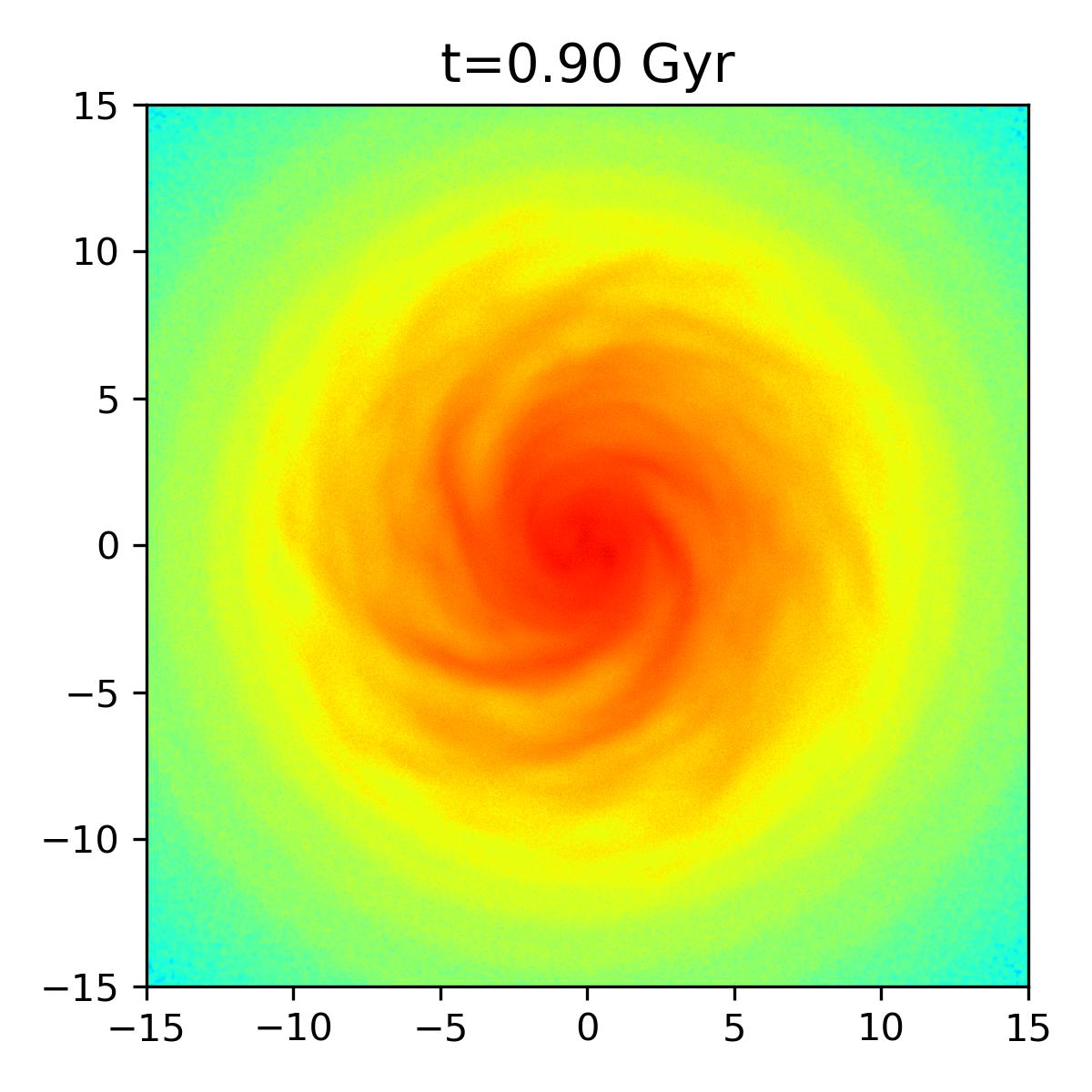} &
        \includegraphics[width=0.26\textwidth]{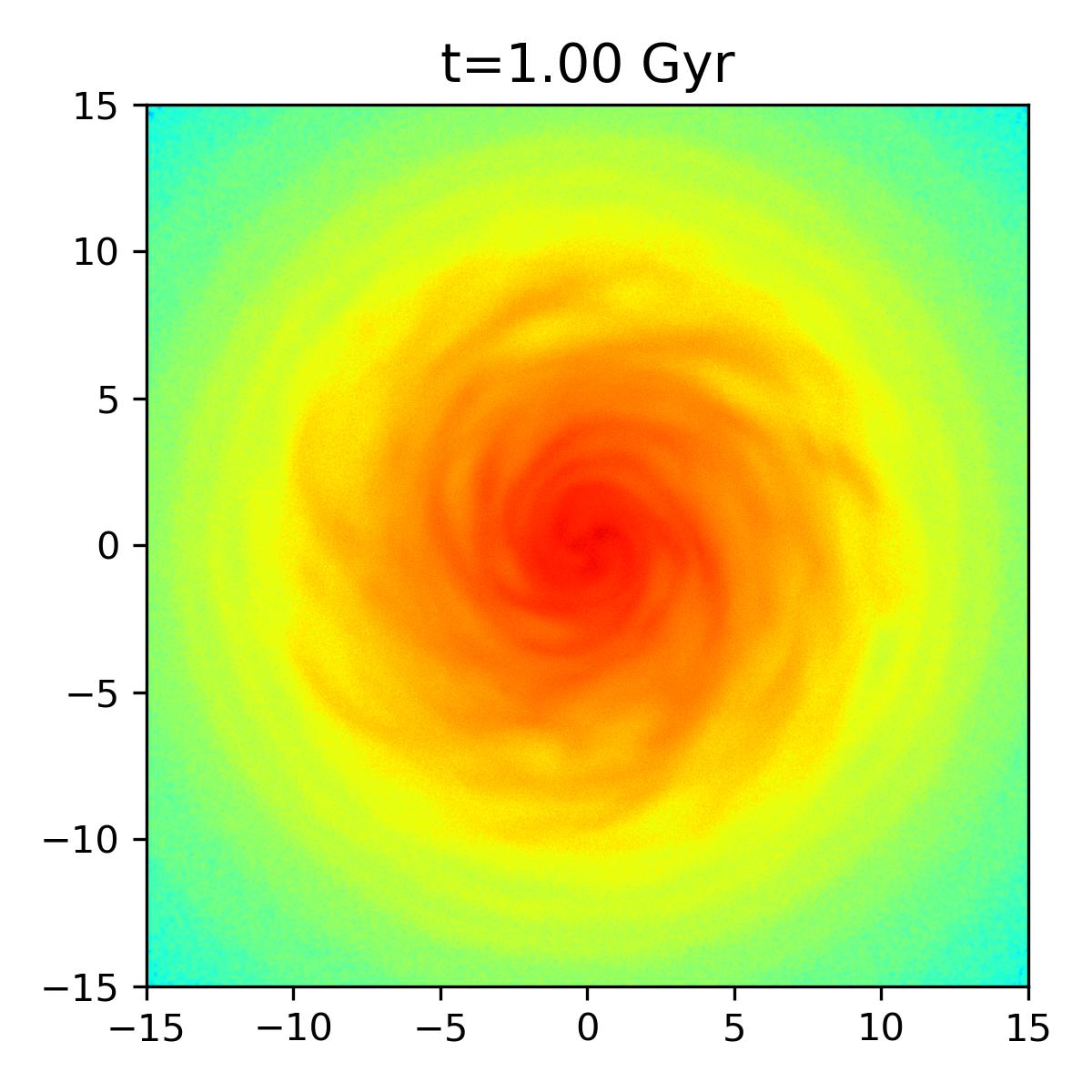} &
        \includegraphics[width=0.26\textwidth]{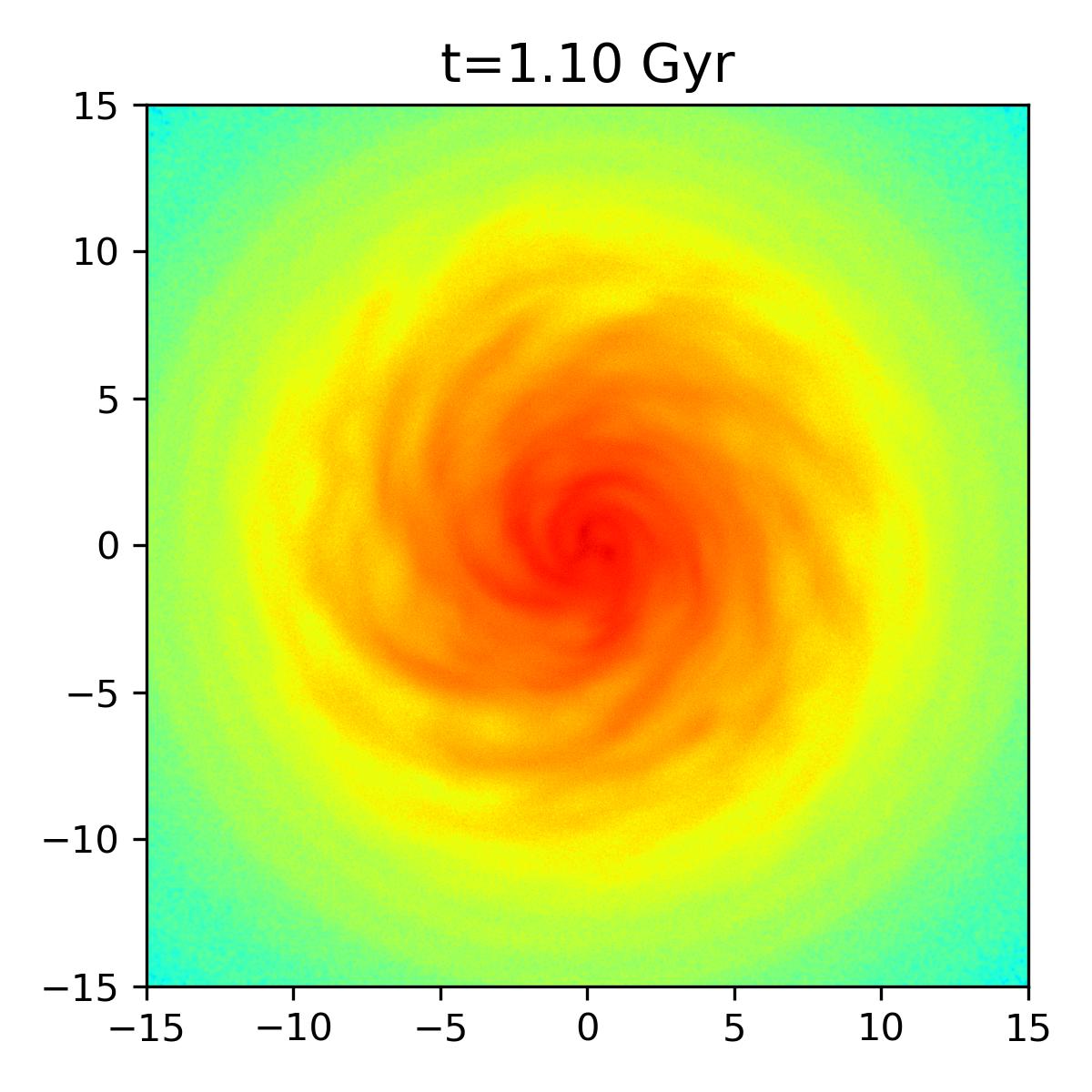} &
        \\[-5pt]

        \includegraphics[width=0.26\textwidth]{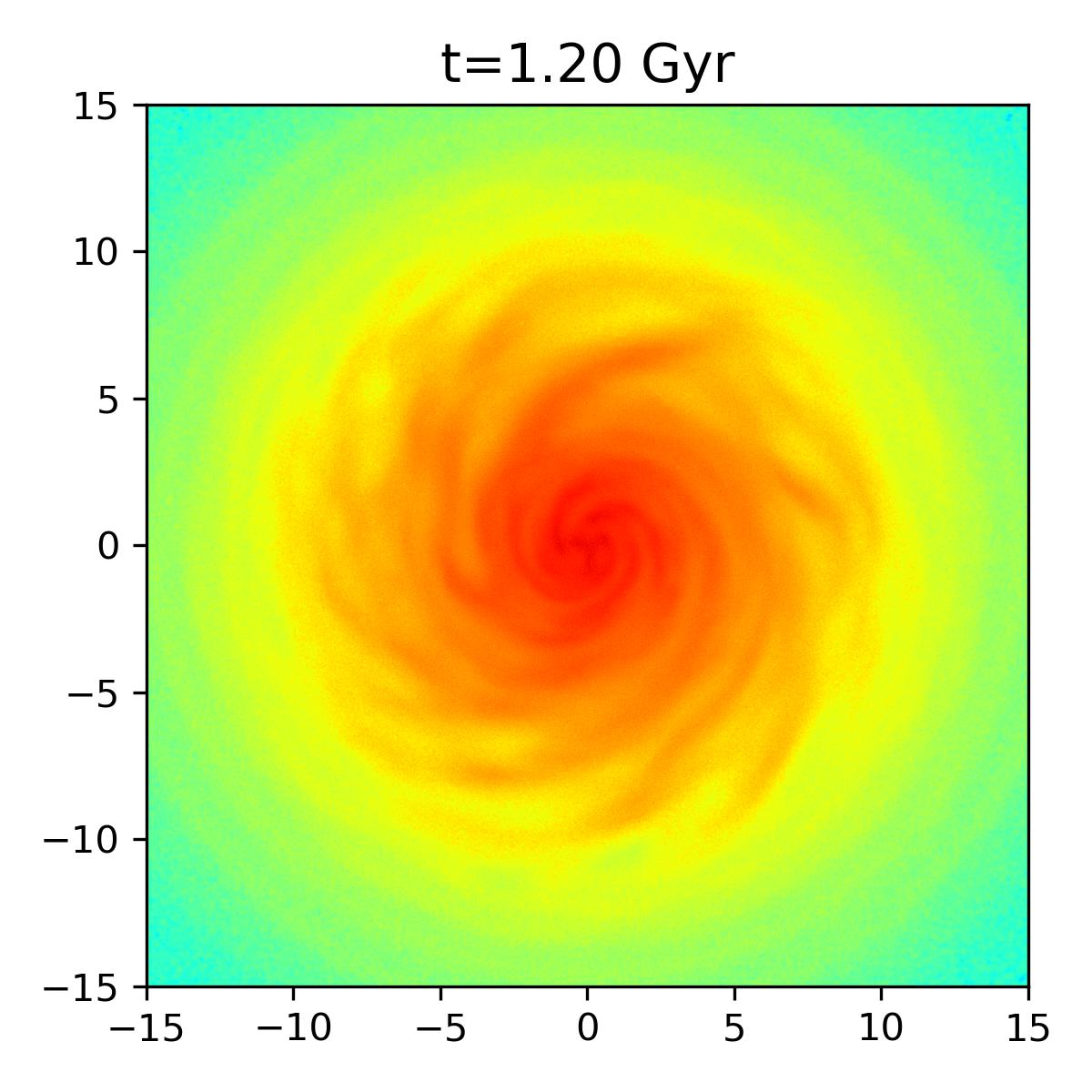} &
        \includegraphics[width=0.26\textwidth]{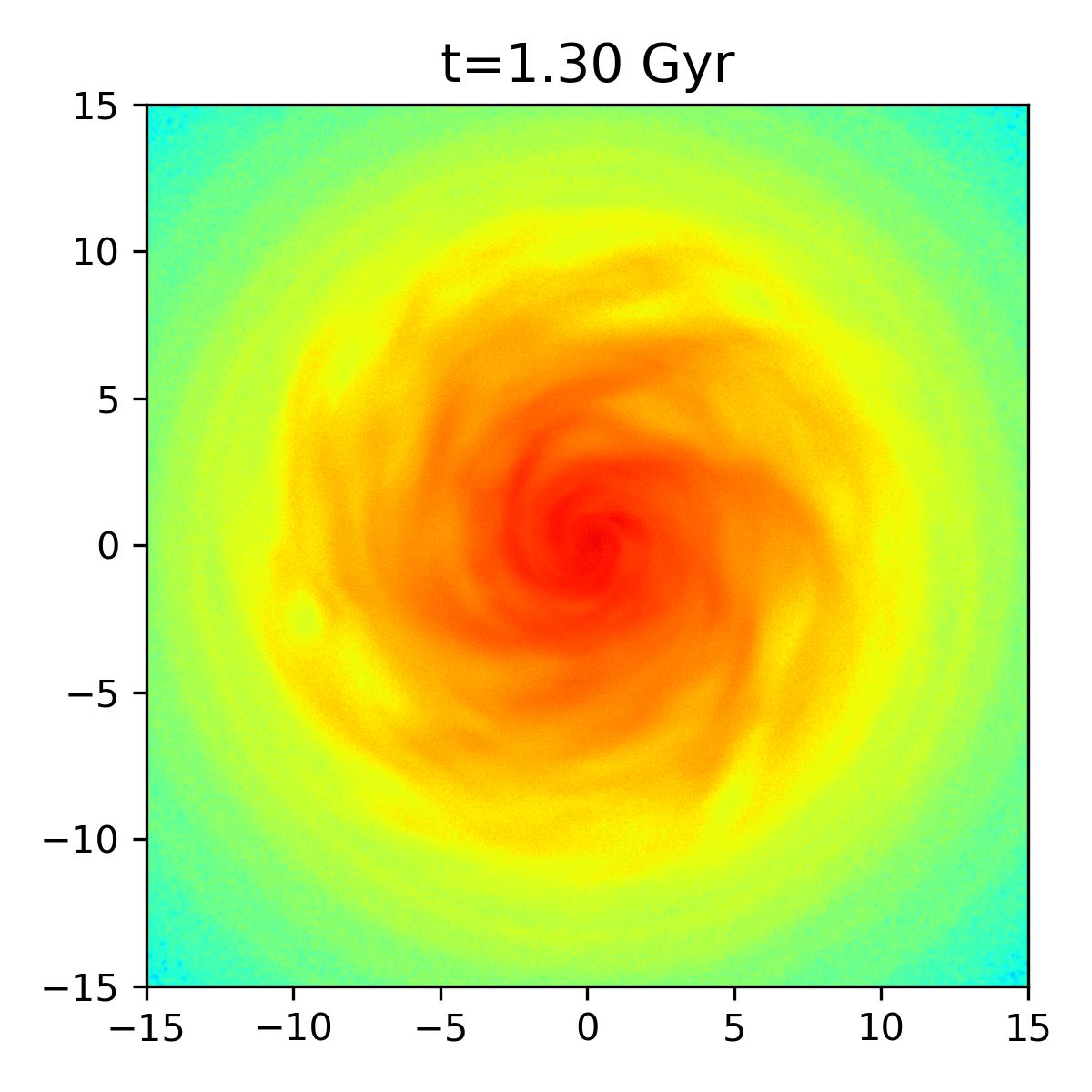} &
        \includegraphics[width=0.26\textwidth]{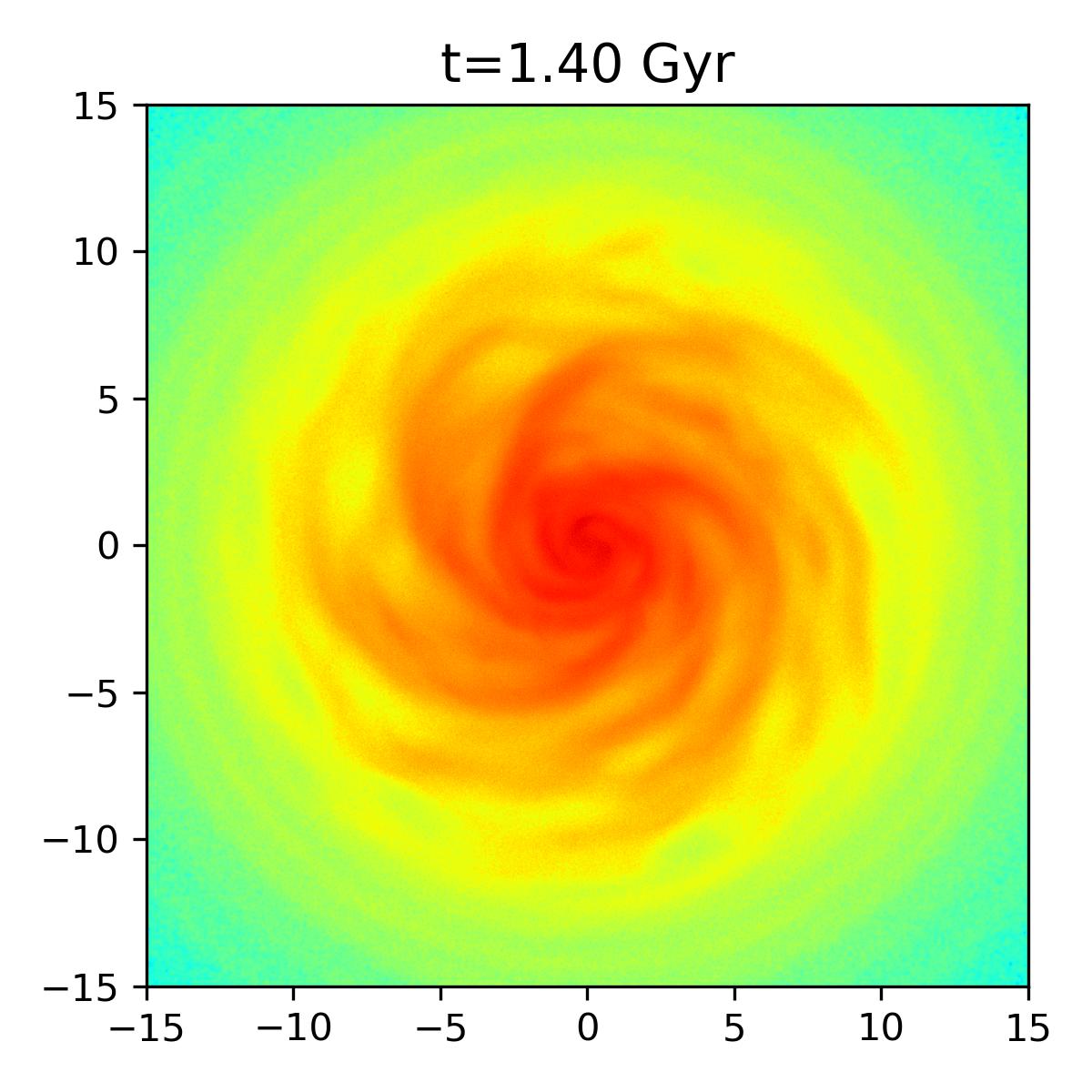} &
        \\[-5pt]

        \includegraphics[width=0.26\textwidth]{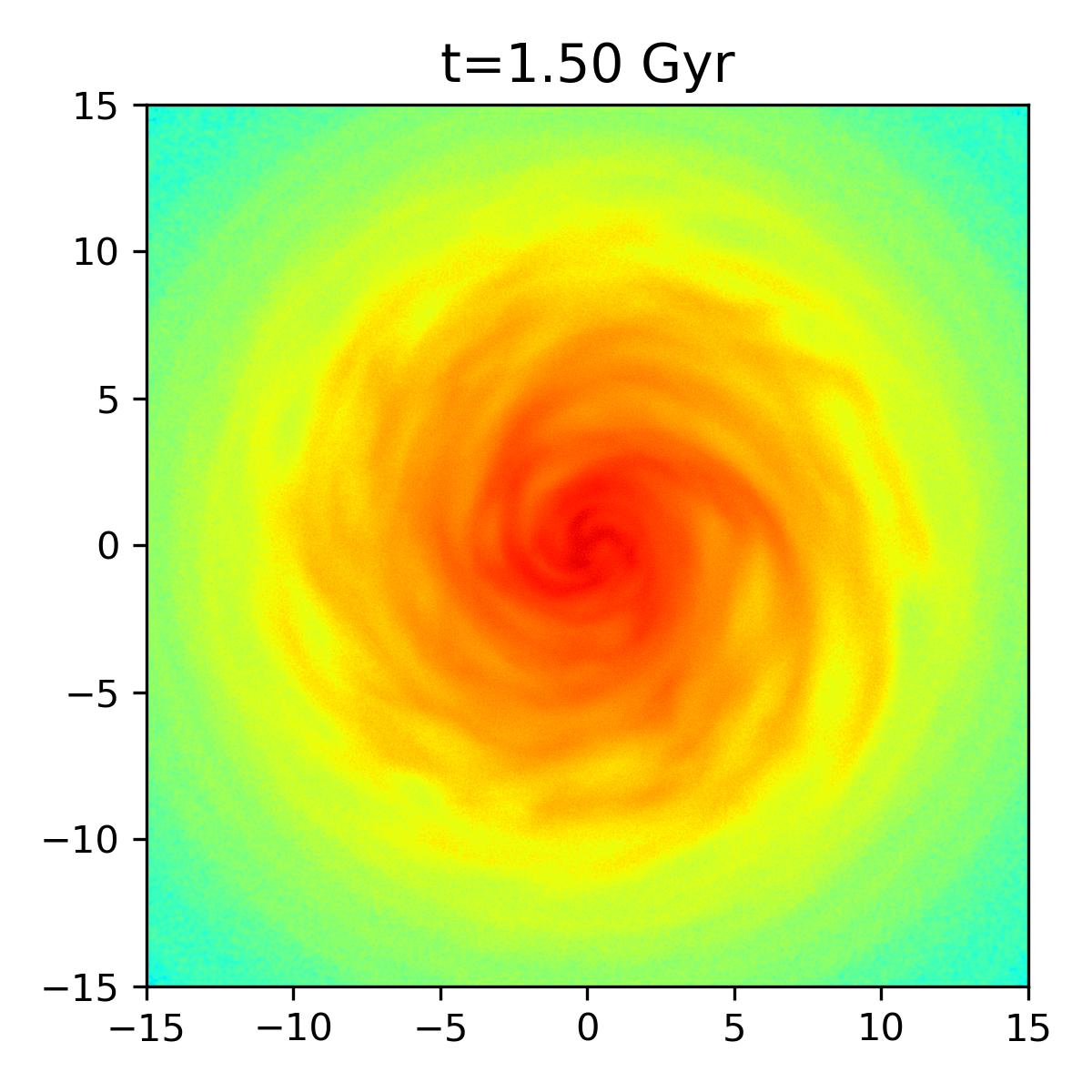} &
        \includegraphics[width=0.26\textwidth]{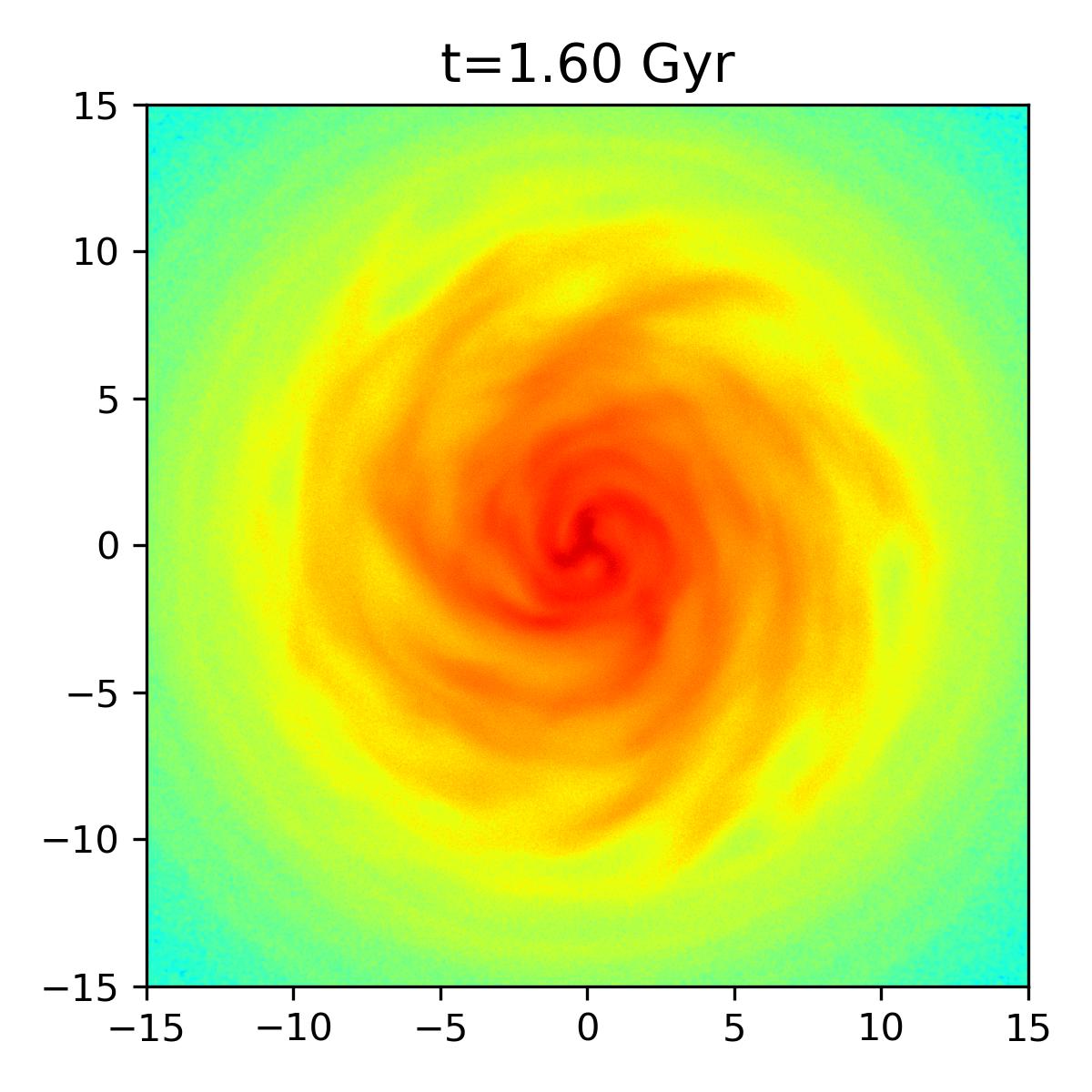} &
        \includegraphics[width=0.26\textwidth]{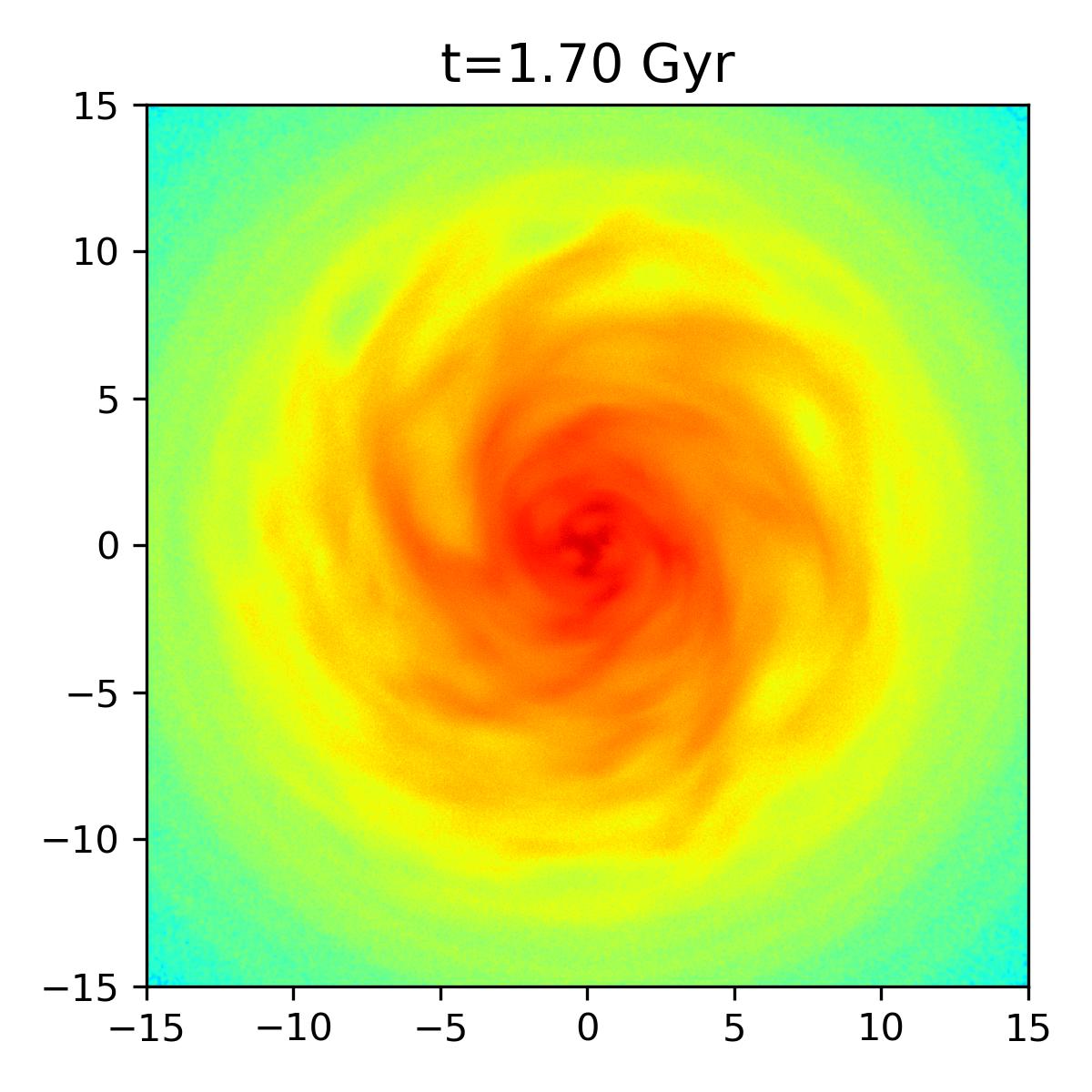} &
        \\[-5pt]

        \includegraphics[width=0.26\textwidth]{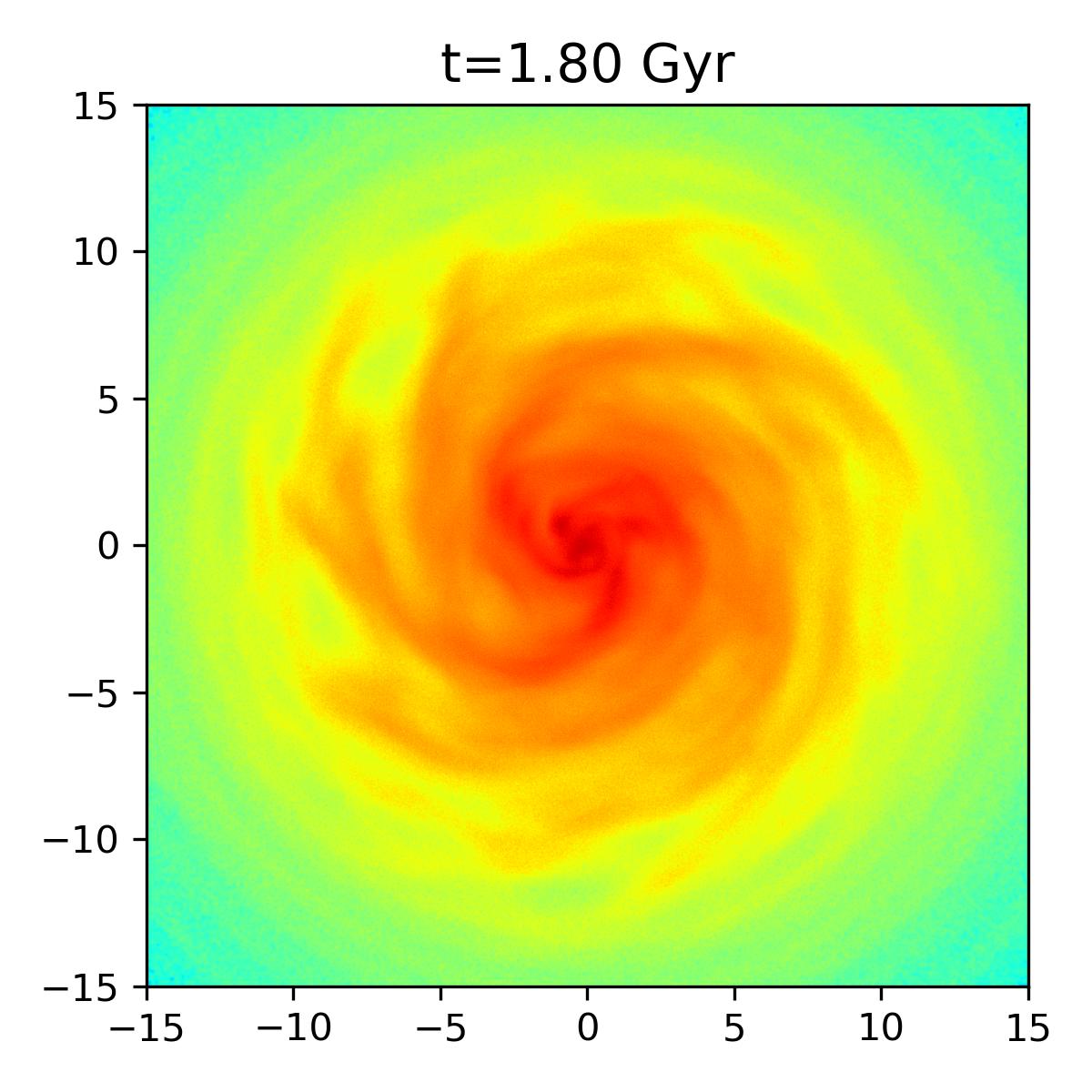} &
        \includegraphics[width=0.26\textwidth]{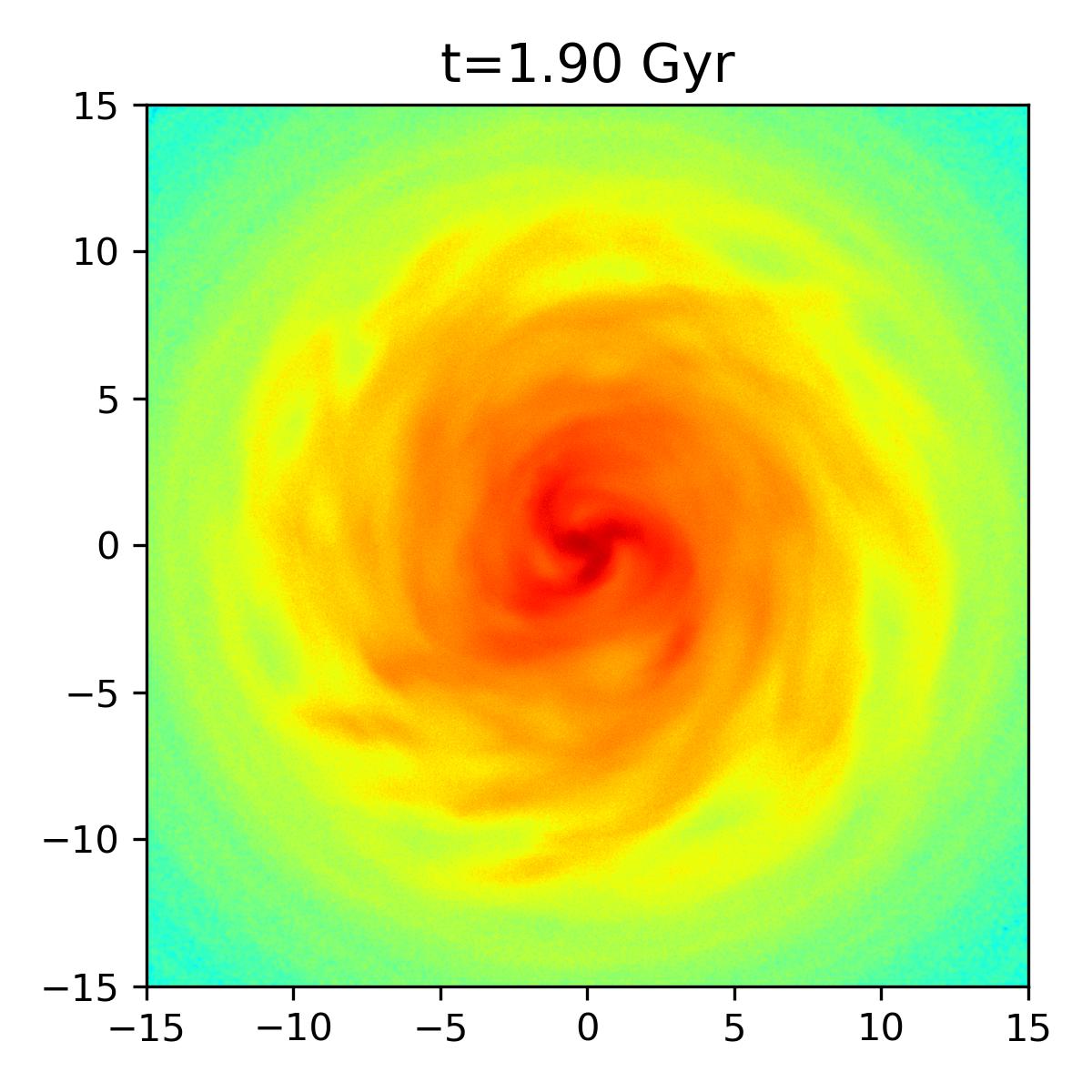} &
        \includegraphics[width=0.26\textwidth]{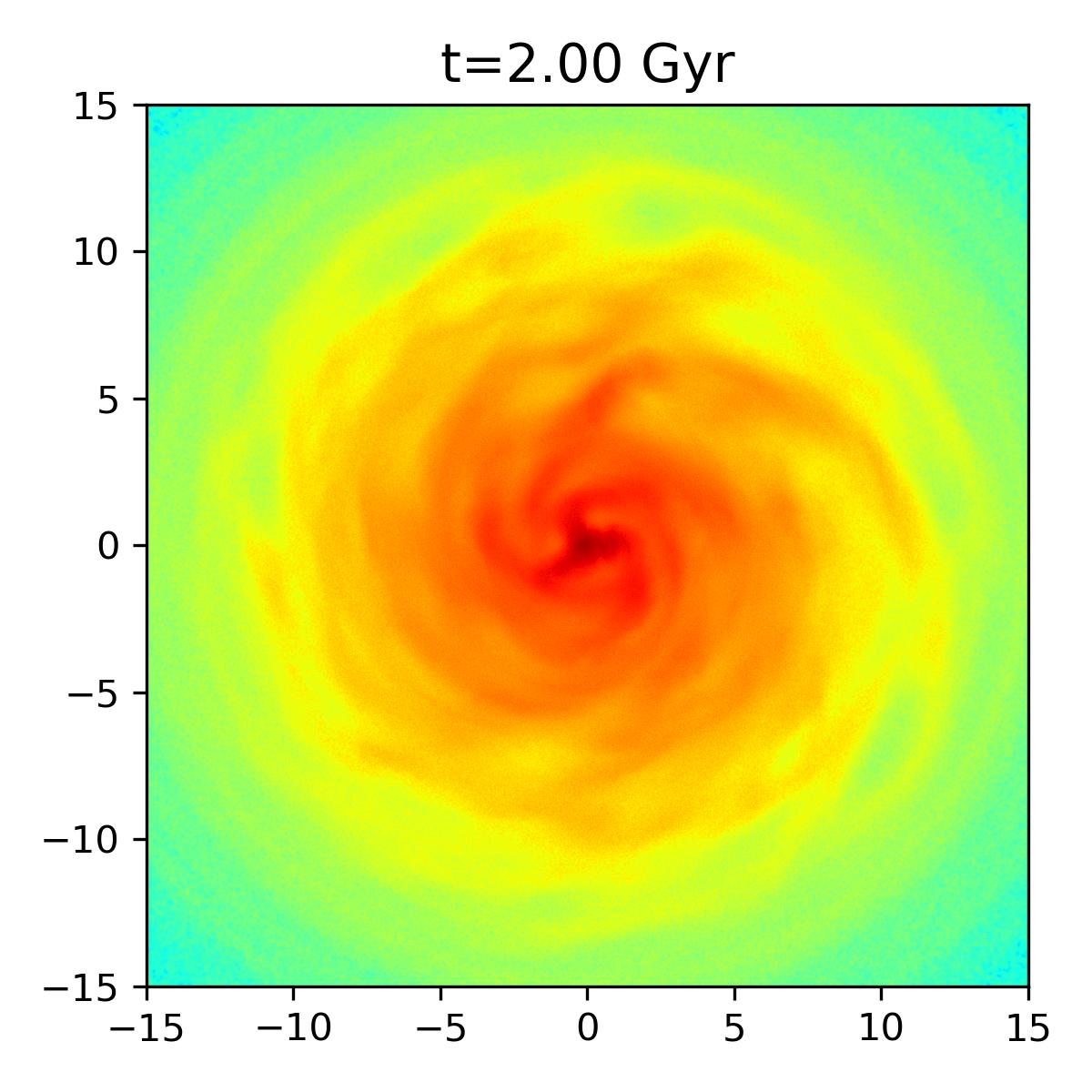} &
        \includegraphics[height=0.25\textwidth]{fig/colorbar.jpg} \\
    \end{tabular}

    \caption{Face-on projections of the stellar surface density distribution in a $30 \times 30$ kpc box from 0.6 to 2.0 Gyr for model r2c16, {\bf in steps of 0.1 Gyr, increasing from left to right and top to bottom}. The color bar indicates the surface density in units of $\Msun\,\mathrm{kpc}^{-2}$.}
    \label{fig:faceon_r2c16_all}
\end{figure*}

\section{Additional Fourier Map}\label{appendix:additional}

\begin{figure*}[htbp]
    \centering
    \begin{tabular}{@{}cccc@{}}
        \textbf{r1c18} & \textbf{r1c20} & \textbf{r1c14fdm} & \textbf{r1c16ldmsf06} \\

        \raisebox{0.1\height}{\rotatebox{90}
        {\textbf{
        $\Fsum \qquad\quad\quad F_6 \qquad\quad\quad\:\: F_5 \qquad\quad\quad\:\:\: F_4 \qquad\quad\quad\:\:\:\: F_3 \qquad\quad\quad\:\:\:\: F_2 \qquad\quad\quad\:\:\:\: F_1$}}}
        \includegraphics[width=0.24\textwidth]{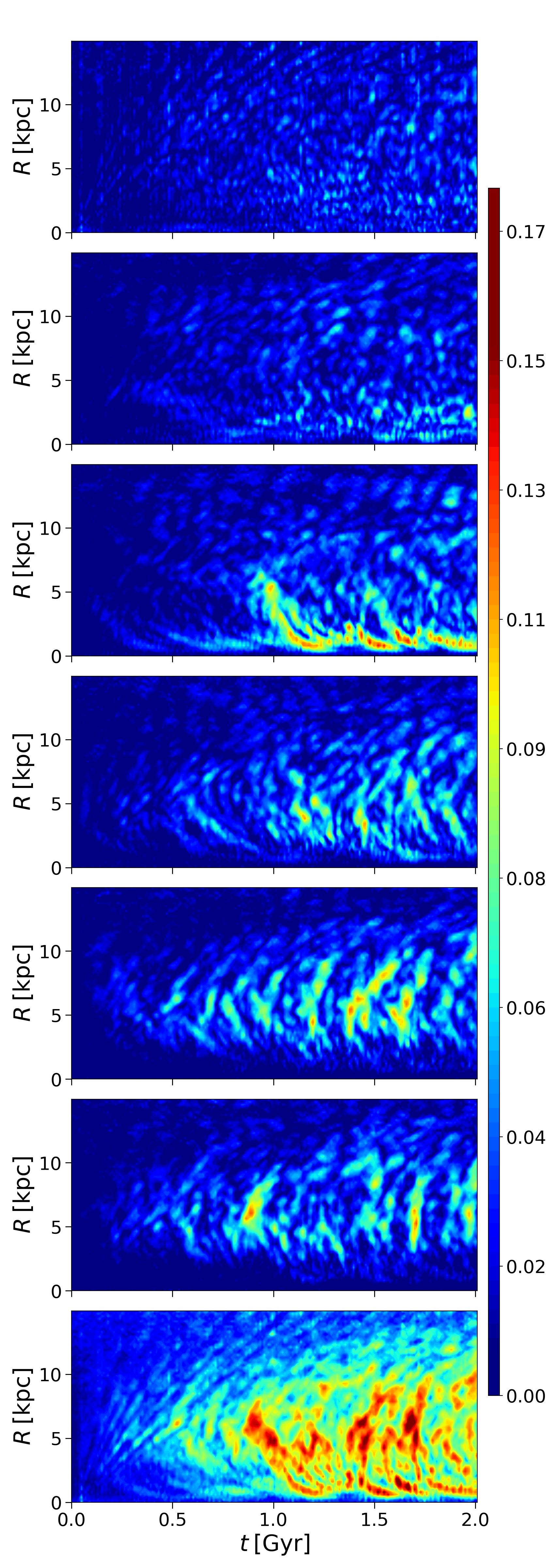} &
        \includegraphics[width=0.24\textwidth]{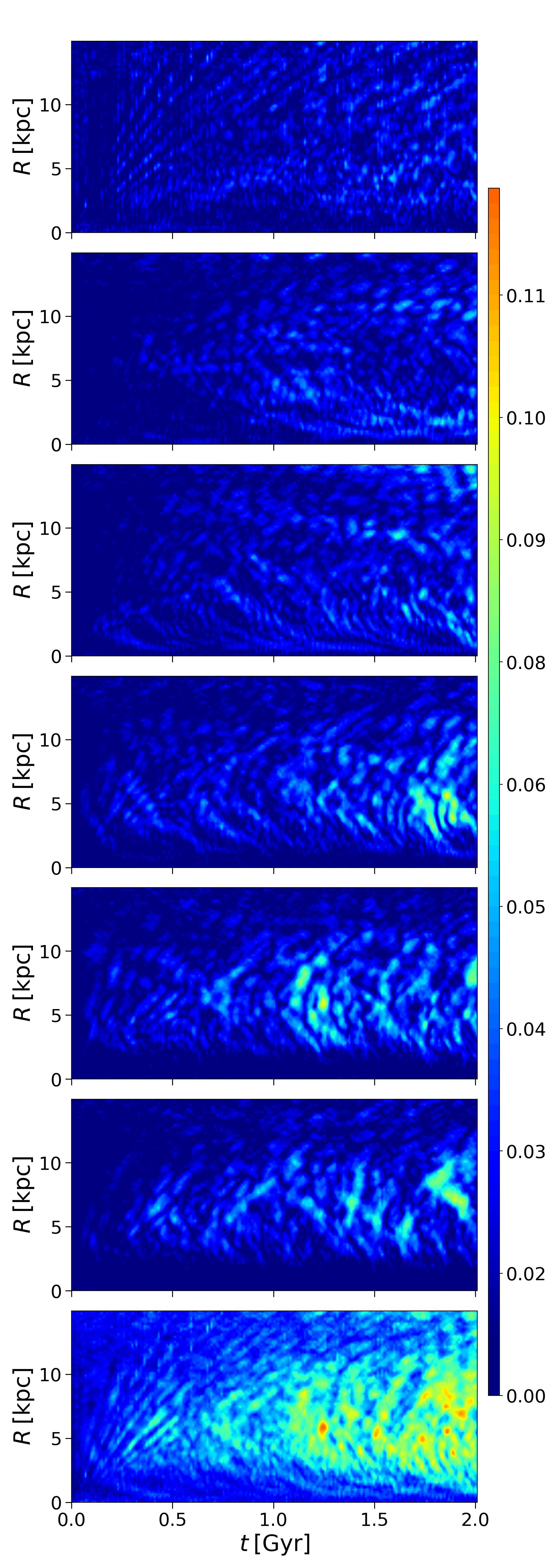} &
        \includegraphics[width=0.24\textwidth]{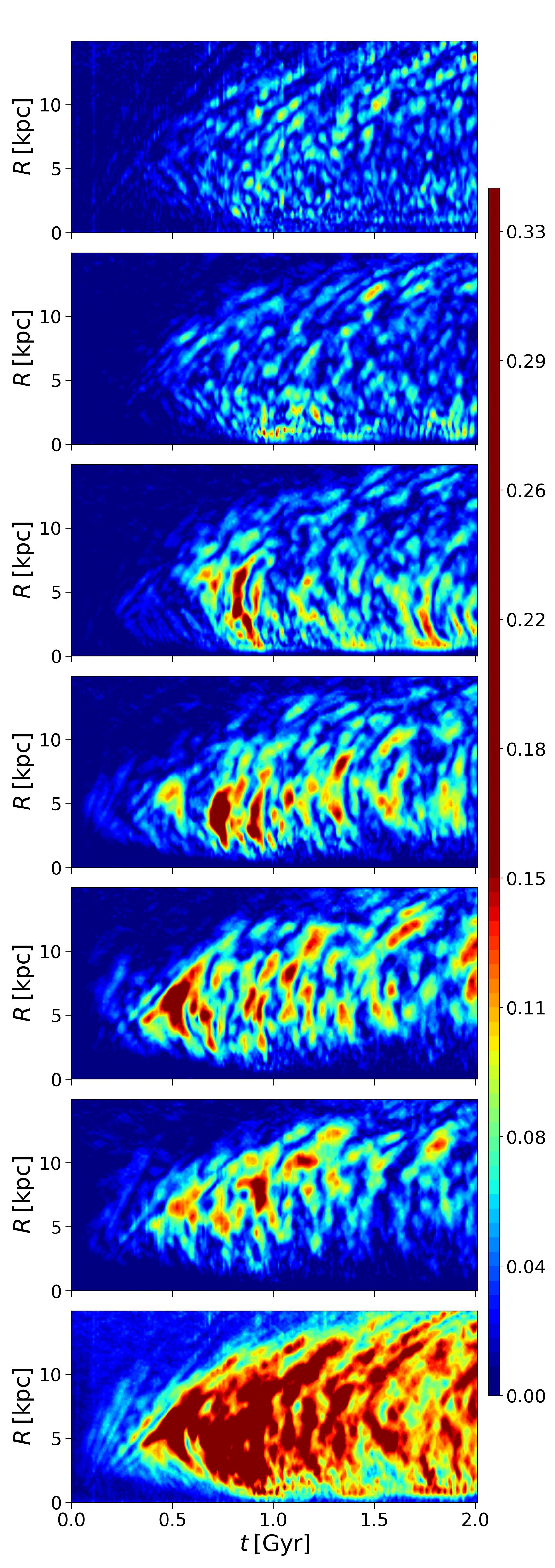} &
        \includegraphics[width=0.24\textwidth]{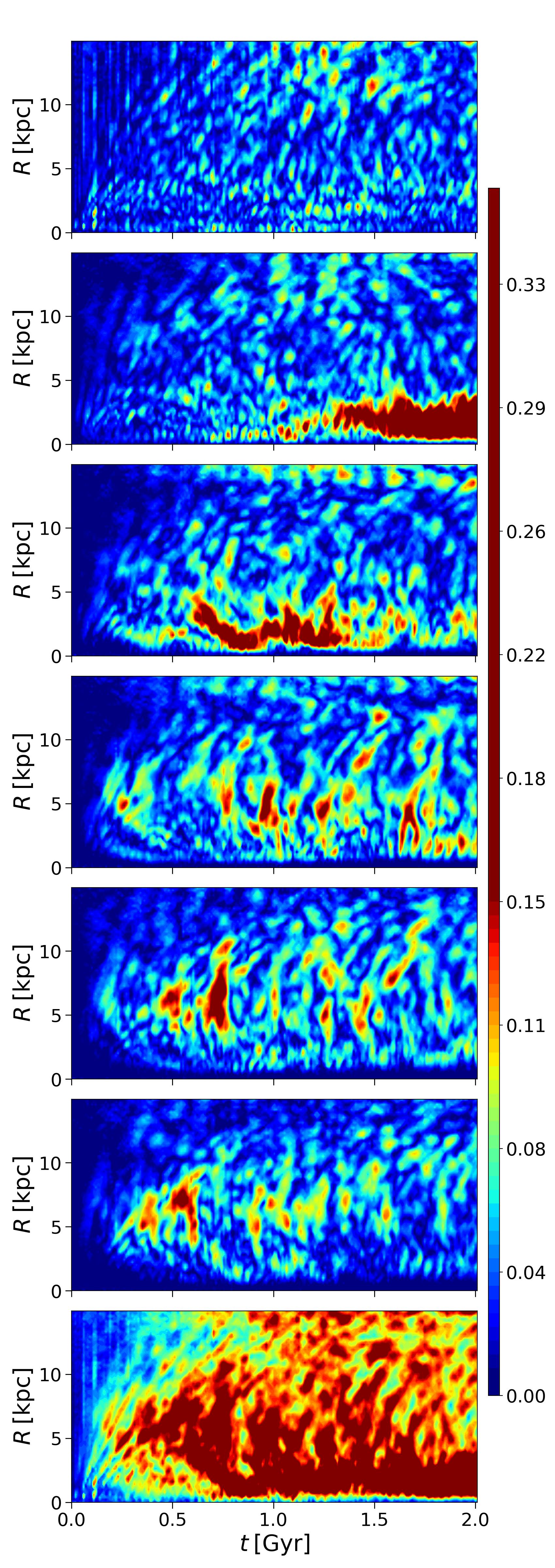}
    \end{tabular}
    \caption{The same as Figure \ref{fig:fmap1}, but for models r1c18, r1c20, r1c14fdm, and r1c16ldmsf06.}
    \label{fig:additional}
\end{figure*}

We present the Fourier maps for models r1c18 and r1c20 to examine the effects of resolution in comparatively more stable systems.
Notably, even in lower-resolution models, the cascading of Fourier modes and spiral epicenters remains visible, as this process unfolds more gradually in more stable models.
In model r1c18, the spirals form faintly and recur in a quasi-periodic manner, without exhibiting visible migration or strong amplification as seen in unstable models.

We additionally simulate model r1c14 using a fixed potential and designate it as r1c14fdm.
Model r1c16fdm does not exhibit continuous amplification of the $m=3$ mode in the central region.
To confirm that this is not due to stability, we simulate a more unstable model and verify that continuous amplification within the central 1 kpc is absent in fixed-potential models, irrespective of stability.

$N$-body simulations often adopt different softening lengths for stellar and DM halo particles, corresponding to their differing mass resolutions, primarily to minimize computational costs for DM halo particles.
Given the substantial deviations observed in model r1c16ldm, we resimulate this model with an increased softening length from 0.03 kpc to 0.6 kpc.
As shown in Figure~\ref{fig:additional}, this adjustment in model r1c16ldmsf06 does not produce notable differences in spiral structures compared to those in model r1c16ldm presented in Figure~\ref{fig:fmap2}.

\section{Extended Evolution and Spectrogram Analysis}\label{appendix:spectrogram}

The central $m=3$ component is visible in the radial Fourier profiles (Fig.~\ref{fig:comparison_grid}), the face-on density maps (Fig.~\ref{fig:faceon}), and the Fourier maps (Figs.~\ref{fig:fmap1} and \ref{fig:fmap2}). Its appearance also coincides with changes in the angular-momentum maps (Figs.~\ref{fig:lzmap1} and \ref{fig:lzmap2}) and with the onset of measurable disk angular-momentum loss before bar formation (Fig.~\ref{fig:t_lz}). To test whether this component behaves as a coherent rotating mode rather than as a short-lived interference feature, we examine extended simulations and compute spectrograms for the $m=2$ and $m=3$ components.

We first extend model r2c16 to 4 Gyr. This model is useful because the mode cascade proceeds gradually: the central $m=3$ component appears around $t\simeq1.5\,\Gyr$, the $m=2$ component strengthens after $t\simeq2\,\Gyr$, and the bar forms at $t=3.56\,\Gyr$ when $F_{2,\rm max}>0.3$. We also analyze an additional, more stable extended model, r1c22, evolved to 10 Gyr. This model allows us to follow the central $m=3$ phase for a longer time before the $m=2$ mode becomes dominant.

The pattern speeds are measured from the time evolution of the complex Fourier coefficient $F_m(R,t)$, defined as in equation~(\ref{eq:fourier}) but evaluated at each snapshot. The amplitude and phase of each mode are
\begin{eqnarray}
\bigl|F_m(R,t)\bigr| &=& \sqrt{\operatorname{Re}(F_m)^2 + \operatorname{Im}(F_m)^2}, \\
\phi_m(R,t) &=& \mathrm{atan2}\bigl(\operatorname{Im}(F_m),\operatorname{Re}(F_m)\bigr).
\end{eqnarray}
The instantaneous pattern speed is then calculated as
\begin{equation}
\Omega_{p,m}(R,t) = -\frac{1}{m} \frac{\mathrm{d}\phi_m}{\mathrm{d}t}(R,t).
\end{equation}
We build spectrograms in the $(R,\Omega_p)$ plane using only regions where the average Fourier amplitude exceeds the adopted threshold, $|F_m|>0.09$. For model r2c16, we use overlapping time windows of $\Delta T=0.20\,\Gyr$ from $t=1.4$ to $4.0\,\Gyr$. For the longer r1c22 run, we use $\Delta T=0.50\,\Gyr$ from $t=3.5$ to $10.0\,\Gyr$, which is better suited to its slower evolution. We overlay the resonance curves $\Omega$ and $\Omega\pm\kappa/m$, where $\Omega$ is the angular velocity and $\kappa$ is the epicyclic frequency. Persistent horizontal ridges in these diagrams indicate density-wave structures that maintain a nearly constant pattern speed, whereas winding or shearing features would not remain confined to a narrow range of $\Omega_p$. We restrict the spectrogram analysis to the epoch in which the central $m=3$ mode and the later $m=2$ bar dominate the inner disk. A mode-by-mode decomposition of the earlier multi-armed phase is beyond the scope of the present paper.

\begin{figure*}[htbp]
    \centering
    \begin{tabular}{@{}cccc@{}}
        \textbf{r2c16}  \\

        \raisebox{0.09\height}{\rotatebox{90}
        {\textbf{
        $\Fsum \qquad\quad\quad\: F_6 \qquad\quad\quad\:\:\:\:\: F_5 \qquad\quad\quad\:\:\:\:\:\: F_4 \qquad\quad\quad\:\:\:\:\:\: F_3 \qquad\quad\quad\:\:\:\:\:\: F_2 \qquad\quad\quad\:\:\:\:\: F_1$}}}
        \includegraphics[width=0.48\textwidth]{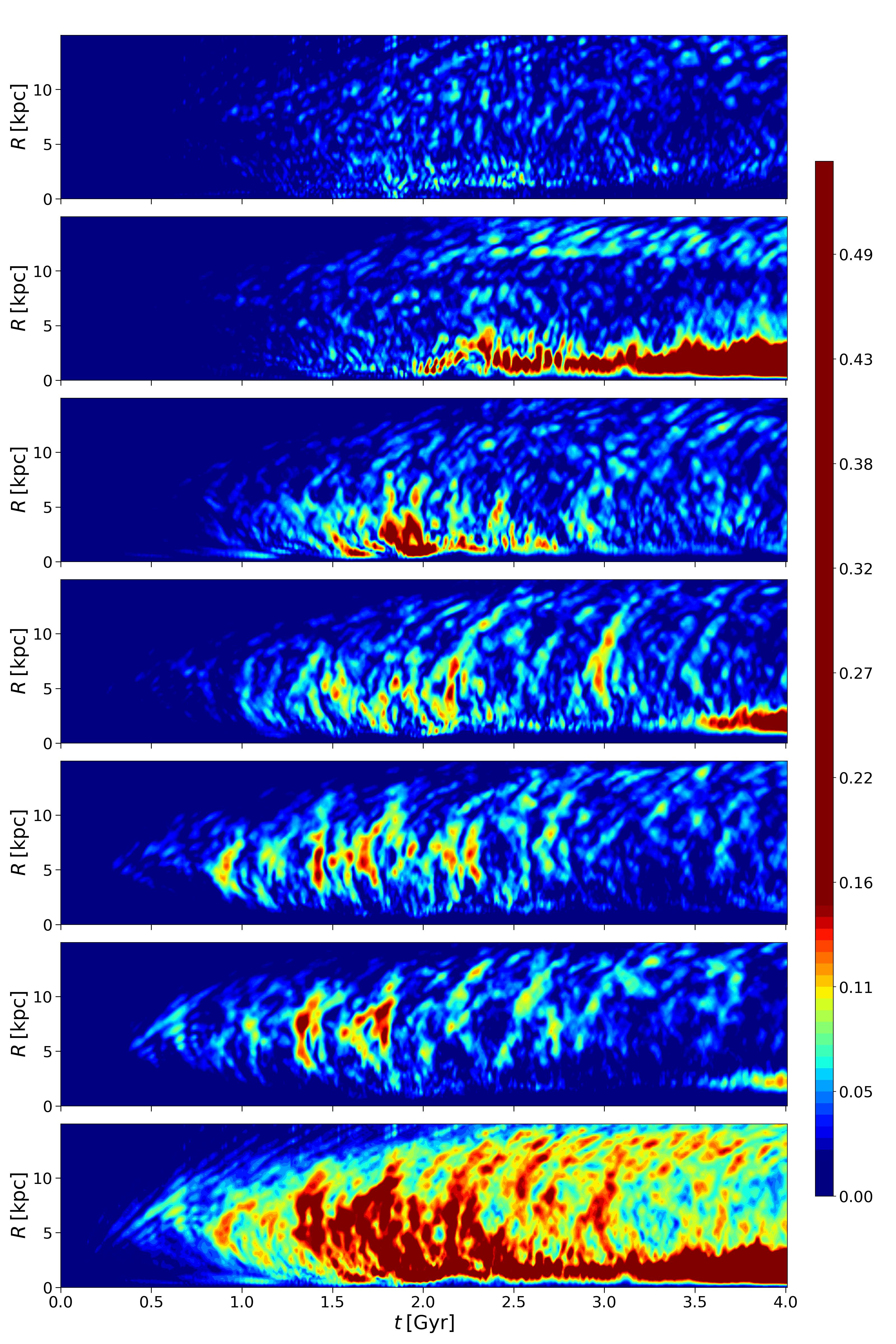} 
    \end{tabular}
    \caption{The same as Figure~\ref{fig:fmap1}, but for model r2c16 evolved for 4 Gyr.}
    \label{fig:fmaplong_r2c16}
\end{figure*}

\begin{figure*}[htbp]
    \centering
    \includegraphics[width=0.83\textwidth]{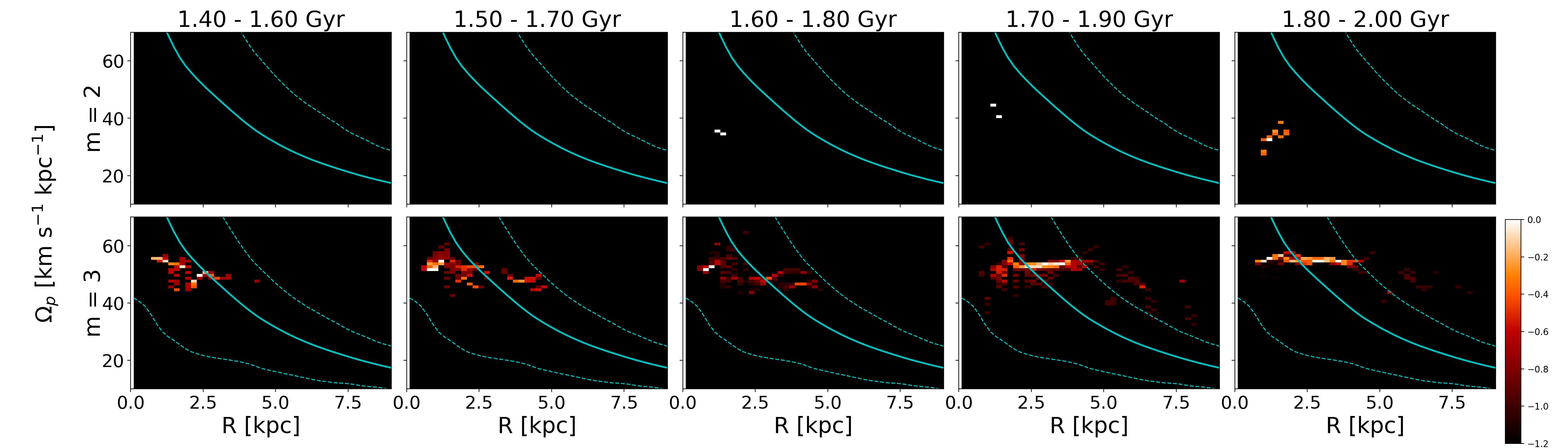} 
    \includegraphics[width=0.83\textwidth]{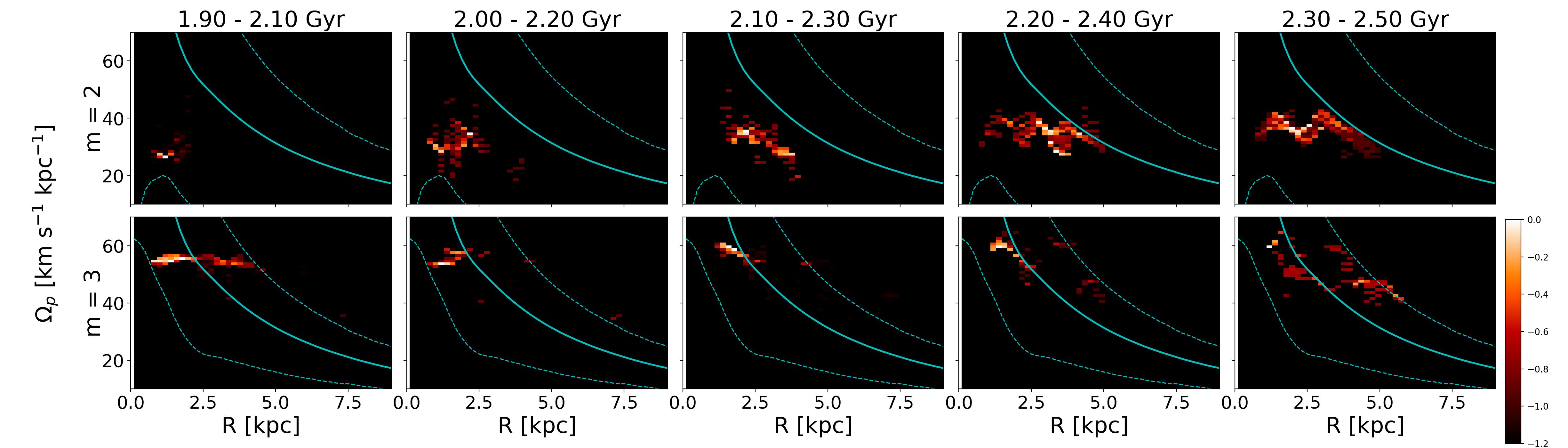}
    \includegraphics[width=0.83\textwidth]{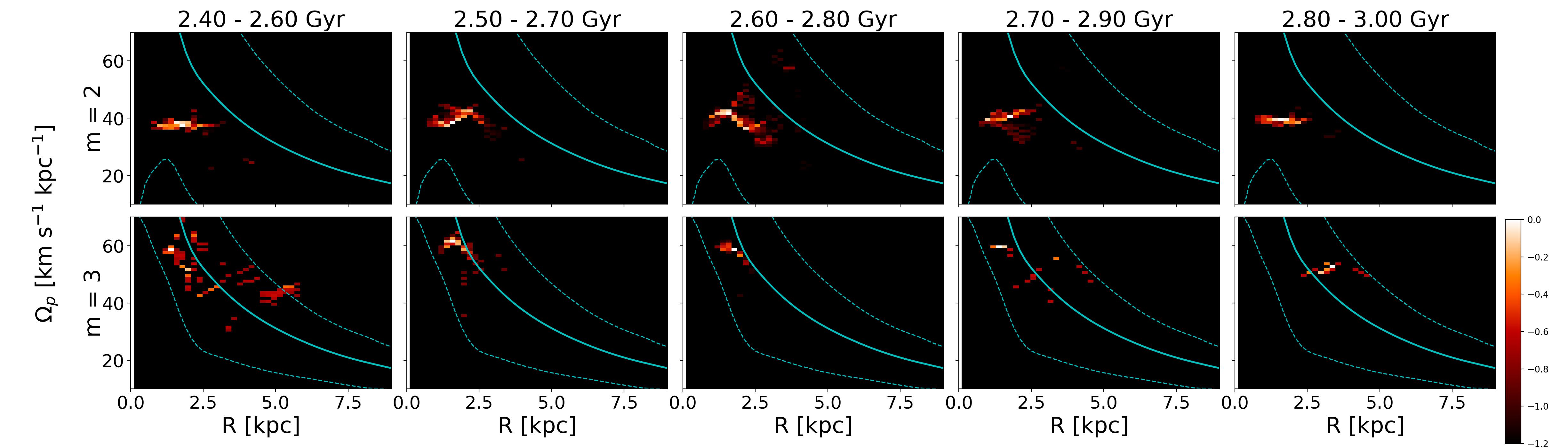} 
    \includegraphics[width=0.83\textwidth]{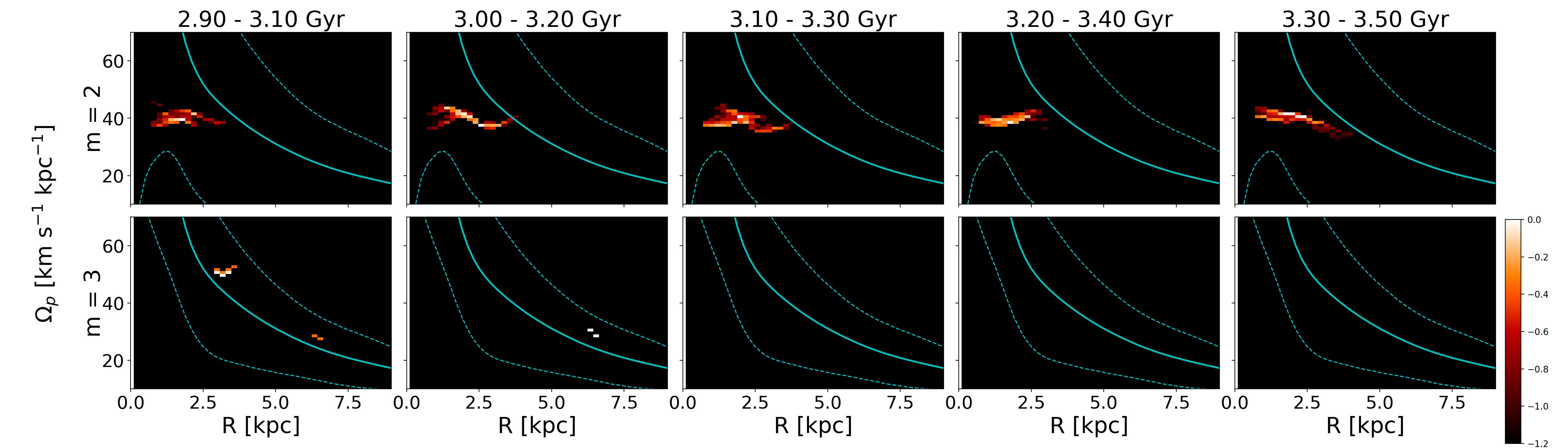}
    \includegraphics[width=0.83\textwidth]{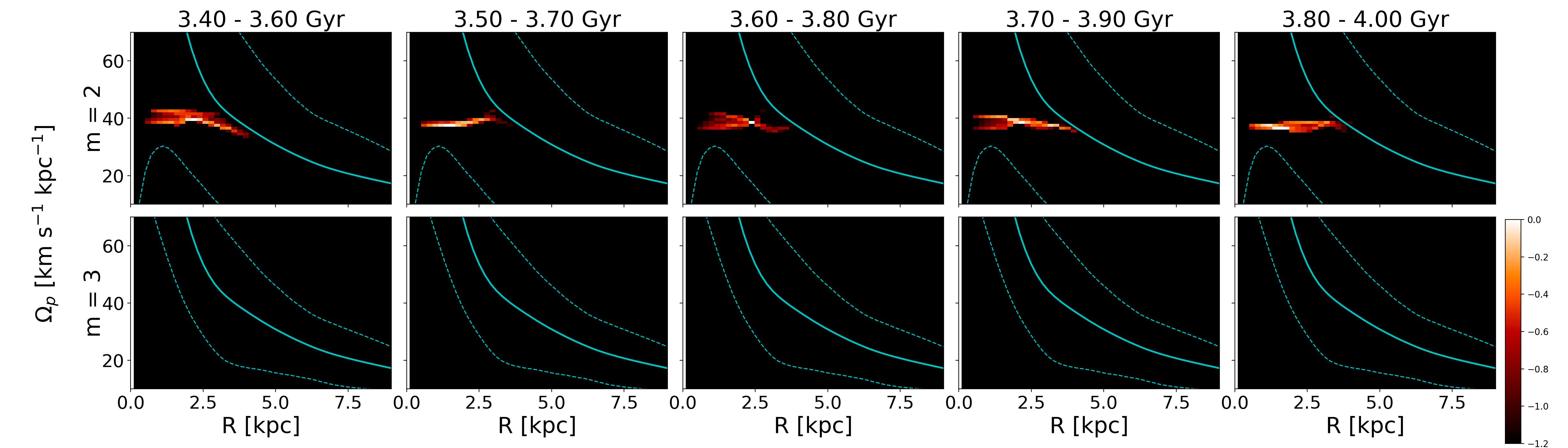} 

    \caption{Spectrograms of the $m=2$ and $m=3$ modes in model r2c16, covering 0.2 Gyr time windows from 1.4 to 4.0 Gyr, with a minimum Fourier amplitude threshold of 0.09 applied for noise filtering. The color bar represents the power in frequencies. Overlaid are the corresponding resonance curves: $\Omega$ and $\Omega \pm \kappa/m$, where $\Omega$ is the angular velocity and $\kappa$ is the epicyclic frequency.}
    \label{fig:spec}
\end{figure*}

Figure~\ref{fig:fmaplong_r2c16} shows the extended Fourier map of model r2c16, and Figure~\ref{fig:spec} shows the corresponding spectrograms. Between $t\simeq1.4$ and $2.1\,\Gyr$, the inner disk is dominated by a compact $m=3$ ridge with $\Omega_p\simeq50$--$60\,{\rm km\,s^{-1}\,kpc^{-1}}$. This ridge is radially confined and remains nearly horizontal in the spectrogram, showing that the central three-armed distortion behaves as a coherent rotating mode rather than as transient interference among neighbouring harmonics. After $t\simeq2\,\Gyr$, the $m=3$ component weakens, while the $m=2$ component appears around $\Omega_p\simeq40\,{\rm km\,s^{-1}\,kpc^{-1}}$ and gradually becomes more prominent. By the time the bar forms at $t=3.56\,\Gyr$, the $m=2$ mode has become the dominant central pattern.

\begin{figure*}[htbp]
    \centering
    \setlength{\tabcolsep}{1pt}
    \renewcommand{\arraystretch}{0.92}
    \begin{tabular}{@{}ccccc@{}}
        \includegraphics[width=0.21\textwidth]{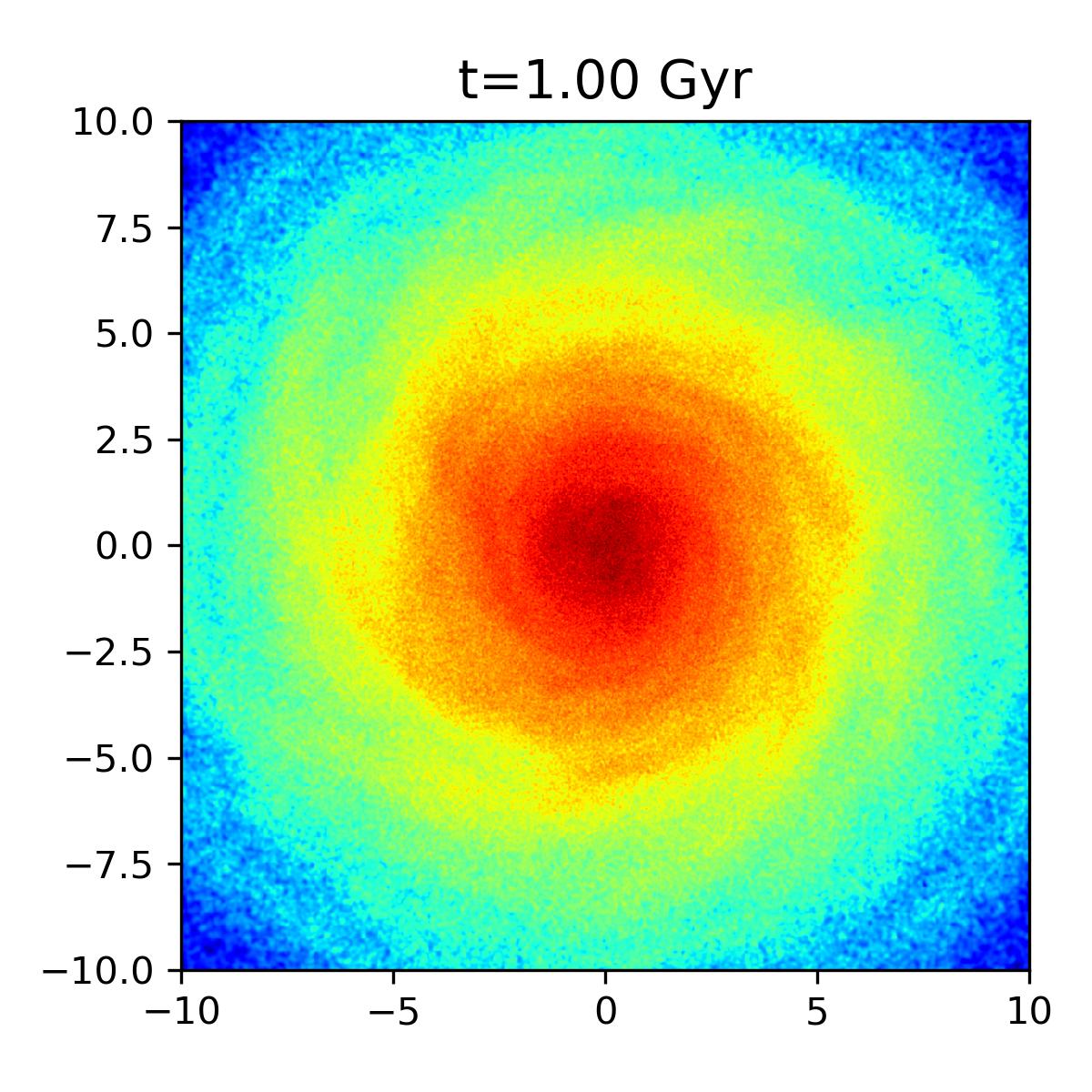} &
        \includegraphics[width=0.21\textwidth]{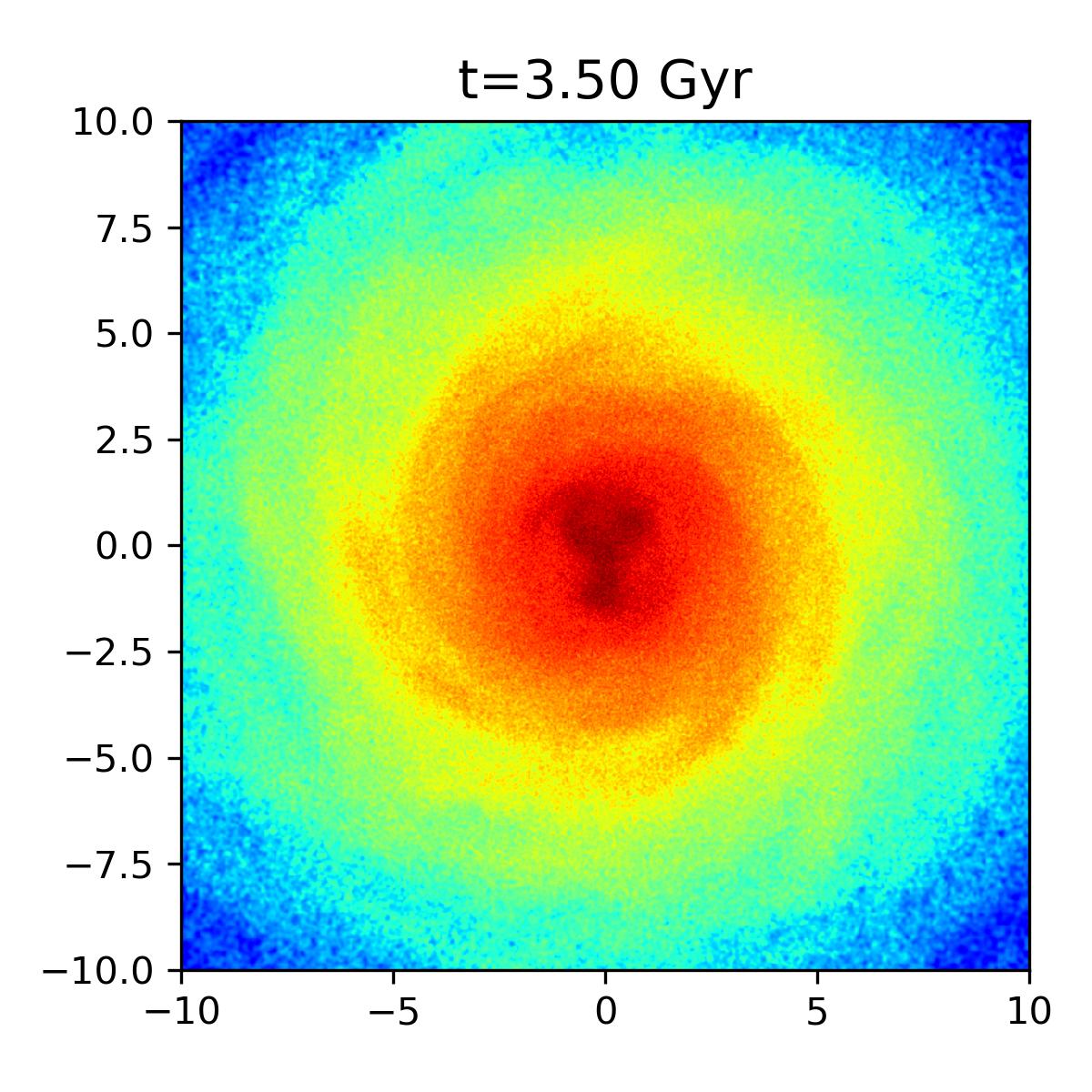} &
        \includegraphics[width=0.21\textwidth]{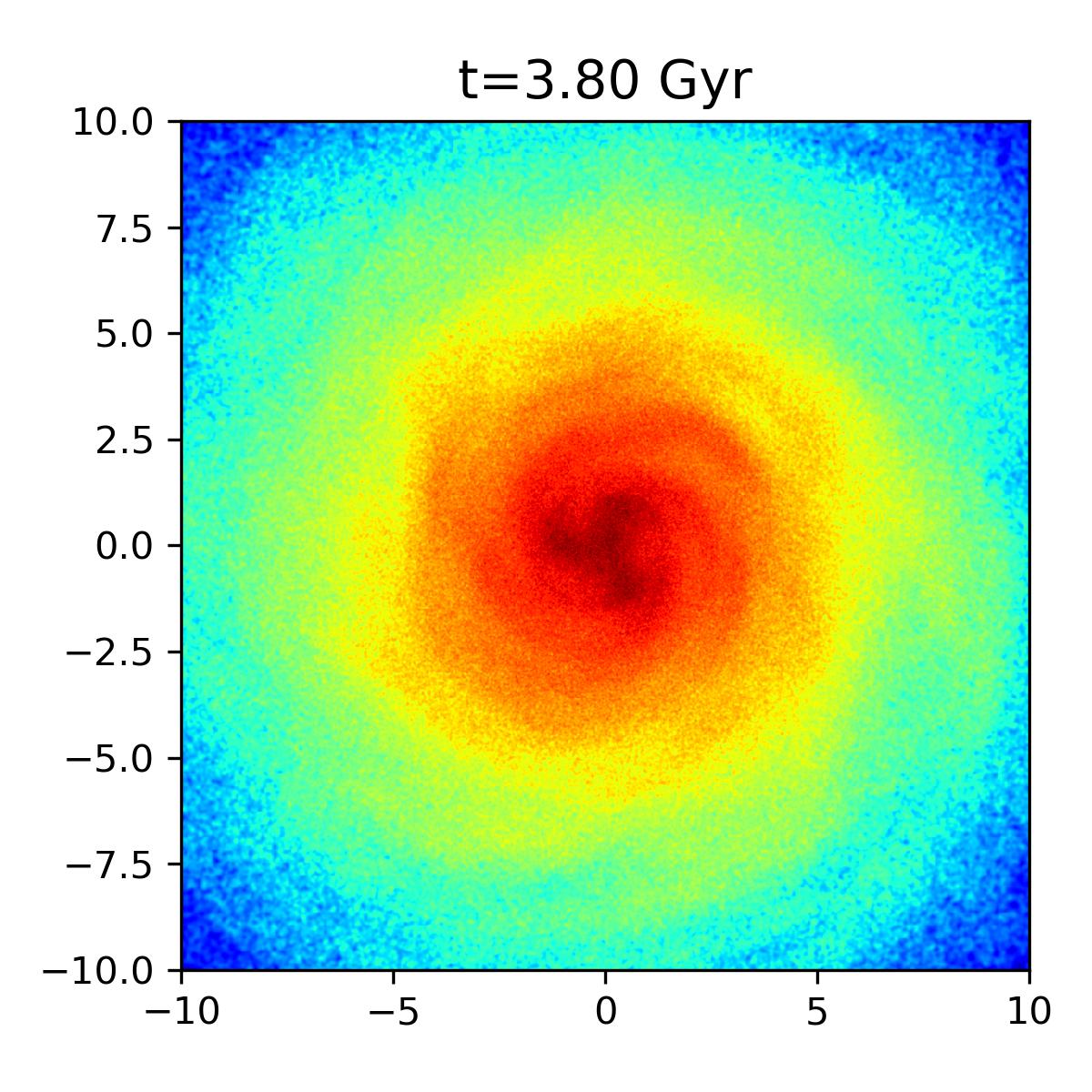} &
        \includegraphics[width=0.21\textwidth]{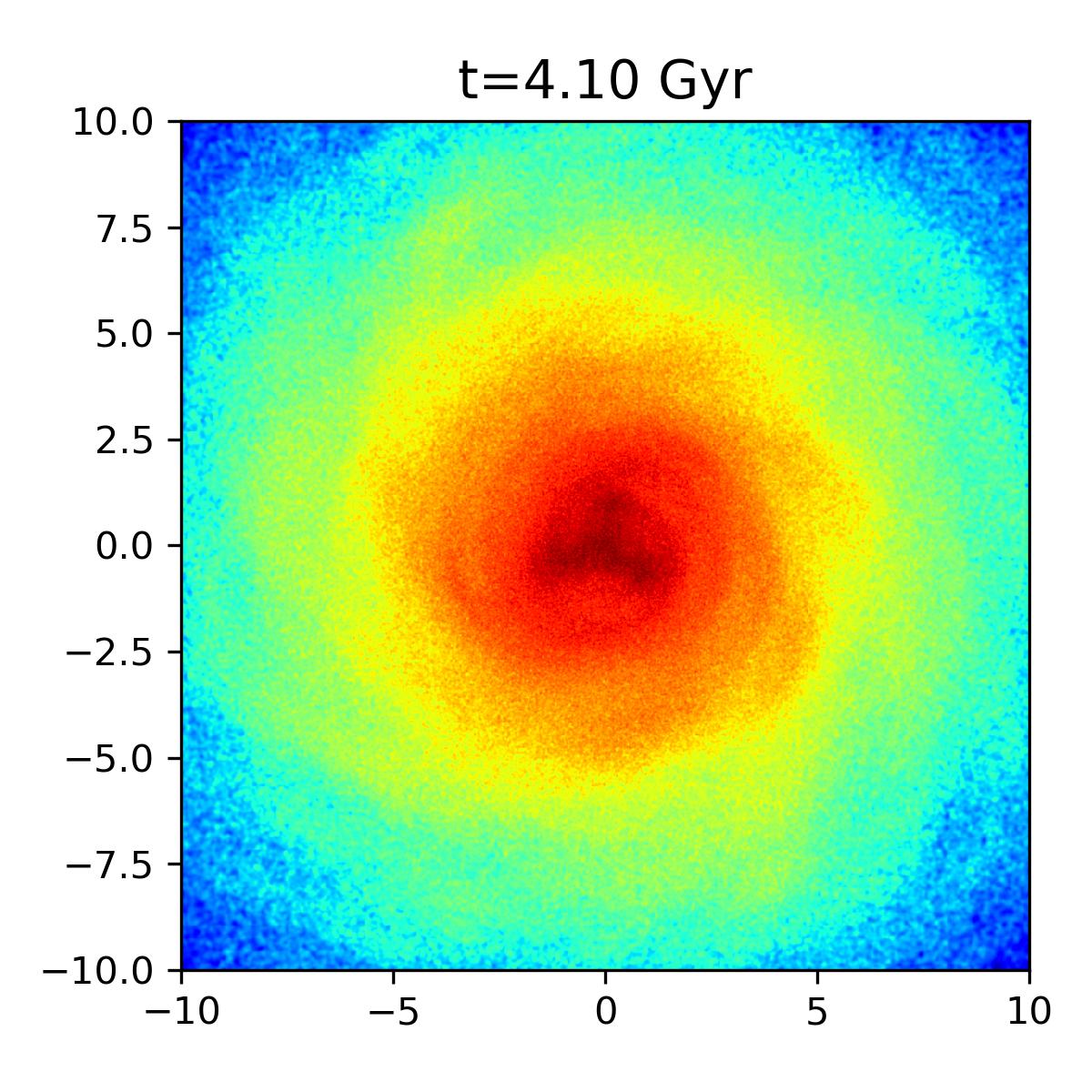} & \\
        \includegraphics[width=0.21\textwidth]{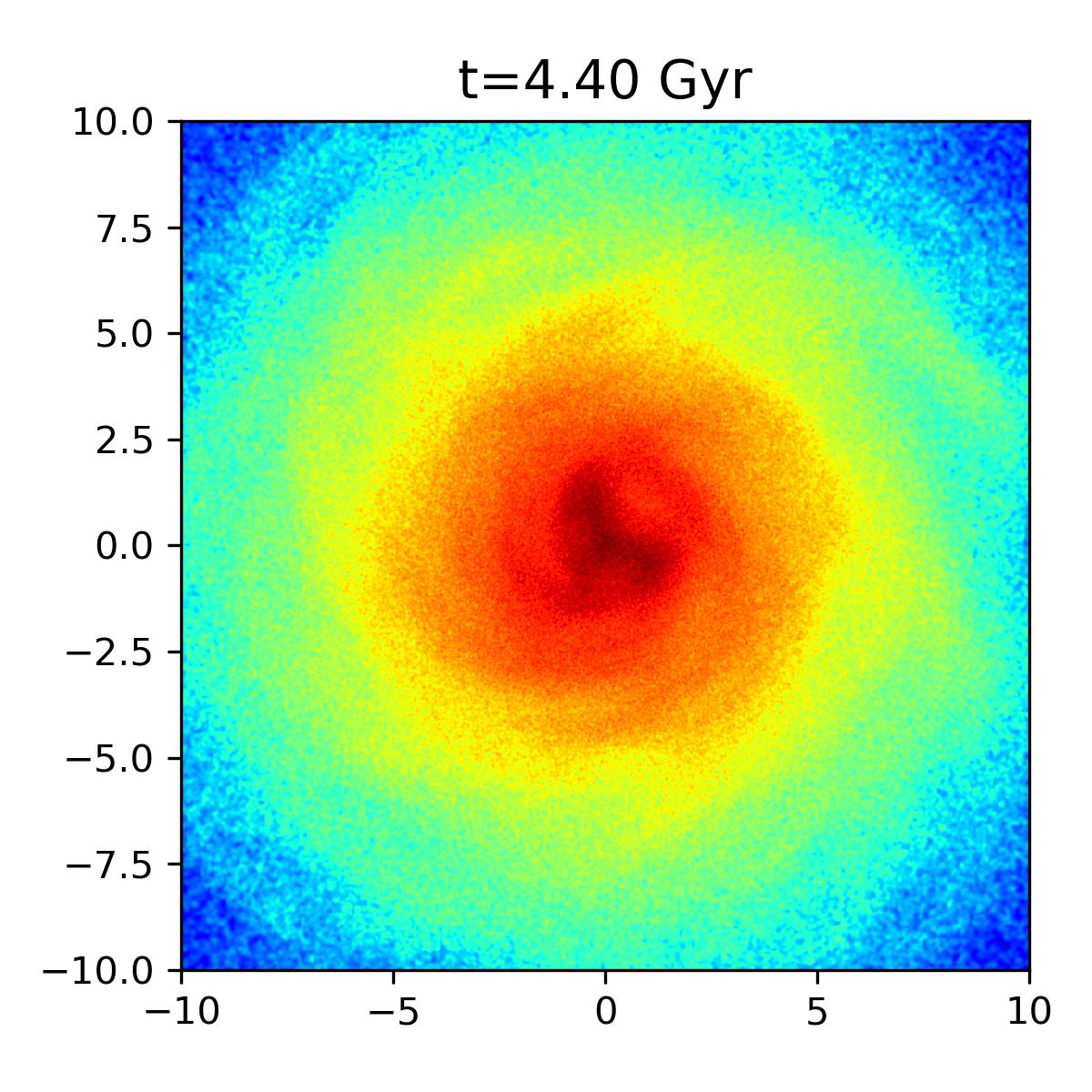} &
        \includegraphics[width=0.21\textwidth]{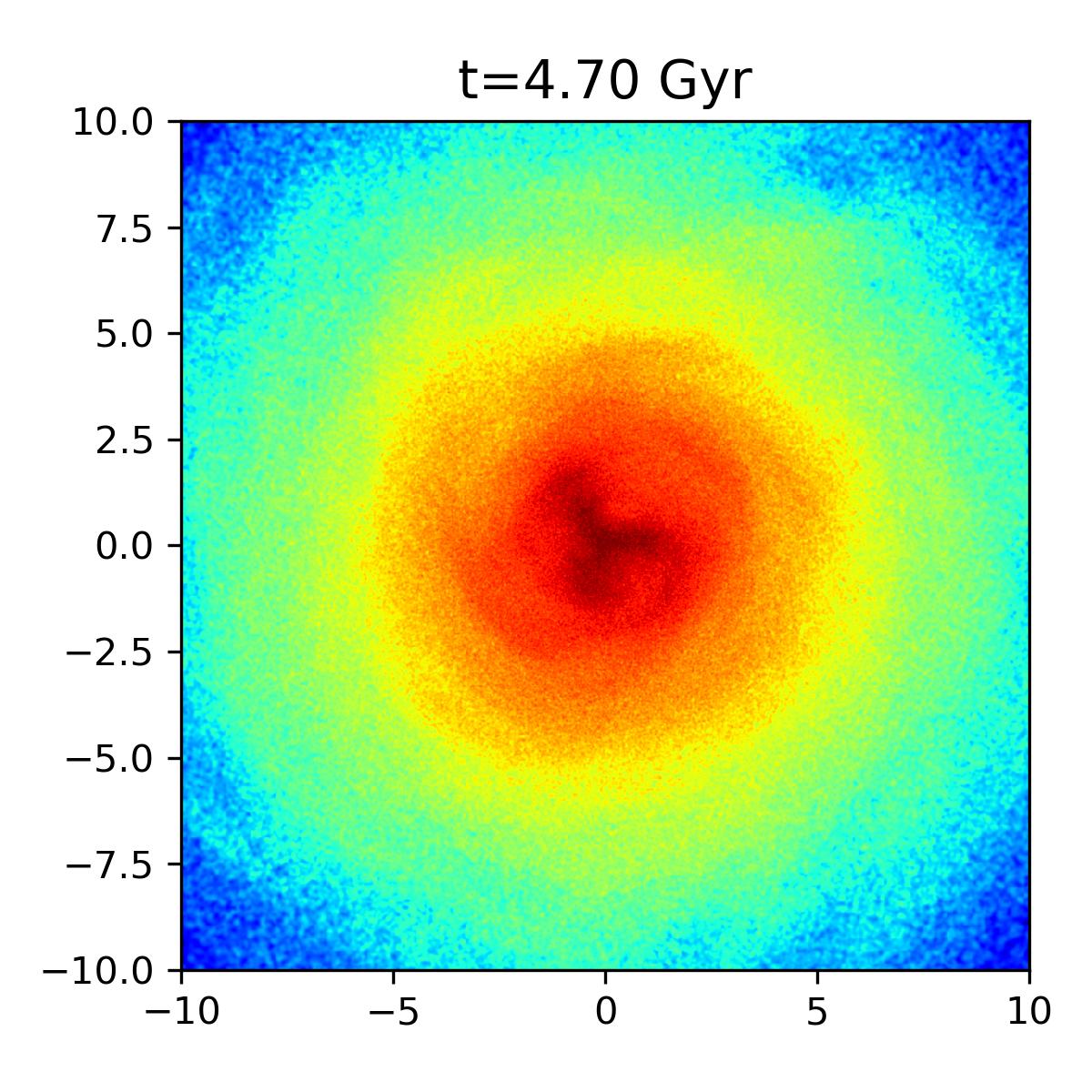} &
        \includegraphics[width=0.21\textwidth]{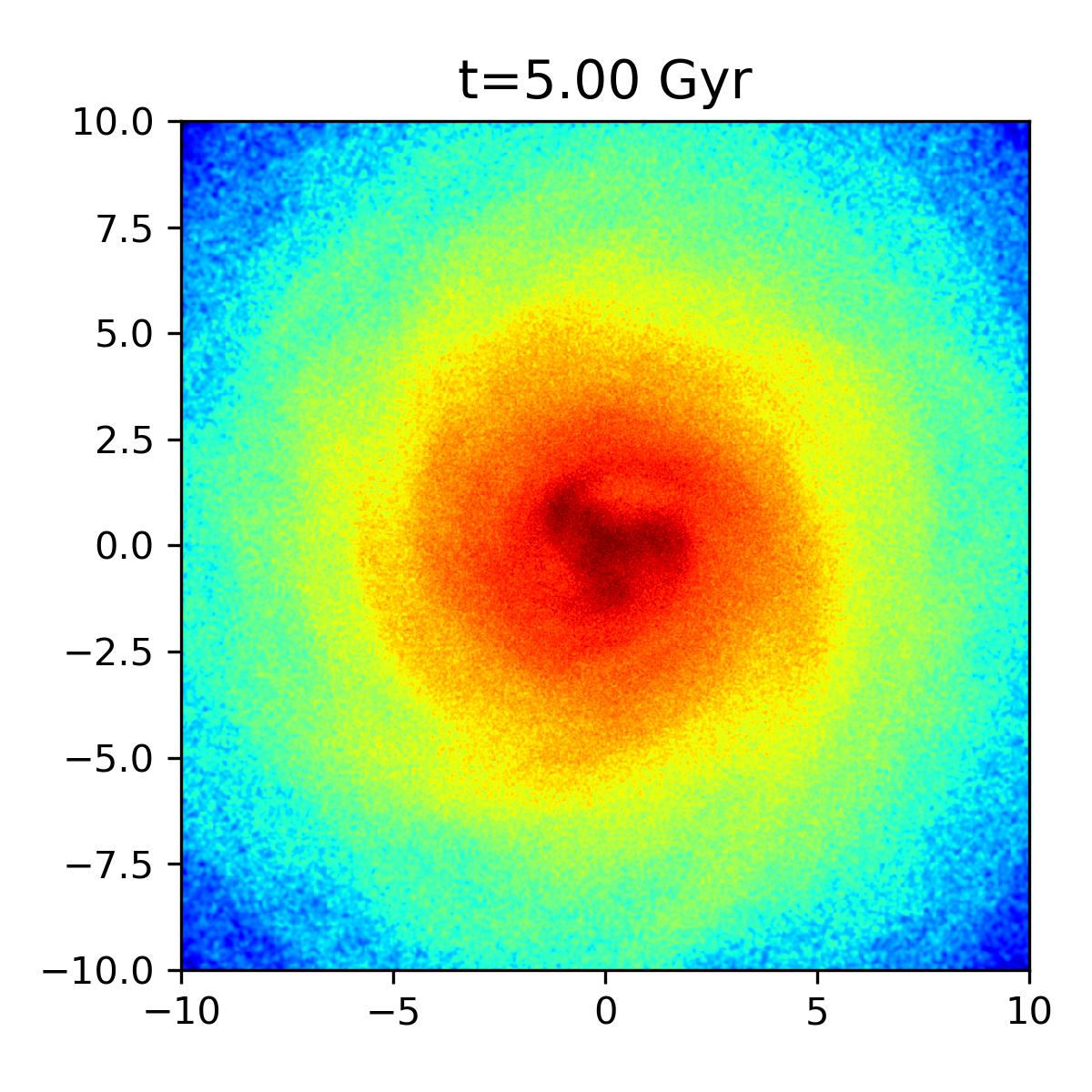} &
        \includegraphics[width=0.21\textwidth]{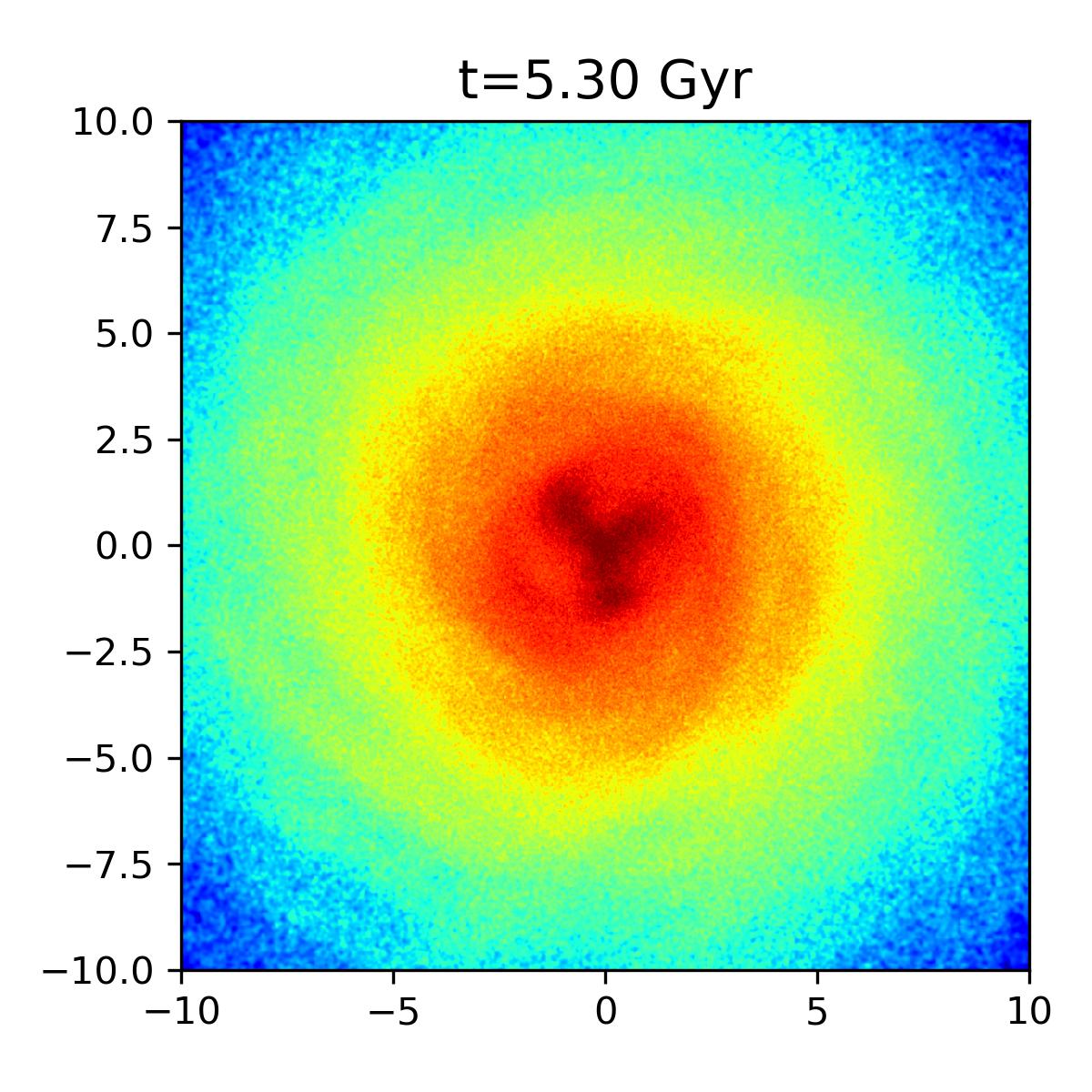} & \\
        \includegraphics[width=0.21\textwidth]{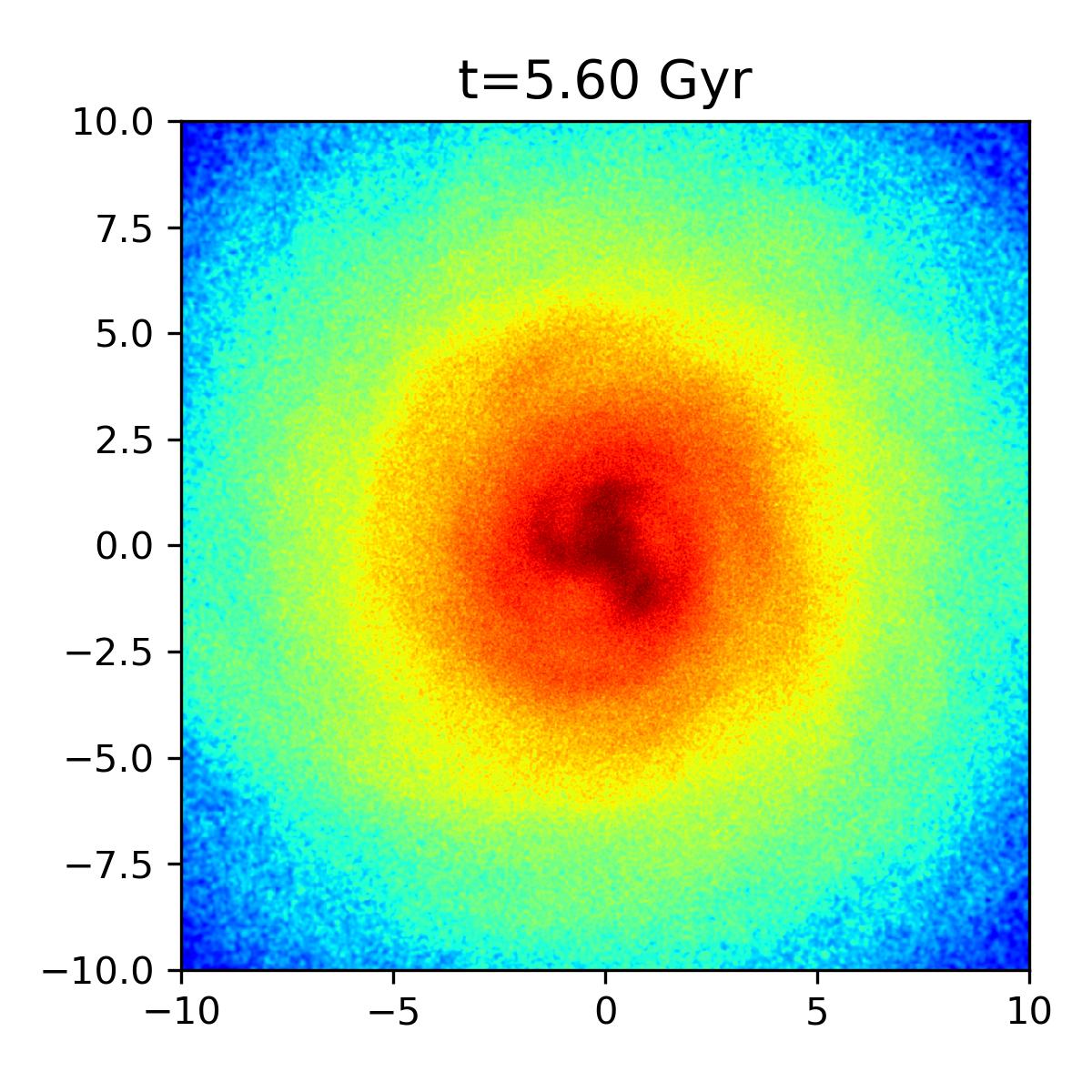} &
        \includegraphics[width=0.21\textwidth]{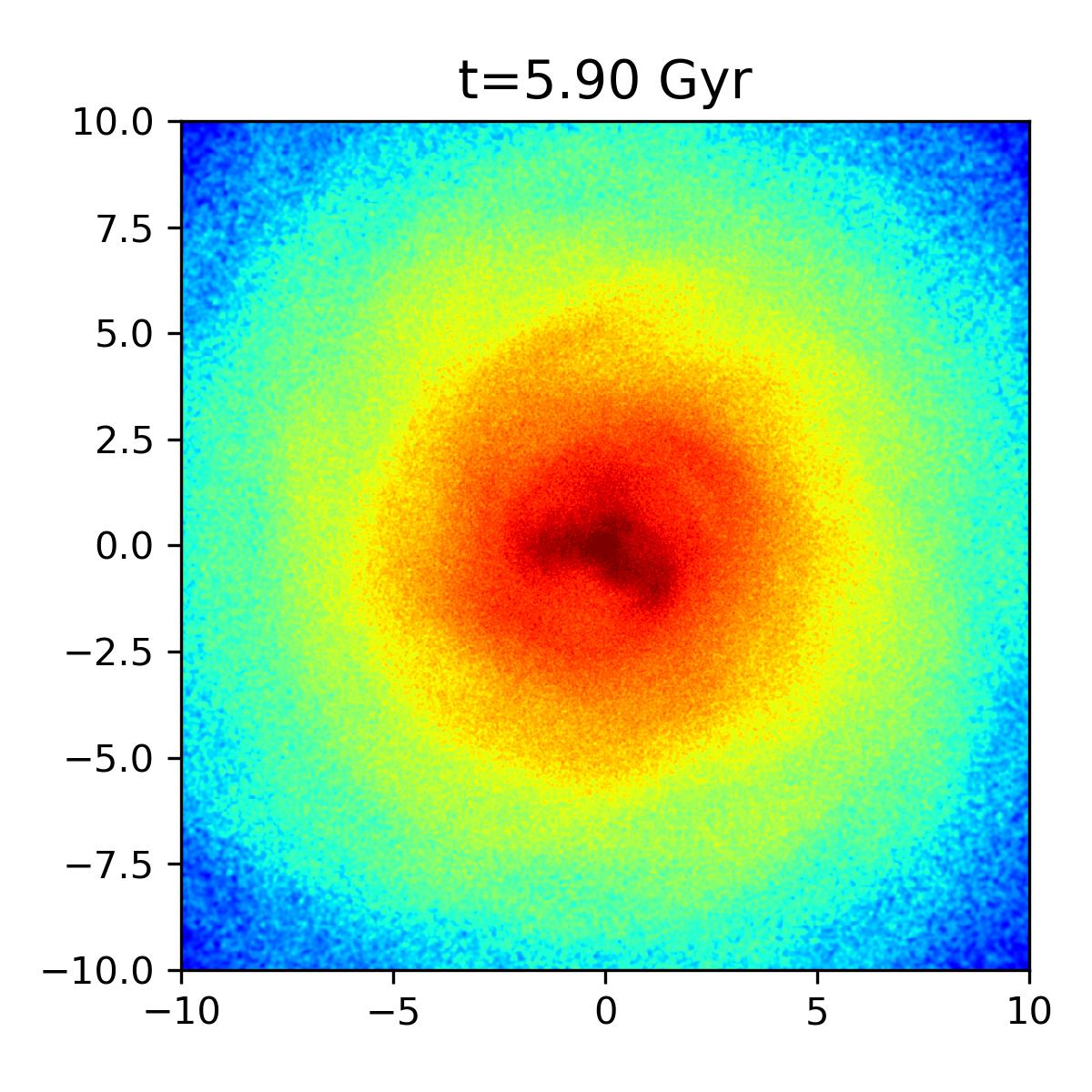} &
        \includegraphics[width=0.21\textwidth]{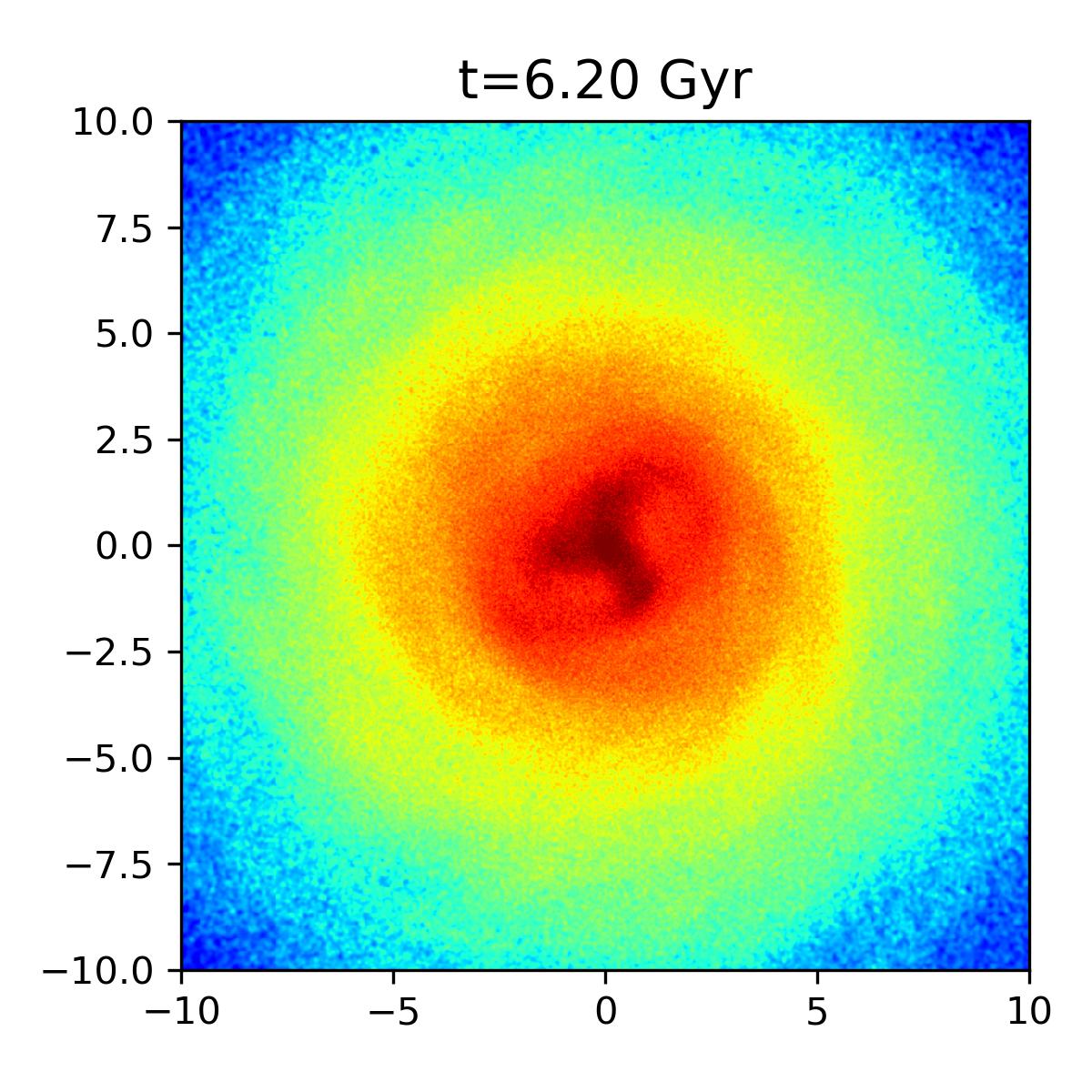} &
        \includegraphics[width=0.21\textwidth]{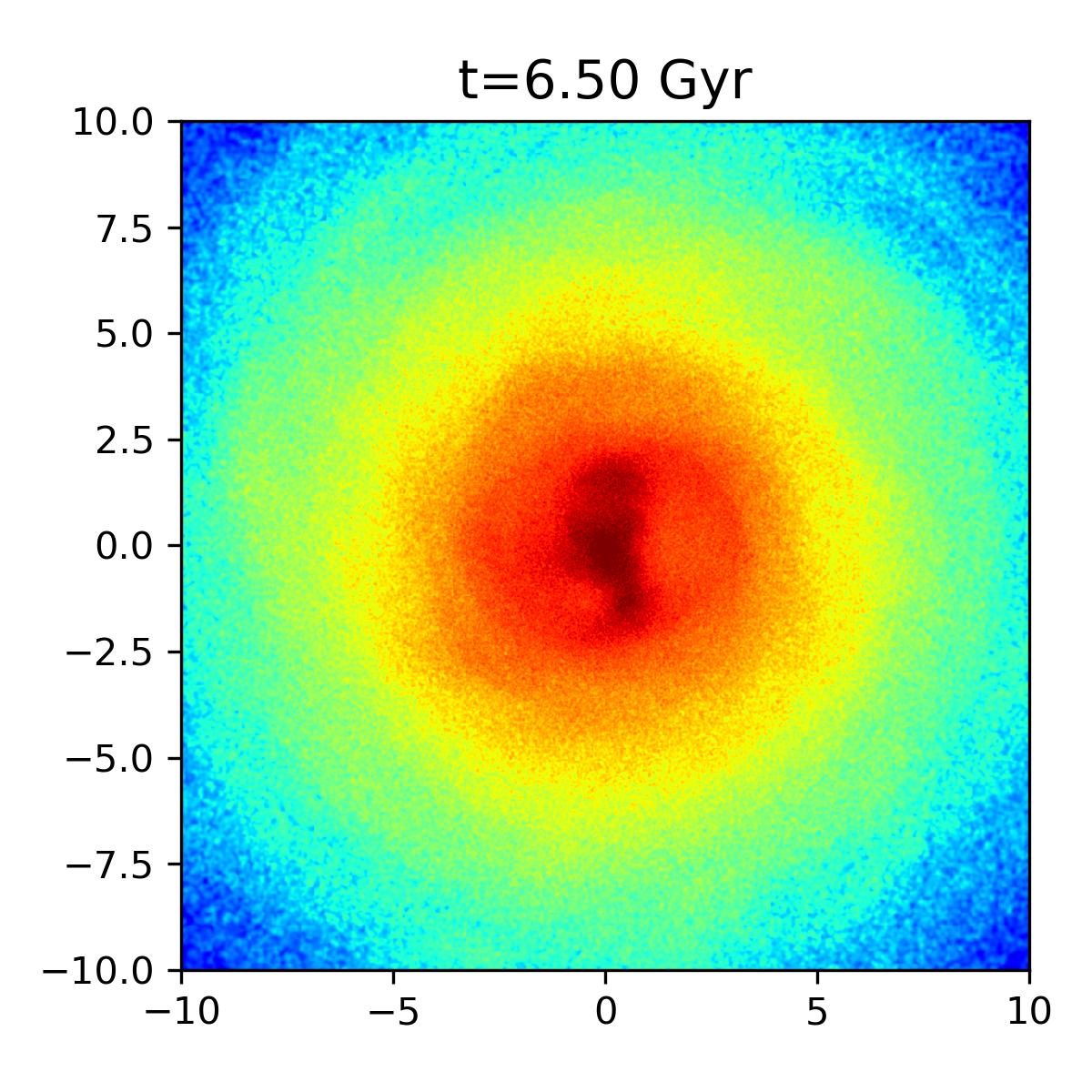} & \\
        \includegraphics[width=0.21\textwidth]{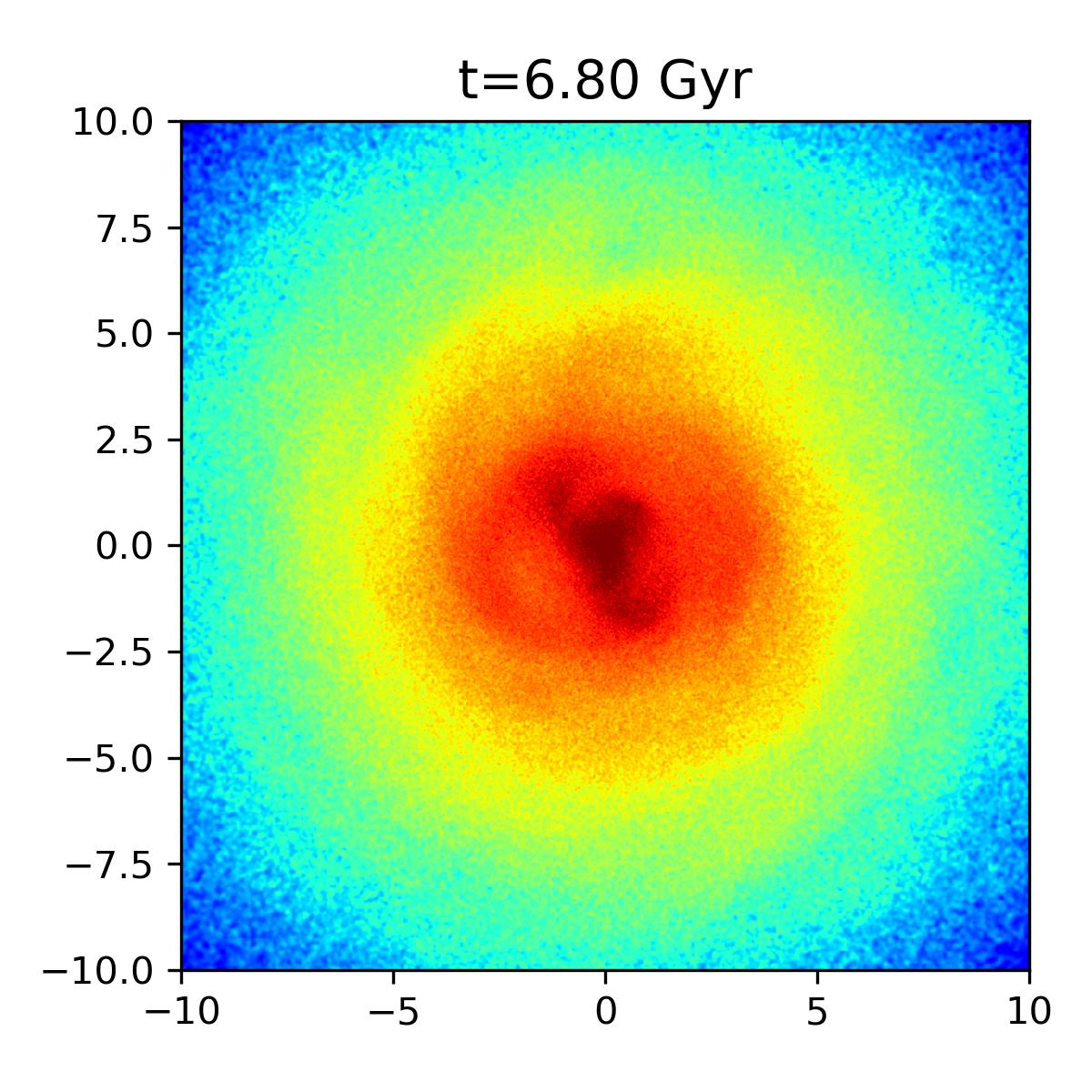} &
        \includegraphics[width=0.21\textwidth]{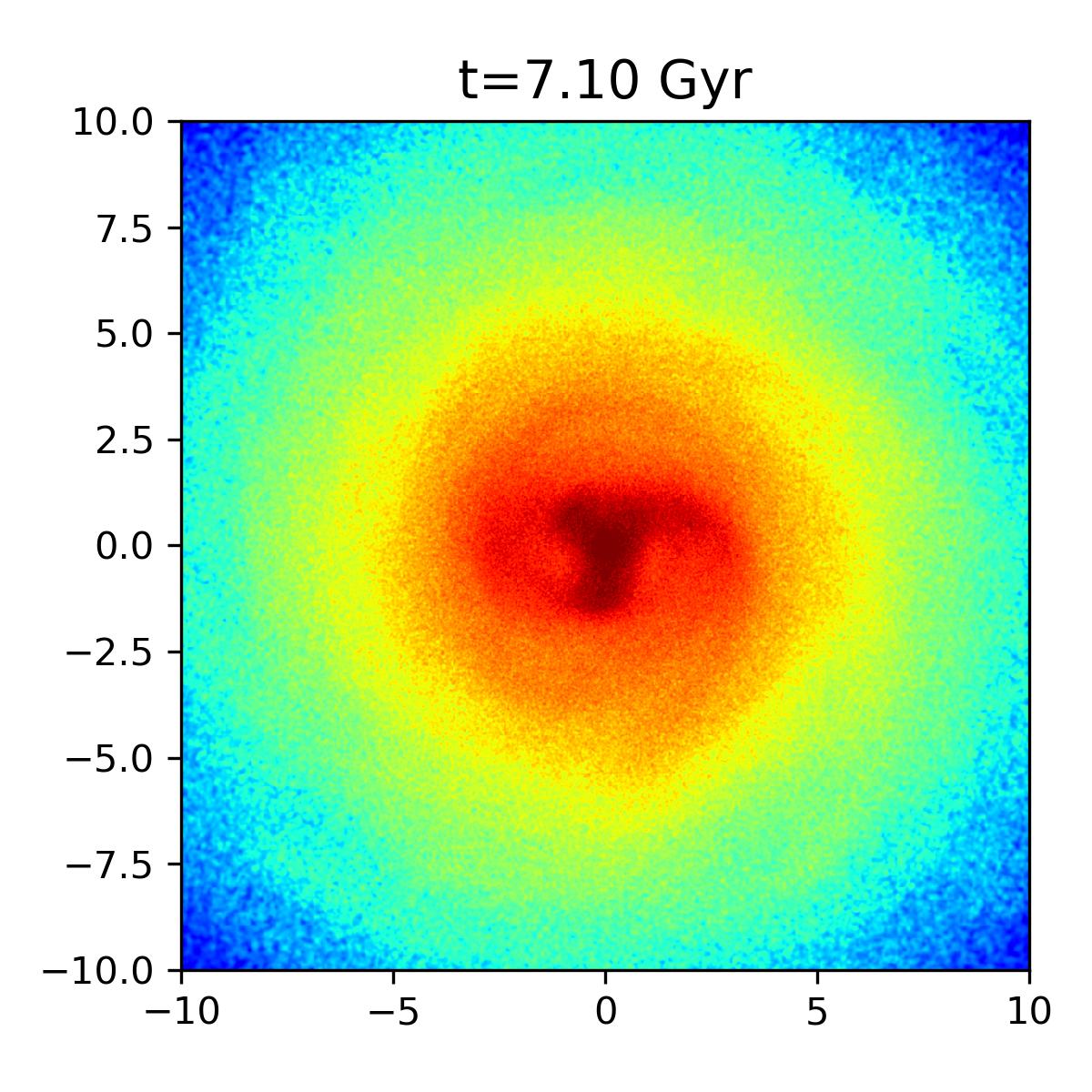} &
        \includegraphics[width=0.21\textwidth]{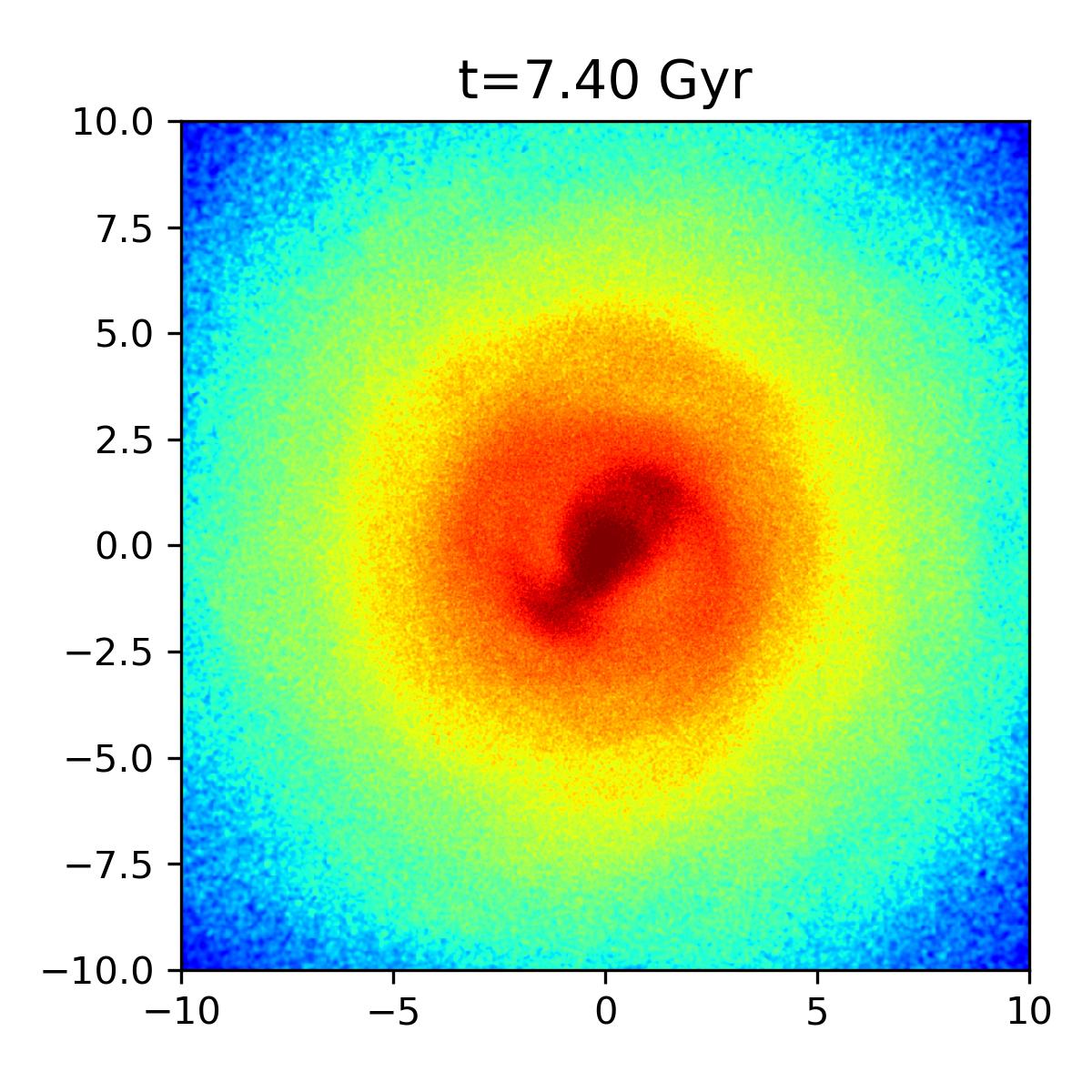} &
        \includegraphics[width=0.21\textwidth]{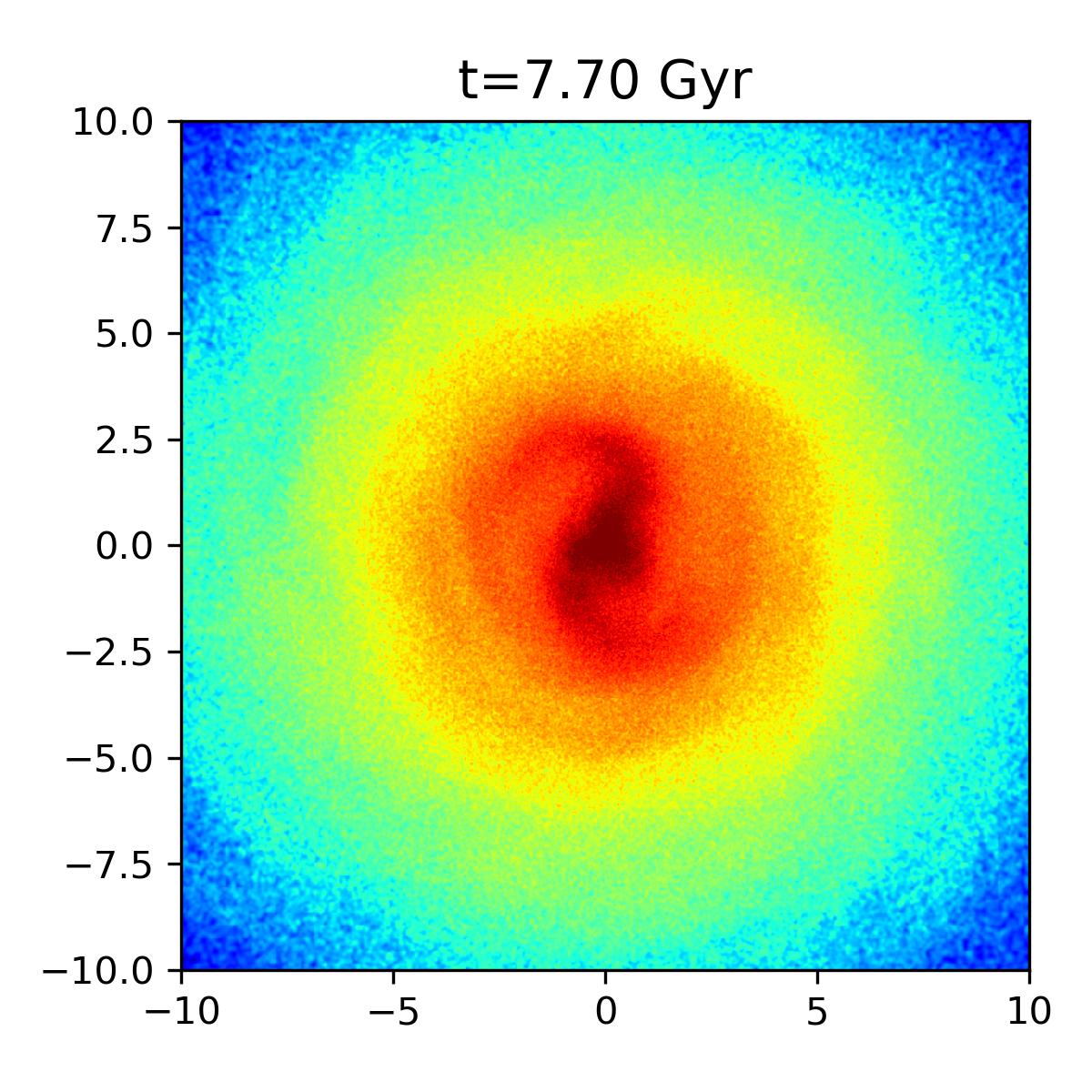} & \\
        \includegraphics[width=0.21\textwidth]{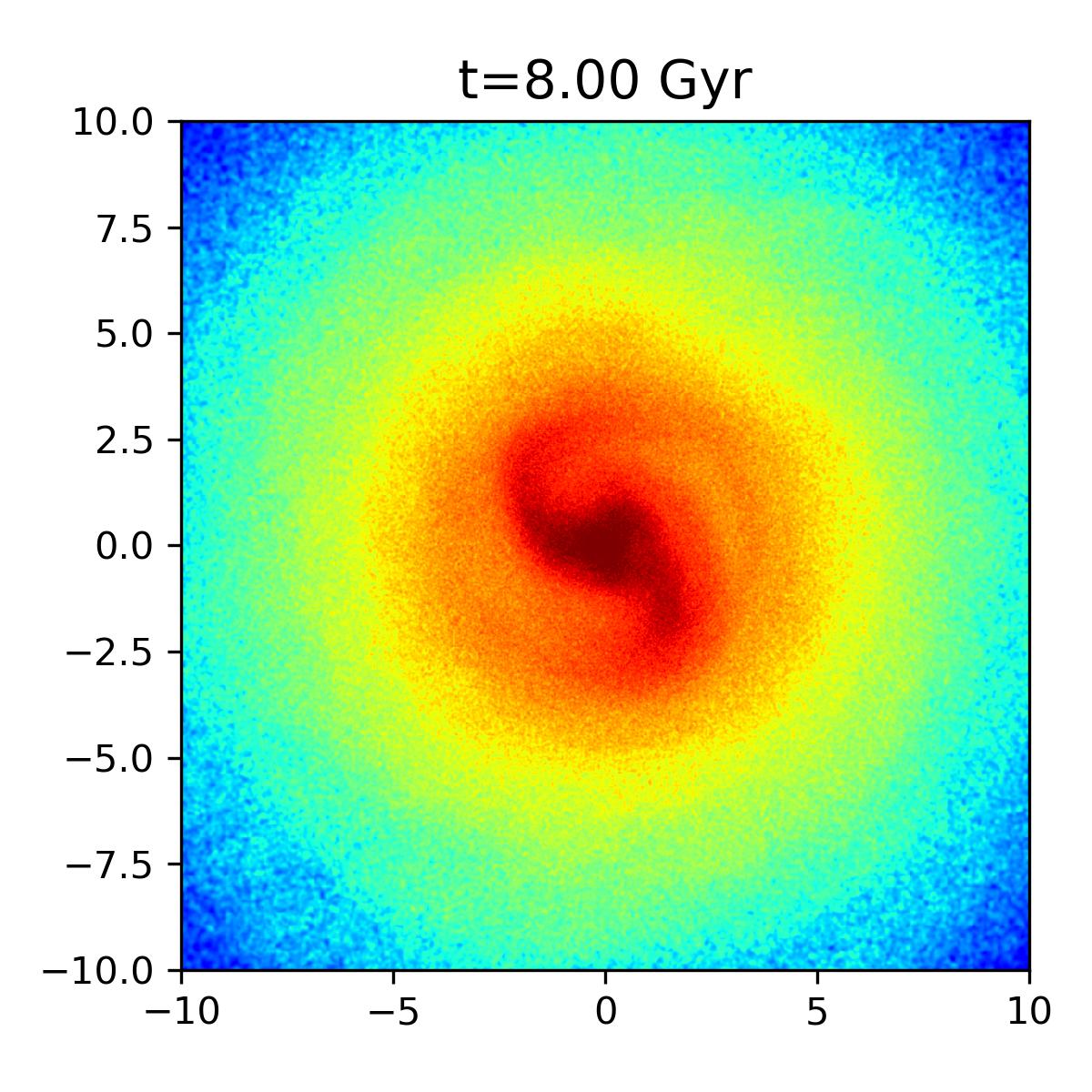} &
        \includegraphics[width=0.21\textwidth]{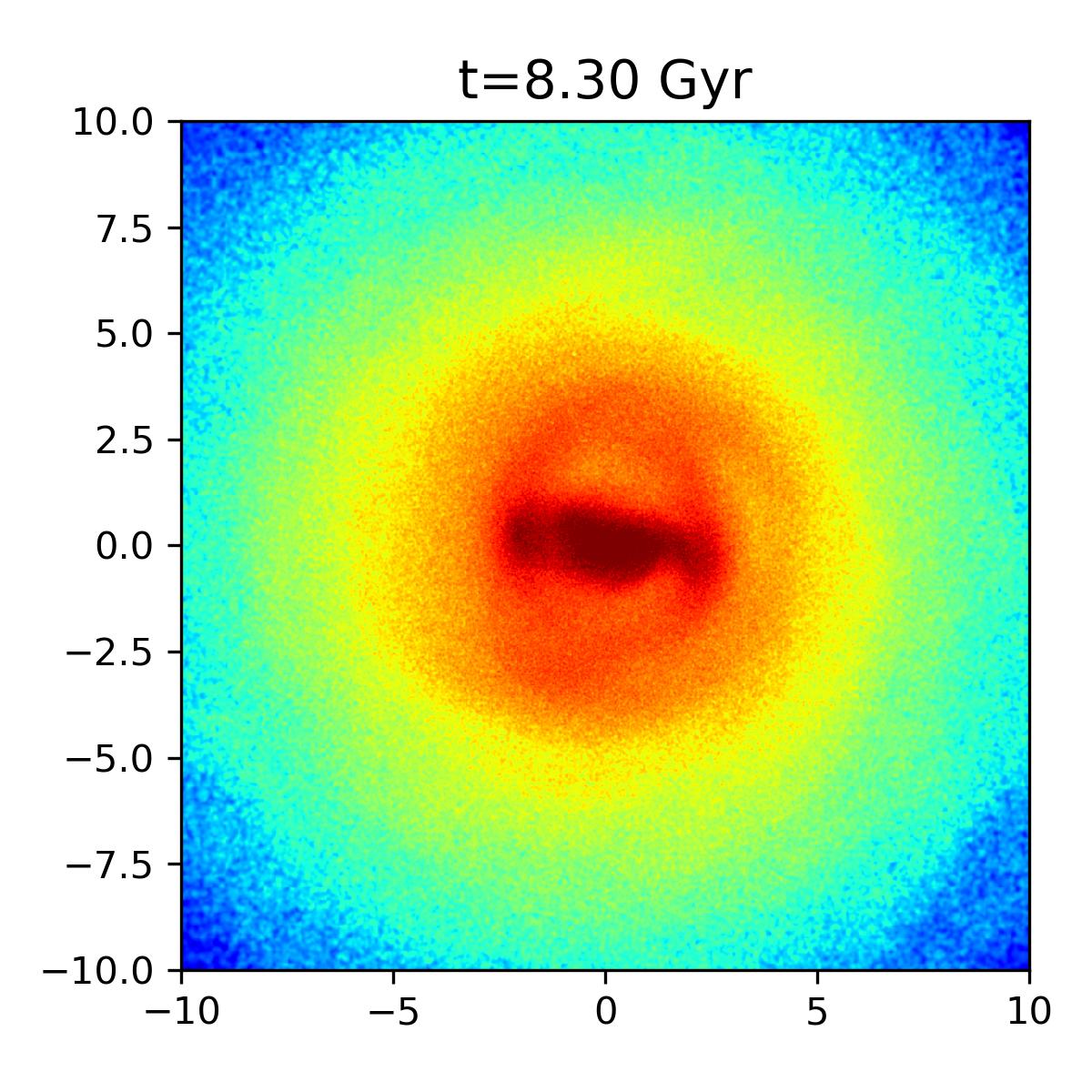} &
        \includegraphics[width=0.21\textwidth]{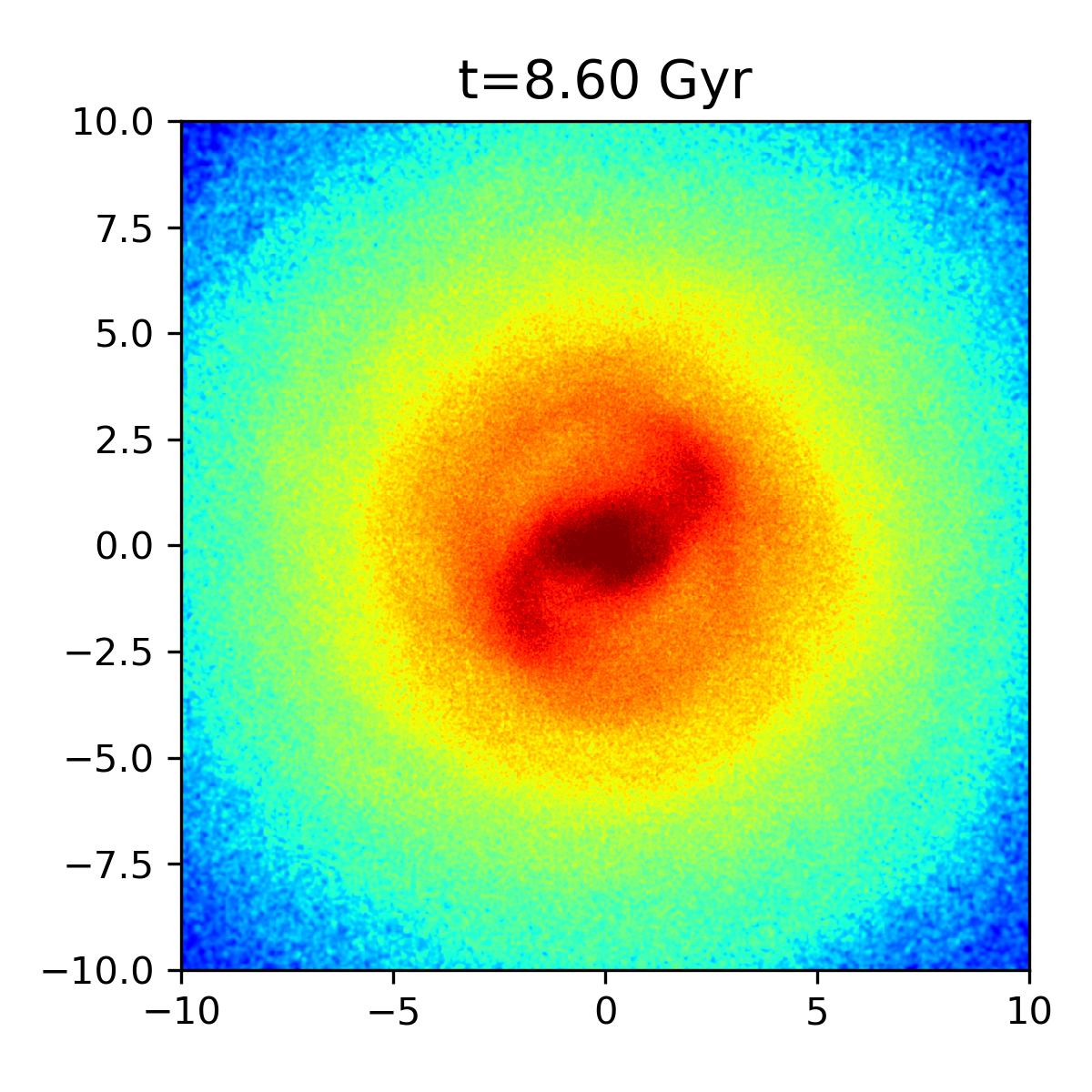} &
        \includegraphics[width=0.21\textwidth]{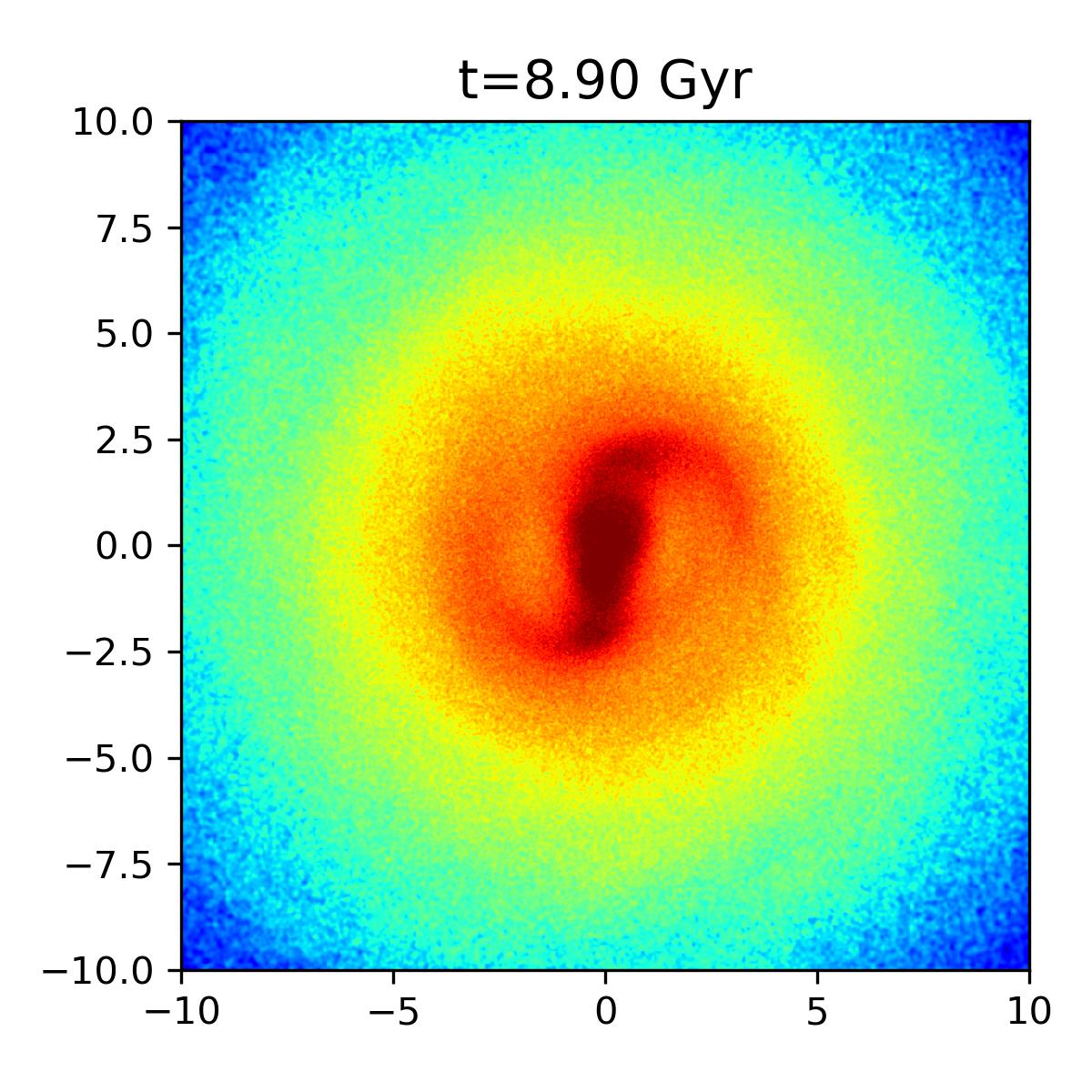} & \\
        \includegraphics[width=0.21\textwidth]{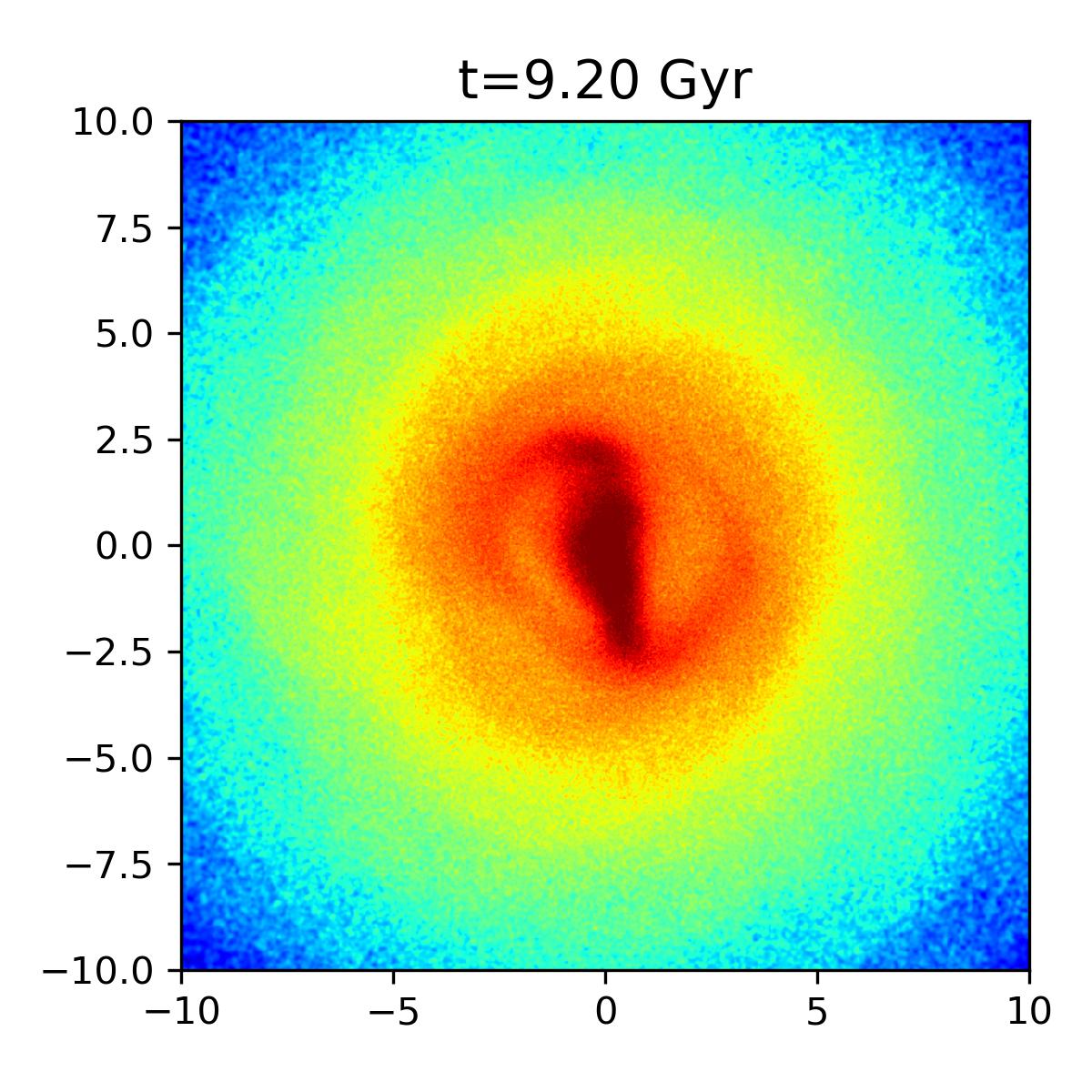} &
        \includegraphics[width=0.21\textwidth]{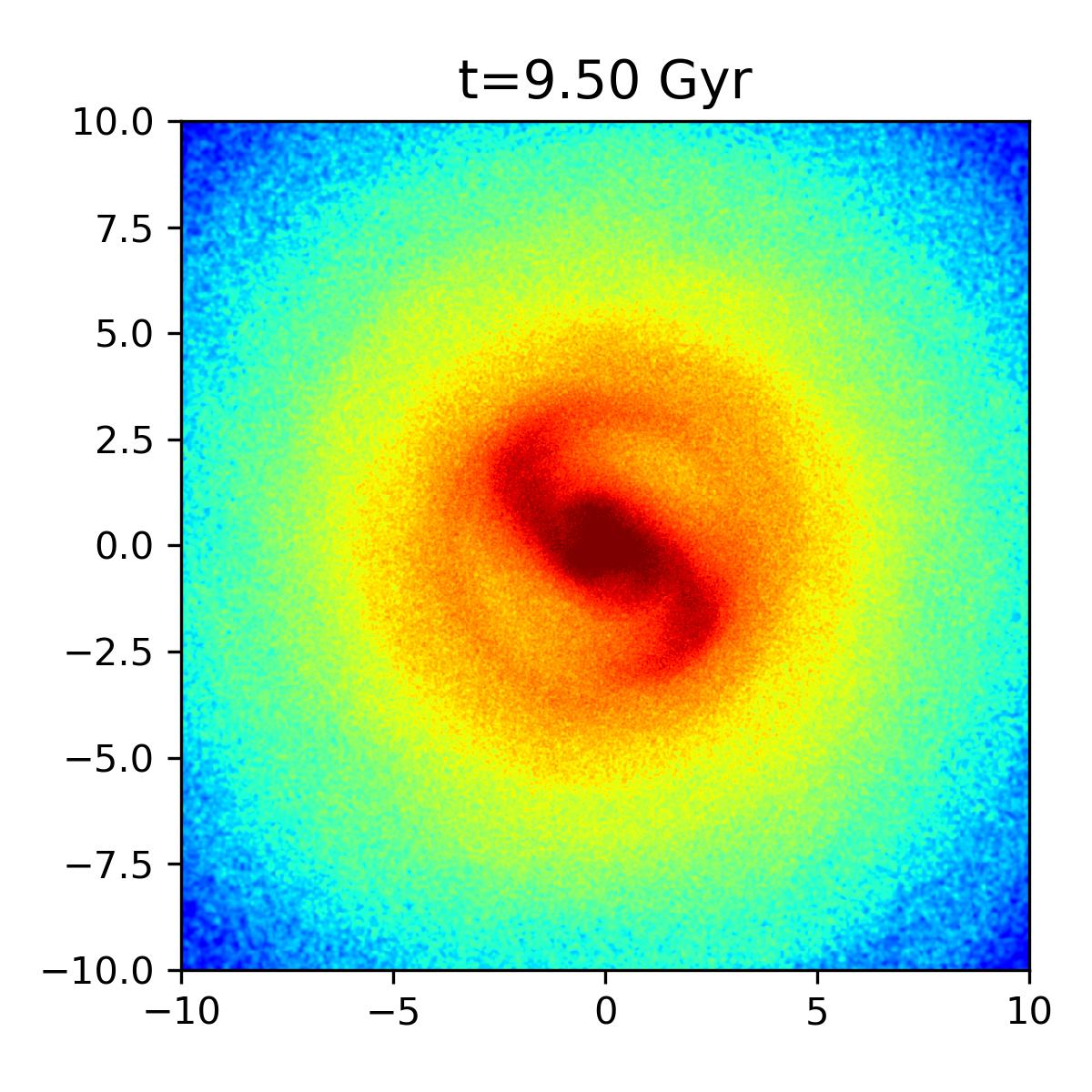} &
        \includegraphics[width=0.21\textwidth]{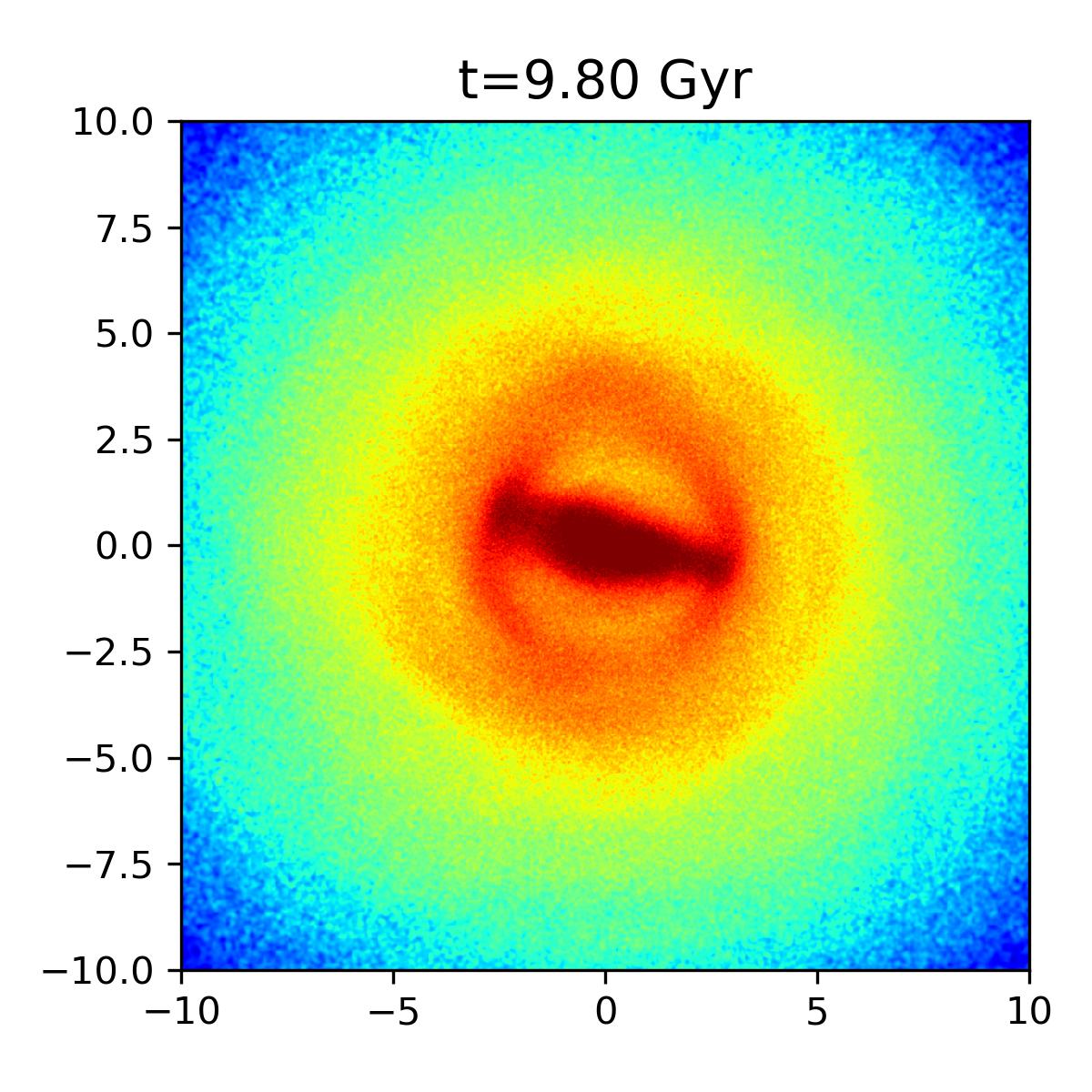} &
        \includegraphics[width=0.21\textwidth]{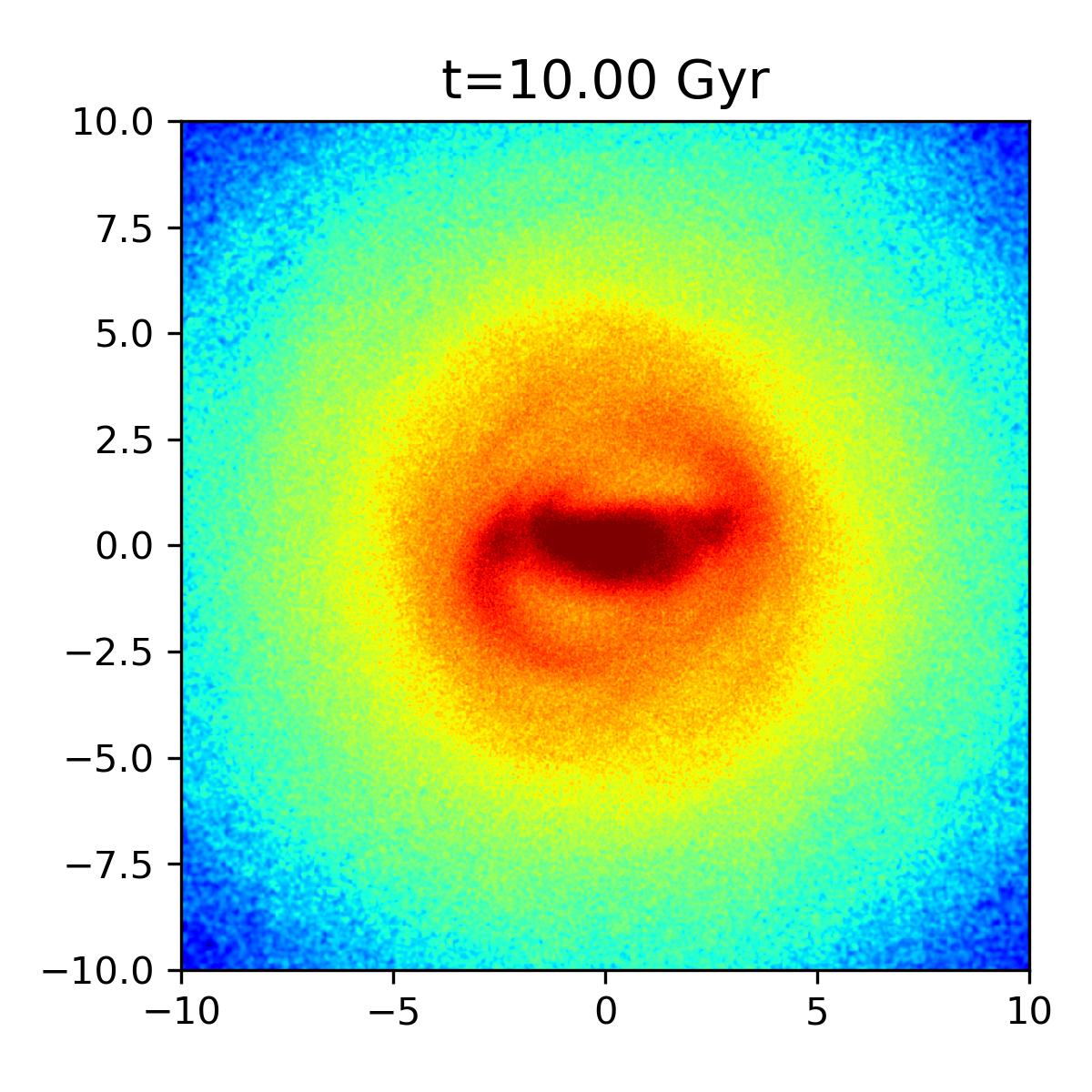} &
        \includegraphics[height=0.19\textwidth]{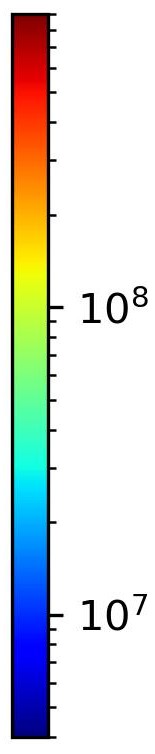} \\
    \end{tabular}
    \caption{Face-on projections in a $20 \times 20$ kpc box for model r1c22 over the same evolutionary period analyzed in Figure~\ref{fig:spec_r1c22}. The snapshots show the {\bf persistent} central three-armed distortion and its subsequent transition toward an $m=2$-dominated bar morphology. The color bar indicates the surface density in units of $\Msun\,\mathrm{kpc}^{-2}$.}
    \label{fig:faceon_r1c22}
\end{figure*}

\begin{figure*}[htbp]
    \centering
    \includegraphics[width=0.83\textwidth]{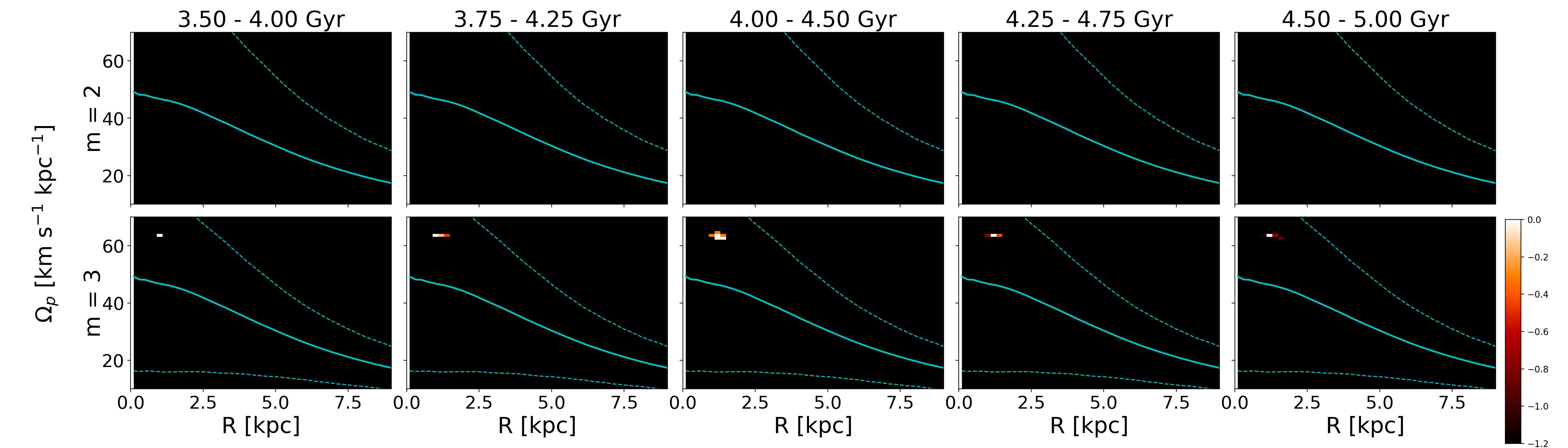} 
    \includegraphics[width=0.83\textwidth]{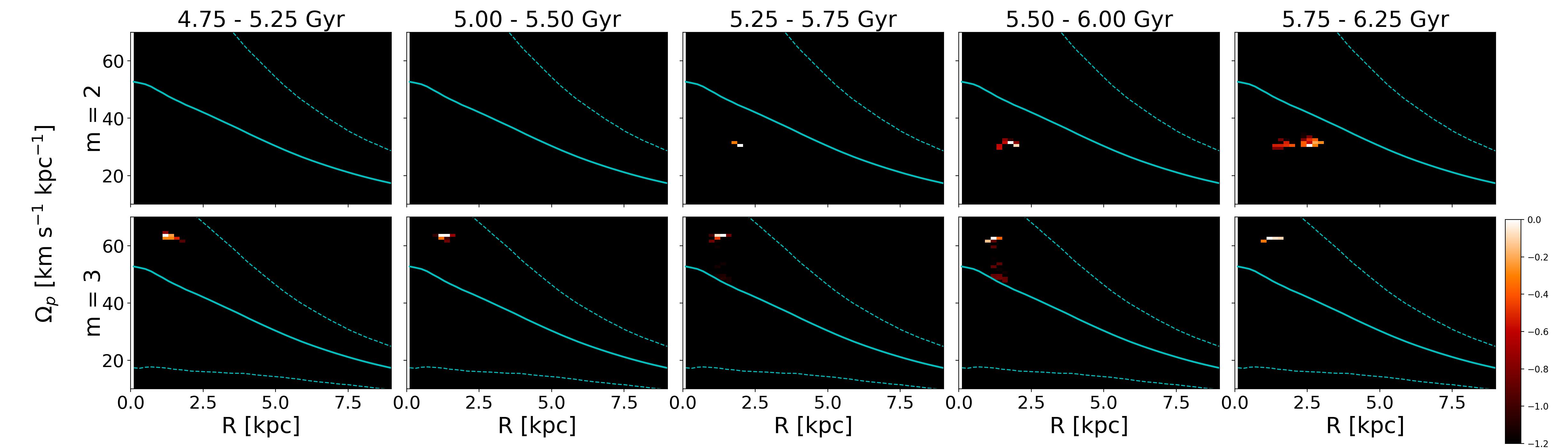}
    \includegraphics[width=0.83\textwidth]{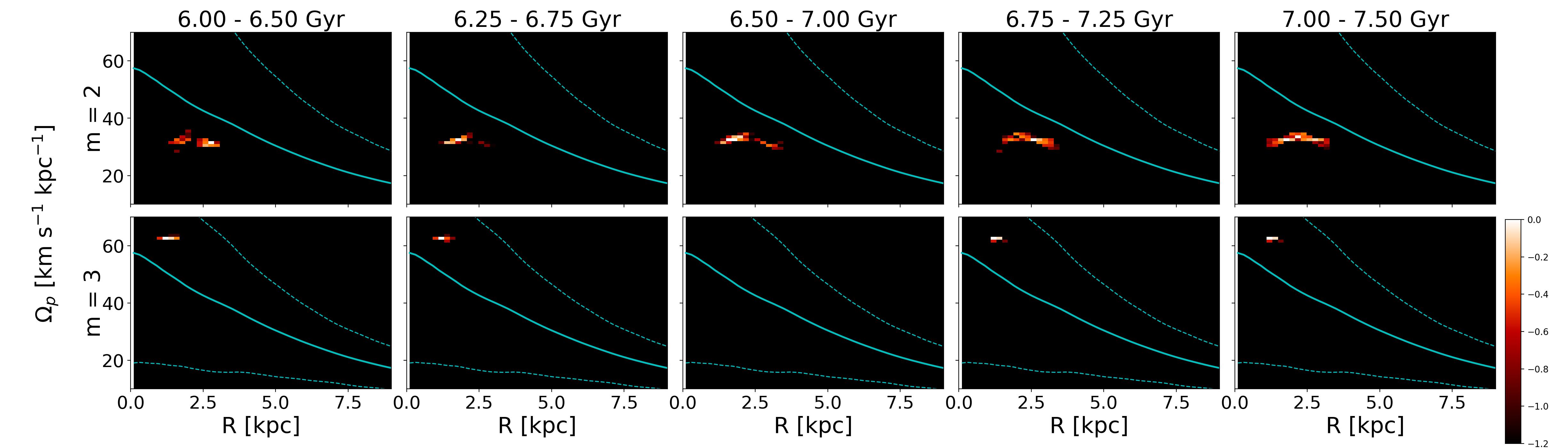} 
    \includegraphics[width=0.83\textwidth]{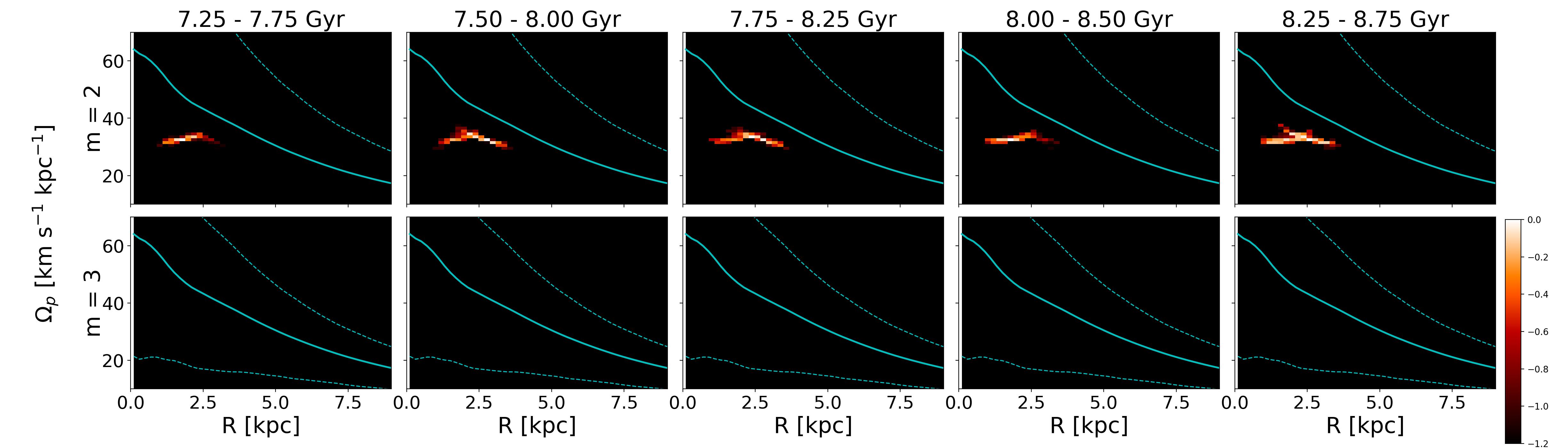}
    \includegraphics[width=0.83\textwidth]{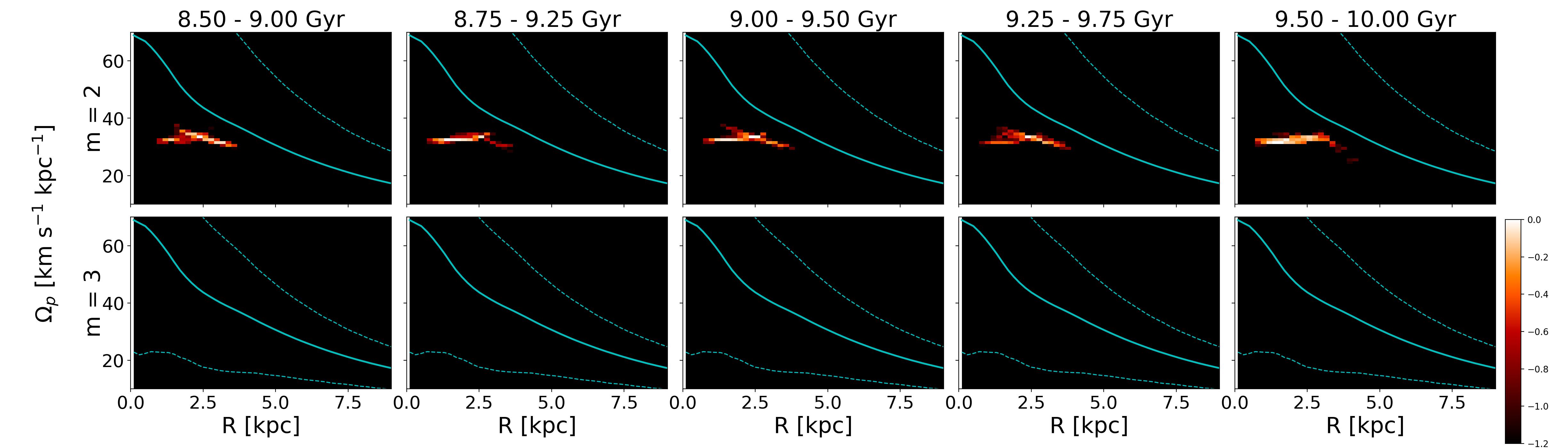} 

    \caption{Spectrograms of the $m=2$ and $m=3$ modes in model r1c22, covering 0.5 Gyr time windows from 3.5 to 10.0 Gyr, with a minimum Fourier amplitude threshold of 0.09 applied for noise filtering. Overlaid are the corresponding resonance curves: $\Omega$ and $\Omega \pm \kappa/m$, where $\Omega$ is the angular velocity and $\kappa$ is the epicyclic frequency.}
    \label{fig:spec_r1c22}
\end{figure*}

Figures~\ref{fig:faceon_r1c22} and \ref{fig:spec_r1c22} show the same type of analysis for model r1c22. The face-on maps show that the central morphology remains dominated by a three-armed distortion for several Gyr. The spectrogram confirms that this $m=3$ component maintains a coherent, nearly constant pattern speed over this extended period. The $m=2$ component strengthens only later and gradually becomes the dominant central mode. Once the $m=2$ bar is fully established, the $m=3$ signal weakens and disappears.

Together, the r2c16 and r1c22 spectrograms support a sequence in which a coherent central $m=3$ mode precedes the later $m=2$ bar. The $m=3$ component can persist for several Gyr when bar growth is delayed, as in r1c22, but it fades once the $m=2$ mode becomes fully established. We therefore interpret the $m=3$ mode as a coherent precursor to bar formation rather than as a momentary interference feature. The main phase of sustained angular-momentum loss, however, occurs only after the $m=2$ bar becomes dominant.

\bigskip

\end{appendix}

\end{document}